\documentclass{mn2e}
\usepackage{times}
\usepackage{amssymb}
\usepackage{epsfig}

\begin{document}
\hsize=6truein

\title[Submm and radio galaxies at high redshift]{A near-infrared morphological comparison of high-redshift submm and radio galaxies: massive star-forming discs vs relaxed spheroids}

\author[T.A.~Targett, et al.]
{Thomas A. Targett$^{1}$\thanks{Email: tat@roe.ac.uk}, James S. Dunlop$^{1}$, Ross J. McLure$^{1}$, Philip N. Best$^{1}$, \and Michele Cirasuolo$^{1}$, Omar Almaini$^{2}$\\
\footnotesize\\
$^{1}$ SUPA\thanks{Scottish Universities Physics Alliance}, 
Institute for Astronomy, University of Edinburgh, 
Royal Observatory, Edinburgh, EH9 3HJ, UK\\
$^{2}$School of Physics \& Astronomy, University of Nottingham, University Park, Nottingham NG7 2RD}
\maketitle

\begin{abstract}
We present deep, high-quality $K$-band images of complete subsamples of powerful radio and sub-millimetre galaxies at redshifts $z \simeq 2$. The data were obtained in the very best available seeing via queue-based observations at the United Kingdom InfraRed Telescope and Gemini North, with integration times scaled to ensure that comparable rest-frame surface brightness levels are reached for all of the galaxies. We fit two-dimensional axi-symmetric galaxy models to these images to determine basic galaxy morphological parameters at rest-frame optical wavelengths $\lambda_{rest} > 4000$\AA, varying luminosity, axial ratio, half-light radius $r_{1/2}$, and S\'{e}rsic index $n$. We find that, while a minority of the images show some evidence of galaxy interactions, $>95$\% of the rest-frame optical light in all the galaxies is well-described by these simple axi-symmetric models. We also find evidence for a clear difference in morphology between these two classes of galaxy; fits to the individual images and to the image stacks reveal that the radio galaxies are moderately large ($\langle r_{1/2} \rangle = 8.4 \pm 1.1$\,kpc; median $r_{1/2}=7.8$ ), de Vaucouleurs spheroids ($\langle n \rangle = 4.07 \pm 0.27$; median $n = 3.87$), while the sub-millimetre galaxies appear to be moderately compact ($\langle r_{1/2} \rangle = 3.4 \pm 0.3$\,kpc; median $r_{1/2}=3.1$\,kpc) exponential discs ($\langle n \rangle = 1.44 \pm 0.16$; median $n = 1.08$). Model fits to the stacked images yield very similar results. We show that the $z \simeq 2$ radio galaxies display a well-defined Kormendy ($\mu_e - r_e$) relation but that, while larger than other recently-studied massive galaxy populations at comparable redshifts, they are still a factor $\simeq 1.5$ times smaller than their local counterparts. The scalelengths of the starlight in the sub-millimetre galaxies are comparable to those reported for the molecular gas, suggesting that the two may be co-located. Their sizes are also similar to those of comparably massive quiescent galaxies at $z > 1.5$, allowing the possibility of an evolutionary connection following cessation/quenching of the observed star-formation activity. In terms of stellar mass surface density, the majority of the radio galaxies lie within the locus defined by local ellipticals of comparable stellar mass. In contrast, while best modelled as discs at the epoch of bright dust-enshrouded star formation, most of the sub-millimetre galaxies have higher stellar mass densities than local galaxies, and appear destined to evolve into present-day massive ellipticals. 

\end{abstract}


\section{INTRODUCTION}
\label{intro}
The statistical and individual properties of galaxies selected at different wavelengths, and at different redshifts must be explained in a self-consistent manner by any successful model of galaxy formation and evolution. 

Driven by multi-wavelength surveys of ever increasing scale, recent years have seen substantial advances in our understanding of the basic demographics of different galaxy populations such as Lyman-break galaxies (e.g. Bouwens et al. 2007), radio galaxies (e.g. Rigby et al. 2010) and sub-millimetre galaxies (e.g. Coppin et al. 2006; Weiss et al. 2009; Austermann et al. 2010). 

The resulting improving information in, for example, number counts and redshift distributions, has been fed back to modify/refine models of galaxy formation. However, important uncertainties and degeneracies in the basic physics generally remain. For example, sub-millimetre galaxies in particular were not a natural prediction of early semi-analytic models (Baugh et al. 2005), and there remains controversy over whether this can be fixed by the adoption of a top-heavy stellar intial mass function (IMF), or whether this tension is symptomatic of a basic inability of current semi-analytic models to produce enough star-formation activity in the young Universe (Bower et al. 2006; Swinbank et al. 2008; Cirasuolo \& Dunlop 2008).

To remove such uncertainties requires higher-level information. Specifically, different galaxy formation models make rather different predictions for galaxy masses, sizes and morphologies. For example, the models of Baugh et al. (2005) predict that sub-millimetre galaxies result from mergers of sub-$L*$ galaxies (with their apparent star-formation rates boosted by a top-heavy IMF). In contrast, the recent cosmological hydrodynamic simulations of Dav\'{e} et al. (2010) predict that sub-millimetre galaxies are massive galaxies fed by smooth infall, are not typically involved in major mergers, and are simply the most extreme examples of normal star-forming galaxies (see also Finlator et al. 2006; Fardal et al. 2007; Dekel et al. 2009).

The estimation of galaxy stellar masses at high reshift has therefore become a major industry in recent years, boosted by the provision of rest-frame near-infrared photometry provided by the {\it Spitzer Space Telescope}. The sensitivity of mass-to-light ratio on star-formation history is minimized at such wavelengths, but uncertainties remain due to choice of IMF or the relative importance of asymptotic giant branch stars. This has motivated attempts to obtain independent confirmation of galaxy masses at $z \simeq 2$ via velocity dispersion measurements derived from emission-line spectroscopy at both millimetre and infrared wavelengths (e.g. Taconni et al. 2006, 2008; van Dokkum, Kriek \& Franx 2009).

While acknowledging the uncertainties, the available photometric and dynamical evidence suggests that, by the epoch of peak sub-millimetre emission at $z \simeq 2-3$, sub-millimetre galaxies were already massive galaxies with estimated stellar masses $M_{star} \simeq 10^{11} M_{\odot}$ (Greve et al. 2004; Tacconi et al. 2006; Borys et al. 2005; Clements et al 2008; Dye et al. 2008; Schael et al. 2010, Michalowski et al. 2010). This suggests that sub-millimetre galaxies represent an important phase in the formation of massive galaxies, and the challenge now is to establish their role in cosmic star-formation history, and the nature of their present-day descendents. Here we attempt to move this subject forward by measuring the basic sizes and morphologies of sub-millimetre galaxies at $z \simeq 2$, performing such measurements at the longest possible wavelength (2.2\,$\mu m$) in order to trace the mass dominant stellar population, and to minimize the potentially confusing effects of dust obscuration at rest-frame ultraviolet wavelengths.

The measurement of galaxy sizes and morphologies at redshifts $z > 1$ has also been the subject of many studies in recent years, and has yielded some surprising and still controversial results. Specifically, a number of studies have now found that massive, passive galaxies at $1.5 < z < 3$ are much more compact than present-day galaxies of comparable mass (Trujillo et al. 2006, 2007; Longhetti et al. 2007; Cimatti et al. 2008; van Dokkum et al. 2008). The resulting inferred high stellar mass densities have been taken to imply that such objects cannot continue to evolve by purely passive evolution into any known type of present-day galaxy. These results are consistent with those of Toft et al. (2007) and Zirm et al. (2007) who, with the aid of {\it Spitzer} MIPS photometry, were in addition able to show that, while passive high-redshift galaxies are surprisingly compact, star-forming galaxies at high redshift appear morphologically similar to normal star-forming disk galaxies. Recently Buitrago et al. (2008), from a {\it Hubble Space Telescope} (HST) NICMOS study of 82 massive galaxies in GOODS North concluded that, for a given stellar mass, disc-like galaxies at $z \simeq 2.3$ were a factor 2.6 $\pm$ 0.3 smaller than present-day systems, while spheroid-like galaxies were a factor 4.3 $\pm$ 0.7 smaller. They also noted that this evolution is at least in qualitative agreement with some hierarchical semi-analytic predictions (e.g. Khochfar \& Silk 2006). However, there is now growing evidence that not all high-mass galaxies at $z > 1$ follow this trend, with some at least appearing to be fully formed (Saracco, Longhetti \& Gargiulo 2010).

At present it is unclear where sub-millimetre galaxies fit into this picture. As star-forming galaxies they might be expected to be extended discs, like Lyman-break galaxies and present-day star-forming galaxies. On the other hand, if they are the progenitors of today's massive ellipticals, their apparently high-density starbursts (Tacconi et al. 2006) might be related to the high stellar densities reported in red, passive galaxies at high redshift ($\sim~2 \times 10^{10} {\rm M_{\odot}\, kpc^{-3}}$, comparable to present-day globular clusters).

To date, genuinely deep, $K$-band imaging has only been obtained for a small number of submm galaxies (e.g. Smail et al. 2003), and generally with inadequate resolution for the reliable extraction of basic morphological parameters. Very recently, Swinbank et al. (2010) reported the results of HST NICMOS+ACS imaging of a sub-sample of sub-millimetre galaxies. However, they did not report reliable measurement of, for example S\'{e}sic indices for individual galaxies, and the wavelength limit of HST prevented them from undertaking observations in the $K$-band, which ensures imaging longward of the 4000\AA/Balmer break out to $z \simeq 4$.

A high-quality, systematic, deep, $K$-band imaging study of a well defined sample of sub-millimetre galaxies is thus desirable, and has the potential to clarify several fundamental issues. In this paper we attempt to take a step towards better quantifying the sizes, morphologies and stellar mass densities of sub-millimetre galaxies via a deep, and high-quality $K$-band imaging study of a well-defined sample of luminous SMGs at $z \simeq 2$. To aid the interpretation of our results we have taken the somewhat unusual step of obtaining images of equivalent depth and resolution for a complete comparison sample of the most luminous radio galaxies known at comparable redshifts ($z \simeq 2$).

The radio galaxy sample is a useful control sample because, given the now-well-established proportionality of black hole and spheroid mass (e.g. McLure \& Dunlop 2002), we can be very confident that these galaxies will end up as massive ellipticals at the present-day. However, given the above-mentioned evidence that many massive galaxies at $z > 1$ are more compact than comparably massive present-day objects, a study of radio galaxy sizes at $z \simeq 2$ is also clearly of interest in its own right. Rather surprisingly, while deep near-infrared imaging of 3CR radio galaxies has been undertaken at $z \simeq 1$ (e.g. Dunlop \& Peacock 1993; Best et al. 1998) very deep, high-quality $K$-band imaging of the most powerful 3CR radio galaxies at $z \simeq 2$ has not in fact been undertaken before (although NICMOS $H$-band images of a sample of even higher redshift radio galaxies were obtained by Penterrici et al. 2001). Consequently, while the morphologies of radio galaxies out to $z \simeq 1$ are rather well defined (McLure \& Dunlop 2000; McLure et al. 2004), their rest-frame optical morphologies and sizes at $z \simeq 2$ have yet to be clarified.

To undertake this study we have used NIRI on Gemini North to image the sub-millimetre galaxies, and UFTI on UKIRT to image the radio galaxies. We have taken great care to ensure that all images are as well matched as possible in $K$-band surface-brightness limit, designing integration times to compensate for the difference in telescope aperture between Gemini North and UKIRT, and for cosmological surface-brightness dimming within our samples. Also crucial to this study has been our ability (due to the queue-based scheduling system in operation at both telescopes) to demand the very best seeing conditions for all of our observations ($FWHM < 0.6$ arcsec). It is only with this powerful combination of consistent high-quality seeing and long exposure times (typically 90 mins on Gemini) that it has proved possible to determine the sizes (i.e. half-light radii) and basic morphologies (i.e. S\'{e}rsic index) of these massive $z \simeq 2$ galaxies via ground-based imaging.

The paper is structured as follows. In Section 2 we summarize the radio galaxy and sub-millimetre galaxy samples, and the existing supporting information. In Section 3 we present the UKIRT and Gemini $K$-band imaging observations, and in Section 4 we explain how these data were reduced. Then in Section 5 we describe how we modelled these new images to extract the basic morphological properties of the galaxies. The results of the individual model fits are presented, summarized and discussed in Section 6, with the images, model fits and residual images provided in Appendix A. We also explore the robustness of our results, through simulations, non parametric tests of galaxy size, and the analysis of radio galaxy and sub-millimetre galaxy image stacks. Finally, in Section 7 we attempt to place our results in the context of studies of other galaxy populations at both high and low redshift. Our main conculsions are summarized in Section 8.

\begin{table*}
 \begin{center}

\caption[Details of the radio and sub-millimetre galaxy $K$-band observations]{Details of the radio and sub-millimetre galaxy $K$-band observations. Column 1 lists the IAU name for the radio galaxies, and the 8-mJy survey name for the sub-millimetre sources. Column 2 gives the original Cambridge and Parkes catalogue names for the radio sources, and the alternative SHADES catalogue names for the Lockman sub-millimetre galaxies (Coppin et al. 2006). Column 3 provides information on the nature of the galaxy identification for the sub-millimetre sources. Column 4 names the telescope with which the data were acquired, and Column 5 lists the total exposure time in seconds for each target. Column 6 provides the measured 6-arcsec diameter aperture $K$-band magnitude (in the Vega magnitude system) for each galaxy, along with the associated error. Finally, Columns 7, 8, and 9 list the adopted redshift for each source, the type of redshift, and the reference for the source from which this redshift was obtained: 
A from Best el al. (1999), 
B from Hewitt \& Burbidge (1991), C from Spinrad et al. (1985), 
a from Aretxaga et al. (2005), 
b from Aretxaga, priv. comm., 
c from Chapman et al. (2003), 
d from Chapman et al. (2005), 
e from Ivison et al. (2005), 
f from Smail et al. (2003).}

\begin{tabular}{llllrclcc}
\hline\hline
Source  &  Radio Catalogue Name  &   &  Telescope & Exp. time (sec)  &  $K$ mag. (Vega) & $z$  &  $z$-type & z-ref\\
\hline
0016$-$12   &  3C 008	     & & UKIRT &  9480    & \phantom{$>$}18.13$\pm$0.07 & 1.589      & $z_{spec}$            &  A\\
0128$-$26   &  PKS 0128$-$26 & & UKIRT &  18840   & \phantom{$>$}17.77$\pm$0.09 & 2.348      & $z_{spec}$            &  A\\
0231$+$31   &  3C 068.2      & & UKIRT &  9720    & \phantom{$>$}17.97$\pm$0.08 & 1.575      & $z_{spec}$            &  B\\
0310$-$15   &  PKS 0310$-$15 & & UKIRT &  8100    & \phantom{$>$}18.35$\pm$0.11 & 1.769      & $z_{spec}$            &  A\\
0851$-$14   &  PKS 0851$-$14 & & UKIRT &  10140   & \phantom{$>$}17.87$\pm$0.08 & 1.665      & $z_{spec}$            &  A\\
1008$+$46   &  3C 239	     & & UKIRT &  11820   & \phantom{$>$}17.89$\pm$0.07 & 1.790      & $z_{spec}$            &  C\\
1019$+$22   &  3C 241        & & UKIRT &  9600    & \phantom{$>$}17.75$\pm$0.08 & 1.617      & $z_{spec}$            &  C\\
1120$+$05   &  3C 257        & & UKIRT &  22560   & \phantom{$>$}17.58$\pm$0.07 & 2.474      & $z_{spec}$            &  B\\
1140$-$11   &  PKS 1140$-$11 & & UKIRT &  13920   & \phantom{$>$}18.79$\pm$0.09 & 1.935      & $z_{spec}$            &  A\\
1422$-$29   &  PKS 1422$-$29 & & UKIRT &  9600    & \phantom{$>$}17.69$\pm$0.07 & 1.632      & $z_{spec}$            &  A\\
1533$+$55   &  3C 322	     & & UKIRT &  9180    & \phantom{$>$}17.86$\pm$0.08 & 1.681      & $z_{spec}$            &  B\\
1602$-$17   &  PKS 1602$-$17 & & UKIRT &  15120   & \phantom{$>$}17.63$\pm$0.09 & 2.043      & $z_{spec}$            &  A\\
2356$+$43   &  3C 470        & & UKIRT &  10260   & \phantom{$>$}18.11$\pm$0.12 & 1.653      & $z_{spec}$            &  B\\
\hline			  	       	   		   			              
\hline
Source      &  SHADES Name   & ID Type     & Telescope & Exp. time (sec)&  $K$ mag. (Vega)    & $z$          & $z$-type & z-ref\\
\hline
N2 850.01   &                & Radio ID    & Gemini N &  4750    & \phantom{$>$}19.90$\pm$0.15 & 2.850      & $z_{spec}$ &  d\\
N2 850.02   &                & Radio ID    & Gemini N &  4500    & \phantom{$>$}20.77$\pm$0.27 & 2.454      & $z_{spec}$ &  d  \\
N2 850.04   &                & Radio ID    & Gemini N &  4625    & \phantom{$>$}18.85$\pm$0.10 & 2.387      & $z_{spec}$ &  f  \\
N2 850.06   &                & ERO/VRO     & Gemini N &  5500    & \phantom{$>$}20.18$\pm$0.16 & 3.0        & $z_{phot}$ &  b  \\
N2 850.07   &                & Radio ID    & Gemini N &  5000    & \phantom{$>$}20.00$\pm$0.16 & 3.1        & $z_{phot}$ &  b  \\
N2 850.08   &                & Radio ID    & Gemini N &  5000    & \phantom{$>$}18.30$\pm$0.09 & 1.190      & $z_{spec}$ &  d  \\
N2 850.09   &                & Radio ID    & Gemini N &  4750    & \phantom{$>$}20.99$\pm$0.37 & 3.3        & $z_{phot}$ &  b  \\
N2 850.12   &                & ERO/VRO     & Gemini N &  5500    & \phantom{$>$}20.31$\pm$0.17 & 2.5        & $z_{phot}$ &  a  \\
N2 850.13   &                & Radio ID    & Gemini N &  5375    & \phantom{$>$}21.61$\pm$0.37 & 2.283      & $z_{spec}$ &  d  \\
LE 850.02   & LOCK850.3      & Radio ID	   & Gemini N &  5500    & \phantom{$>$}19.66$\pm$0.11 & 3.041      & $z_{spec}$ &  d  \\
LE 850.03   & LOCK850.17     & Radio ID	   & Gemini N &  5500    & \phantom{$>$}19.66$\pm$0.12 & 2.239      & $z_{spec}$ &  d/e\\
LE 850.04   & LOCK850.6      & ERO/VRO     & Gemini N &   5500    & \phantom{$>$}20.74$\pm$0.14 & 3.6       & $z_{phot}$ &  b  \\
LE 850.06   & LOCK850.14     & Radio ID	   & Gemini N &  5625    & \phantom{$>$}19.16$\pm$0.17 & 2.610      & $z_{spec}$ &  c/e\\
LE 850.07   & LOCK850.16     & Radio ID	   & Gemini N &  5500    & \phantom{$>$}18.87$\pm$0.10 & 1.9        & $z_{phot}$ &  b  \\
LE 850.08   & LOCK850.41     & Radio ID	   & Gemini N &  5500    & \phantom{$>$}20.38$\pm$0.16 & 3.4        & $z_{phot}$ &  b  \\
\hline
\end{tabular}
\label{tabsample}

 \end{center}
\end{table*}

\section{Samples and supporting data}
\label{samples}
The target sample for this study consisted of 29 sources, and comprised a complete sample of the 14 most radio-luminous radio galaxies at $z \simeq 2$, and 15 of 
the brightest 850\,${\rm \mu m}$ sources in the SCUBA 8-mJy survey. 

\subsection{The radio galaxy sample}
Of the 14 objects in the original radio galaxy sample, 8 were drawn from the 3CRR catalogue (Laing, Riley \& Longair 1983) and 6 from the equatorial sample of Best et al. (1999). The 3CRR sample is complete to a 178-MHz flux-density limit of $S_{\rm 178\,MHz} > 10.9$\,Jy over an area of 4.2 steradians, and contains a total of 170 steep-spectrum sources over the redshift range $0.0<z<2.5$. These data are essentially doubled by the addition of the Best et al. sample, which was designed to have a roughly equivalent selection criterion of $S_{\rm 408\,MHz} > 5$\,Jy, using the Molonglo Reference Catalogue. From this combined survey we have defined a complete (\textit{cf} Pentericci et al. 2001) sample of the 14 most powerful, low-frequency-selected, radio sources in the redshift range $1.5<z<2.5$. One of the sources in this high-power sample, 3C\,294, transpired to lie too close on the sky to a nearby star for deep $K$-band imaging to be successfully secured. Thus, for the present study, the parent sample was trimmed to the 13 sources listed in Table 1.

\subsection{The sub-millimetre galaxy sample}

The sub-millimetre galaxies were drawn from the SCUBA 8-mJy survey (Scott et al. 2002). The SCUBA 850\,${\rm \mu m}$ imaging in the 8-mJy survey was split between the Lockman Hole East and Elais N2 fields, and covered a total area of 260 arcmin$^{2}$ to a typical rms noise level of $\sigma_{850}\simeq2.5$\,mJy\,beam$^{-1}$. From the resulting source catalogue, all 25 sub-millimetre sources with a signal-to-noise $S/N > 3.5$ at 850\,${\rm \mu m}$ were selected as potential targets for deep $K$-band imaging in the present study. In Table 1 we list the 15 sub-millimetre sources which were actually observed with sufficient signal:noise in $K$-band to undertake a meaningful morphological analysis.

\subsection{Identification of ${\bf K}$-band counterparts}

All of the radio galaxies in the 13-source sample have unambiguous near-infrared galaxy counterparts, as a result of the sub-arcsec positional accuracy provided by the VLA detections of their radio cores.

However, as is well known, the identification of the $K$-band counterparts of the sub-millimetre sources is more complicated, due to the large (FWHM $\simeq 15$ arcsec) size of the Airy disc delivered by the 15-m JCMT at $\lambda \simeq 850\,{\rm \mu m}$. However, for 12 of the 15 sub-millimetre sources listed in Table 1, robust radio identifications (and hence positions accurate to $\simeq$ 1 arcsec) were successfully deduced by Ivison et al. (2002), via deep 1.4\,GHz imaging with the Very Large Array (VLA) reaching a typical 5$\sigma$ detection limit of $S_{1.4GHz} \simeq 25 {\rm \mu Jy}$.

As indicated in Table 1, identifications for the remaining 3 sub-millimetre sources in our sub-sample were therefore by necessity determined on the basis of optical-infrared information only. For the 2 outstanding ELAIS N2 sources, the choice of identification was based on the $I-K$ colour derived from the existing (and now public) $I$-band HST ACS imaging of the ELAIS N2 field (PI: Almaini, PID: 9761), and the pre-existing UKIRT $K$-band imaging described by Ivison et al. (2002). Specifically, following Pope et al. (2005) the criterion that sub-millimetre galaxies are typically very red in $I-K$ colour (see also Smail et al. 2004) was used to select the most likely $K$-band counterpart in the absence of a secure radio detection. In the case of LE\,850.04, where no optical imaging was available at the commencement of this study, the $K$-band source closest to the sub-millimetre position was selected. It should be noted that although LE\,850.04 is therefore potentially the most uninformed identification, no other possible $K$-band counterpart is found within the SCUBA positional error circle for this source.

While it is to some extent desirable to partially remove the biases introduced by insisting on a robust radio identification, it is nonetheless true that these three non-radio selected galaxy identifications could be regarded as less secure than their radio-selected counterparts. Because of this, they are explicitly flagged as a specific subset in much of the subsequent analysis presented in this paper.

\subsection{Redshifts}
Secure spectroscopic redshifts exist for all 13 radio galaxies (generally aided by bright emission lines) and are given in Table 1.

As is well documented in the literature (e.g. Ivison et al. 2005), obtaining secure spectroscopic redshifts for sub-millimetre galaxies has proved much more difficult, not only because the identifications are harder to establish, but also because the correct galaxy identifications are often very red, providing little optical light for optical spectroscopy. Where secure identifications exist, spectroscopic redshifts are thus typically only available for sub-millimetre sources with optically bright counterparts or strong-lined AGN emission (Chapman et al. 2005). Nevertheless secure spectroscopic redshifts have now been secured for 8 of the 15 sub-millimetre sources, and these are listed in Table 1.

In the absence of secure spectroscopic redshifts, photometric redshifts based on the observed shape of the sub-millimetre--radio spectral energy distribution have been adopted from Aretxaga et al. (2007). These redshift estimates have been found to be accurate to $\Delta z < 0.5$ when compared to existing spectroscopic values. We note that in the general high-redshift range of interest here, derived physical galaxy scalelength is relatively insensitive to the precise redshift.

\section{OBSERVATIONS}

\subsection{UKIRT observations of the radio galaxies}
The $z \simeq 2$ radio galaxies were imaged in the $K$-band with the 3.8-m United Kingdom Infrared Telescope (UKIRT) on Mauna Kea. All observations used the UKIRT Fast-Track Imager (UFTI), a 1-2.5\,${\rm \mu m}$ camera with a 1024$^2$ HgCdTe array, a pixel scale of 0.091 arcsec, and hence a 1.5$^{\prime}$ $\times$ 1.5$^{\prime}$ field of view. All data were taken in $<0.6$-arcsec seeing and photometric conditions. Integration times were set to obtain a signal-to-noise of 5 at twice the anticipated half-light radius in the azimuthally binned luminosity profile. Note that even passive evolution of the stellar population offsets to some extent the impact of surface brightness dimming with increasing redshift. In the design of the observations, galaxy scalelengths of 10\,kpc were assumed (comparable to those found by McLure \& Dunlop (2000) for 3CR galaxies at $z=1$), and the total anticipated $K$-band magnitudes were estimated using the infrared Hubble diagram of De Breuck et al. (2002).

\subsection{Gemini observations of the sub-millimetre galaxies}
The sub-millimetre sources were imaged in the $K-$band with the 8-m Gemini North Telescope on Mauna Kea. All observations used the Near-Infrared Imager (NIRI), a 1-5\,${\rm \mu m}$ camera with a 1024$^2$ ALADDIN InSb array. Camera f/6 was used, delivering a pixel scale of 0.1171 arcsec, and hence a 2.0$^{\prime}$ $\times$ 2.0$^{\prime}$ field of view. All data were taken in $<0.6$ arcsec seeing and photometric conditions. Integration times were near constant over all observations at $\simeq 1.5$ hours per source, a choice made to yield images of comparable depth to those obtained with the longest exposures used on UKIRT for the most distant ($z>2$) radio galaxies in the comparison sample (see Table 1). The resulting $K$-band imaging reaches a $3\sigma$ limiting magnitude of $K=22.5$ (Vega) in a 4-arcsec diameter aperture.

\section{Data reduction}
Data reduction was performed using standard IRAF packages. Dark frames of equal integration time to the science data were taken on all nights. Each dark-subtracted image was divided by a normalized flat-field, derived though the scaled sigma-clipped median combination of neighbouring science frames. Registering of the offset frames was performed using the brightest stars in the images. A map of the bad pixels in the array was obtained and used to exclude these pixels during image combination, followed by cosmic-ray rejection. The initial reduced image was processed to create a mask of all source flux detected in the mosaiced data. With all known source flux excluded, the science frames were then reprocessed to create an improved flat-field, which was used in a re-reduction of the data to produce the final science frames.

\subsection{Astrometry and Photometry}
Given the numerous faint sources visible in these deep data, accurate astrometry was essential for associating the Ivison et al. (2002) radio positions with the correct $K$-band counterparts. Owing to uncertainties on scales of a few arcsec in the telescope pointing, it was therefore necessary to re-determine the astrometry of the images. Using GAIA, finding-charts from the Two Micron All Sky Survey (2MASS) were obtained, and the co-ordinates of stars present in both the 2MASS data and the $K$-band imaging were used to determine the plate solution for each image. After correction, the offsets between coordinates of 2MASS sources and their counterparts in the $K$-band data were consistently $<0.1$ arcsec.

The aperture photometry listed in Table \ref{tabsample} was obtained using the IRAF APPHOT package. Background sky counts were measured and subtracted using a curve-of-growth analysis. Faint standard stars were observed for each target object for photometric calibration. These data were reduced via the process applied to the science data. The observations of these standards consistently yielded zero-points of $22.3\pm0.06$ for UFTI and $23.68\pm0.08$ for NIRI, consistent with the values given in the literature.

\begin{table*}
 \begin{center}

\caption[Results from two-dimensional of $z \simeq 2$ radio galaxies and submm hosts]{Results from two-dimensional modelling of the $K$-band images of the radio galaxies (upper section) and the sub-millimetre galaxies (lower section). Column 1 gives the source name. Column 2 lists the semi-major axis scalelength (half-light radius) of the host galaxy fit in kiloparsec. Column 3 lists the mean surface brightness in $K$-band magnitude (Vega) per square arcsec. Column 4 lists the value of the S\'{e}rsic index ($n$). Column 5 gives the axial ratio of the host galaxy. Column 6 lists the total host-integrated $K$-band magnitude (Vega). Column 7 gives the value of reduced $\chi^{2}_{\nu}$ for each model fit, and finally Column 8 gives the \% of $K$-band flux from the object which is attributed to the smooth, axi-symmetric model.}

\begin{tabular}{lccccccc}
\hline\hline
Source & $r_{1/2}$ & $<\mu_{e}>$ & S\'{e}rsic $n$ & Axial ratio & $K-$mag & $\chi^{2}_{\nu}$ & \% flux in model\\
\hline	       				   
0016$-$129 &  7.58 &  19.03  &  5.23  &  0.52  &  17.84  &  1.01 & 98.3 \\
0128$-$264 & 15.66 &  20.64  &  3.28  &  0.45  &  17.37  &  0.94 & 98.5 \\
0231$+$313 & 11.54 &  19.49  &  2.73  &  0.69  &  17.70  &  0.95 & 99.3 \\
0310$-$150 &  7.75 &  19.51  &  4.51  &  0.33  &  18.07  &  0.82 & 98.2 \\
0851$-$142 &  9.72 &  19.78  &  3.85  &  0.68  &  18.19  &  0.80 & 96.9 \\
1008$+$467 &  4.24 &  17.44  &  3.76  &  0.85  &  17.81  &  0.94 & 99.4 \\
1019$+$222 &  5.09 &  17.97  &  4.59  &  0.60  &  17.72  &  0.96 & 97.7 \\
1120$+$057 & 11.85 &  16.69  &  6.26  &  0.42  &  17.44  &  1.14 & 88.1 \\
1140$-$114 &  4.20 &  19.39  &  3.05  &  1.00  &  18.90  &  1.02 & 95.6 \\
1422$-$297 &  9.29 &  19.16  &  5.74  &  0.45  &  17.40  &  0.97 & 99.5 \\
1533$+$557 &  5.95 &  18.90  &  4.21  &  1.00  &  17.82  &  0.93 & 98.9 \\
1602$-$174 &  0.90 &  14.99  &  4.67  &  0.88  &  17.84  &  0.88 & 99.8 \\
2356$+$438 & 11.02 &  20.11  &  3.87  &  0.71  &  18.47  &  0.99 & 97.3 \\ 
\hline								    
N2 850.1   &  2.09 &  18.61  &  1.74  &  0.76  &  19.77  & 0.99  & 97.7 \\
N2 850.2   &  1.15 &  17.78  &  3.33  &  0.47  &  20.83  & 1.06  & 95.9 \\
N2 850.4   &  3.24 &  18.2   &  1.79  &  0.66  &  18.65  & 0.85  & 99.3 \\
N2 850.6   &  1.95 &  18.14  &  3.29  &  0.37  &  20.19  & 0.90  & 96.1 \\
N2 850.7   &  2.77 &  19.16  &  1.08  &  0.75  &  19.67  & 0.72  & 96.2 \\
N2 850.8   &  3.70 &  18.09  &  1.35  &  0.72  &  18.21  & 0.79  & 99.4 \\
N2 850.9   &  3.05 &  19.74  &  1.08  &  0.41  &  20.65  & 0.79  & 96.0 \\
N2 850.12  &  4.65 &  19.11  &  2.17  &  0.19  &  20.06  & 0.93  & 99.5 \\
N2 850.13  &  5.78 &  20.34  &  2.75  &  0.22  &  20.76  & 0.79  & 96.9 \\
LE 850.2   &  3.88 &  18.24  &  1.07  &  0.19  &  19.48  & 0.82  & 97.2 \\
LE 850.3   &  2.56 &  17.53  &  0.97  &  0.53  &  18.75  & 0.58  & 96.7 \\
LE 850.4   &  2.87 &  18.76  &  0.93  &  0.20  &  20.49  & 0.96  & 96.1 \\
LE 850.6   &  3.60 &  19.3   &  1.99  &  0.89  &  19.16  & 0.90  & 97.8 \\
LE 850.7   &  3.10 &  17.89  &  0.86  &  0.55  &  18.71  & 0.80  & 97.7 \\
LE 850.8   &  3.04 &  20.04  &  0.94  &  0.75  &  20.16  & 0.72  & 97.2 \\
\hline
\end{tabular}
\label{tabmodel}

 \end{center}
\end{table*}

\subsection{Point spread function}
\label{psf}
A point spread function (PSF) for each science frame was built using the IRAF package PSF for use during two-dimensional modelling. Stars of comparable brightness to the target were selected and fitted with a range of alternative statistical distributions to determine the best fit. The resulting models were then centroided, averaged, and processed with look-up tables to reproduce any asymmetries in the PSF.

\section{Host galaxy analysis}
\label{analysis}

\subsection{Image modelling}
A tried and tested two-dimensional modelling code (previously developed for the more demanding task of studying quasar host galaxies) was used to extract basic galaxy morphological parameters from the $K$-band images of the sub-millimetre and radio galaxies. Full details of the two-dimensional modelling code can be found in McLure et al. (2000). In brief, the code fits the image with an axially symmetric model (convolved with the observed PSF) with 5 free parameters (luminosity, half-light radius, position angle, axial ratio, and S\'{e}rsic index $n$), minimising $\chi^2$ using a downhill-simplex technique. In order to ensure that the solution was stable (and not a local minimum), the modelling code was repeatedly restarted from a range of positions in the parameter space close to the minimum $\chi^2$ solution. The results are summarized in Table 2, and discussed further below in Section 6.

\subsection{Reliability and limitations of modelling results}

The accuracy of scalelength and S\'{e}rsic index recovery was examined via two-dimensional modelling of synthetic galaxies.
Model galaxies at $z \simeq 2$ were constructed with a scalelength $r_{1/2}$ of 1, 1.5, 2, 3, 4, 5, 6, 7, and 8\,kpc, and with S\'{e}rsic index values of $n=1$ (exponential disc) and $n=4$ (de Vaucouleurs spheroid). The axial ratio and position angle were set to be constant, with luminosity and redshift adopted from the mean values of the radio-galaxy sample. The synthetic sources were inserted into an image, with noise levels based upon the average rms-per-pixel values derived from the real images. Background counts were also added to each simulated model using the IRAF routine MKNOISE. To ensure a realistic level of uncertainty in the accuracy of the PSF used to model a given galaxy, synthetic data were convolved with a PSF derived from a science image, and then the modelling was undertaken using a new PSF constructed from a selection of stars elsewhere in the same field. 

Example results of these tests are given in Tables \ref{tabsersic1} and \ref{tabsersic4}. These tests were repeated using PSFs and sky noise from other targets in the UKIRT and Gemini samples with similar results, demonstrating that host galaxy scalelength and S\'{e}rsic index can be recovered to within a typical accuracy of 10 per cent. It was, however, also found that S\'{e}rsic index can only be usefully recovered when $r_{1/2}>2$\,kpc; for smaller galaxies the limitations of ground-based seeing render elliptical and disc galaxies indistinguishable.

\section{Results}
\subsection{Scalengths and luminosity profiles}

The original $K$-band images, the resulting best fitting axi-symmetric model, and the residual images are presented in Appendix A (Fig. A1 for the radio galaxies, and Fig. A2 for the sub-millimetre galaxies). The best-fitting values for the model parameters are summarized in Table 2, where it can be seen that statistically acceptable axisymmetric fits were achieved for all galaxies. 

\begin{figure}
\centerline{\epsfig{file=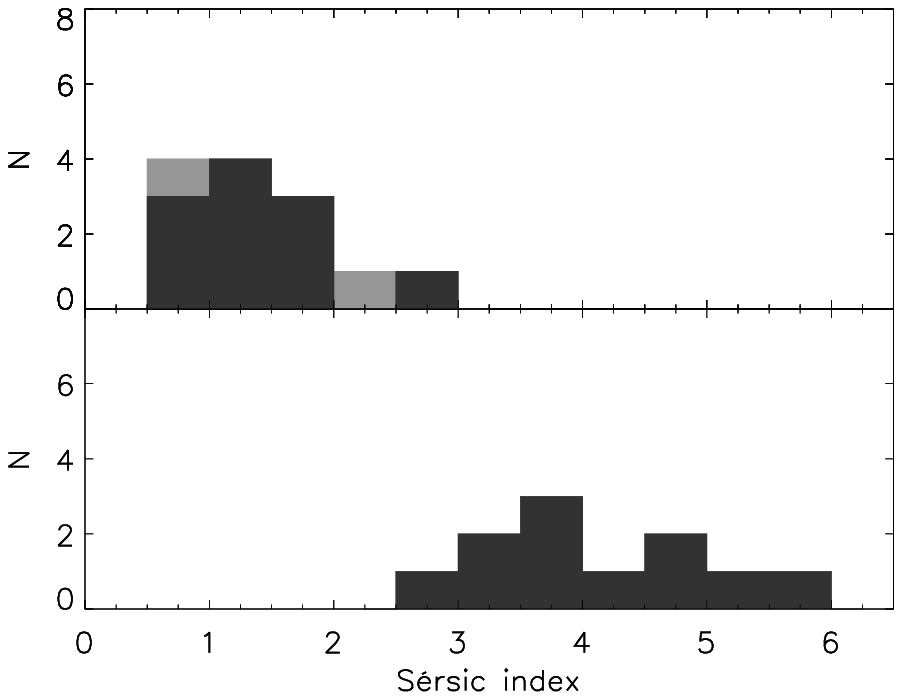,width=9.0cm,angle=0}}
\caption[The distribution of S\'{e}rsic index $n$ for the sub-millimetre and radio galaxies]{The distribution of S\'{e}rsic index $n$ for 
the sub-millimetre galaxies (upper panel) and radio galaxies (lower panel) as derived from the two-dimensional modelling of the $K$-band
images.}
\label{histsersic}
\end{figure}

\begin{figure}
\centerline{\epsfig{file=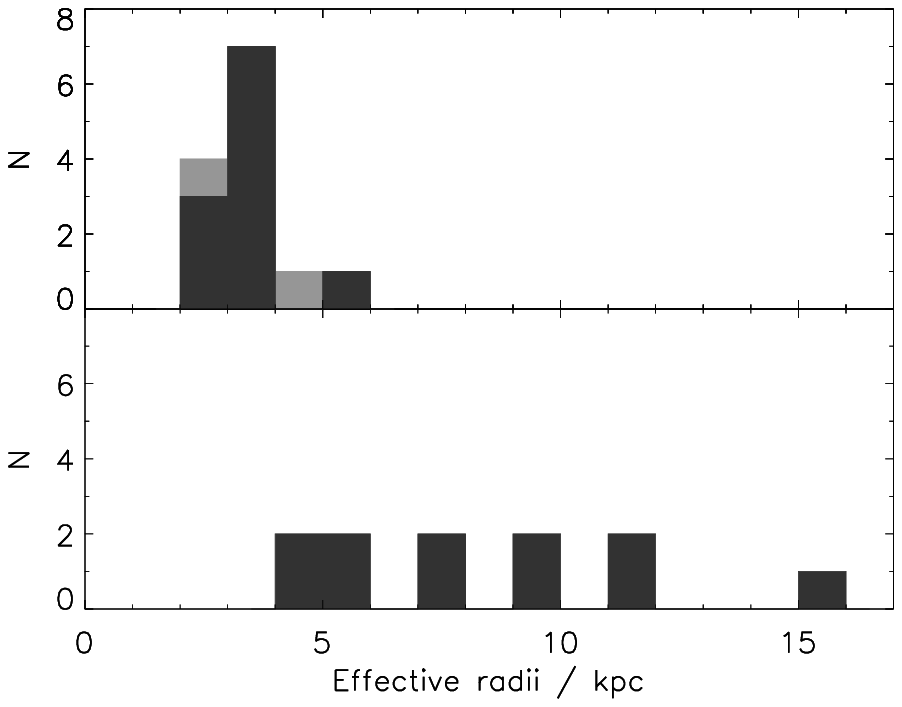,width=9.0cm,angle=0}}
\caption[The distribution of half-light radius, $r_{1/2}$, for sub-millimetre and radio galaxies]{The distribution of the half-light radius, $r_{1/2}$, for the sub-millimetre galaxies (upper panel) and radio galaxies (lower panel), as derived from the two-dimensional modelling of the $K$-band images.}
\label{histhalf}
\end{figure}

A rather clear separation is found in the sizes and luminosity profiles of these two classes of galaxy, as can be seen from inspection of Table \ref{tabmodel}. The high-redshift radio galaxies emerge as moderately-large dynamically-relaxed ellipticals, while the sub-millimetre sources appear better described as disc galaxies. A comparison of the S\'{e}rsic indices is presented in Fig. \ref{histsersic}. The 13/15 sub-millimetre galaxies with reliable modelling results (i.e. excluding N2\,850.2 and N2\,850.6 which have $r_{1/2} < 2$\,kpc) possess S\'{e}rsic values peaked at $n=1$, and confined to the range $1<n<2.5$. The S\'{e}rsic indices of the 11/13 reliably-modelled radio galaxies (i.e. excluding 1120+057 and 1602-174; see below) are in the range $2.5<n<5.75$ and peaked around $n=4$, corresponding to a de Vaucouleurs luminosity profile. The distribution of half-light radii is shown in Fig. \ref{histhalf}. Sub-millimetre galaxy scale-lengths are clearly peaked around $r_{1/2}\sim3$\,kpc, while the radio galaxies are spread over a larger range, and are systematically larger (with an average $r_{1/2}=8$\,kpc). These results are discussed further in Section 7.

\subsection{Notes on individual sources}

\subsubsection{Radio galaxies}

As explained in Section 2, all the $z \simeq 2$ radio galaxy targets were initially identified with known radio positions in the literature (Laing, Riley \& Longair 1983, Best et al. 1999). In the light of our deep $K$-band imaging, we have revised one of these radio-galaxy identifications. This is the case of 0851$-$14, where Best et al. (1999) selected a discy object to the east of the radio position as seen in their $R$-band imaging. Our deep $K$-band data reveals a multi-component merging system closer to the radio centroid. The larger component of this merger (which also displayed a redder $R-K$ colour) was adopted as the true galaxy counterpart, and it is this object which is shown and modelled in Fig. A1.

In the case of 1120$+$05, the highest-redshift radio galaxy in the 3CR catalogue, we find a partially-obscured point-like nucleus which is clearly shown in the two-dimensional modelling residual (see Appendix A). A nuclear component is also clearly present in 1602$-$17. In this case, the $R$-band image from Best et al. (1999) shows a distinctly different morphology (with no sign of activity in the nucleus) to that revealed in our $K$-band data, which is dominated by a very compact source suggestive of obscured AGN emission. As we do not believe our data are of sufficient quality to reliably separate a significant nuclear contribution from the host, and still derive robust galaxy scalelengths and S\'{e}rsic indices, 1120$+$05 and 1602$-$17 have been excluded from further discussion of the statistical distribution of galaxy morphological parameters for the radio-galaxy sample. In all other cases no evidence 
of significant AGN contamination was found, with combined nuclear+host model fits introducing no significant nuclear component. 

\subsubsection{Sub-mm galaxies}

Due in part to their early redshift determinations by Chapman et al. (2003; 2005), two of the ELAIS N2 8-mJy sources, N2\,850.2 and N2\,850.4, have previously been imaged in some detail at sub-millimetre, optical and near-infrared wavelengths (Smail et al. 2003; Tacconi et al. 2006; Tacconi et al. 2008). In particular, Tacconi et al. (2008) report that, as viewed via high-resolution mm-wave CO line emission, N2\,850.2 is a very compact object, with $r_{1/2} = 0.8 \pm 0.5$\,kpc, while N2\,850.4 is somewhat more extended, with $r_{1/2} = 2.4 \pm 1$\,kpc. These figures agree remarkably well with the completely independent (and, at least in principle, physically distinct) measures of the half-light radius of the starlight in these two galaxies, as derived here from our modelling of the $K$-band imaging (N2\,850.2 has the smallest value of stellar half-light radius listed in Table 2, with $r_{1/2} = 1.1$\,kpc, while for N2\,850.4 the derived value is $r_{1/2} = 3.2 \pm 0.3$\,kpc). This suggests that the basic distribution of the stars and molecular gas in these galaxies may be rather similar.

Tacconi et al. (2008) also present a colour optical-infrared image for N2\,850.4, derived from HST ACS and NICMOS imaging of this sub-mm galaxy. This shows a complex morphology, but as we demonstrate in sub-section 6.6, this is {\it not} inconsistent with our finding that the $K$-band light is well-described by a rather straightforward, axi-symmetric smooth galaxy model. Rather it provides further evidence of the dramatic wavelength dependence of morphology in at least some sub-mm galaxies (due to localized star-formation and/or patchy obscuration), and re-emphasizes the importance of moving to the longest possible near-infrared wavelength to uncover the basic morphological parameters of the mass-dominant host-galaxy stellar population.

\begin{table}
\begin{center}
\caption[Example results of two-dimensional modelling tests using synthetic disc galaxies ($n=1$)]{Example results of two-dimensional modelling tests using synthetic disc galaxies (i.e. with a S\'{e}rsic index $n=1$). Column 1 lists the actual input scale-lengths of the simulated sources in parsecs. Columns 2 and 3 list the recovered scale-lengths and S\'{e}rsic indices as derived from the two-dimensional modelling code. $r_{1/2}$ values are given in parsecs at $z = 2$.}

\begin{tabular}{ccc}
\hline\hline
Simulated $r_{1/2}$ (pc) & Recovered $r_{1/2}$ (pc) & Recovered $n$\\
\hline
1000  &  878  & 2.12\\
1500  & 1389  & 1.97\\
2000  & 1886  & 1.52\\
3000  & 3089  & 1.12\\	
4000  & 4181  & 1.07\\
5000  & 5246  & 1.04\\
6000  & 6375  & 1.06\\	
7000  & 7180  & 1.02\\
8000  & 8238  & 1.00\\
9000  & 9138  & 1.10\\
10000 & 10192 & 1.00\\
\hline	   
\end{tabular}
\label{tabsersic1}

 \end{center}
\end{table}

\begin{table}
 \begin{center}

\caption[Example results of two-dimensional modelling tests using synthetic elliptical galaxies ($n=4$)]{Example results of two-dimensional modelling tests using synthetic elliptical galaxies (i.e. with a S\'{e}rsic index $n=4$). Column 1 lists the actual scalelengths of the simulated sources in parsecs. Columns 2 and 3 list the recovered scalelengths and S\'{e}rsic indices as derived from the two-dimensional modelling code. $r_{1/2}$ values are given in parsecs at $z = 2$.}

\begin{tabular}{ccc}
\hline\hline
Simulated $r_{1/2}$ (pc) & Recovered $r_{1/2}$ (pc) & Recovered $n$\\
\hline
1000  &  882 & 2.64\\
1500  & 1402 & 3.29\\
2000  & 1920 & 3.51\\
3000  & 3256 & 3.89\\
4000  & 4115 & 3.74\\
5000  & 5220 & 3.92\\
6000  & 5936 & 4.05\\
7000  & 6185 & 3.94\\
8000  & 8223 & 3.79\\
9000  & 9082 & 3.82\\
10000 &10076 & 3.90\\
\hline	   
\end{tabular}
\label{tabsersic4}

 \end{center}
\end{table}

\subsection{Non-parametric tests of ${\bf r_{1/2}}$}

The Petrosian radius and $r_{1/2}$ were determined from a curve-of-growth analysis for comparison with the modelling results. Surface-brightness profiles for each object were derived by using the IRAF package PHOT to measure aperture magnitudes through progressively larger apertures out to a radius of 12 arcsec. Upon determining the extent of the total flux from the source, the value of $r_{1/2}$ was read from the profile. The Petrosian radius was defined as described by Blanton et al. (2001) and Yasuda et al. (2001), taken to be the point at which the ratio of the surface brightness in an annulus at radius $r$ to the mean surface brightness within an aperture of radius $r$ equals $0.2$. 

The values of $r_{1/2}$ as determined from the two-dimensional modelling, the curve-of-growth, and Petrosian techniques were found to be consistent with within 10 per cent. Typically, the results from two-dimensional modelling are systematically smaller than those derived from curve-of-growth techniques. This is expected given that the two-dimentional modeling effectively deconvolves the PSF (typically $\sim4$\,kpc) from the data, yielding an unbiased measurement of true galaxy size.

\subsection{Non-parametric tests of galaxy morphology}
\begin{figure}
\centerline{\epsfig{file=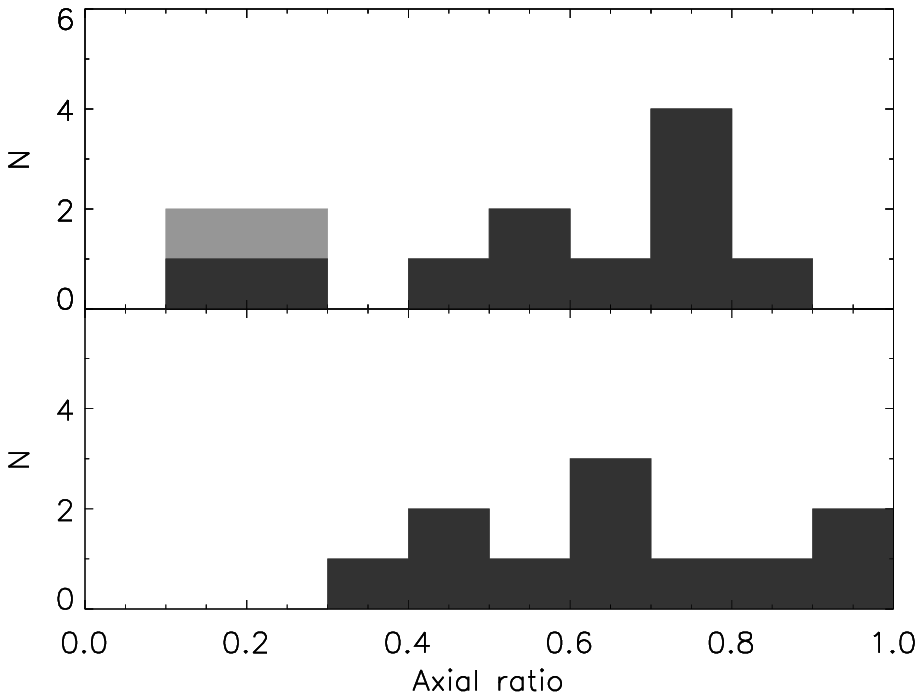,width=9.0cm,angle=0}}
\caption[Distribution of axial ratios for sub-millimetre and radio galaxies]{The distribution of axial ratios for sub-millimetre galaxies (above) and radio galaxies (below) as derived from the two-dimensional modelling of the $K$-band images. The secure ({\it i.e.} radio-identified) sub-millimetre galaxies are shown with dark grey shading, while the less secure optical-infrared identifications are indicated by light grey shading.}
\label{axrat}
\end{figure}

The distributions of the axial ratios and concentration indices for the two samples are plotted as a model-independent test of galaxy morphology. The axial ratios are plotted in Fig. \ref{axrat}. Although detailed statistical comparison is limited by small sample size, it can be seen that the distribution of axial ratios for the sub-millimetre galaxies is very broad and flat, extending to values $\simeq 0.1$, as expected for a population of disc galaxies (Sandage et al. 1970). In contrast, the axial ratios of all but one of the radio galaxies are confined to $> 0.45$, and the mean value of axial ratio is $\simeq 0.7$, more consistent with that expected from a sample drawn from a population of elliptical galaxies (Sandage et al. 1970).  

\begin{figure}
\centerline{\epsfig{file=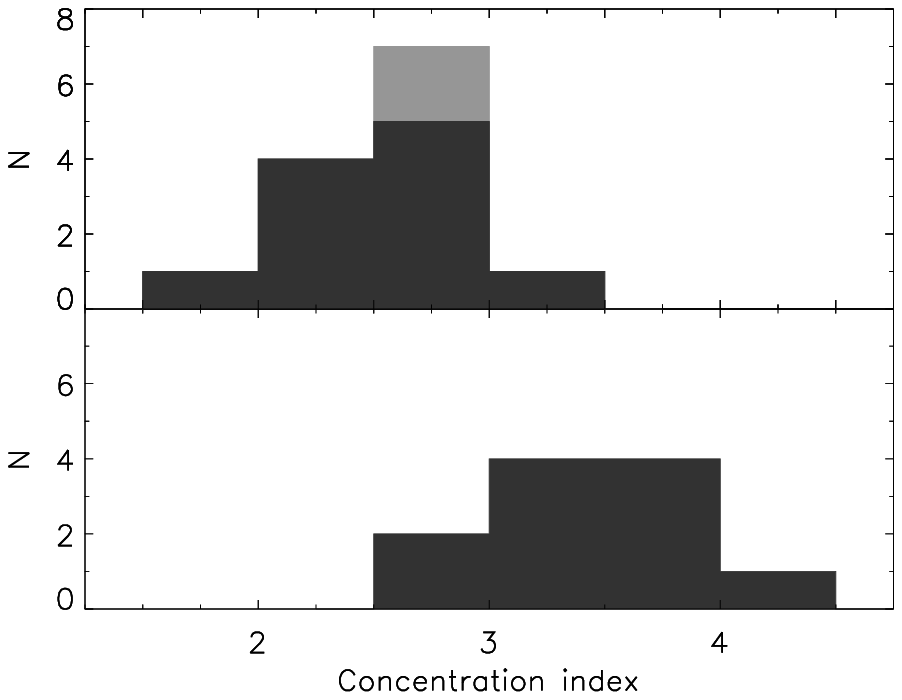,width=9.0cm,angle=0}}
\caption[Distribution of concentration index for sub-millimetre and radio galaxies]{The distribution of concentration index for sub-millimetre galaxies (above) and radio galaxies (below) as derived from the two-dimensional modelling of the $K$-band images. The secure ({\it i.e.} radio-identified) sub-millimetre galaxies are shown with dark grey shading, while the less secure optical-infrared identifications are indicated by light grey shading.}
\label{concentration}
\end{figure}

The distribution of source concentration index is shown in Fig. \ref{concentration}. Concentration index represents the fraction of stars in the bulge component of a galaxy (Conselice 2003), and was calculated using Equation \ref{eq:conseq}, where $r_{80}$ and $r_{20}$ are the radii containing 80 per cent and 20 per cent of the total light from the galaxy.

\begin{equation}
\label{eq:conseq}
C=5 \times \log(r_{80} / r_{20})
\end{equation}

The lower values found for the sub-millimetre sources are consistent with those expected for late-type galaxies, while the higher values for the radio galaxies suggest early-type hosts. A KS test finds a significant separation between the two samples, yielding a significance level of only $p=0.01$ that the two subsamples are drawn from the same distribution. It should be noted that constraining source morphology from concentration index typically requires higher-resolution data than that used in this study. However, the relatively clean separation seen here between these two sub-samples suggests a difference in morphology completely consistent with that implied by the S\'{e}rsic index distributions produced by the 2-dimensional modelling.

The ability of the two-dimensional modelling to recover accurate morphological information from simulated data, and the consistency of these results with multiple non-parametric tests of $r_{1/2}$ and basic morphology, all support the conclusion that the $z \simeq 2$ radio galaxies are already relaxed ellipticals, while the sub-millimetre galaxies are much better described as massive discs.

\subsection{Image stacking}
\label{stack}
\begin{figure*}
 \begin{center}

\begin{tabular}{ccc}
\epsfig{file=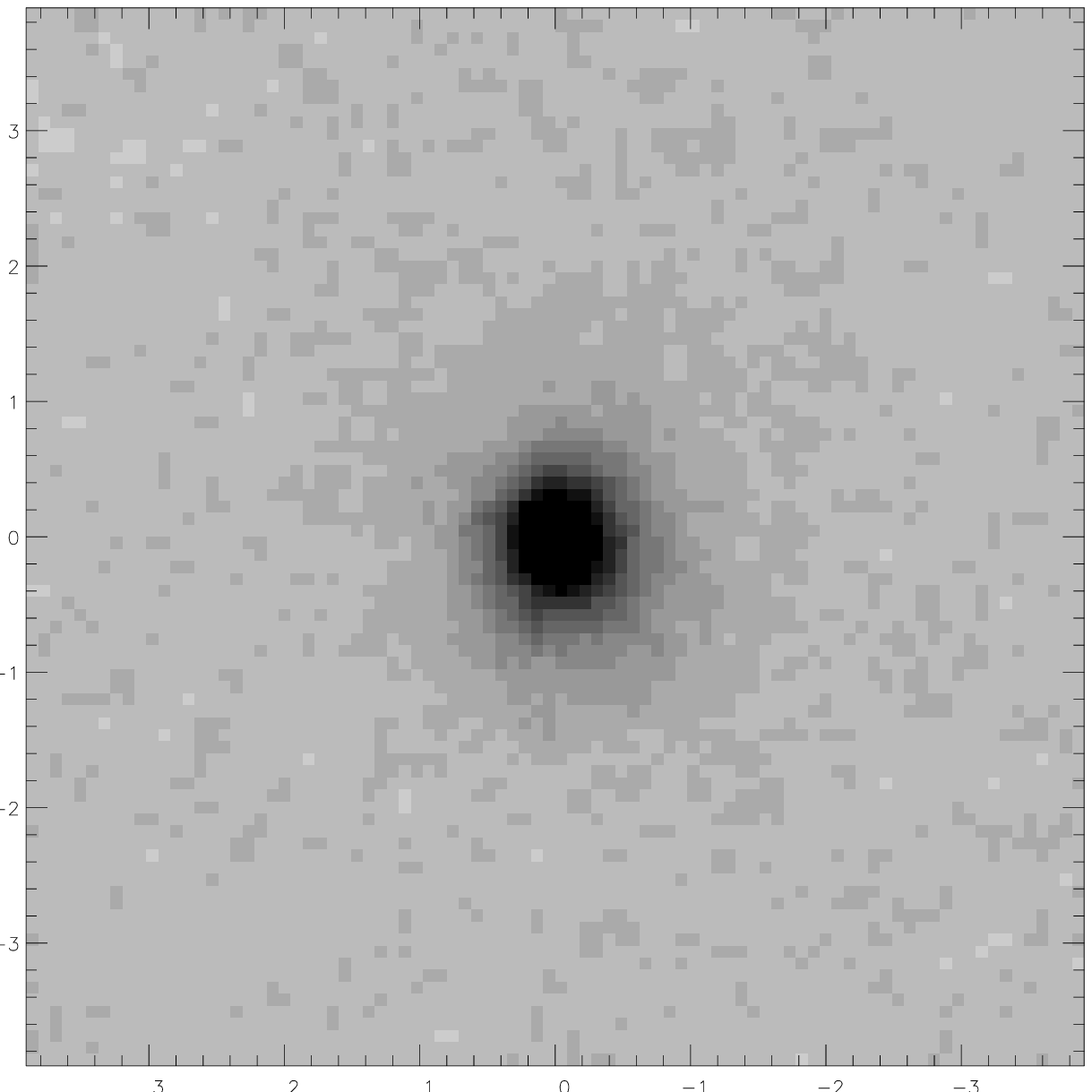,width=0.29\textwidth}&
\epsfig{file=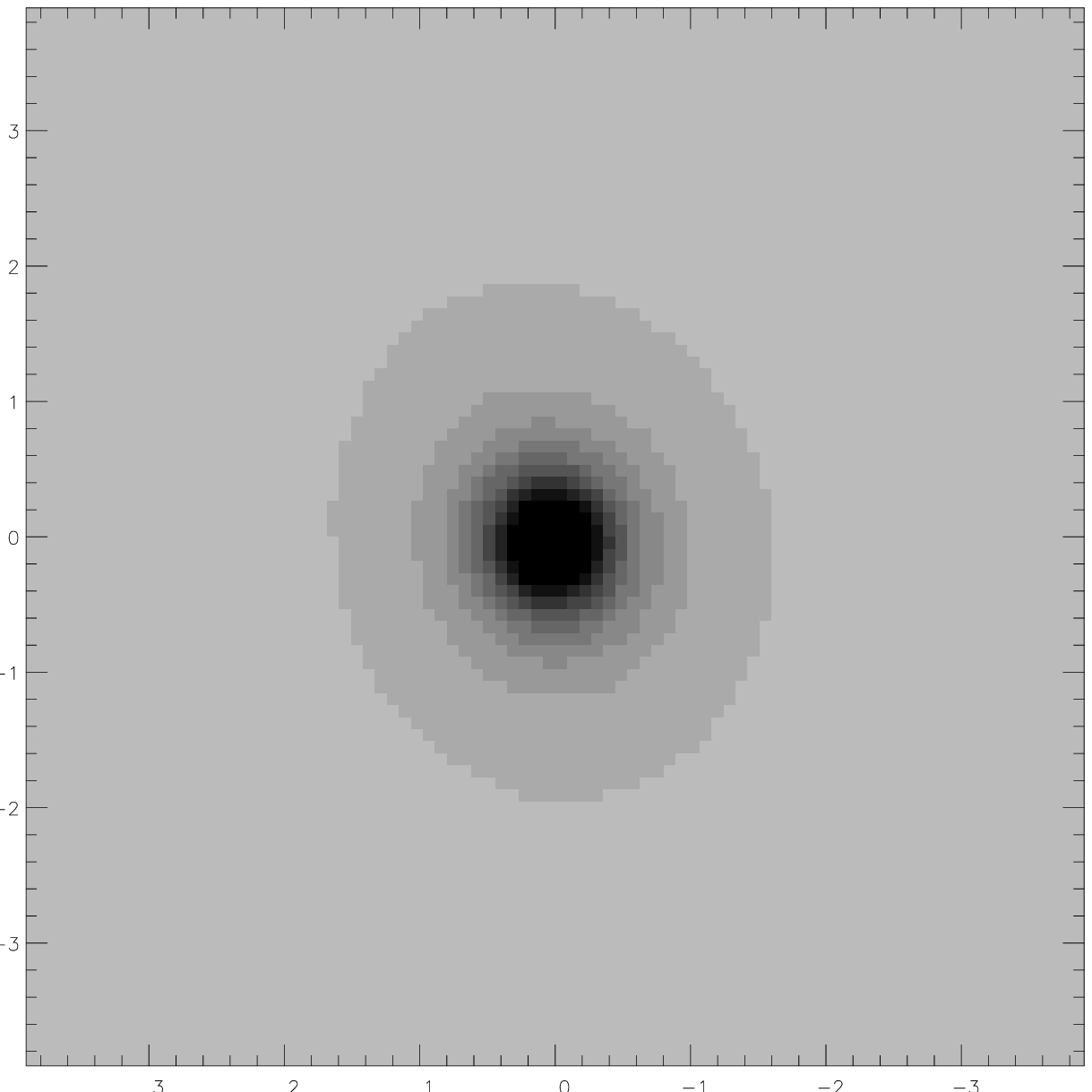,width=0.29\textwidth}&
\epsfig{file=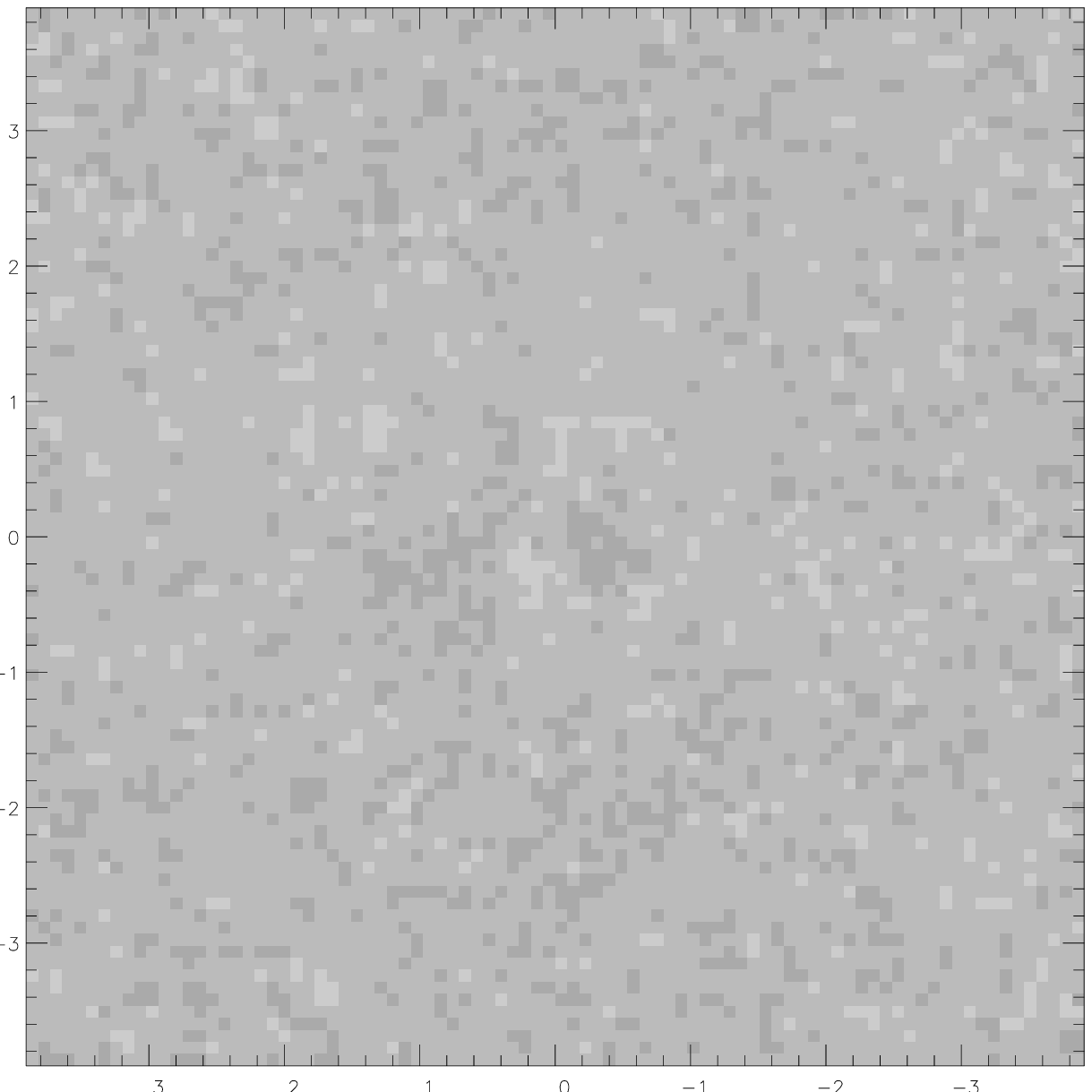,width=0.29\textwidth}\\
\\
\epsfig{file=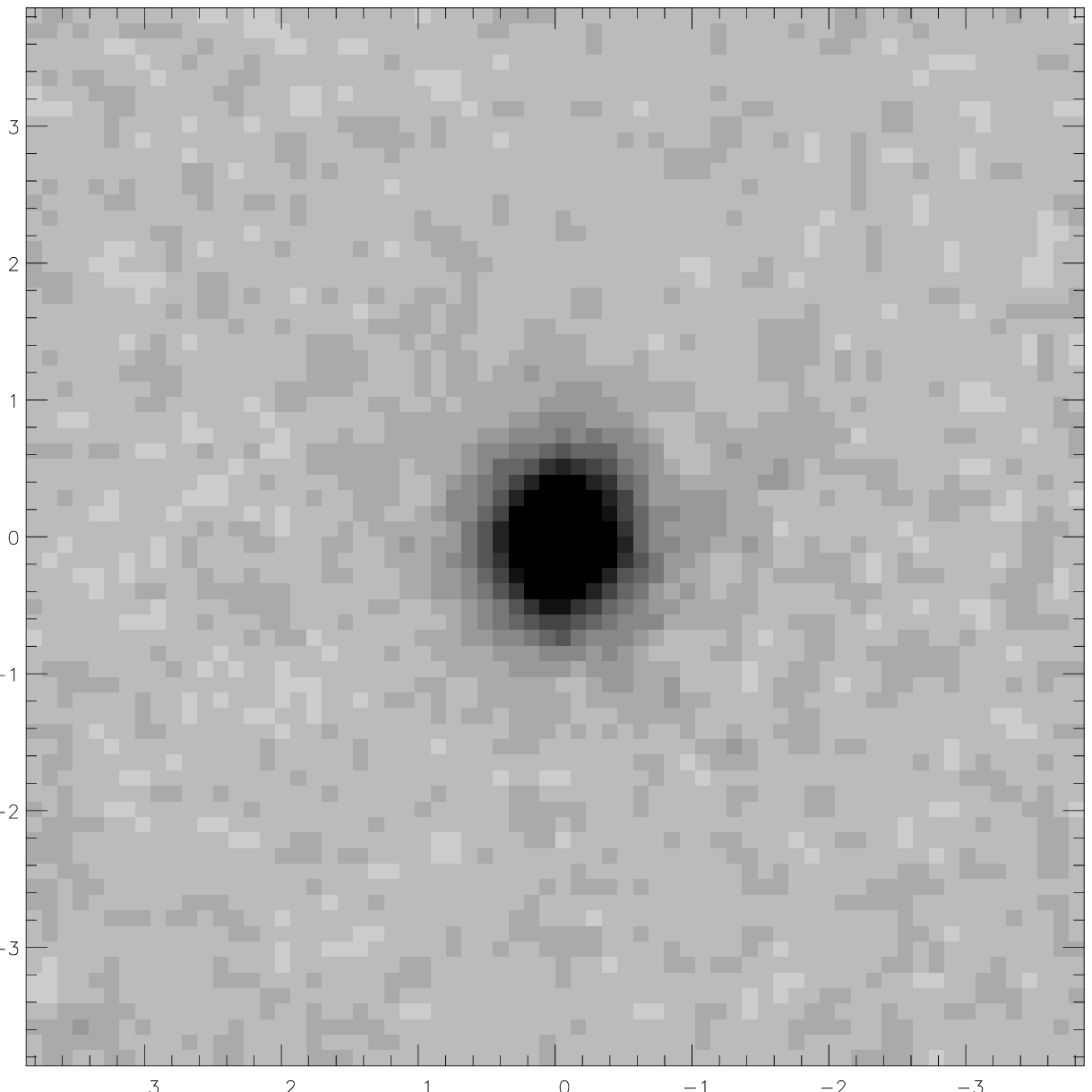,width=0.29\textwidth}&
\epsfig{file=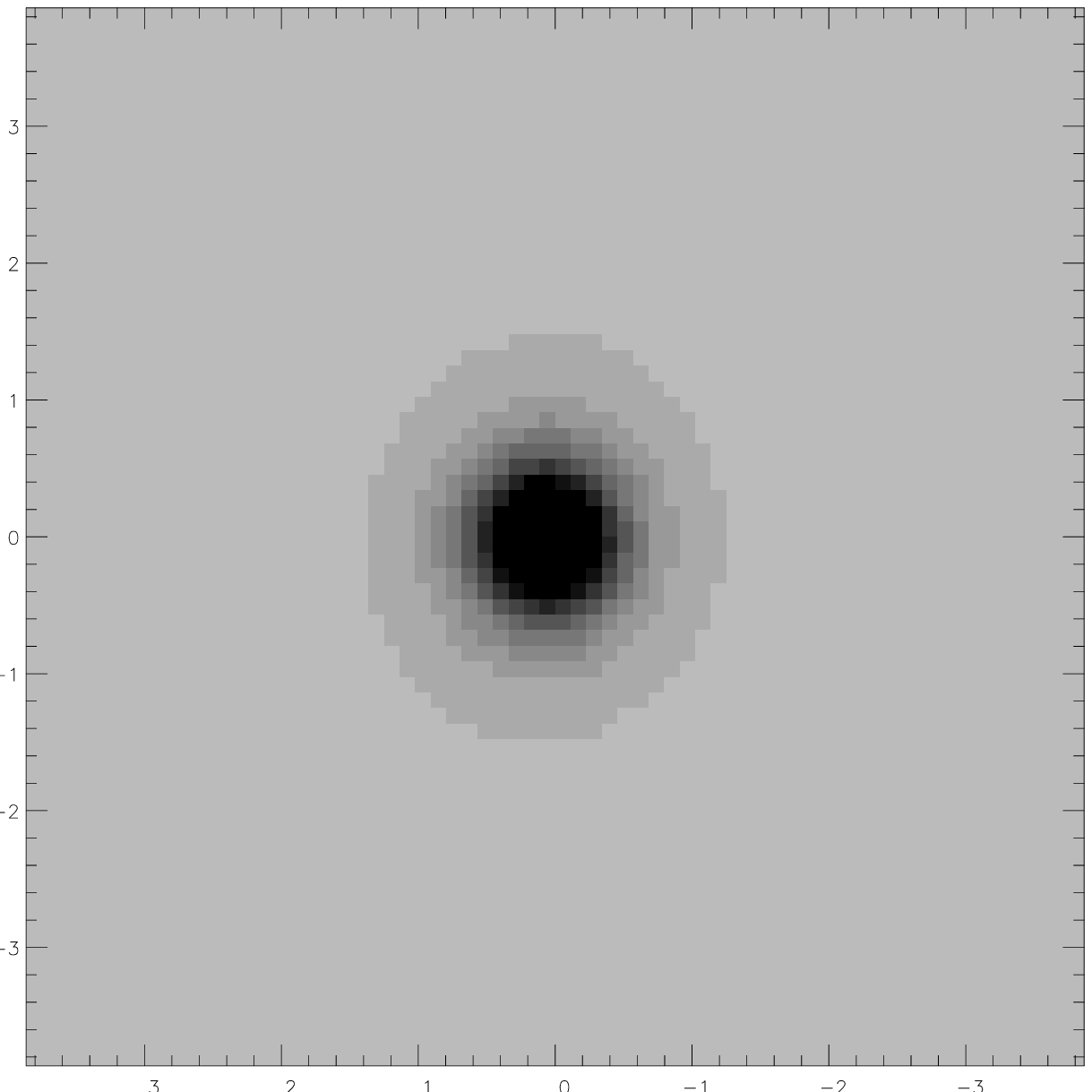,width=0.29\textwidth}&
\epsfig{file=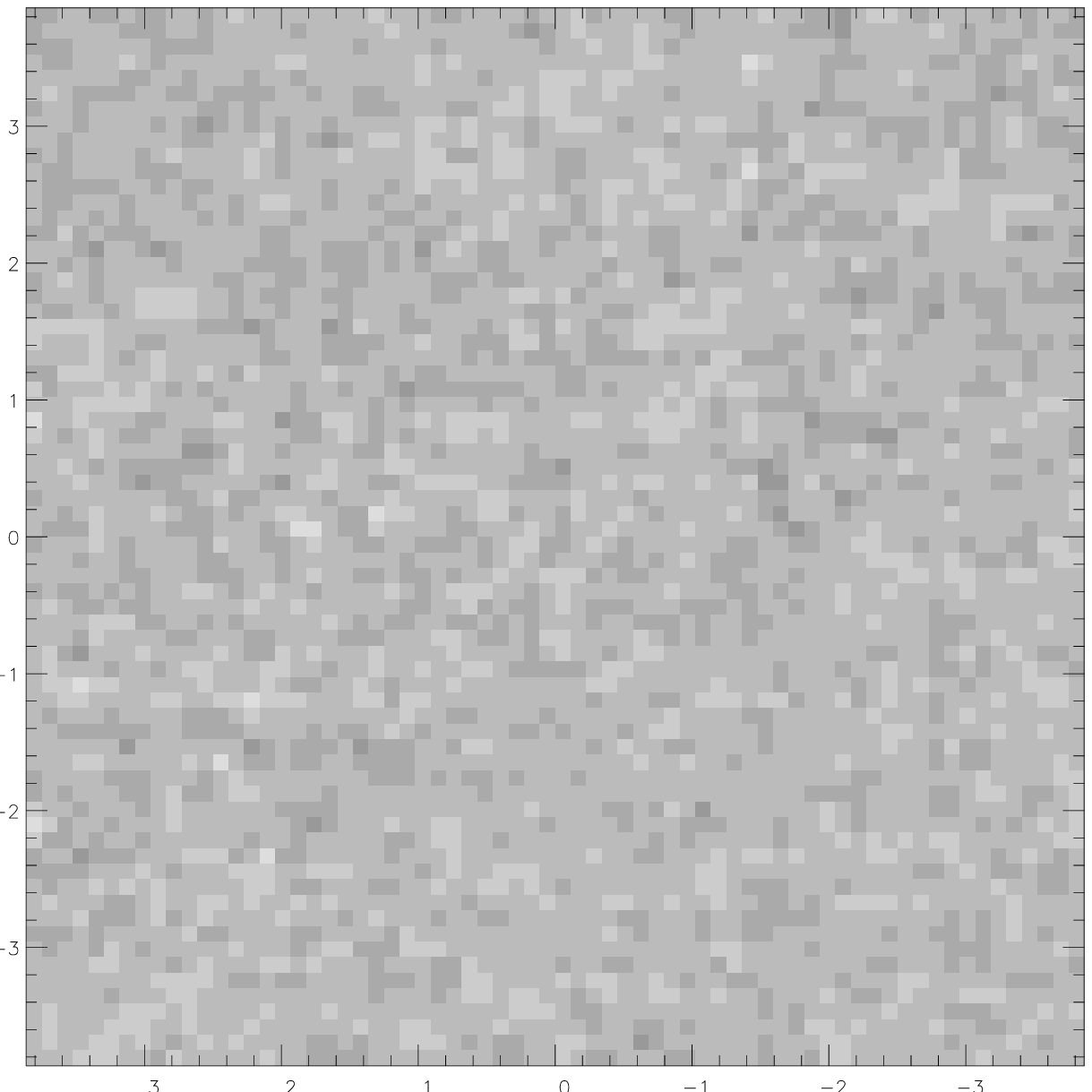,width=0.29\textwidth}\\
\end{tabular}

\caption[Two-dimensional modelling of stacked radio and sub-millimetre galaxies]{The stacked $K$-band images of the radio galaxies (above) and sub-millimetre galaxies (below), along with the results of fitting these stacked images with the axi-symmetric two-dimensional galaxy models. The left-hand panels show the stacked images. The middle panels show the best-fitting models. The right-hand panels show the residual images produced by subtracting the best-fitting models from the stacked data. All panels are $8.0^{\prime \prime} \times 8.0^{\prime \prime}$, and the images are shown with a linear greyscale in which black corresponds to $2.5\sigma$ above, and white to $1\sigma$ below the median sky value in the stacked images.}
\label{figstack}

 \end{center}
\end{figure*}

As described in Section 3, the exposure times adopted in this study were selected to achieve a signal-to-noise of 5 at twice the anticipated $\simeq 10$\,kpc scalelength of the radio galaxies, in order to allow accurate two-dimensional modelling of galaxy profiles. To explore the extent to which extended low-surface brightness emission could still have remained undetected, the image data for both galaxies sub-classes were stacked to create images with effective exposure times of 33.5 hours on UKIRT for the radio galaxies, and 21.7 hours on Gemini for the sub-millimetre galaxies ($\equiv$80 hours on UKIRT). All the individual galaxy images were centroided, scaled to the average luminosity, aligned to the same position angle, and mean-combined using the IRAF package IMCOMBINE. The resulting images were then modelled using PSFs created from a scaled stack of all individual PSFs. The stacked images, and the results of the two-dimensional model fits are presented in Fig. \ref{figstack} and Table \ref{tabstack}.

\begin{figure*}
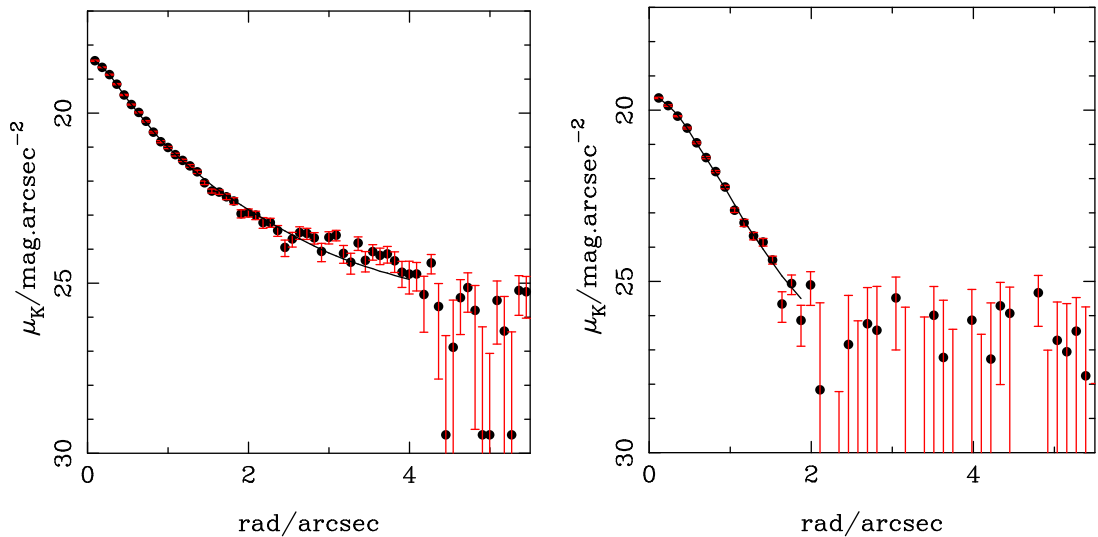

 \begin{center}

\begin{tabular}{cc}
\epsfig{file=TAT2010A_FIG6_1.PS,width=0.4\textwidth,angle=270}&
\epsfig{file=TAT2010A_FIG6_2.PS,width=0.4\textwidth,angle=270}\\
\end{tabular}

\caption[Luminosity profiles of stacked $z \simeq 2$ radio and submm galaxies]{The luminosity profiles (data and model fits) derived from the stacked $K$-band images of the $z \simeq 2$ radio (left) and sub-millimetre galaxies (right). Model profiles are plotted out to the radius at which the signal:noise level in the stacked images allows them to be constrained.}
\label{profile}

 \end{center}
\end{figure*}

The luminosity profiles of these two stacked images are shown in Fig. \ref{profile}. It is clear that there is a significant amount of extended emission around the radio galaxies which can be traced out to a radius of $\sim5$ arcsec, consistent with a large relaxed elliptical. In contrast, the profile of the sub-millimetre galaxy stack confirms the less extended, power-law luminosity profiles of these objects, with no signal detected beyond a radius of 2 arcsec. This comparison demonstrates very clearly that the NIRI imaging of the sub-millimetre galaxies is deep enough to reveal extended emission as displayed by the radio galaxies if it was present at a comparable level. The parameter values derived from the two-dimensional modelling of these two image stacks are given in Table \ref{tabstack}. The results are reassuringly consistent with the average values of the individual object sub-samples which have been incorporated into the stacks, reinforcing the evidence that these two classes of galaxy are very different in terms of size and basic morphology.

\begin{table}
 \begin{center}

\caption[Two-dimensional modelling results of stacked radio and SCUBA galaxies]{The results of the two-dimensional modelling of the stacked radio and sub-millimetre galaxy $K$-band images.}

\begin{tabular}{lccccc}
\hline
Stack &  $r_{1/2}$  &  $<\mu_{e}>$  &  S\'{e}rsic $n$ & $K-$mag & $\chi^{2}_{\nu}$\\
\hline
Radio galaxy  & 6.12 & 19.18 & 4.6 & 17.88 & .97\\
Sub-mm galaxy & 2.19 & 18.65 & 1.4 & 19.48 & .94\\
\hline
\end{tabular}
\label{tabstack}

 \end{center}
\end{table}

\subsection{Analysis of degraded HST ACS $I$-band imaging}

The absence of significant morphological asymmetries in the $K$-band sub-millimetre galaxy modelling seems at first glance to be in conflict with the known presence of asymmetries seen in high-resolution space-based HST imaging of sub-millimetre galaxies ({\it e.g.} Tacconi et al. 2008; Swinbank et al. 2010). The Swinbank et al. (2010) data reveals a population of often multi-component sources, typically not well described by standard "Hubble tuning fork" morphologies. One interpretation is, of course, that the galaxies do indeed appear less disturbed when viewed at $\lambda_{obs} \simeq 2.2\,{\rm \mu m}$ than at shorter wavelengths. However, an obvious concern is that such strong asymmetries are in fact still present, but are simply not apparent in our $K$-band data due to the limitations of our (albeit good) ground-based seeing (typically $\simeq 0.5$ arcsec FWHM). 

To attempt to distinguish between these alternatives, we selected the four objects in our ELAIS N2 sub-millimetre galaxy sample which were imaged at $I$-band with HST ACS (PI: Almaini, PID: 9761), and which yielded optical detections of sufficient significance to warrant image modelling (namely ELAIS N2\,850.1, N2\,850.4, N2\,850.7, and N2\,850.8). We degraded these HST data to the same pixel-scale and seeing quality as our ground based observations by convolving the ACS images with an averaged PSF from our $K$-band imaging programme. We then performed two-dimensional modelling of these ``pseudo ground-based'' red optical images in the same way as described above for the real $K$-band imaging.

These degraded images, and the results of trying to model them with axi-symmetric galaxy models are shown in Appendix A, Fig. A3. Unlike the $K$-band images, we found that the degraded ACS images of these four objects could not be so well-described by a simple axi-symmetric galaxy model, with much more significant asymmetries present in the model-subtracted residual images (with only $60-90$\% of the light accounted for by the axi-symmetric models). This favours the morphological k-correction interpretation, supporting the view that the galaxies really are smoother when observed at $K$-band, longward of the rest-frame 4000\AA\ or Balmer break. This conclusion is consistent with the results of Swinbank et al (2010), who found that the $I$-band ACS images of their sub-millimetre galaxies generally displayed larger degrees of asymmetry than suggested by the NICMOS data. 

In conclusion, our modelling of the available data suggests that while at least some $z \simeq 2$ sub-millimetre galaxies may appear complex when viewed at rest-frame ultra-violet wavelengths (perhaps due to a mixture of patchy obscuration and/or localised starburst activity), when viewed at $2.2\,{\rm \mu m}$ a single dominant galaxy is generally seen, the vast majority ($> 95$\% in every case) of whose observed rest-frame optical emission can be well described by a simple axi-symmetric exponential disc. 

\section{Discussion}

\subsection{Galaxy morphologies}

One of the cleanest results from this study is that the radio galaxies, even at $z \simeq 2$, appear to rather well described as de Vaucouleurs spheroids ({\it i.e.} S\'{e}rsic index $n = 4$) whereas, in contrast, the sub-millimetre galaxies are much better described as exponential discs ({\it i.e.} S\'{e}rsic index $n = 1$). This does not of course preclude the possibility that the sub-millimetre galaxies could evolve into present-day ellipticals, but it is clear that, when selected as bright sub-millimetre emitters, the luminosity profiles of these galaxies are not well described by a de Vaucouleurs law (either individually, or in the stacked image shown in Fig. 5). The axial ratio distributions displayed by the two samples are consistent with this morphological division, although they are not formally statistically different.

It is often argued that the extreme starbursts required to produce easily detected sub-millimetre galaxies are the result of major mergers. For example, in the simulations of Narayan et al. (2009), the merging of two galaxies with stellar masses $\simeq 3 \times 10^{11}\,{\rm M_{\odot}}$ is required to produce a sub-millimetre flux density $S_{850} \simeq 5$\,mJy. In addition, direct observational evidence that at least some sub-millimetre galaxies display complex, disturbed gas motions has been presented by Tacconi et al. (2008). 

However, it is not yet clear that major mergers are the dominant cause of high-redshift sub-millimetre galaxies. At least some sub-millimetre galaxies in fact display CO velocity gradients consistent with a compact, rotating disc (Tacconi et al. 2008), and Dav\'{e} et al. (2010) have argued that the sub-millimetre galaxies predicted by current cosmological hydrodynamic simulations are massive galaxies being fed by smooth infall and gas-rich satellites at rates comparable to their star-formation rates. In this scenario the fraction of sub-millimetre galaxies expected to show clear evidence of a major merger is predicted to be small, and most sub-millimetre galaxies would be expected to display relatively undisturbed morphologies.

From Fig. A2 it can be seen that while some of the sub-millimetre galaxies imaged here show evidence of a recent galaxy-galaxy interaction ({\it e.g.} N2850.4, N2850.8, LE850.3, LE850.7), in all cases the vast majority ($>$95\%) of the $K$-band light from these galaxies is well described by a single, simple axi-symmetric galaxy model (see final column in Table 2 for individual object results). 

\begin{figure}
\centerline{\epsfig{file=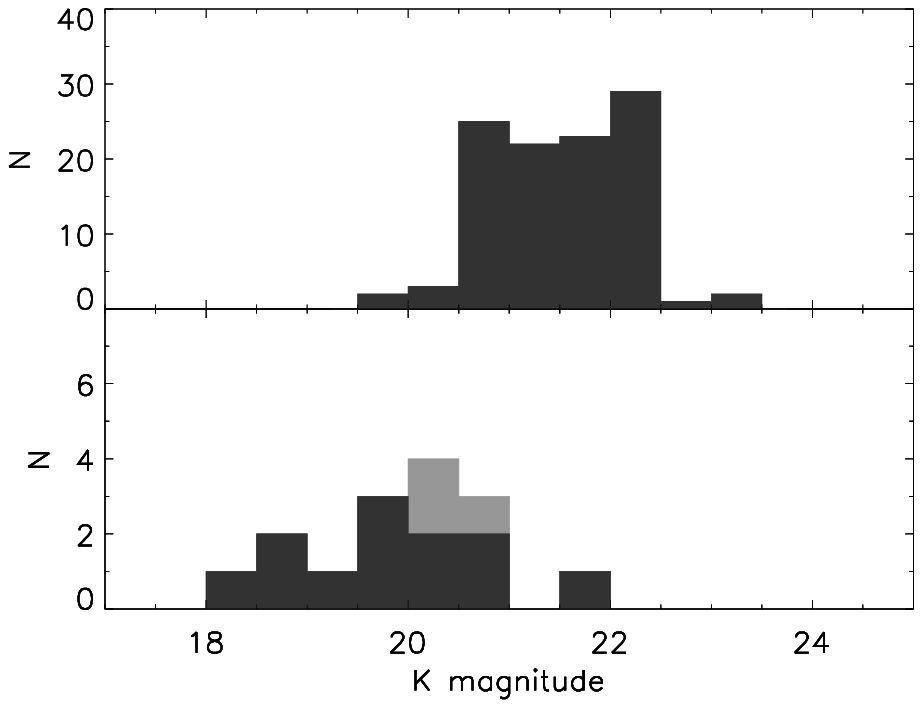,width=9.0cm,angle=0}}
\caption[Magnitude distribution of Lyman-break and submm galaxies from two-dimensional modelling]{The $K$-band magnitude distribution of Lyman-break galaxies at $z \simeq 2 - 3$ (upper panel) from Shapley et al. (2001), compared with that derived here for sub-millimetre galaxies (lower panel) at comparable redshifts. The radio-identified sub-millimetre galaxies are shown with dark grey shading, while the optical-infrared identifications are indicated by light grey shading.}
\label{histlyman}
\end{figure}

The average, deboosted 850\,${\rm \mu m}$ flux density of the sub-millimetre galaxies imaged here is $\langle S_{850} \rangle \simeq  6$\,mJy. Our results indicate that, at least at this flux-density level, $z \simeq 2$ sub-millimetre galaxies are massive, star-forming disc galaxies which generally display little obvious evidence of a recent massive galaxy-galaxy merger. The concept that sub-millimetre galaxies are simply the high-mass tail of the general star-forming galaxy population at $z \simeq 2-3$ gains at least qualitative support from the comparison of the $K$-band magnitudes measured here for sub-millimetre galaxies with the $K$-band magnitudes measured for Lyman-break galaxies at comparable redshifts by Shapley et al. (2001), shown in Fig. \ref{histlyman}. How high-redshift sub-millimetre galaxies compare to present-day star-forming discs is discussed further below.

\subsection{Galaxy sizes}

The sub-millimetre galaxies are fairly compact, with stellar half-light radii $\simeq 3$\,kpc. As already mentioned for individual sources above, this typical size is comparable to the gas sizes of sub-millimetre galaxies measured by Tacconi et al. (2008) using IRAM/PdB ($r_{1/2} \simeq 2$\,kpc), a result which is at least consistent with the idea that both the stars and molecular gas are co-located in these star-forming discs. 

Until recently, such small sizes inferred for massive galaxies ($M_{stars} \simeq 2 \times 10^{11}\,{\rm M_{\odot}}$; see Section 7.3 below) at high-redshift might have been regarded as unexpected. However, recent work has shown that other populations of massive galaxies at $z = 1.5 - 3$ are generally similarly compact, with $r_{1/2} \simeq 2 - 3$\,kpc ({\it e.g.} Daddi et al. 2005; Trujillo et al. 2006, 2007; Zirm et al. 2007; Cimatti et al. 2008; van Dokkum et al. 2008), with both star-forming and quiescent galaxies apparently a factor of $2-5$ times smaller than present-day counterparts of comparable mass (Toft et al. 2009, but see also Mancini et al. 2010, and Saracco, Longhetti \& Gargiulo 2010). The controversial question of how such compact galaxies might evolve into present-day massive ellipticals is discussed further in the next sub-section, but observationally the key point is that the sizes of the sub-millimetre galaxies measured here are not unexpectedly small in the context of those measured for other comparably massive galaxies at comparable redshifts.

The radio galaxies are markedly larger, with stellar half-light radii $\simeq 8$\,kpc. However, they are also typically $\simeq 3$ times more massive. Interestingly, the average size of the $z > 1.5$ radio galaxies measured here is very similar to that displayed by the brightest cluster galaxies at $z \simeq 1.5$ recently studied by by Collins et al. (2009). Further morphological studies of purely mass-selected high-mass ($M_{star} > 5 \times 10^{11}\,{\rm M_{\odot}}$) galaxies at $z \simeq 2$ are required to establish whether these larger scale sizes are typical of this extreme mass regime at these redshifts, or whether they are seen in only certain subclasses of object which may have special merger histories (radio selection, or brightest cluster galaxy selection may preferentially select the products of early major mergers). 

However, while clearly larger than the sub-millimetre galaxies, the $z \simeq 2$ radio galaxies are nonetheless undoubtedly more compact than comparably radio-luminous galaxies at $z < 1$. Specifically, the $z \simeq 2$ radio galaxies studied here are found to be, on average, a factor $\simeq 1.5$ smaller than the $0<z<1$ 3CR radio galaxies studied by McLure et al (2004). Motivated by the fact that the $z \simeq 2$ radio galaxies appear to be well described as dynamically relaxed spheroids, we have attempted to explore further whether they follow the scaling relations displayed by elliptical galaxies in general, and radio galaxies in particular, at lower redshifts.

\subsubsection{The Kormendy relation}
\begin{figure}
\centerline{\epsfig{file=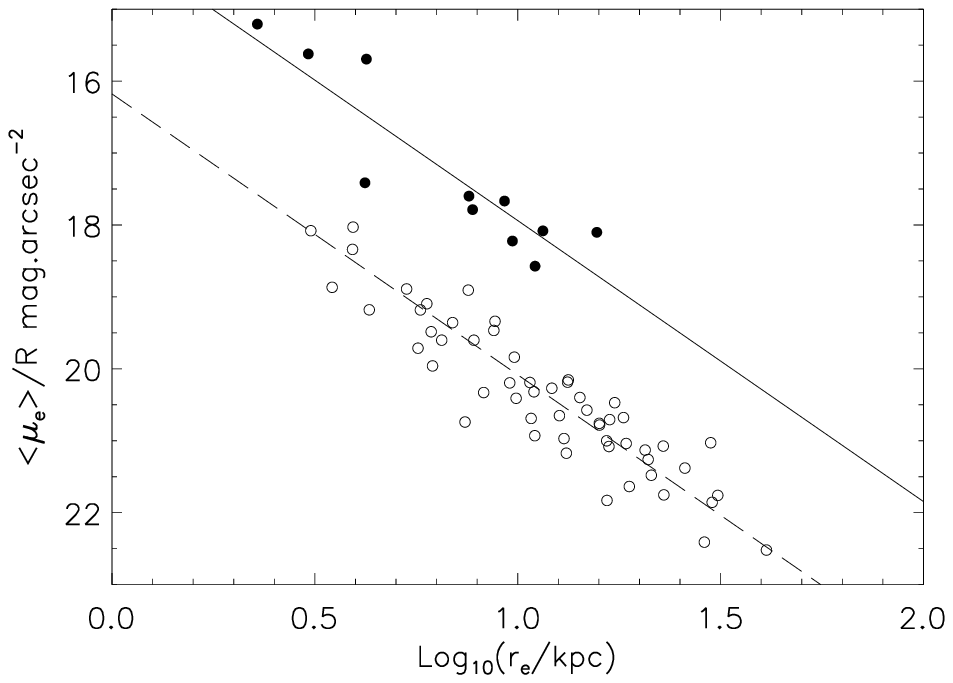,width=9.0cm,angle=0}}
\centerline{\epsfig{file=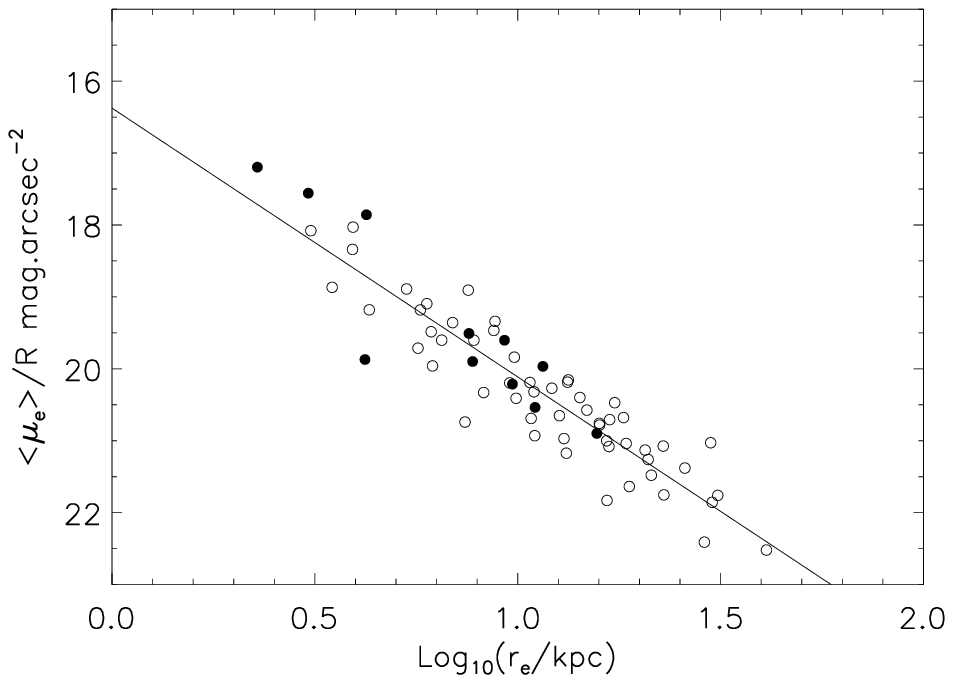,width=9.0cm,angle=0}}
\caption[Kormendy relations for the $z \simeq 2$ radio galaxy sample, 3C-type galaxies, and submm hosts]{The top panel shows the best-fitting fixed-slope Kormendy relations for the $z \simeq 2$ radio galaxy sample (solid line) and the $z \simeq 0$ 3CR-type galaxies from McLure et al. (2004) (dashed line). In the bottom panel the $z \simeq 2$ radio galaxies have been corrected for passive evolution of their stellar populations between $z \simeq 2$ and $z \simeq 0$ as described in the text, and the solid line indicates the best-fitting Kormendy relation to the combined sample.}
\label{kormendy}
\end{figure}

Early-type galaxies are known to exist on a two-dimensional manifold (the ``fundamental plane''), defined by effective scalelength, mean surface brightness, and the central stellar velocity dispersion in three-dimensional parameter space (e.g. Dressler et al. 1987; Djorgovski \& Davis 1987). In the absence of velocity dispersion measurements, the Kormendy or $\mu_{e}-r_{e}$ relation (Kormendy 1977) allows examination of the photometric projection of the fundamental plane using only the scalelength and surface-brightness parameters. The $R-$band Kormendy relation deduced for the $z \simeq 2$ radio galaxies studied here is compared with that deduced for $z=0$ 3CR-type radio galaxies by McLure et al. (2004) in the upper panel of Fig. \ref{kormendy}. The $<\mu_{e}>$ values of the $z \simeq 2$ radio galaxies have been corrected for cosmological surface-brightness dimming, and $k$-corrected into the $R-$band. The data points from the McLure et al. sample have been corrected to $z \simeq 0$ assuming passive evolution derived from the Bruzual \& Charlot (2003) galaxy-evolution models. The best-fitting form of the Kormendy relation for the non-evolutionary corrected $z \simeq 2$ radio galaxy sample is

\begin{equation}
<\mu>_{e}=3.96(\pm1.19) \log r_{e} + 14.39(\pm1.03)
\label{eq:korsep}
\end{equation}

To within the errors, the slope of this relation is indistinguishable from that displayed by the $z \simeq 0$ radio galaxy sample. If the two samples are fitted with a Kormendy relation of fixed slope 3.87 (intermediate between the slopes of their two independent fits), as shown in the upper panel of Fig. \ref{kormendy}, then the relations appear to be offset by a (vertical) shift in rest-frame $R$-band luminosity equivalent to 1.76 magnitudes. This is in very good agreement with the expected passive evolutionary correction as derived from a $z_{form}=3$ burst model, using the evolutionary synthesis models of Bruzual \& Charlot (2003) ($\Delta R \simeq 1.65$). Motivated by this, we have then applied a passive evolution correction, galaxy-by-galaxy, to the $z \simeq 2$ radio-galaxy sample, to move them to $z \simeq 0$ assuming no evolution in scalelength. The results are over-plotted on the $z \simeq 0$ data in the lower panel of Fig. \ref{kormendy}, and a fit to the combined dataset yields

\begin{equation}
<\mu>_{e}=3.74(\pm0.44) \log r_{e} + 16.38(\pm0.47)
\label{eq:korfit}
\end{equation}

In summary, not only are the $z \simeq 2$ radio galaxies well-described by a de Vaucouleurs $r^{1/4}$ law, but they also display a Kormendy relation which is indistinguishable from that displayed by low-redshift radio galaxies (and other radio-quiet massive ellipticals) after simply correcting their luminosities for anticipated passive evolution of their stellar populations. This suggests that any further increase in the typical sizes of these objects since $z \simeq 2$, must be accompanied by sufficient increase in mass to preserve the form of the Kormendy relation.

In concluding this discussion of the radio galaxy sizes and morphologies, it is interesting to assess how our results compare with those of Pentericci et al. (2001), who attempted to fit models to 14 of the 19 radio galaxies in the redshift range $z = 1.7 - 3.0$ which they imaged with HST NICMOS in the $H$-band. They reported that this did not, in general, work well and that many of the highest-redshift radio galaxies had complex, clumpy morphologies. But they did successfully fit de Vaucouleurs models to 5 of their objects, all of which lay at the low-redshift end of their sample, at $z \simeq 1.8 - 2.2$ (more comparable to the redshifts of the radio galaxies studied here). The average size of these five radio galaxies was found to be 5.5\,kpc, again consistent with our findings in the present study.

These results, now reinforced by our own findings, lend support to the general idea that the most massive galaxies at redshift $z \simeq 2$ have evolved into relaxed spheroids. However, we note that it would be interesting to extend the ultra-deep $K$-band imaging of radio galaxies out to $z \simeq 3$, to explore the extent to which the NICMOS imaging of the higher redshift end of the Pentericci et al. sample could have been affected by emission-line contamination, the movement of the 4000\AA/Balmer break through the F160W filter, and the rapidly decreasing sensitivity to low-surface brightness emission.

\subsection{Surface mass densities}
\label{smd}
Finally we use our $K$-band galaxy model fits to explore where the $z \simeq 2$ radio galaxies and sub-millimetre galaxies lie in the stellar surface-mass-density versus stellar mass plane.

The stellar masses ($M_{star}$) of the galaxies were estimated by matching the measured total ({\it i.e.} 6-arcsec diameter aperture, excluding companions) $K$-band magnitudes of the galaxies to the predictions of the evolutionary synthesis models of Bruzual \& Charlot (2003). Given the evidence that the radio galaxies are already evolving essentially passively, a uniform burst-model age of 2 Gyr was adopted in establishing their $K$-band mass-to-light ratio. The inferred average stellar mass of the $z \simeq 2$ galaxies is $\simeq 7 \times 10^{11}\,{\rm M_{\odot}}$, consistent with the findings of McLure et al. (2004). 

For the sub-millimetre galaxies, which are by definition {\it not} passively evolving, such a simple assumption is clearly not justified. However, for the majority of these galaxies, multi-wavelength (optical--mid-infrared) photometry is now available from Subaru and {\it Spitzer} which, when combined with the $K$-band photometry presented here, allows detailed fitting to the predictions of the evolutionary synthesis models (e.g. Dye et al. 2008). The stellar masses derived from these fits were used to calibrate the typical mass-to-light ratio for the sub-millimetre galaxy sample. This was then used to estimate the masses of the remaining sub-millimetre galaxies for which less complete multi-frequency photometry was available. The inferred average stellar mass of the $z \simeq 2$ sub-millimetre galaxies is $\simeq 2.2 \times 10^{11}\,{\rm M_{\odot}}$, consistent with the results of Dye et al. (2008) and Schael et al. (2010).

The average stellar mass surface density, $\sigma_{50}$, is then defined as

\begin{equation}
 \label{eq:surface}
 \sigma_{50}=\frac{0.5M_{star}}{\pi r_{1/2}^{2}}
\end{equation}

where $r_{1/2}$ is the half-light radius of the starlight, as derived from our axi-symmetric two-dimensional fits to the $K$-band images (see Table 2). This value is morphology idependant, and allows direct comparison of different classes of galaxy.

\begin{figure*}
 \begin{center}
\epsfig{file=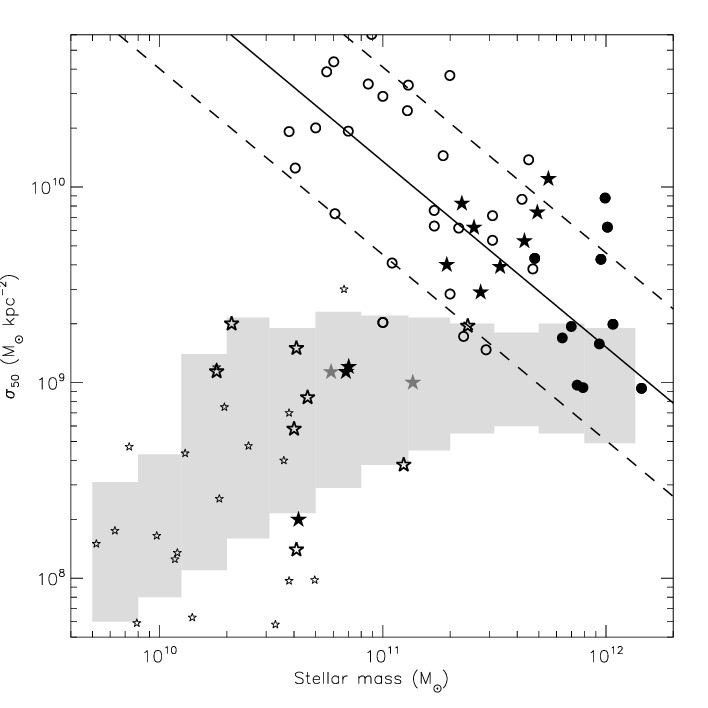,width=0.58\textwidth,angle=0}
\caption[Average surface mass density versus stellar mass]
{Average surface mass density within the half-light radius plotted versus stellar mass. The $z \simeq 2$ radio galaxies studied here are indicated by the filled circles, while the sub-millimetre galaxies are indicated by the solid stars (black for radio-identified galaxies, grey for optical-infrared identifications). Open circles indicate the location of passively evolving massive early-type galaxies in the redshift range $1.4<z<2.6$ from Daddi et al. (2005), Longhetti et al. (2007), Zirm et al. (2007), and van Dokkum et al. (2008) are shown. The measurements for star-forming distant red galaxies (DRGs) and Lyman-break galaxies given by Zirm et al. (2007) are also plotted as large open stars and small open stars respectively. The solid line traces the projected trend of surface density with mass expected for dissipationless mergers (Nipoti, Londrillo, \& Ciotti 2003) normalised to the quiescent DRGs from Zirm et al. (2007). The dotted lines show the same trend if the stellar mass estimates from Zirm et al. (2007) are systematically too high or low by a factor of 3. Light-grey shading represents local values for early- and late-type galaxies from Shen et al. (2003).}
\label{surface1}

 \end{center}
\end{figure*}

The results are plotted in Fig. \ref{surface1}, following Zirm et al. (2007), alongside comparable measurements made for other high-redshift populations taken from a number of recent studies (Daddi et al. 2005, Longhetti et al. 2007, Zirm et al. 2007, van Dokkum et al. 2008).  
 
As discussed by Zirm et al. (2007), the star-forming DRGs and Lyman-break galaxies at $z \simeq 2$ appear to lie within the locus defined by local galaxies, as indicated by the grey shaded region in Fig. \ref{surface1} (Shen et al. 2003). However, the more quiescent high-redshift galaxies appear to lie well above this relation. This is simply a graphical restatement of the fact that, given their inferred stellar masses, these objects are surprisingly compact compared to low-redshift galaxies. The question of how such objects can have evolved onto the local galaxy locus has been considered by several authors. For example, it has been suggested that while secular evolution would exceed the Hubble time, the necessary relaxation of $r_{1/2}$ could arise via dissipationless 'dry' merging (e.g. van Dokkum 2005), with a linear decrease in the surface density with accumulated mass (Nipoti, Londrillo, \& Ciotti 2003). A fiducial normalisation of this power-law trend is shown in Fig. \ref{surface1}. The implication is that dynamical evolution via dry mergers would allow the quiescent DRGs to reach local galaxy densities, but only at very high masses ($\sim10^{12} M_{\odot}$).

What can be learned about the $z \simeq 2$ radio galaxies and sub-millimetre galaxies from their inferred location on this diagram? First, we note that most (but not all) of the $z \simeq 2$ radio galaxies already lie on the local galaxy locus, with project stellar mass densities of $\sigma_{50} \simeq 10^9\,{\rm M_{\odot} kpc^{-2}}$. This is interesting, but perhaps not unexpected given their moderately large sizes and apparently dynamically relaxed stellar populations. Four of the radio galaxies have higher surface mass densities. This might be telling us that their stellar masses have been over-estimated, or their half-light radii under-estimated, but the two obvious cases of an AGN contribution have already been removed and are not plotted in Fig. \ref{surface1}. Alternatively these objects could really be destined to evolve further, in which case the dry-merging tracks shown in Fig. \ref{surface1} would imply that they will land on the present day locus with stellar masses $\simeq 2-3 \times 10^{12} \,{\rm M_{\odot}}$ . This is in fact not unreasonable, given that they are already among the most massive galaxies known at $z \simeq 2$, and yet still appear somewhat more compact than the most massive elliptical galaxies in the present-day Universe.

Interestingly, and unlike the lower-luminosity star-forming Lyman-break galaxies, most of the sub-millimetre galaxies also appear to lie above the local galaxy locus, in the same general regime as the moderately compact quiescent galaxies. Thus, while the morphological evidence suggests the sub-millimetre galaxies are star-forming discs, in terms of surface mass density they can clearly evolve into high-redshift quiescent galaxies without significant size evolution. Also, if they subsequently follow the dry-merger evolutionary tracks indicated in Fig. \ref{surface1}, they are ultimately destined to evolve into present-day massive elliptical galaxies with $M_{star} > 5 \times 10^{11}\,{\rm M_{\odot}}$.

\looseness+ 1There are 5 sub-millimetre galaxies which appear to have lower surface mass densities, and thus lie within the locus defined by present-day galaxies and high-redshift DRGs and Lyman-break galaxies. Both of the radio unidentified sub-millimetre galaxies lie within this subsample, suggesting either that they have been mis-identified, or (not unreasonably) that strong radio emission is an indicator for the densest starbursts.

In summary, Fig. \ref{surface1} supports the conclusion that, while the sub-millimetre galaxies appear (at the epoch of observation) to be best described as star-forming discs, their stellar masses and stellar mass surface densities indicate that they are capable of evolving rapidly into dense, quiescent galaxies at $z > 1.5$, and ultimately into present-day massive ellipticals.

\section{Conclusions}

We have obtained deep, high-quality $K$-band images of complete subsamples of powerful radio and sub-millimetre galaxies at redshifts $z \simeq 2$. The data were obtained in the very best available seeing (FWHM $\simeq$ 0.5\,arcsec) through the queue-based observing systems at the United Kingdom InfraRed Telescope and Gemini North, with integration times scaled to ensure that comparable rest-frame surface brightness levels are reached for all of the galaxies. We have analysed the resulting images by fitting two-dimensional axi-symmetric galaxy models to determine basic galaxy morphological parameters at rest-frame optical wavelengths $\lambda_{rest} > 4000$\AA, varying luminosity, half-light radius $r_{1/2}$, and S\'{e}rsic index $n$. We have also undertaken an extensive set of simulations which show that, for galaxies with scalelengths $r_{1/2} > 2$\, kpc, the modelling of our deep, high-quality ground-based images should yield half-light radii and S\'{e}rsic indices accurate to $\simeq 10$\%.

We find that, while a minority of the images show some evidence of galaxy interactions, $>95$\% of the rest-frame optical light in all the galaxies is well-described by these simple axi-symmetric models. We also find evidence for a clear difference in morphology between the radio galaxies and the sub-millimetre galaxies. Specifically, the fits to the individual images reveal that the radio galaxies are moderately large ($\langle r_{1/2} \rangle = 8.4 \pm 1.1$\,kpc; median $r_{1/2}=7.8$ ), de Vaucouleurs spheroids ($\langle n \rangle = 4.07 \pm 0.27$; median $n = 3.87$), while the sub-millimetre galaxies appear to be moderately compact ($\langle r_{1/2} \rangle = 3.4 \pm 0.3$\,kpc; median $r_{1/2}=3.1$\,kpc) exponential discs ($\langle n \rangle = 1.44 \pm 0.16$; median $n = 1.08$). Model fits to the stacked images yield very similar results.
  
We find that the $z \simeq 2$ radio galaxies display a well-defined Kormendy ($\mu_e - r_e$) relation, which is offset in surface brightness from the local relation by $\simeq 1.7$ mag. This luminosity offset is consistent with that expected due to purely passive evolution between $z \simeq 2$ and $z \simeq 0$ for a stellar population formed at $z > 3$. However, while larger than other recently-studied massive galaxy populations at comparable redshifts, the $z \simeq 2$ radio galaxies are still on average a factor $\simeq 2$ times smaller than their local counterparts, suggesting that they have still to undergo further dynamical evolution.

The scalelengths we have derived for the starlight in the sub-millimetre galaxies are consistent with those reported for the molecular gas (e.g. Tacconi et al. 2008), suggesting that the two may be co-located. Such scalelengths are also similar to those of comparably-massive quiescent galaxies at $z > 1.5$, allowing the possibility of an evolutionary connection following cessation/quenching of the observed star-formation activity (Ricciardelli et al. 2010). 

Finally, in terms of stellar mass surface density, we find that the majority of the radio galaxies lie within the locus defined by local ellipticals of comparable stellar mass. In contrast, while best modelled as discs at the epoch of bright dust-enshrouded star formation, most of the sub-millimetre galaxies have higher stellar mass densities than local galaxies, and appear destined to evolve into present-day massive ellipticals. 

With the advent of {\it Herschel} and SCUBA-2, coupled with the wide-field near-infrared imaging provided by VISTA and WFC3/IR on {\it HST}, it should soon be possible to extend this type of study to larger samples of sub-millimetre galaxies which include even more extreme starbursts at $z \simeq 2$. It will then be possible to test whether the dense disc morphologies of sub-millimetre galaxies uncovered in the present study are a universal feature of dust-enshrouded starbursts at high redshift, or whether they are specific to the mass/star-formation rate regime sampled in the early sub-millimetre surveys.

\section*{ACKNOWLEDGEMENTS}
JSD acknowledges the support of the Royal Society via a Wolfson Research Merit award, and also the support of the European Research Council via the award of an Advanced Grant. RJM acknowledges the support of the Royal Society via a University Research Fellowship. PNB is grateful for support from the Leverhulme Trust. The United Kingdom Infrared Telescope is operated by the Joint Astronomy Centre on behalf of the UK Science \& Technology Facilities Council. This work was based in part on observations obtained at the Gemini Observatory, which is operated by the Association of Universities for Research in Astronomy, Inc., under a cooperative agreement with the NSF on behalf of the Gemini partnership: the National Science Foundation (United States), the Science \& Technology Facilities Council (United Kingdom), the National Research Council (Canada), CONICYT (Chile), the Australian Research Council (Australia), CNPq (Brazil) and CONICET (Argentina). This work is based in part on observations made with the NASA/ESA {\it Hubble Space Telescope}, which is operated by the Association of Universities for Research in Astronomy, Inc, under NASA contract NAS5-26555. This work is based in part on observations made with the {\it Spitzer Space Telescope}, which is operated by the Jet Propulsion Laboratory, California Institute of Technology under NASA contract 1407. 

{}

\appendix

\section{Science images, galaxy model fits, and residuals}

In this appendix we present greyscale plots of the $K$-band images, the axi-symmetric model fits, and then the residual images after subtraction of the PSF-convolved model. Plots for the radio galaxies are shown in Fig. A1, with the plots for the sub-millimetre galaxies presented in Fig. A2.

In addition, in Fig. A3, we show comparable greyscale plots of the HST ACS $I$-band images of the sub-milimeater galaxies N2850.1, N2850.4, N2850.7, and N2850.8, again accompanied by the axi-symmetric model fits, and the model-subtracted residual images. 
 
\begin{center}
\begin{figure*}
\begin{tabular}{ccc}
\epsfig{file=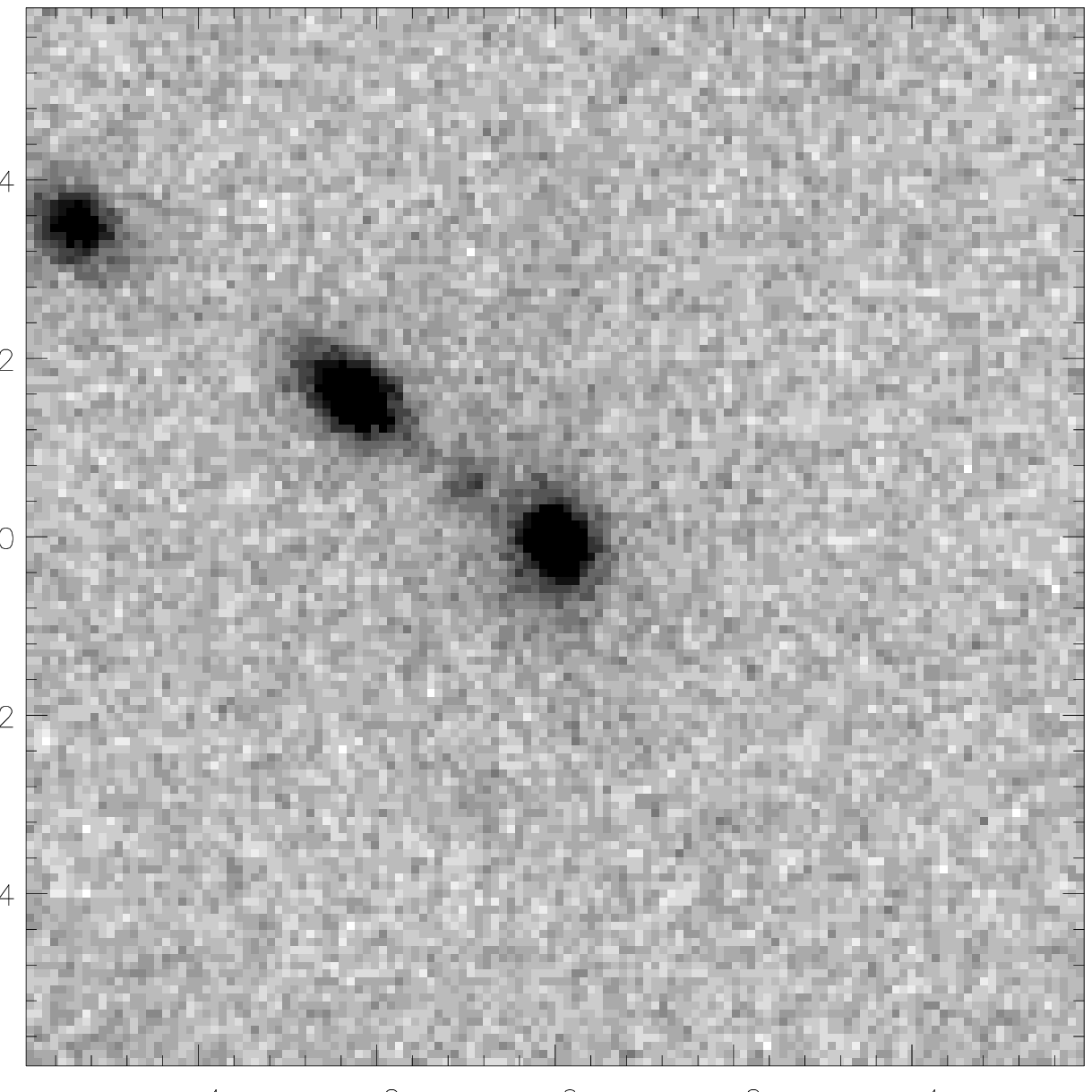,width=0.3\textwidth}&
\epsfig{file=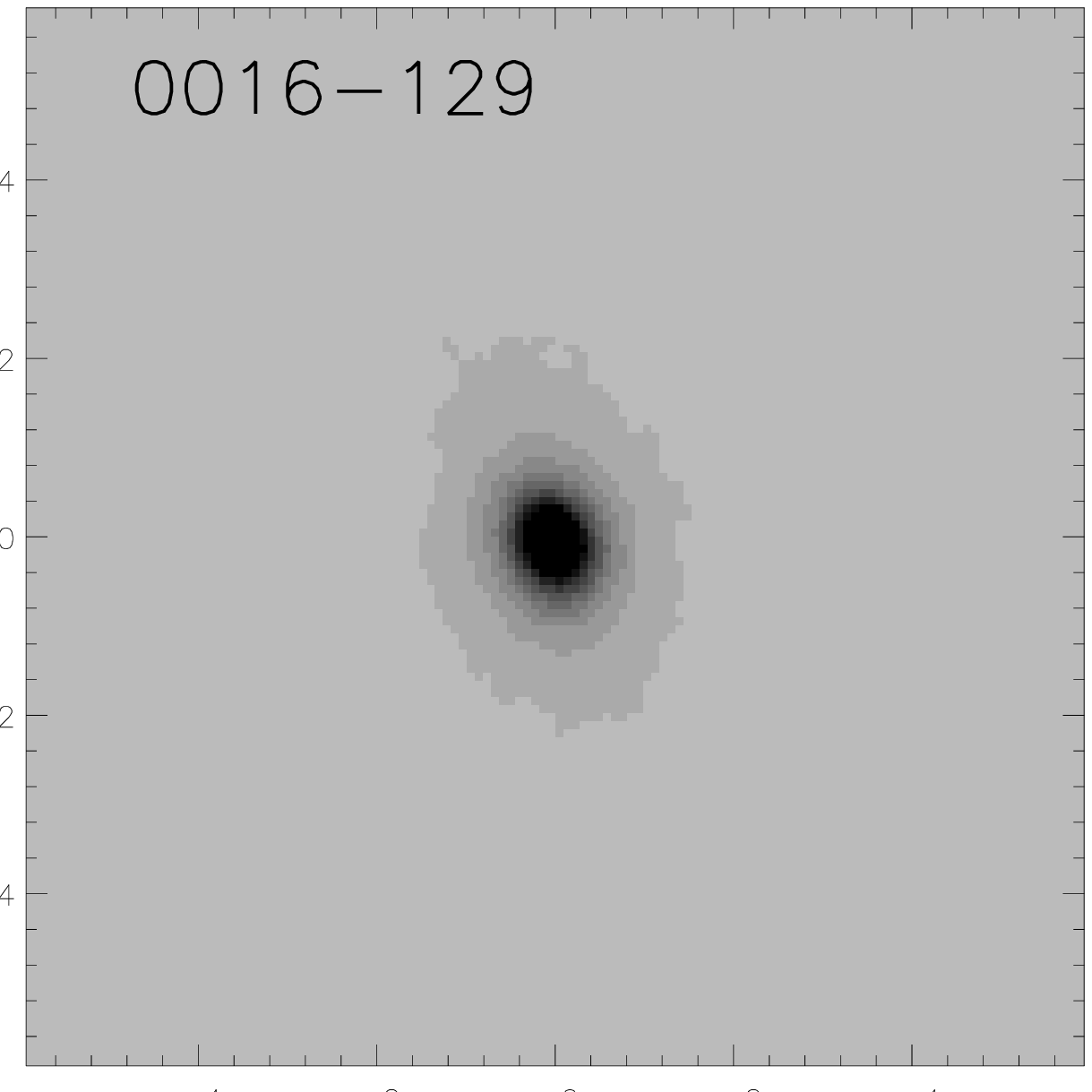,width=0.3\textwidth}&
\epsfig{file=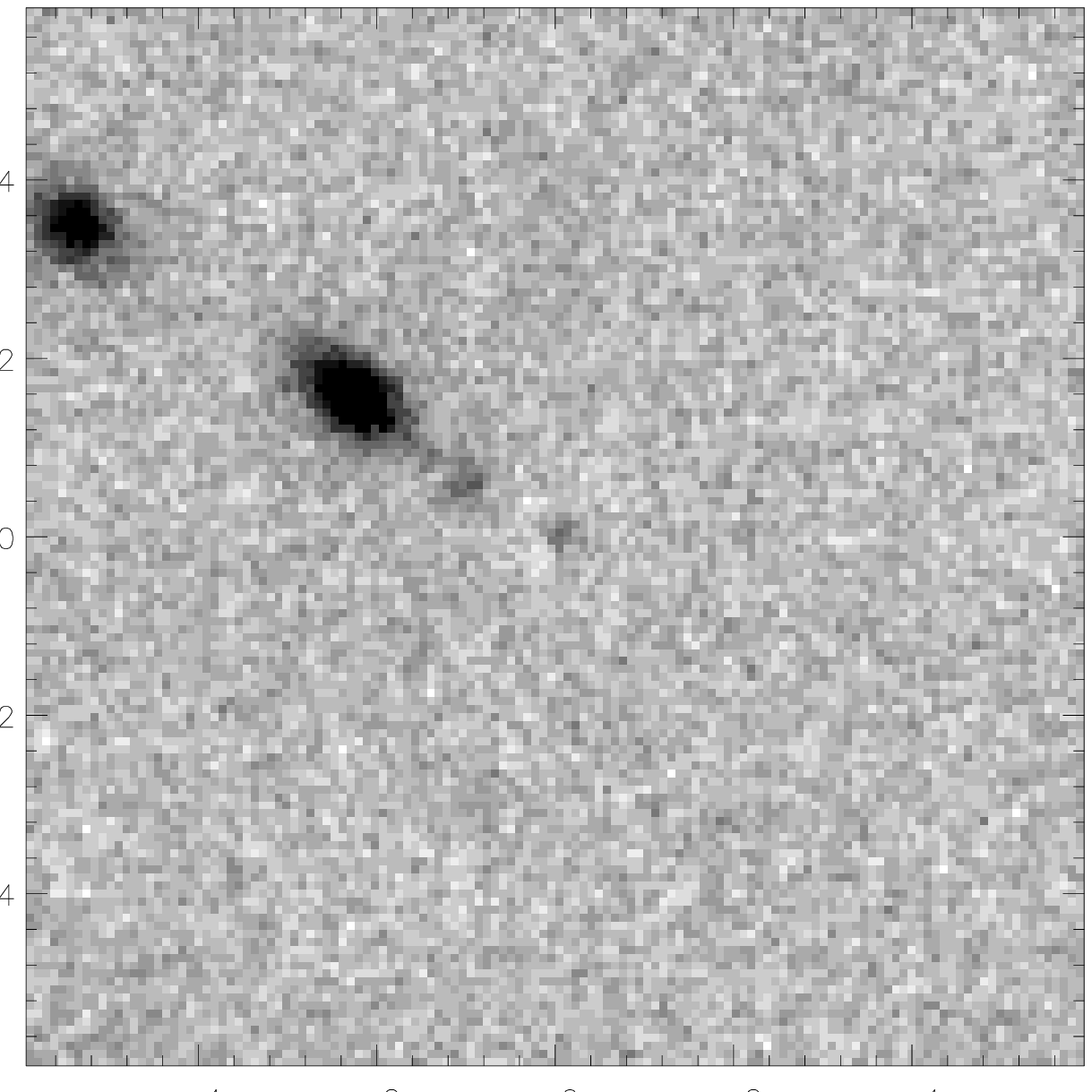,width=0.3\textwidth}\\
\\
\epsfig{file=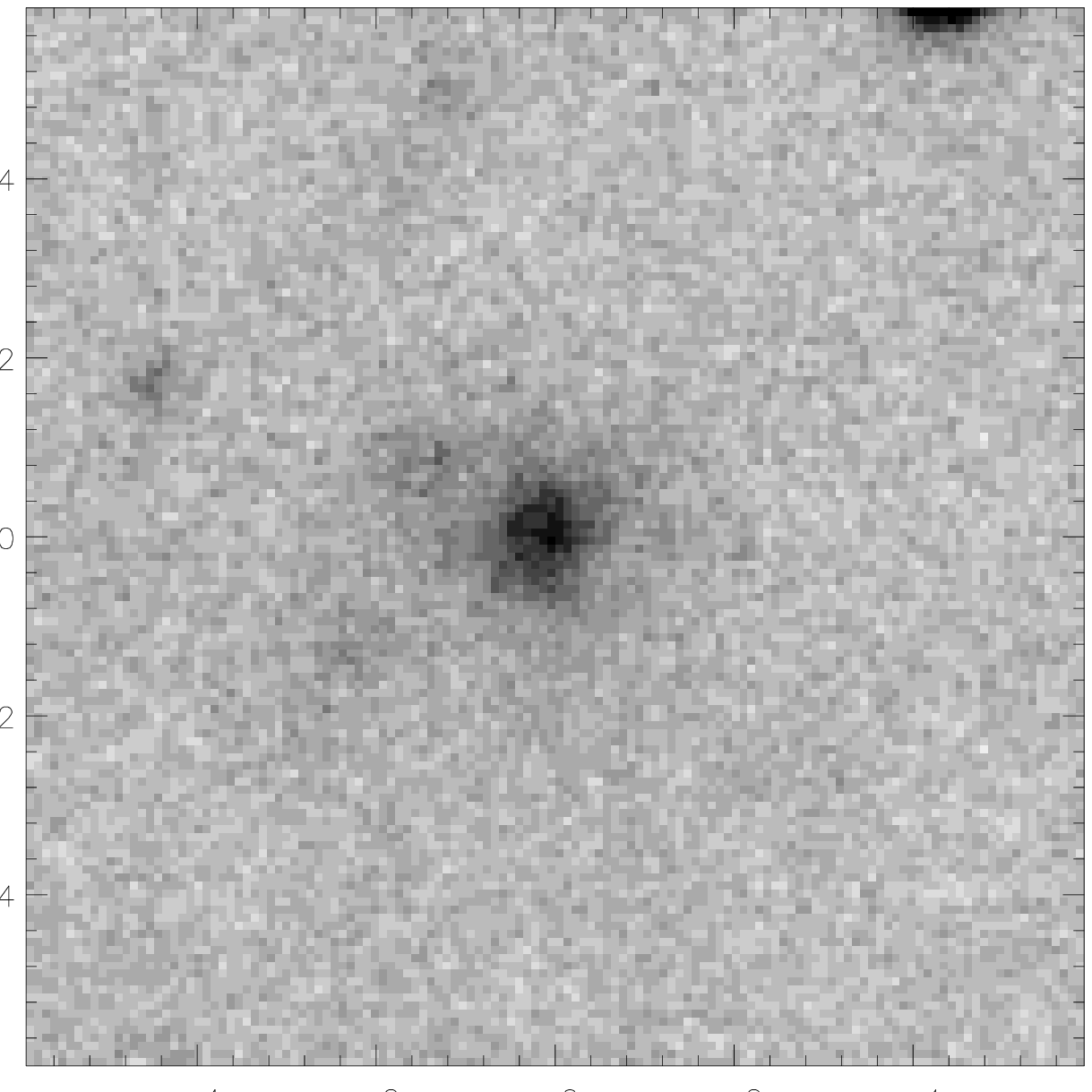,width=0.3\textwidth}&
\epsfig{file=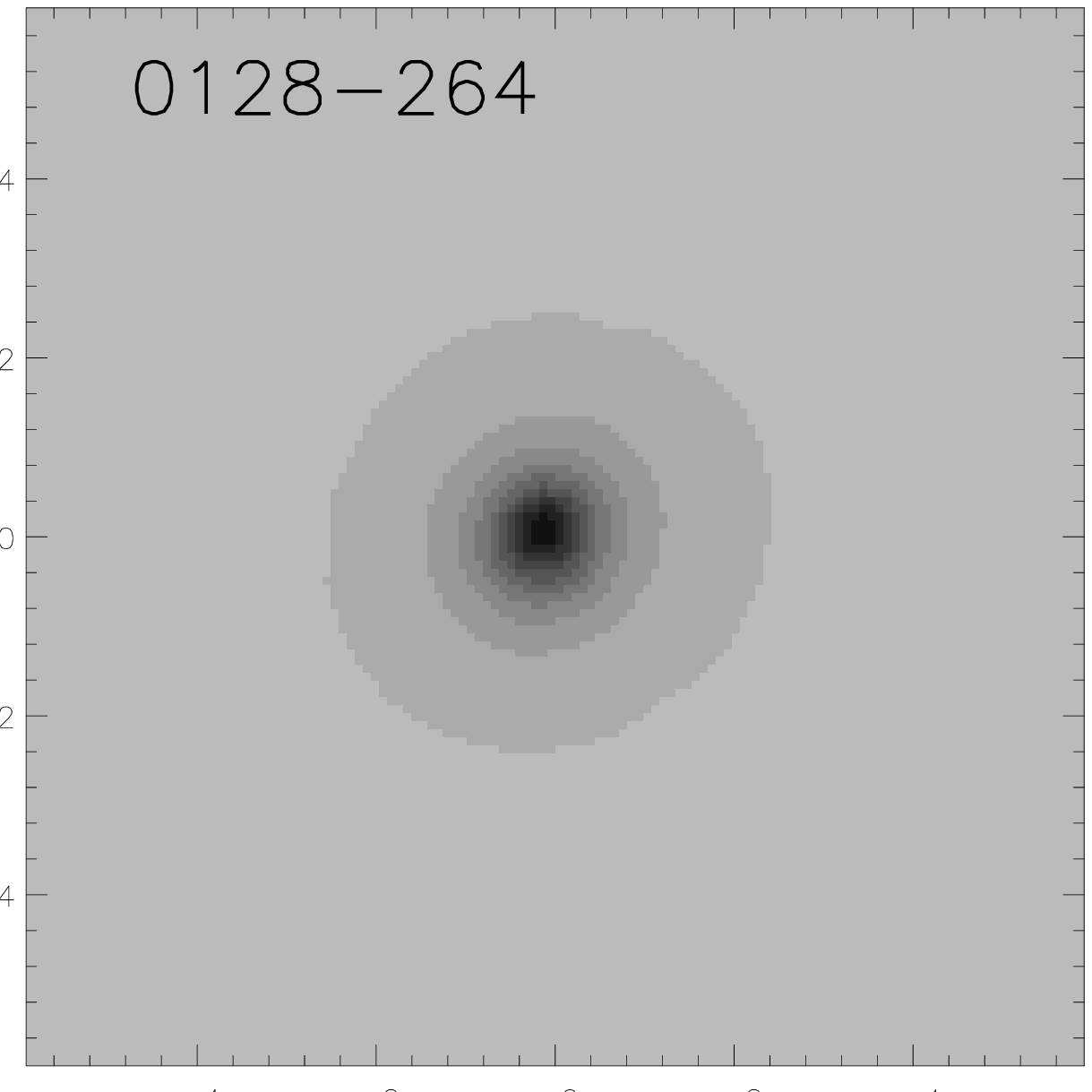,width=0.3\textwidth}&
\epsfig{file=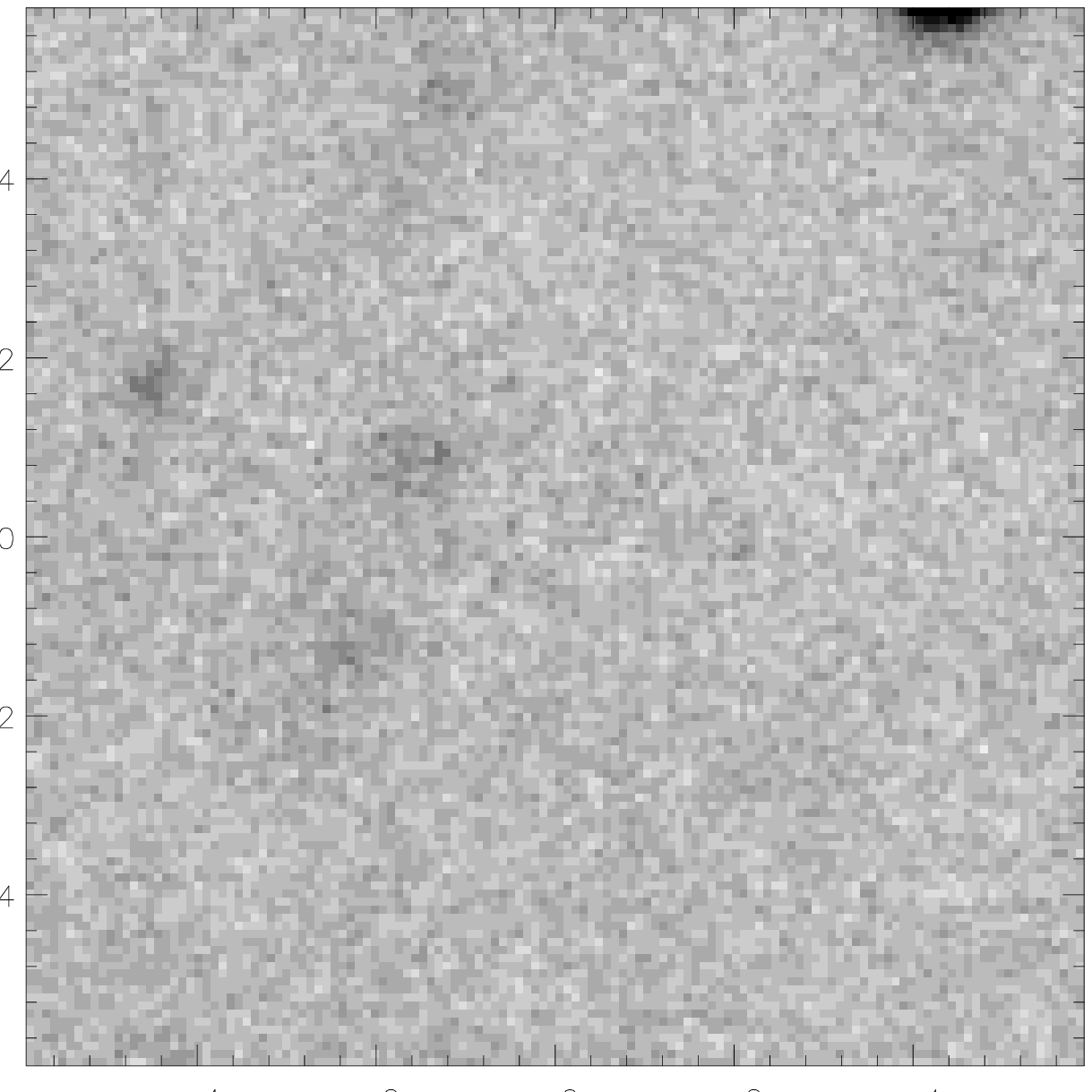,width=0.3\textwidth}\\
\\
\epsfig{file=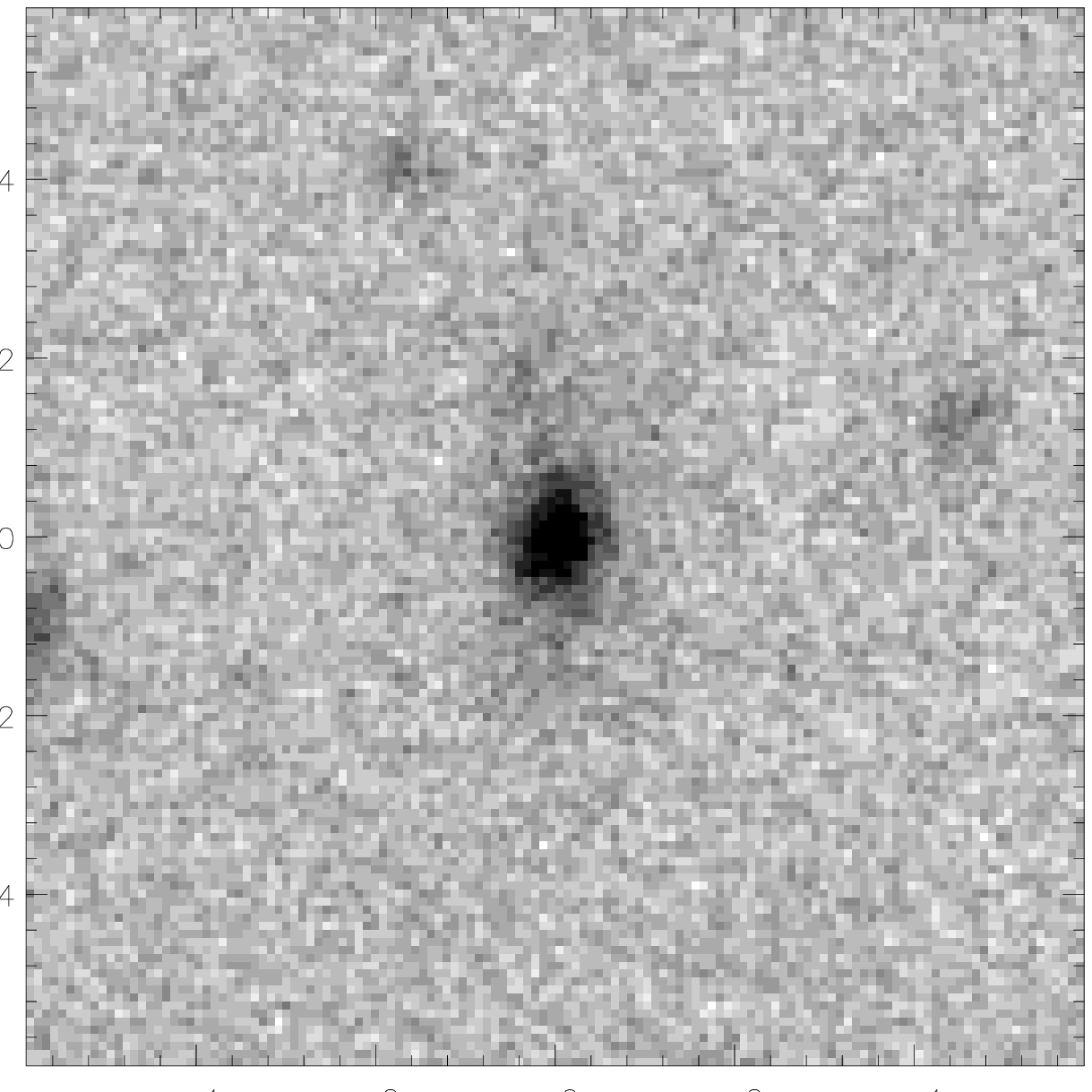,width=0.3\textwidth}&
\epsfig{file=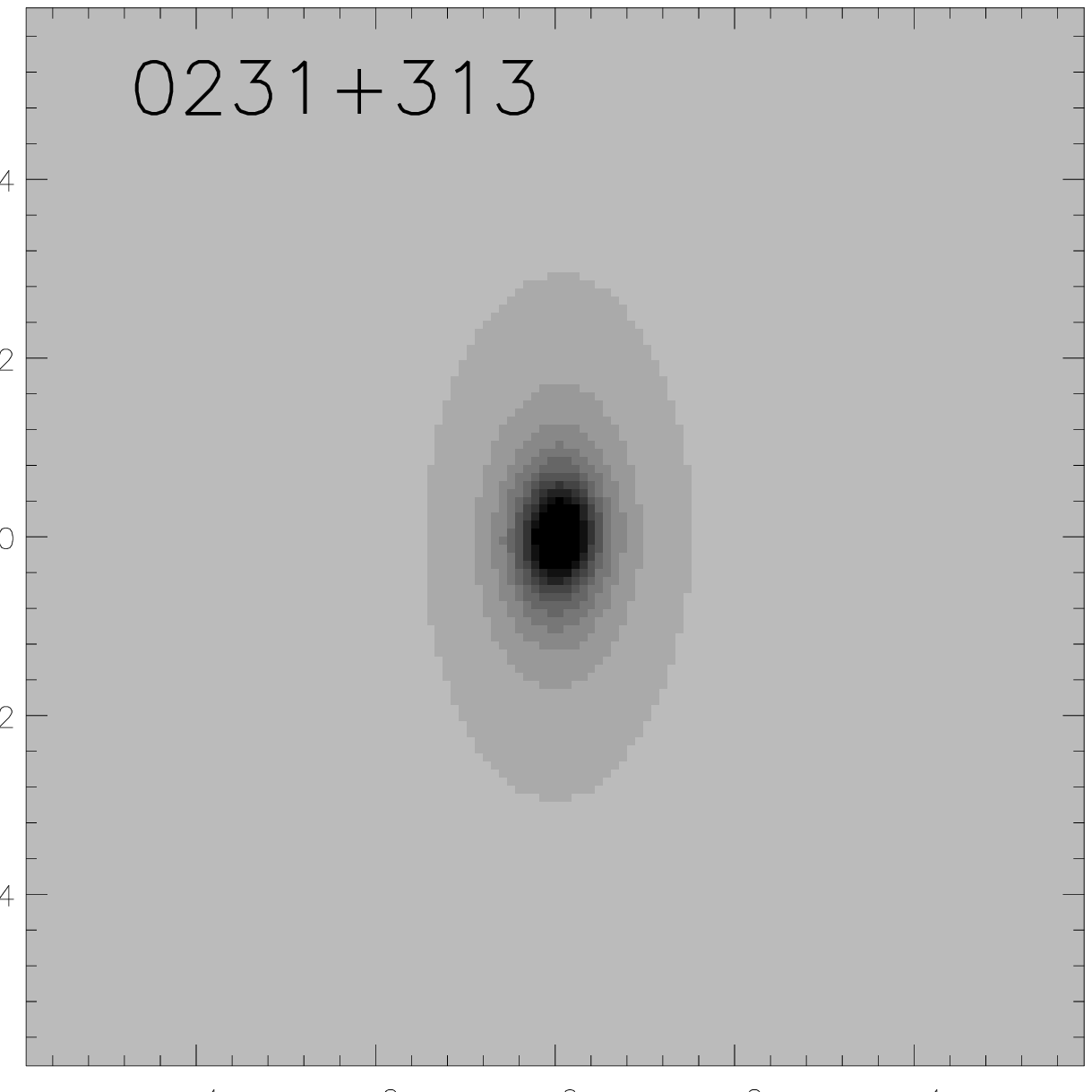,width=0.3\textwidth}&
\epsfig{file=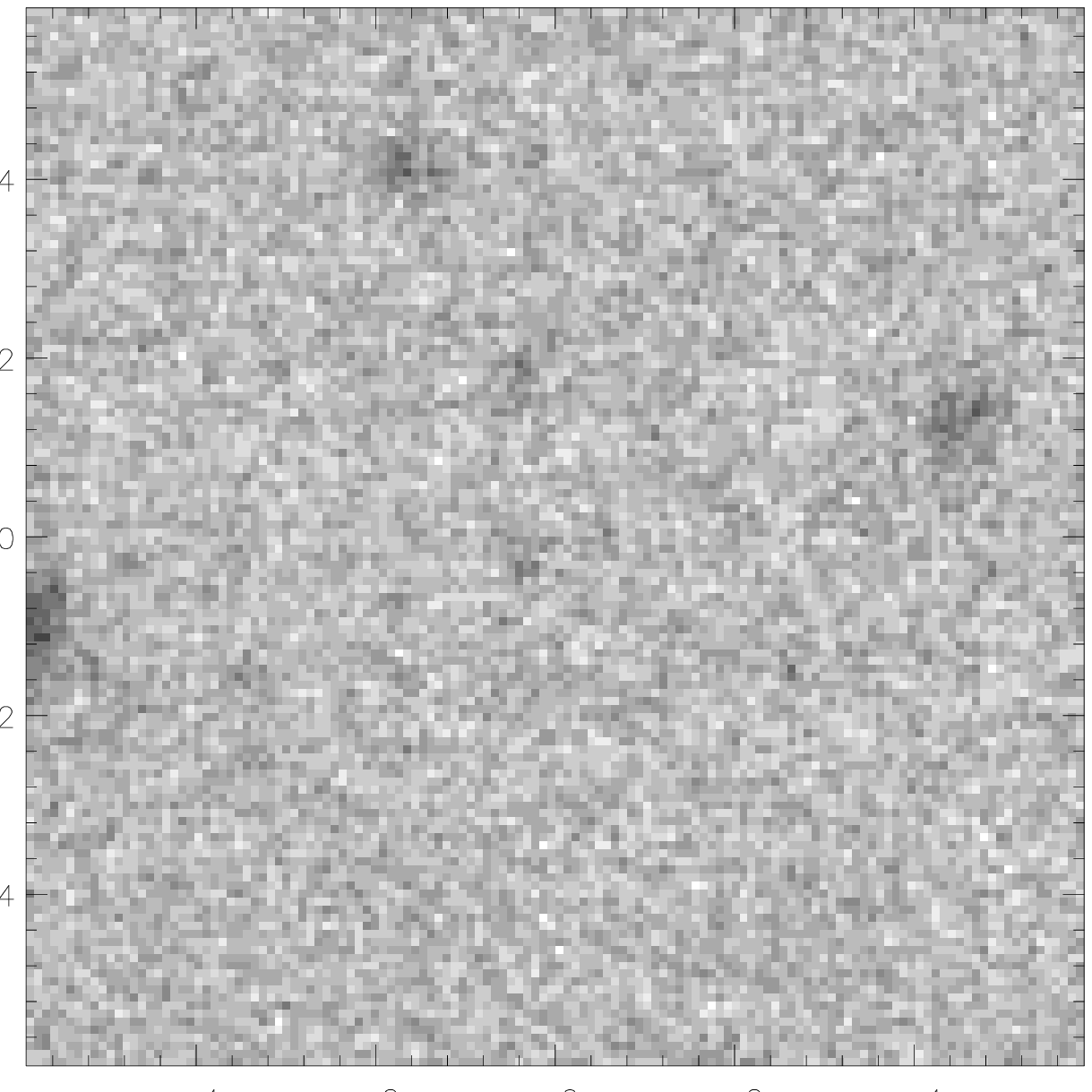,width=0.3\textwidth}\\
\\
\epsfig{file=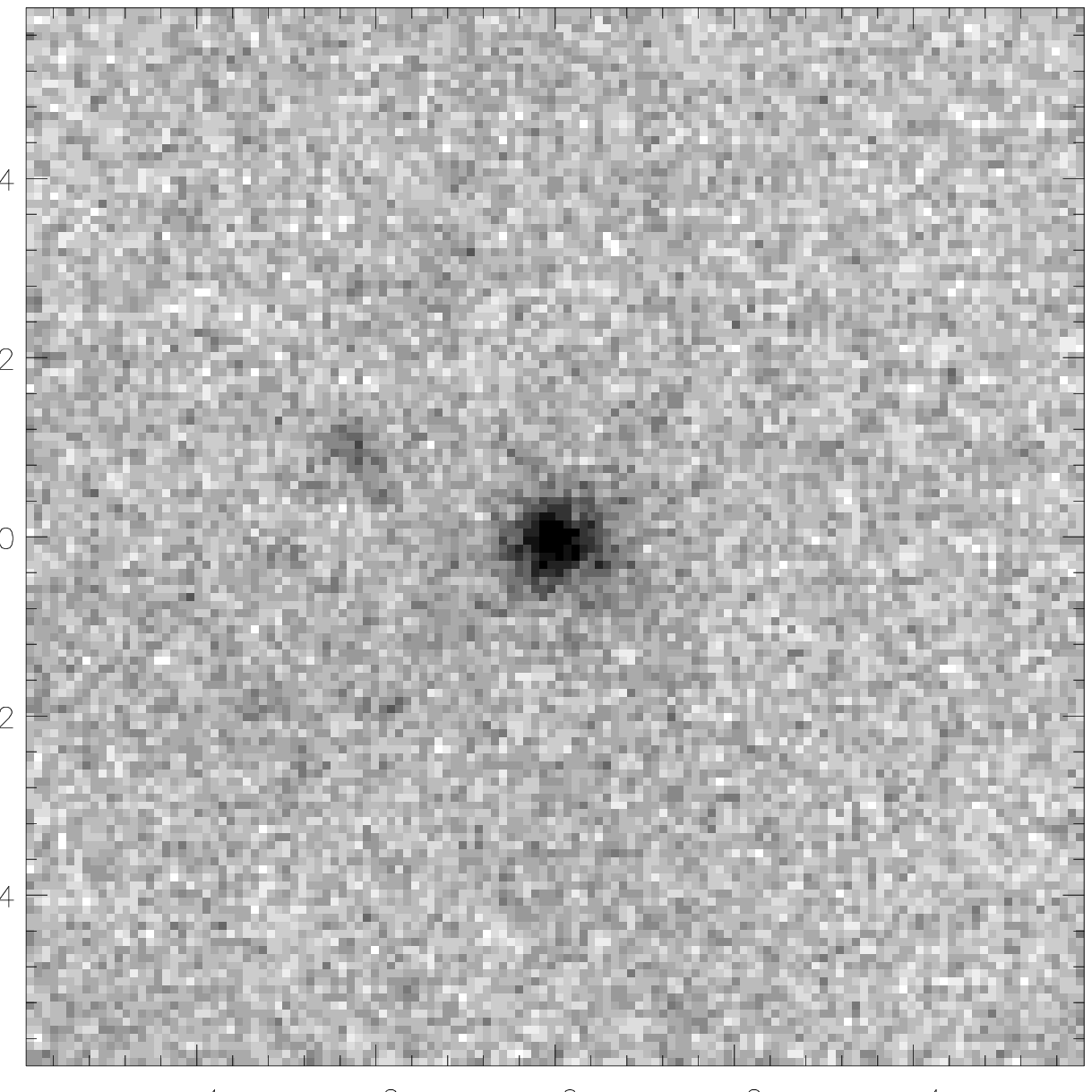,width=0.3\textwidth}&
\epsfig{file=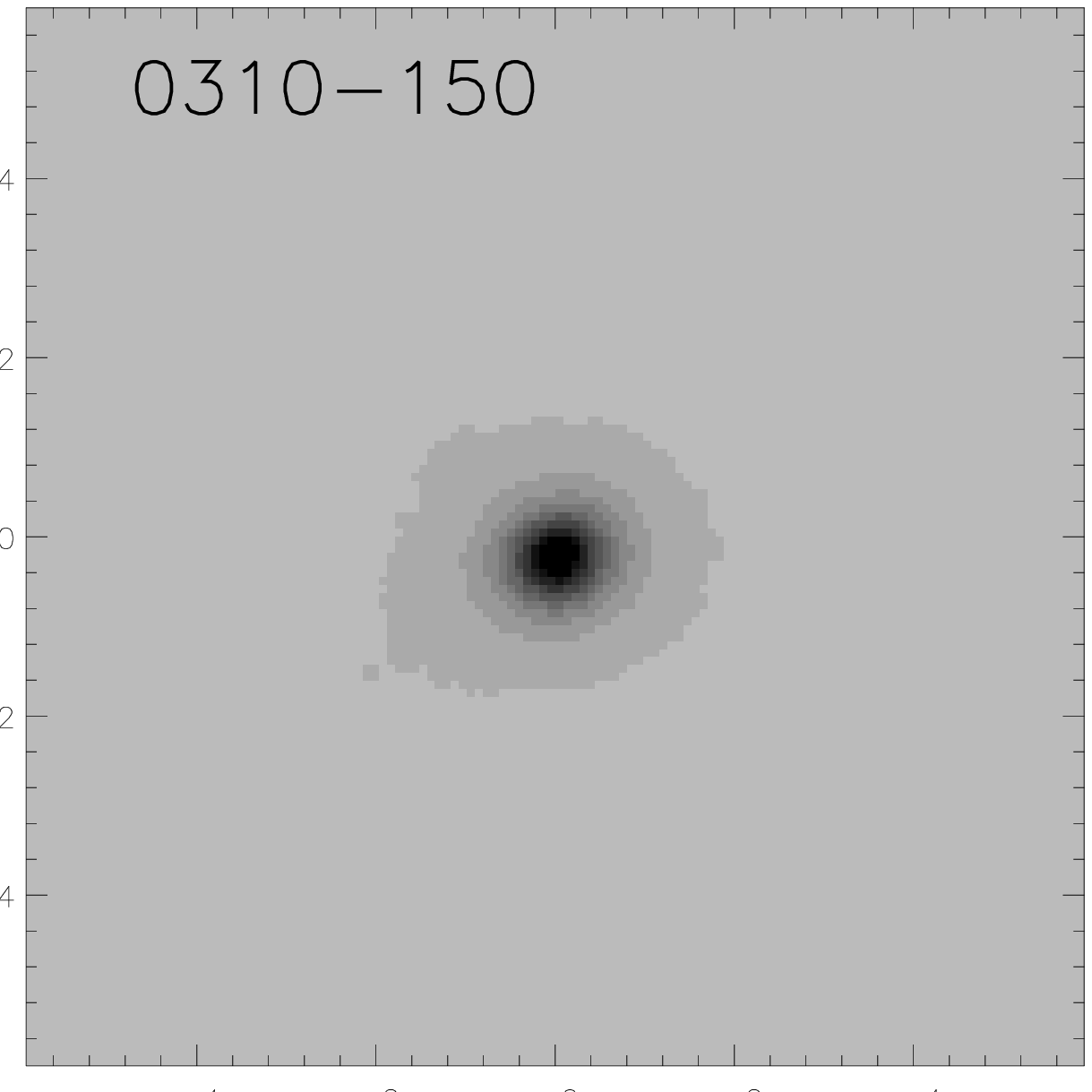,width=0.3\textwidth}&
\epsfig{file=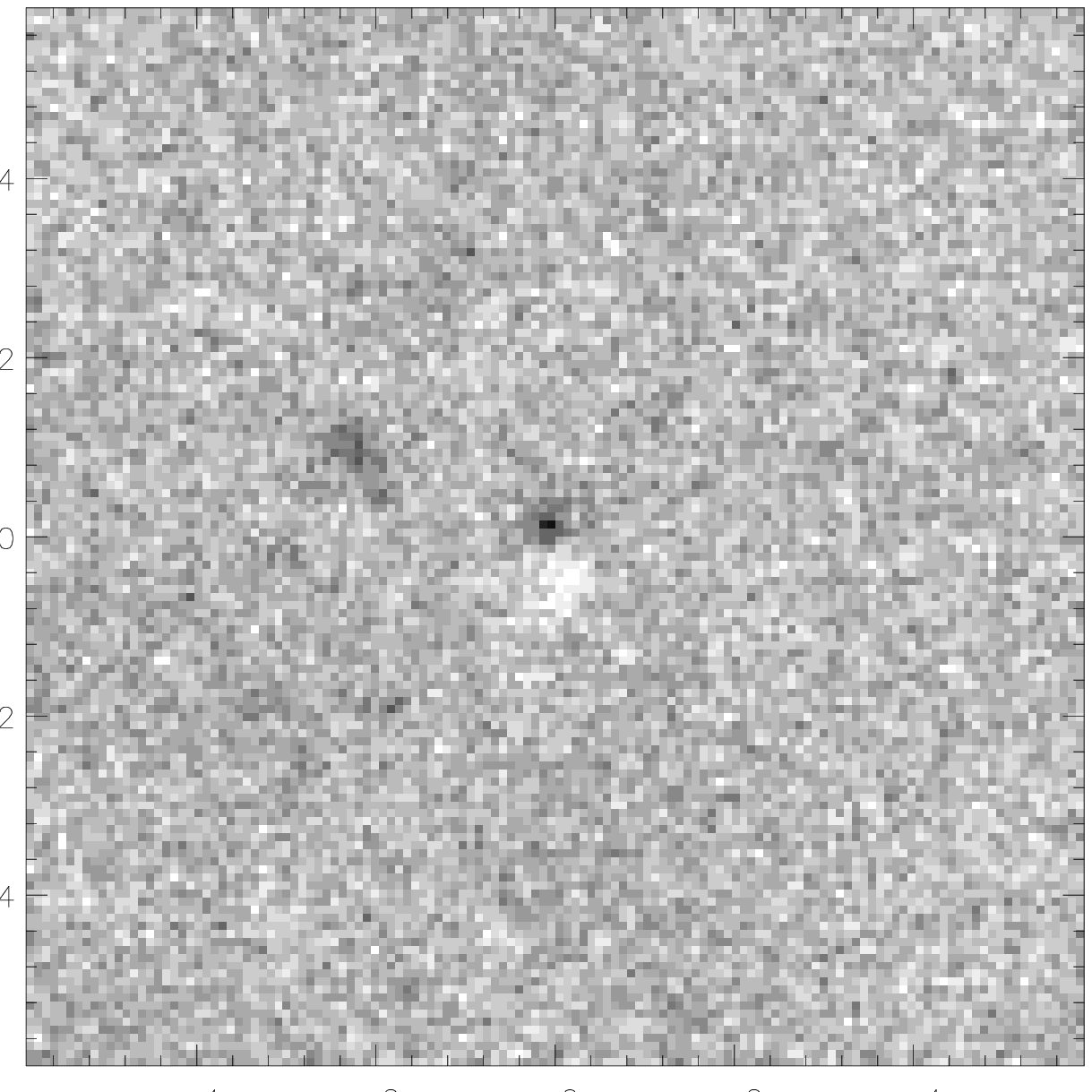,width=0.3\textwidth}\\
\end{tabular}
\caption[Two-dimensional modelling of the $z \simeq 2$ radio galaxies]{Two-dimensional modelling of the $z \simeq 2$ radio galaxies. The left-hand panel shows the $K$-band image centred on the radio galaxy. The middle panel shows the best-fitting two-dimensional model. The right-hand panel shows the residual image after subtraction of the model from the data. All panels are 12.0$^{\prime \prime}$ $\times$12.0$^{\prime \prime}$, and the images are shown with a linear greyscale in which black corresponds to $2.5\sigma$ above, and white to $1\sigma$ below the median sky value.}
\label{ukirtmodel1}
\end{figure*}
\end{center}

\begin{center}
\begin{figure*}
\begin{tabular}{ccc}
\epsfig{file=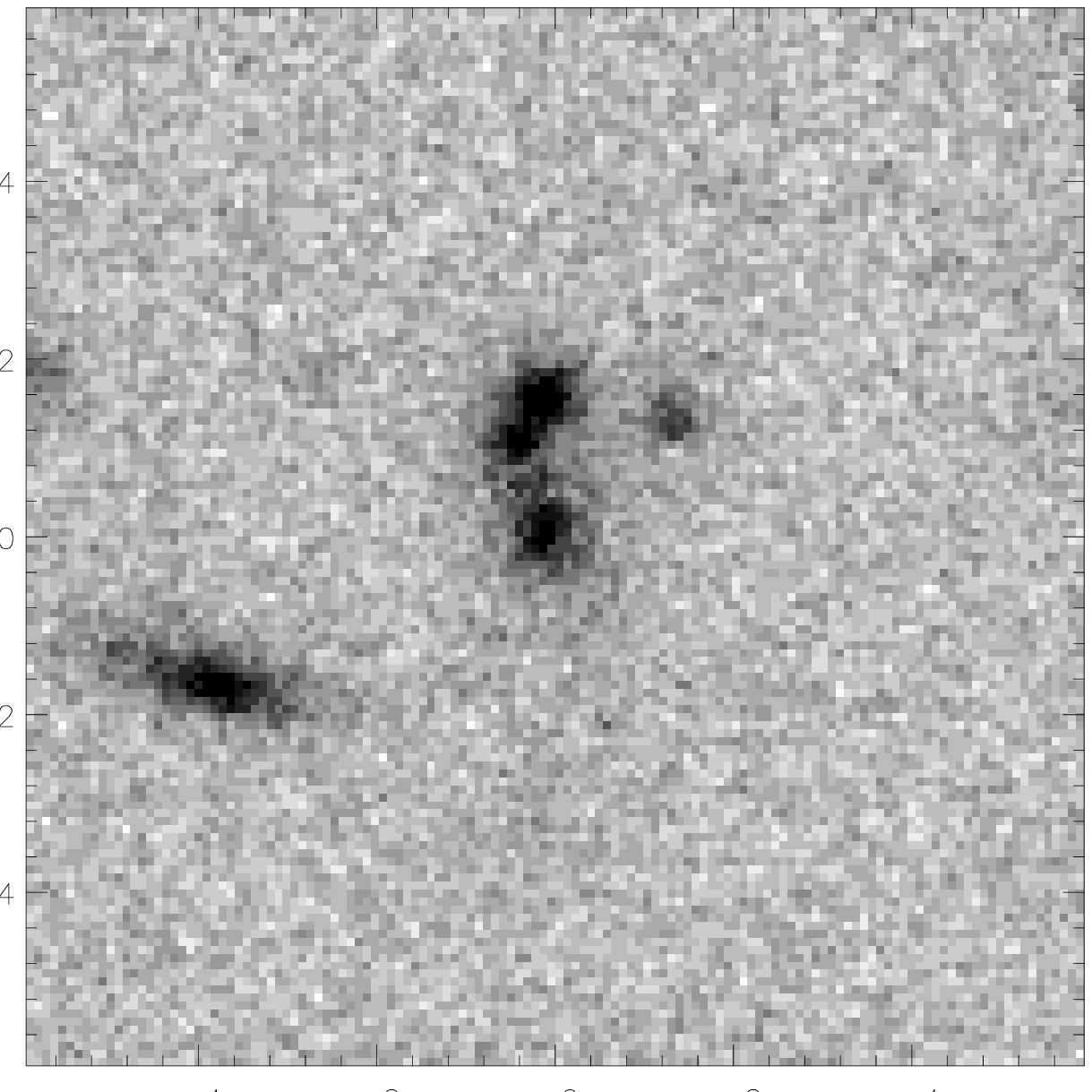,width=0.3\textwidth}&
\epsfig{file=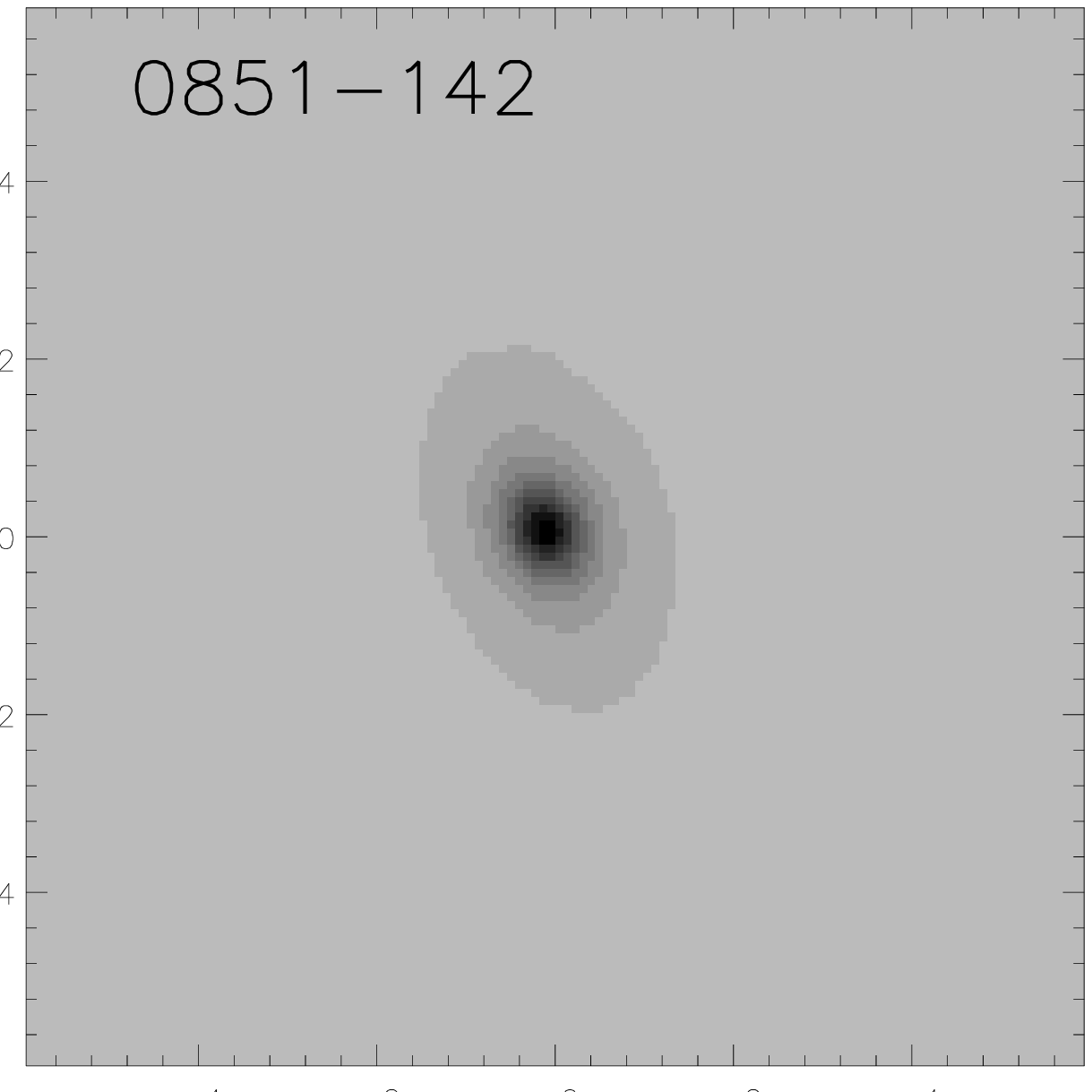,width=0.3\textwidth}&
\epsfig{file=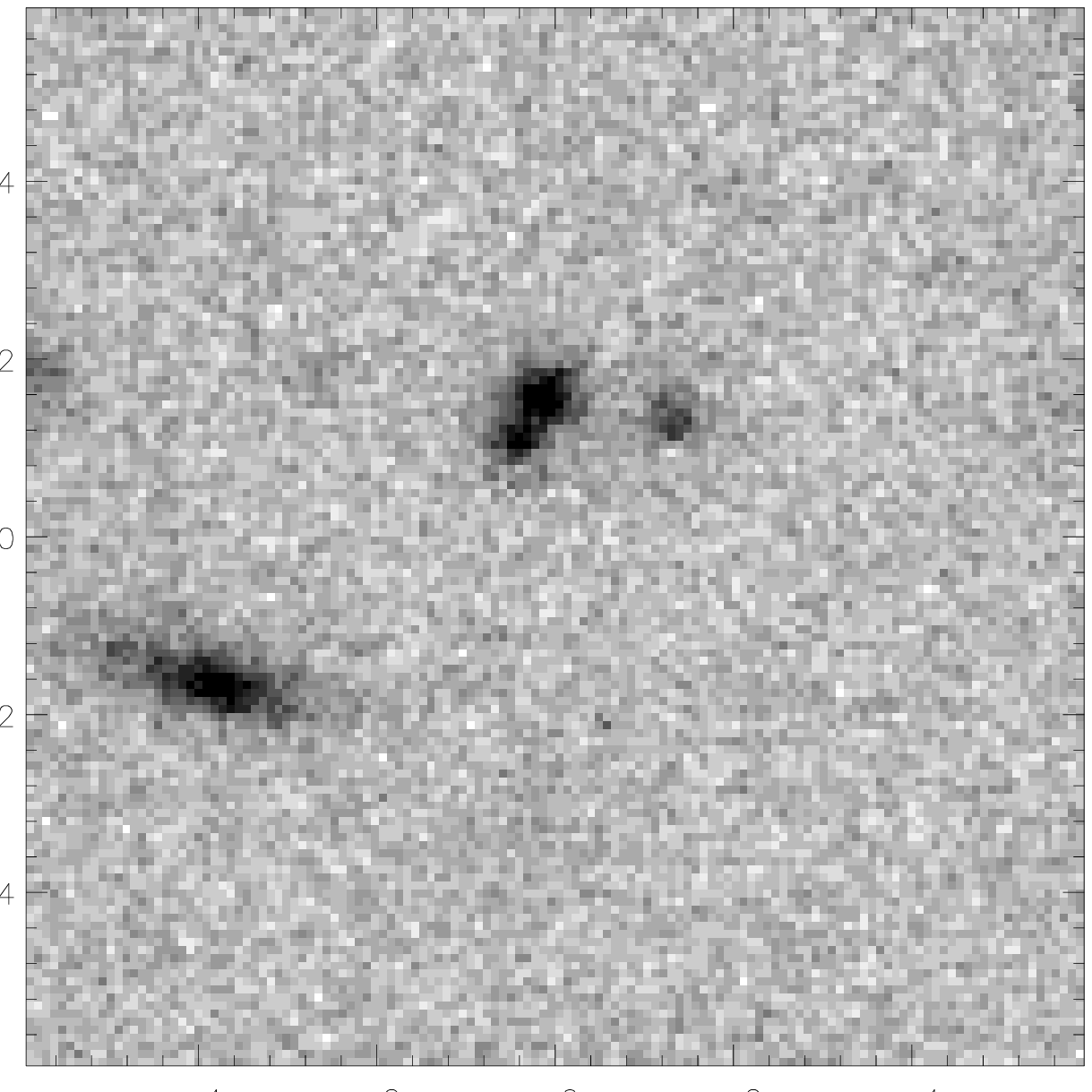,width=0.3\textwidth}\\
\\
\epsfig{file=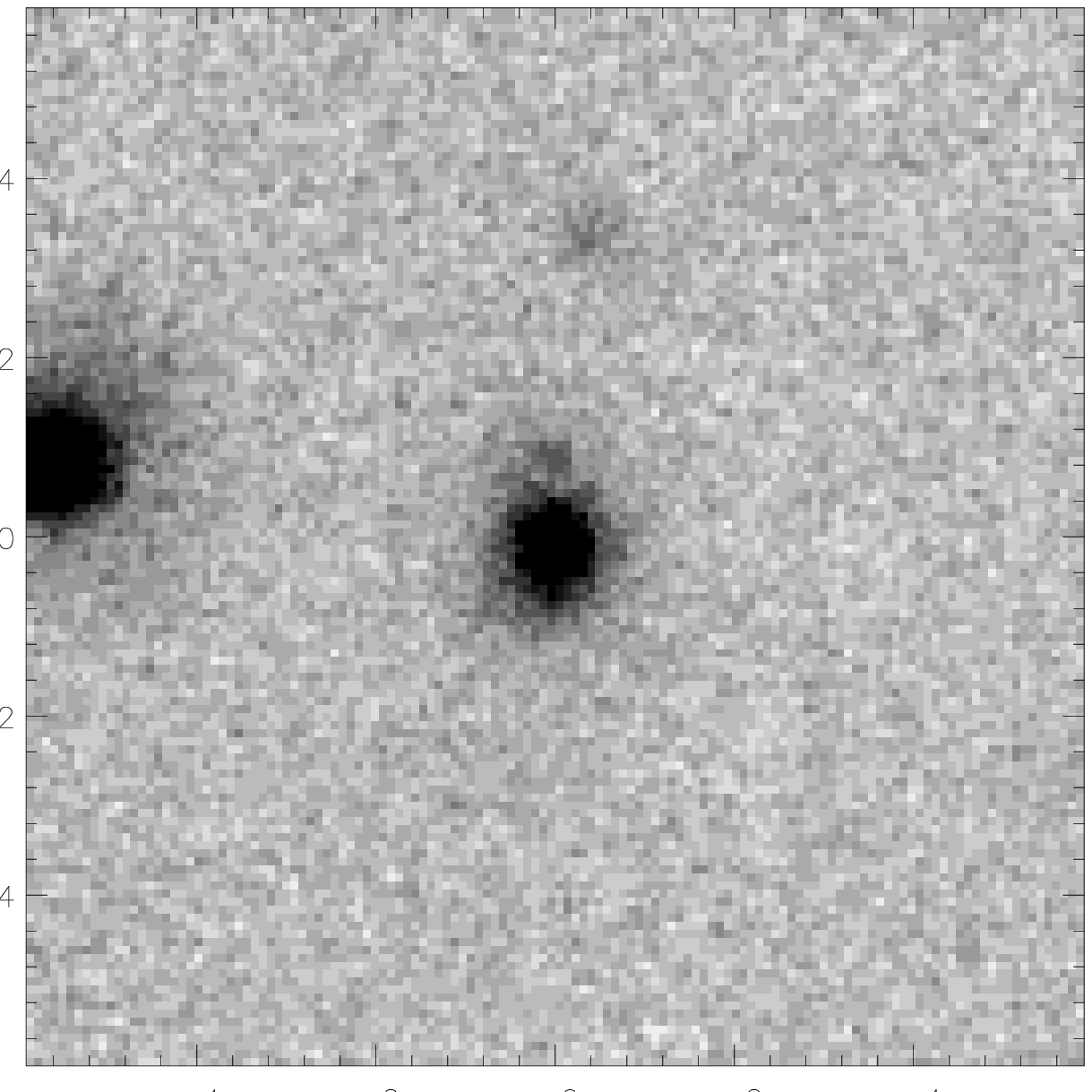,width=0.3\textwidth}&
\epsfig{file=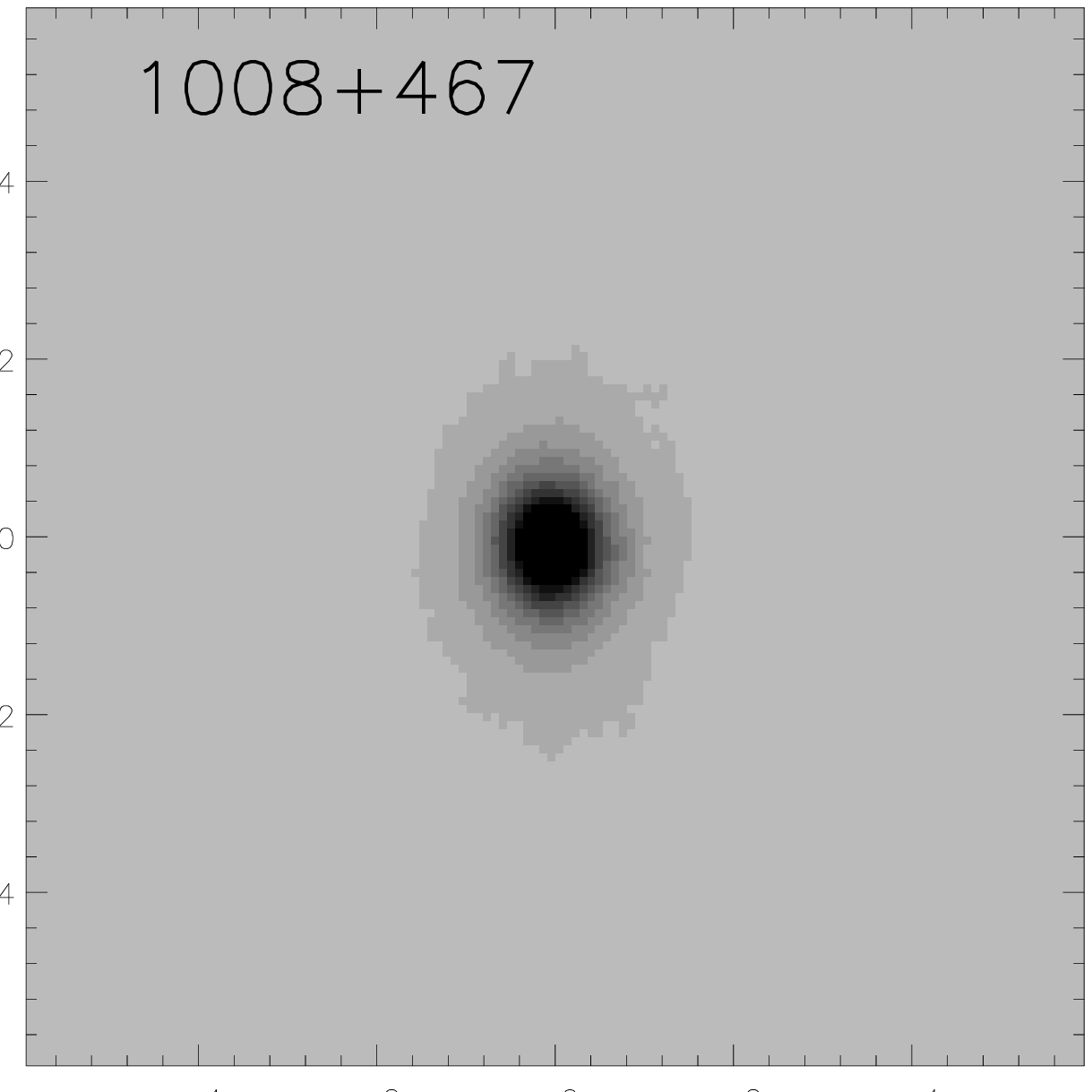,width=0.3\textwidth}&
\epsfig{file=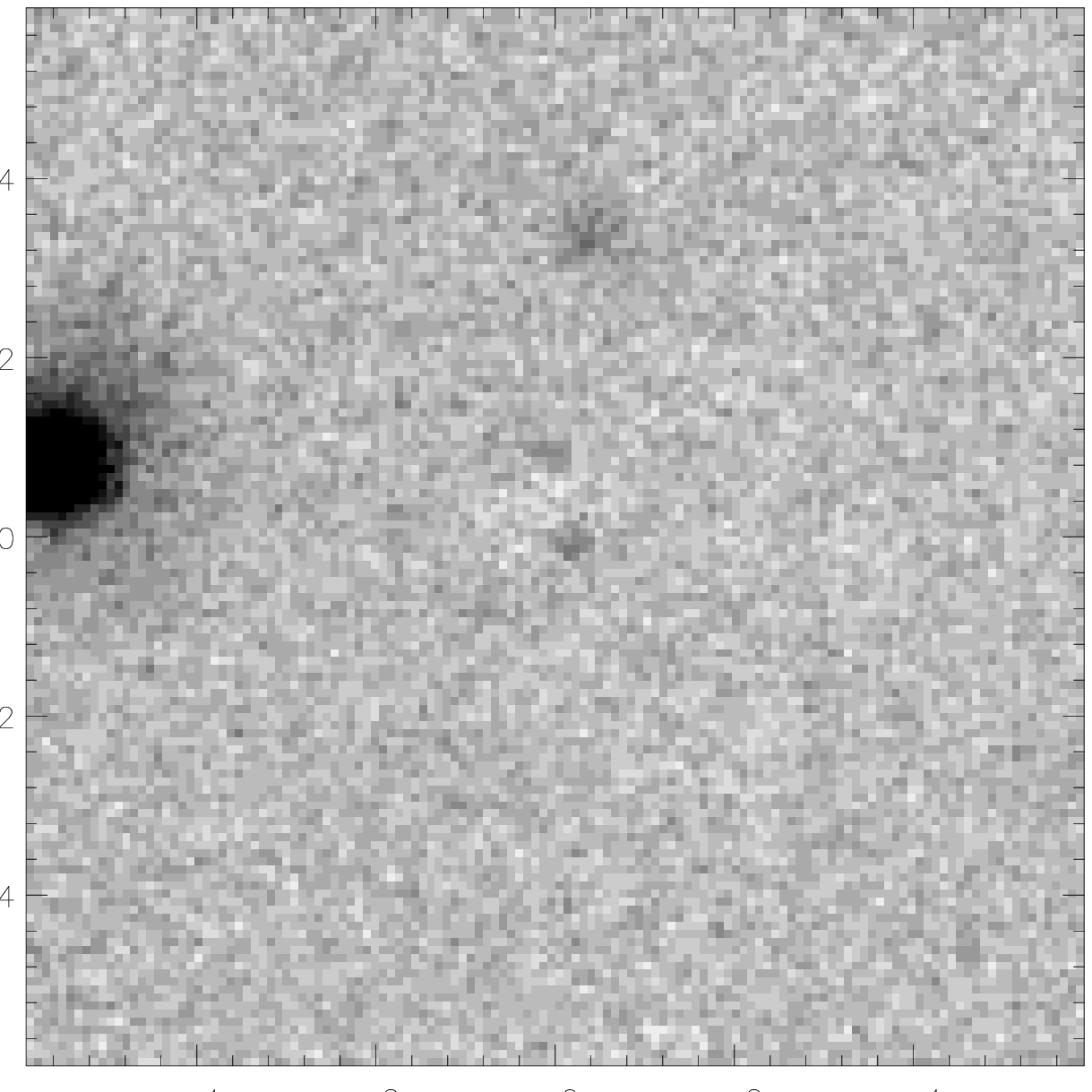,width=0.3\textwidth}\\
\\
\epsfig{file=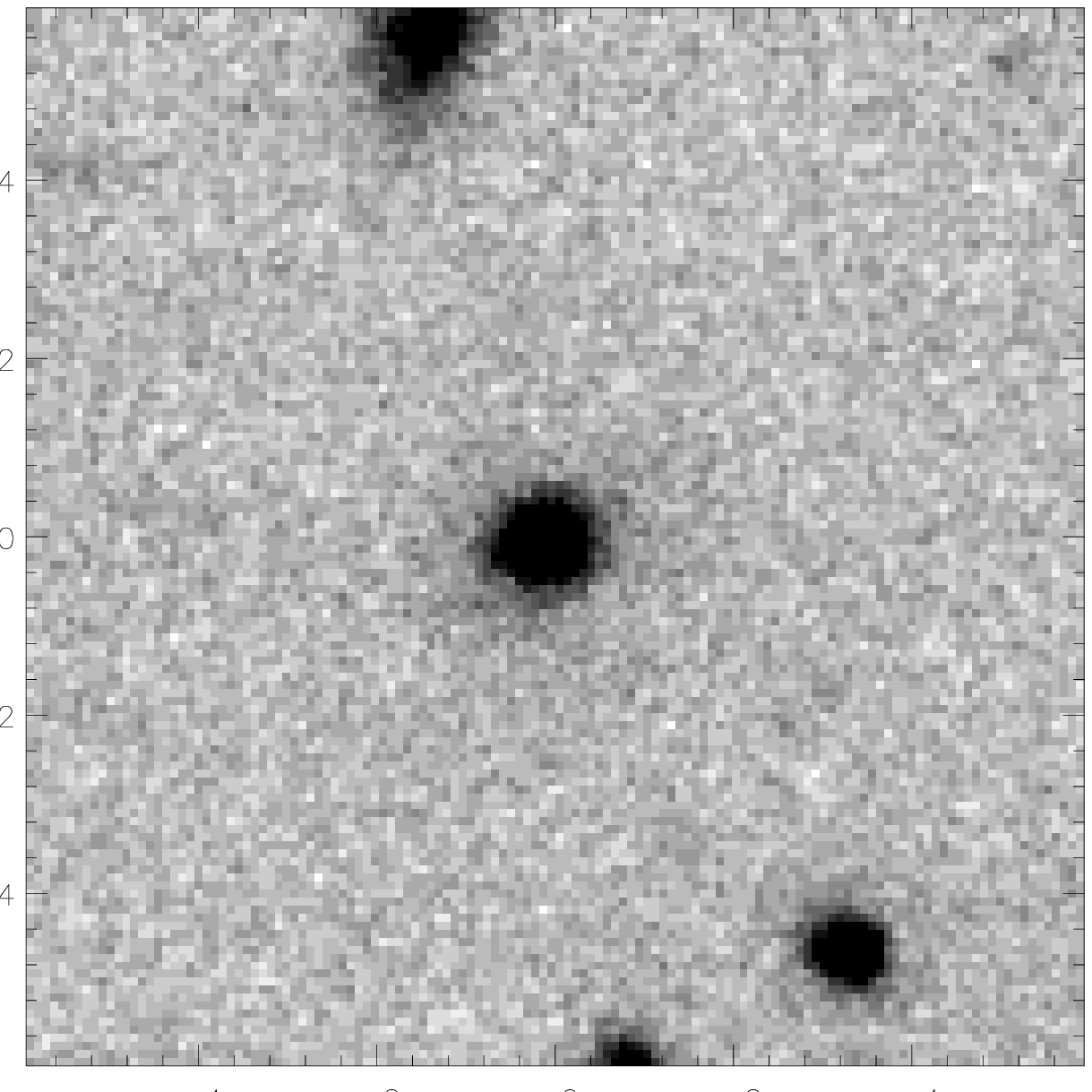,width=0.3\textwidth}&
\epsfig{file=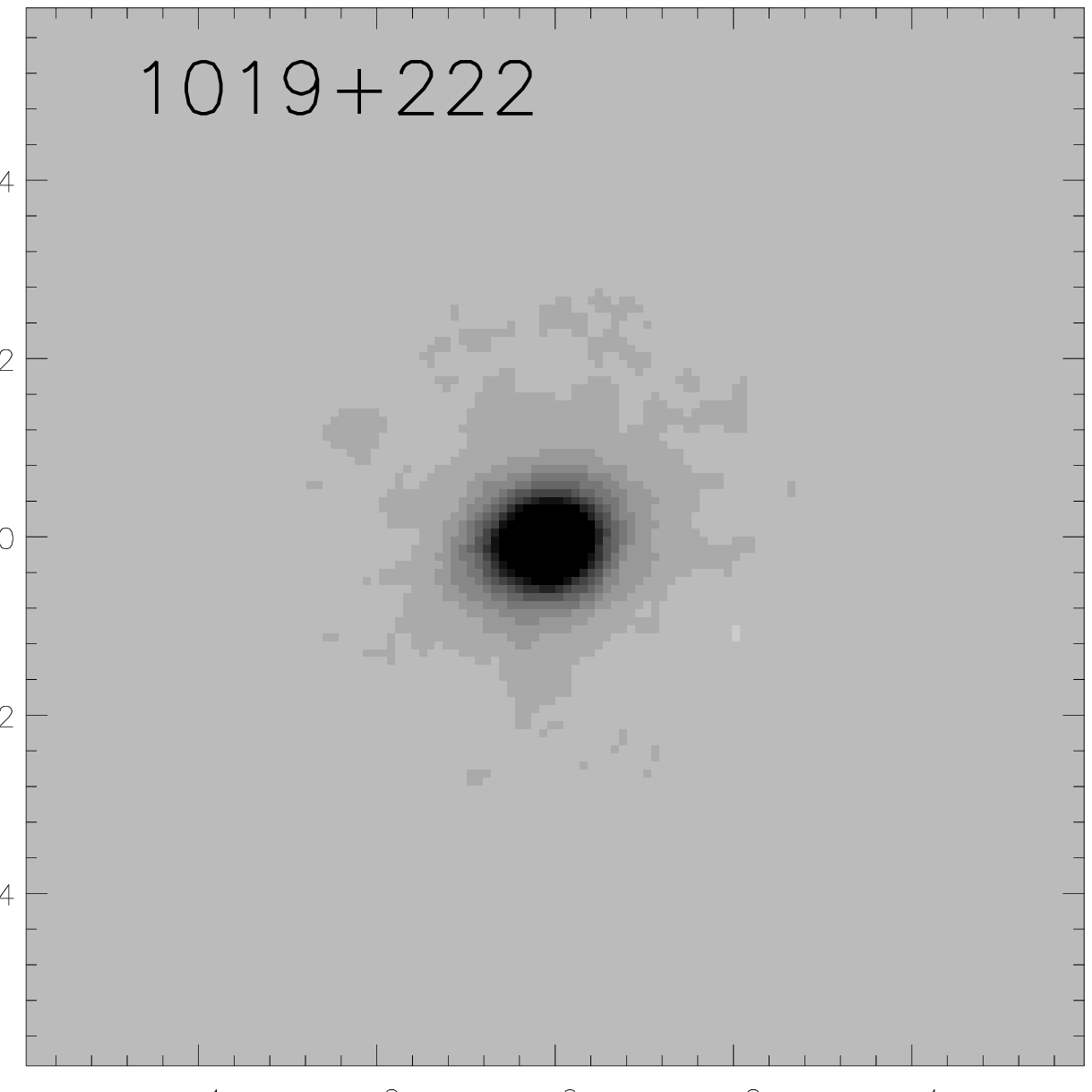,width=0.3\textwidth}&
\epsfig{file=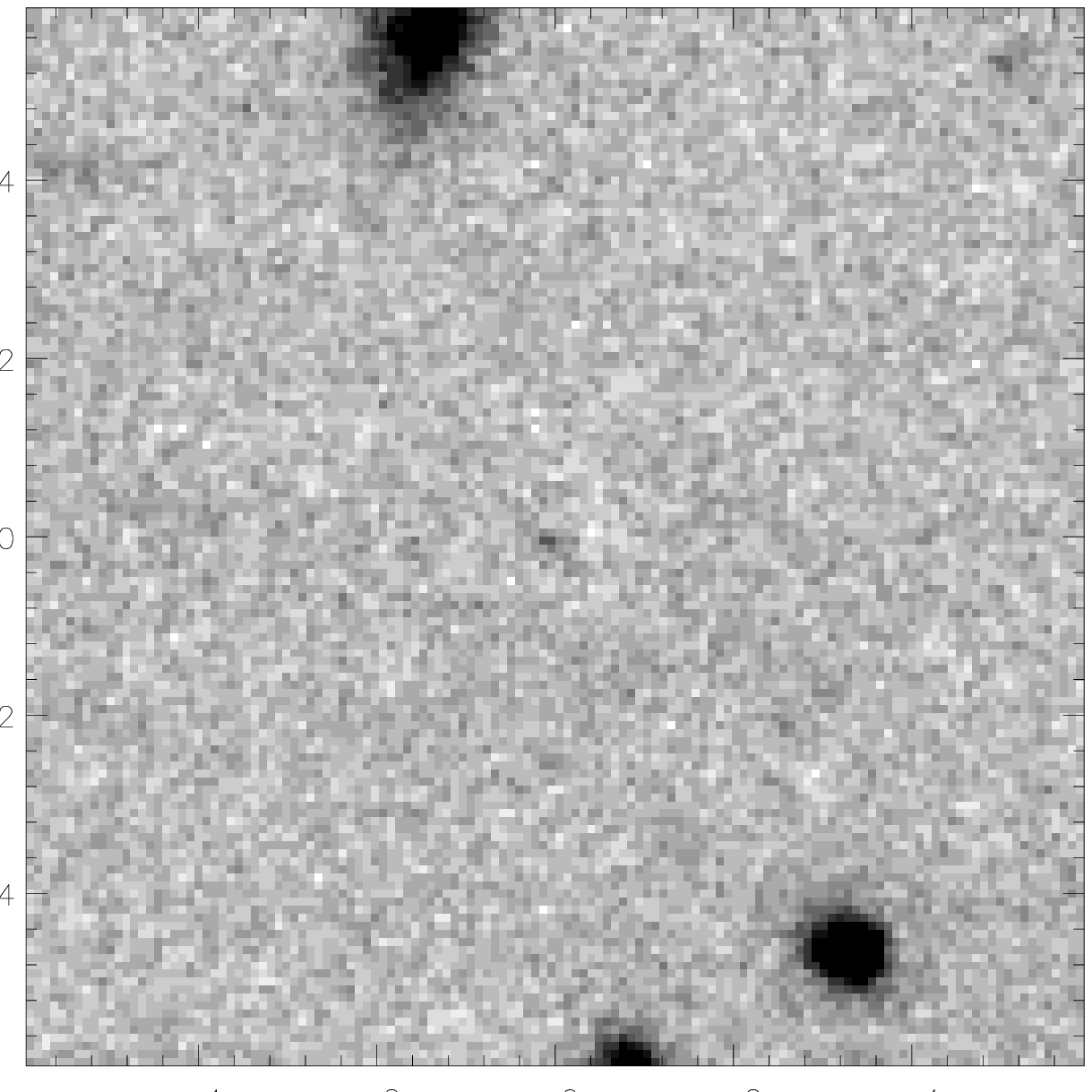,width=0.3\textwidth}\\
\\
\epsfig{file=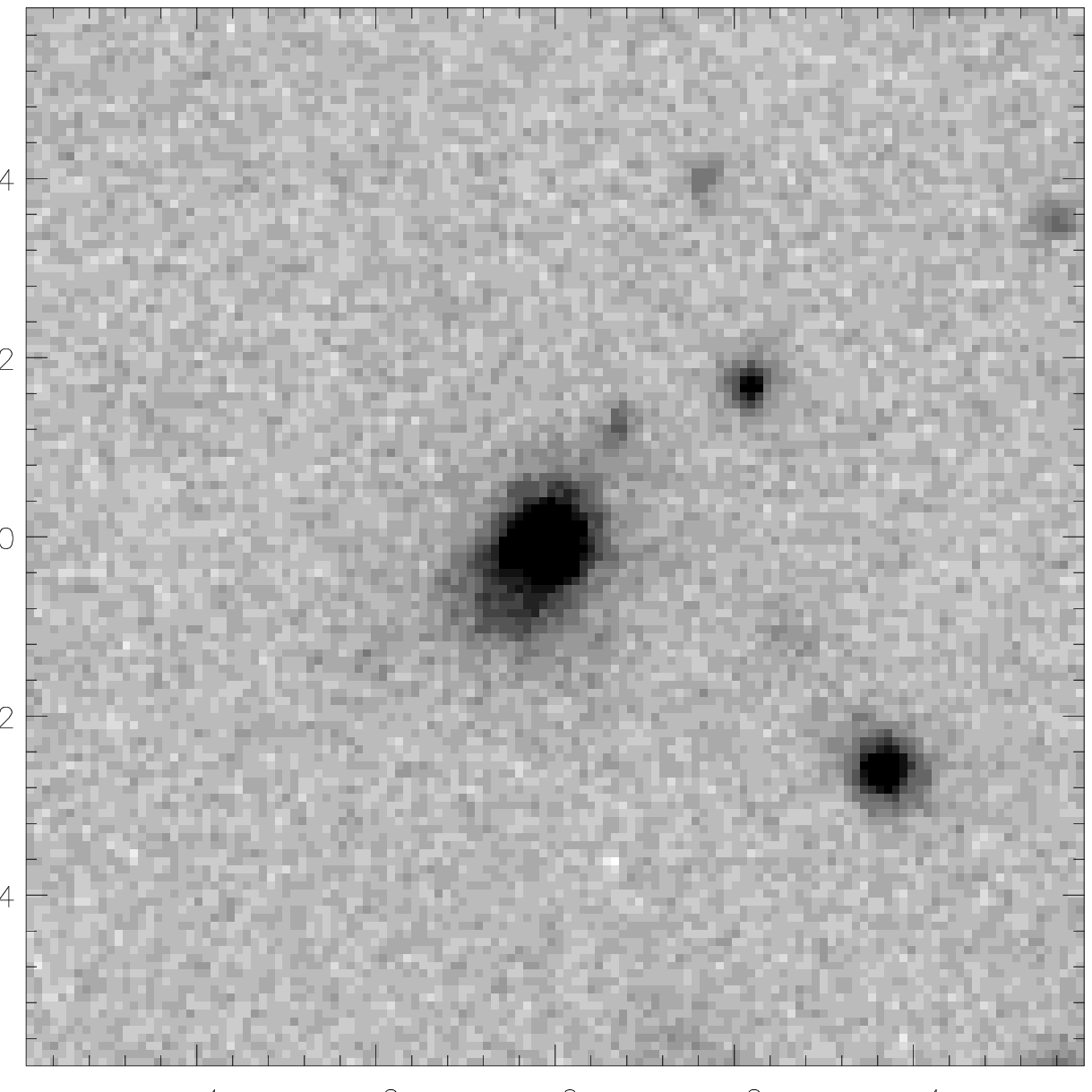,width=0.3\textwidth}&
\epsfig{file=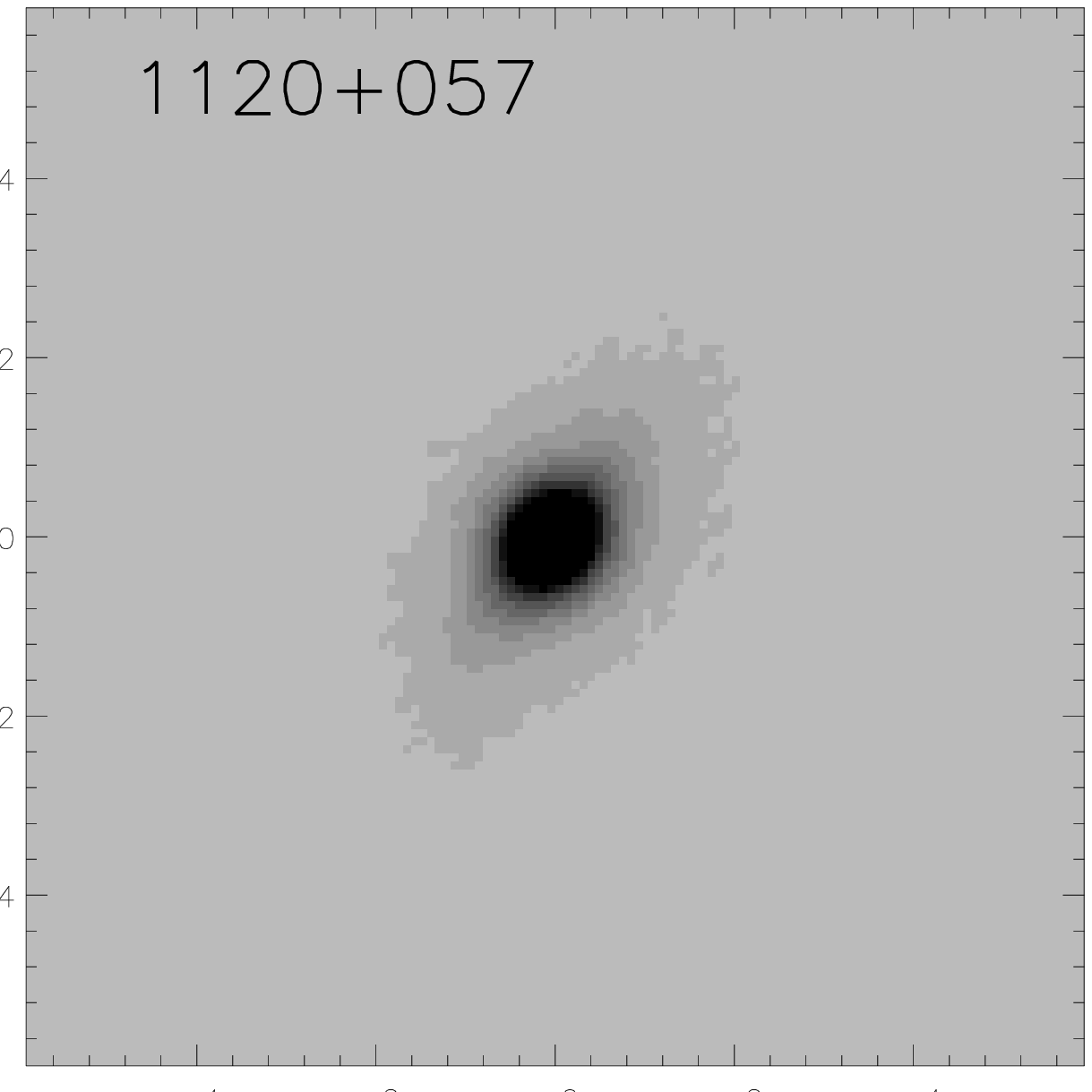,width=0.3\textwidth}&
\epsfig{file=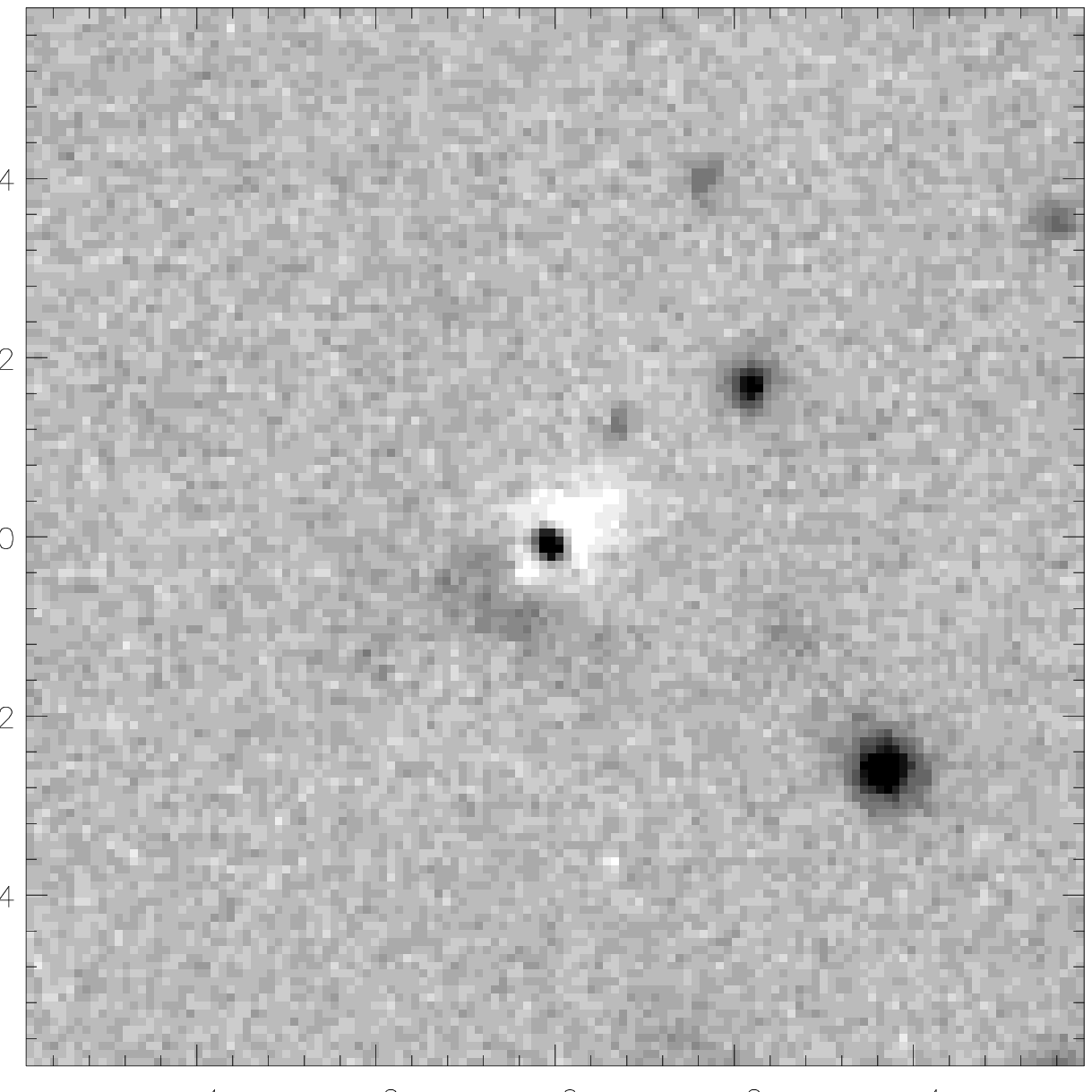,width=0.3\textwidth}\\
\end{tabular}
\addtocounter{figure}{-1}
\caption{- continued}
\label{ukirtmodel1}
\end{figure*}
\end{center}

\begin{center}
\begin{figure*}
\begin{tabular}{ccc}
\epsfig{file=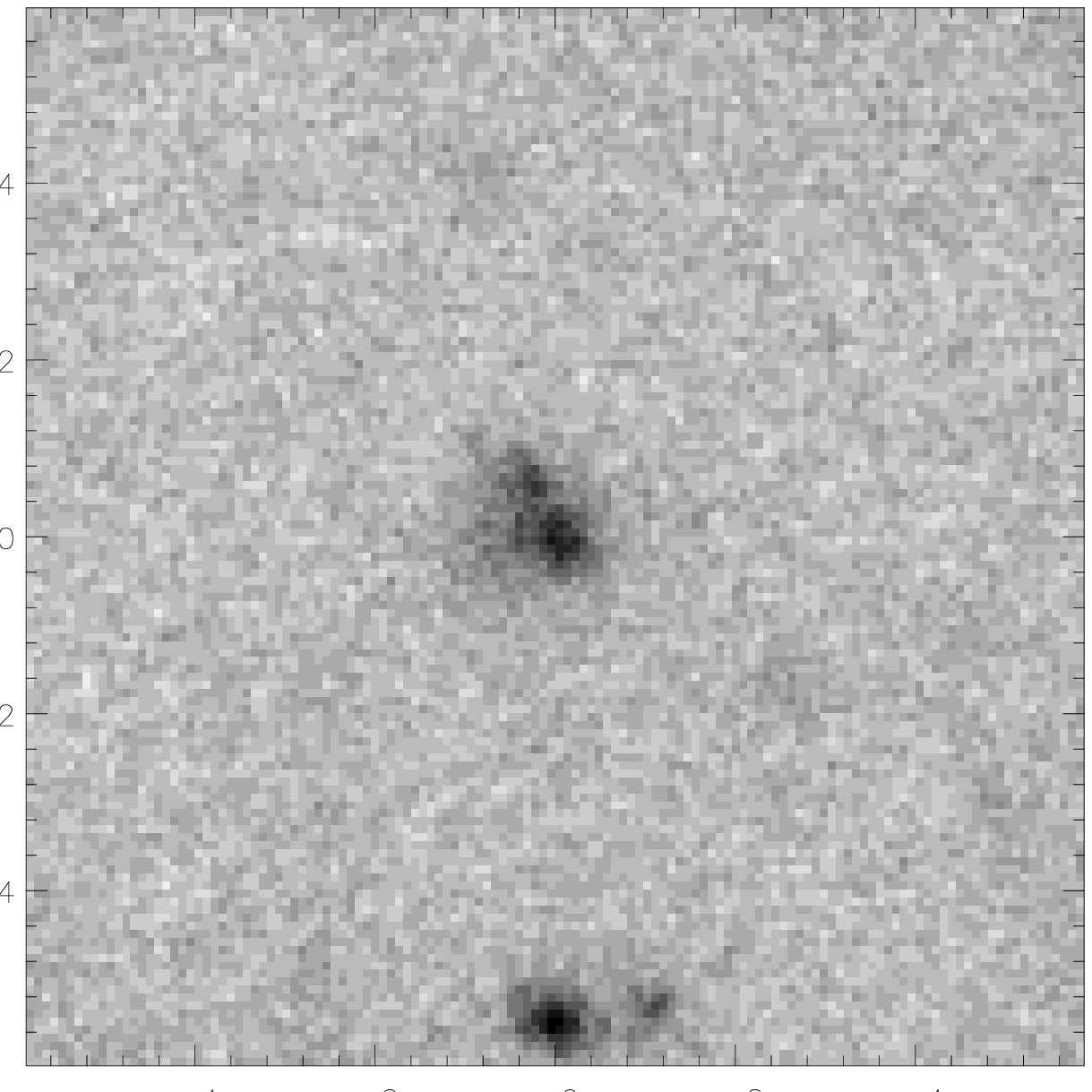,width=0.3\textwidth}&
\epsfig{file=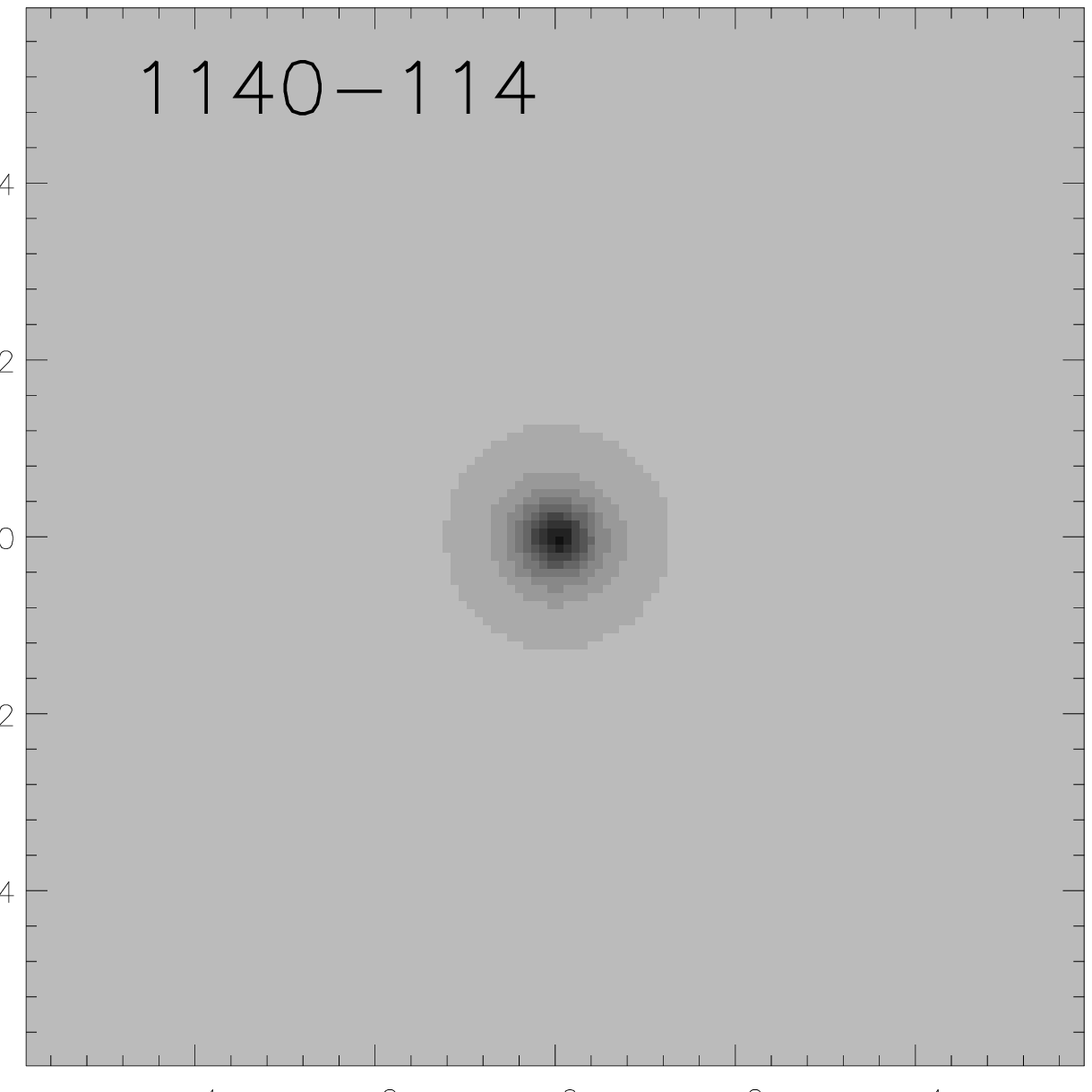,width=0.3\textwidth}&
\epsfig{file=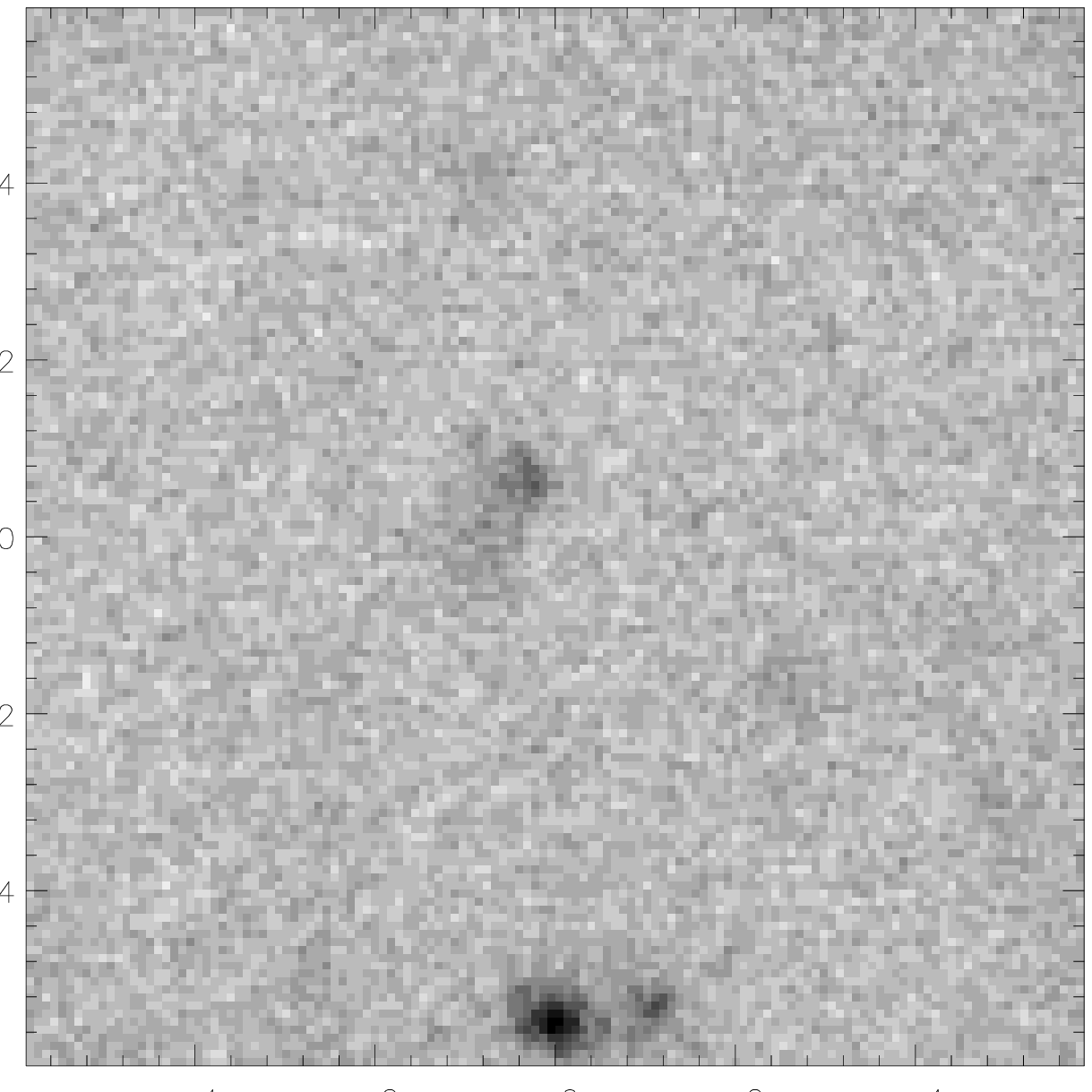,width=0.3\textwidth}\\
\\
\epsfig{file=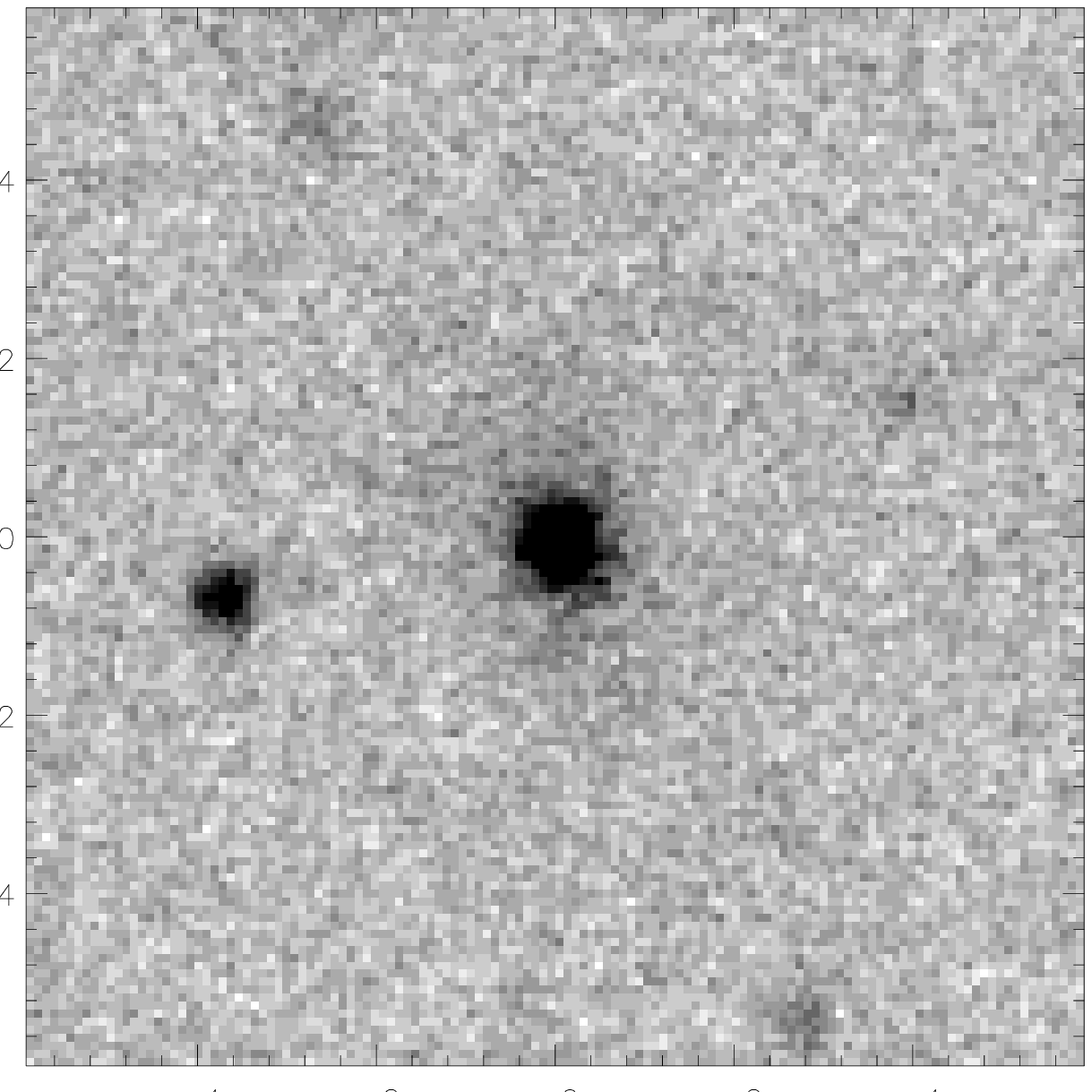,width=0.3\textwidth}&
\epsfig{file=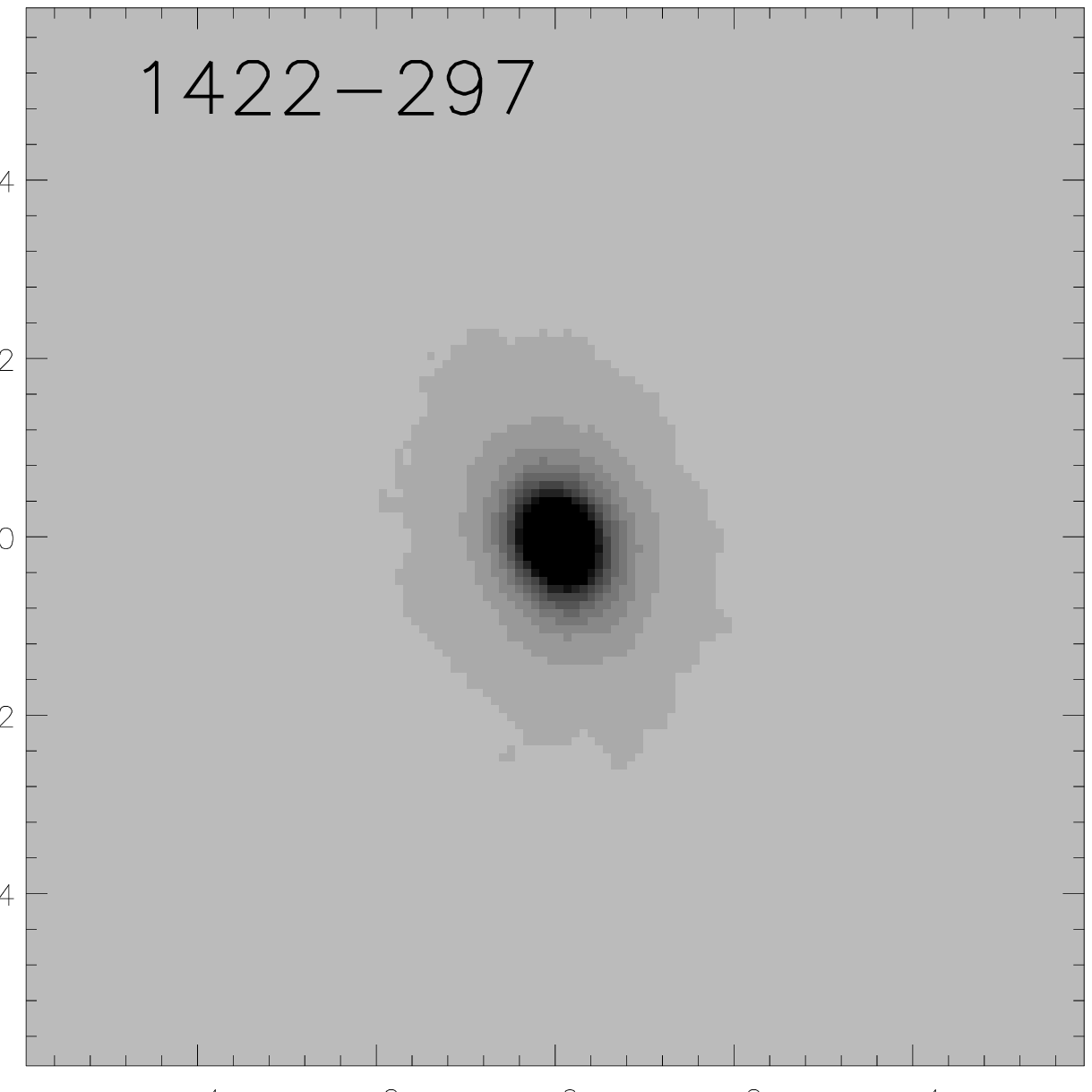,width=0.3\textwidth}&
\epsfig{file=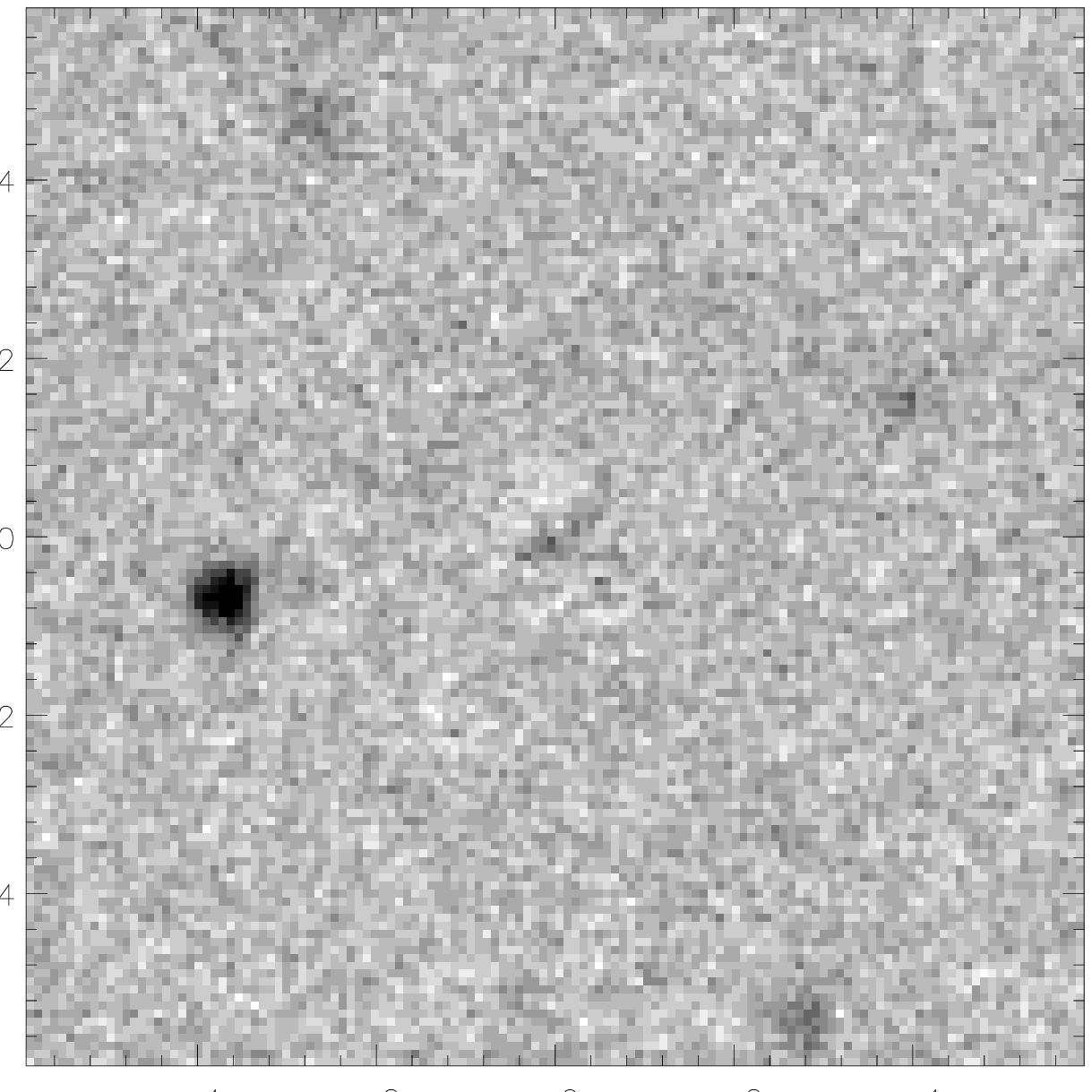,width=0.3\textwidth}\\
\\
\epsfig{file=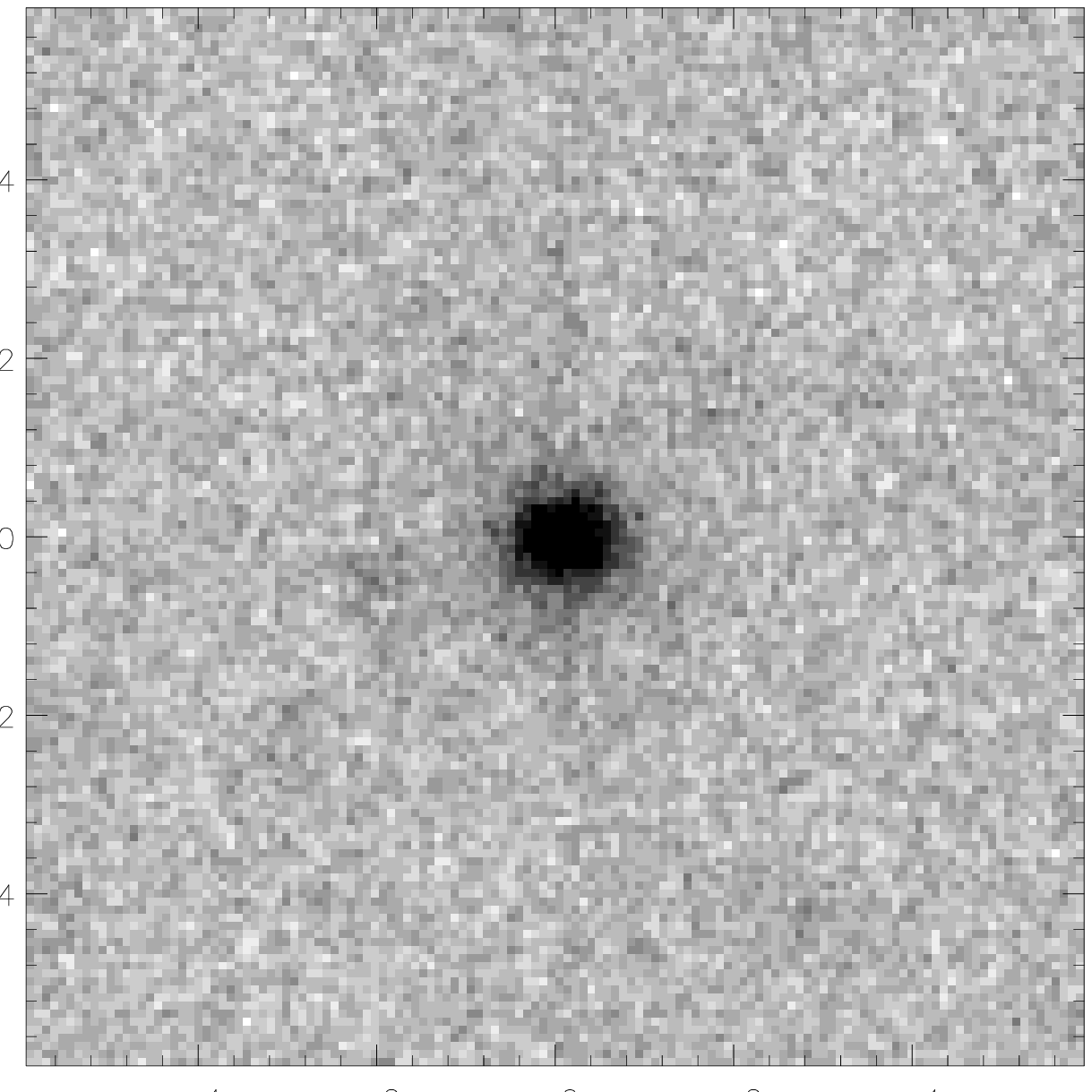,width=0.3\textwidth}&
\epsfig{file=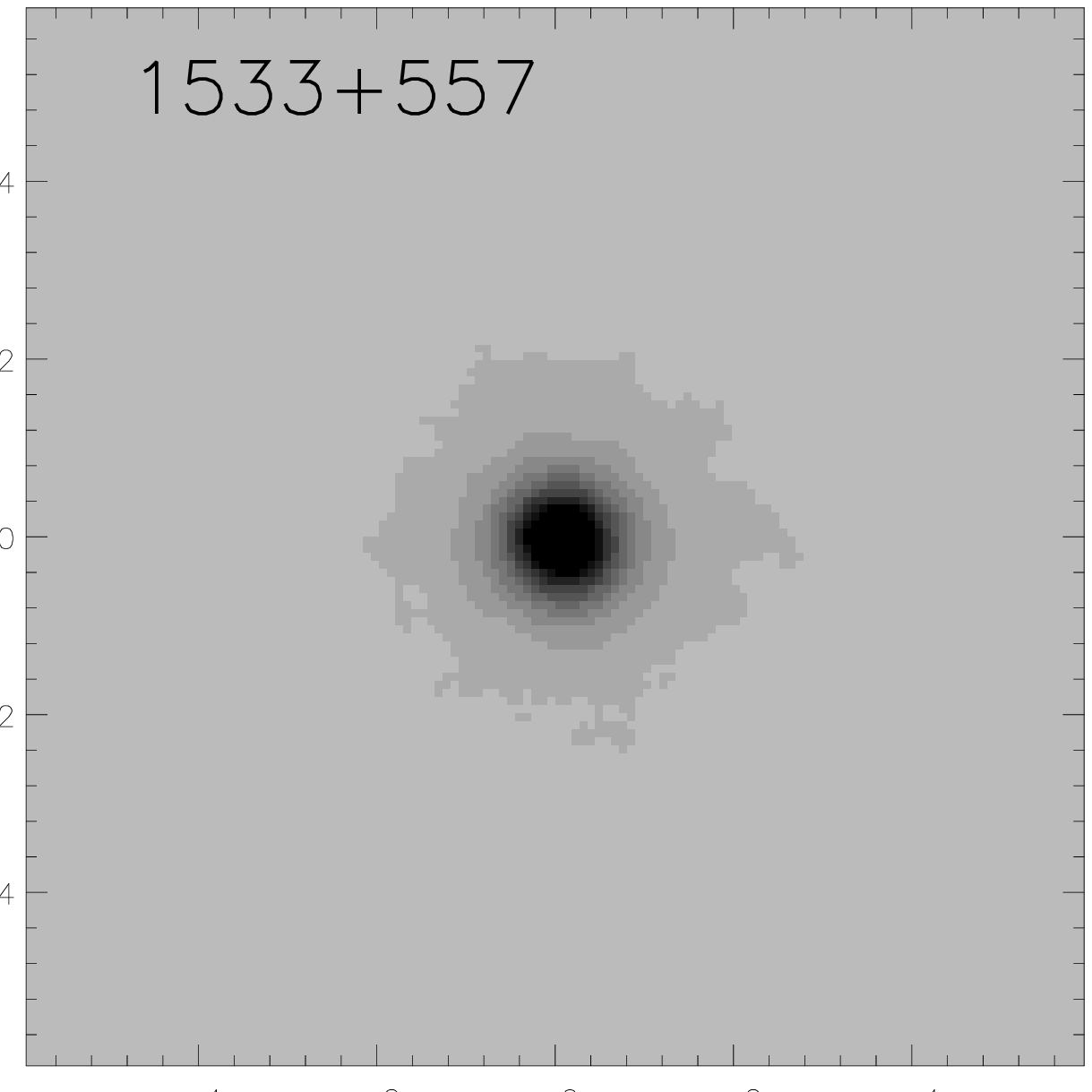,width=0.3\textwidth}&
\epsfig{file=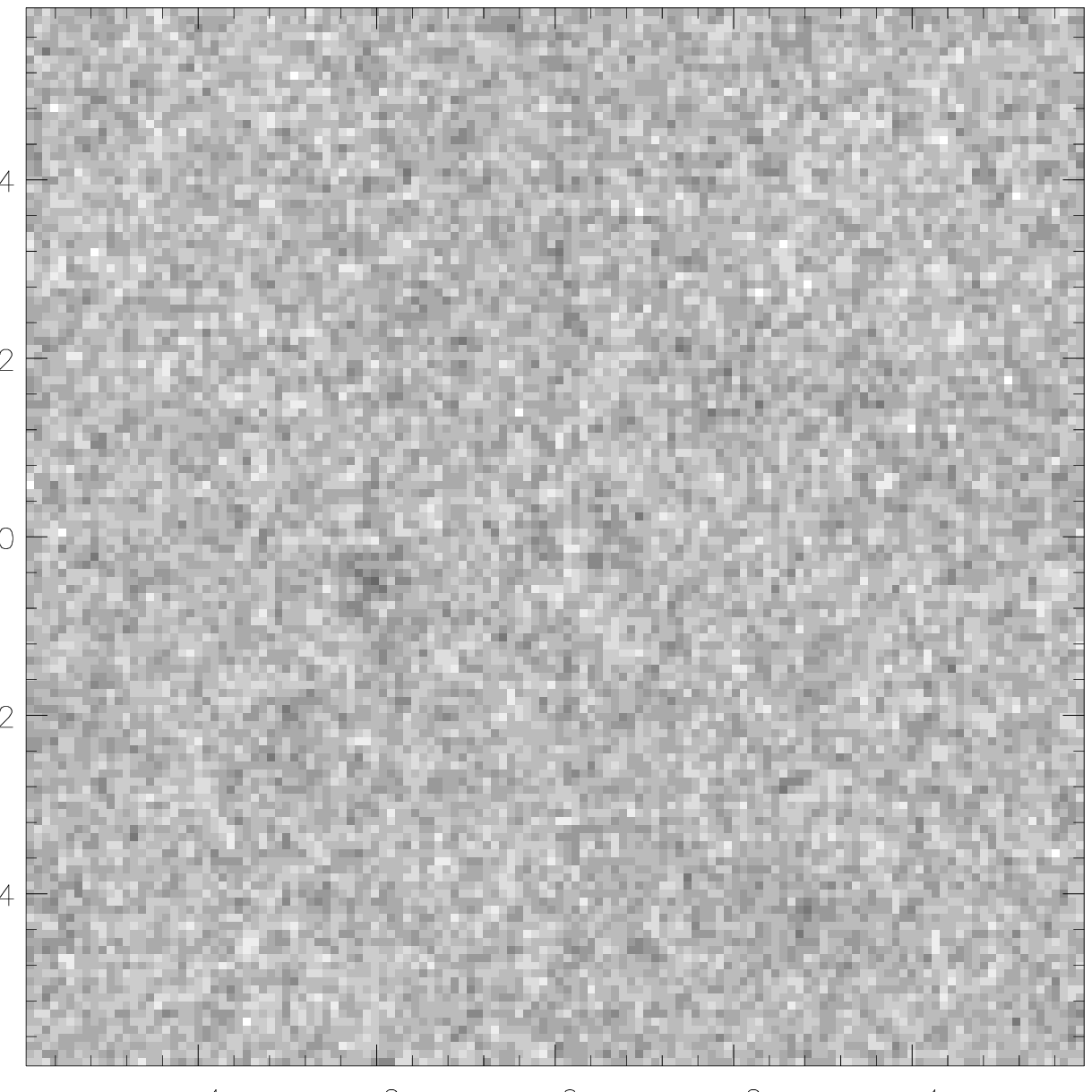,width=0.3\textwidth}\\
\\
\epsfig{file=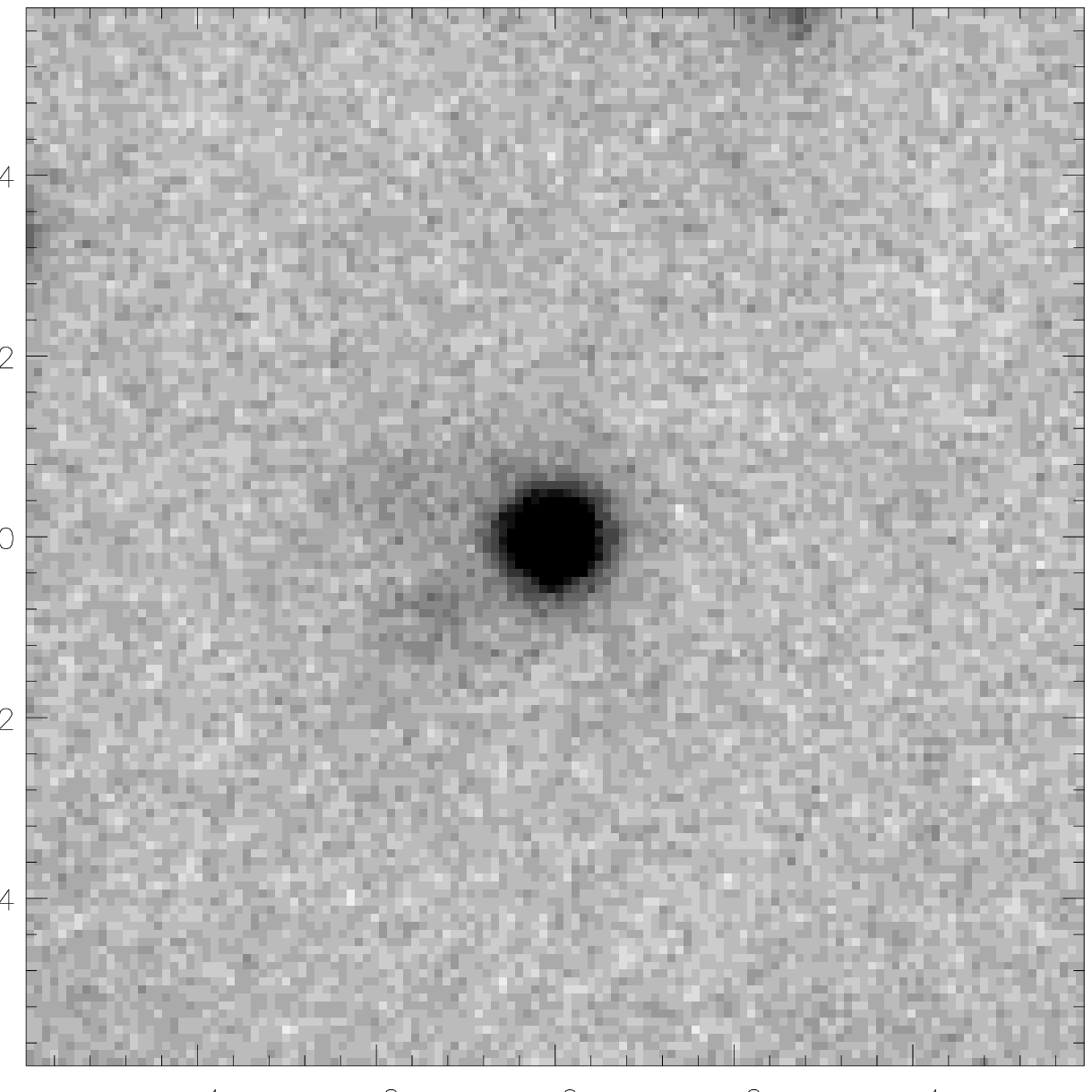,width=0.3\textwidth}&
\epsfig{file=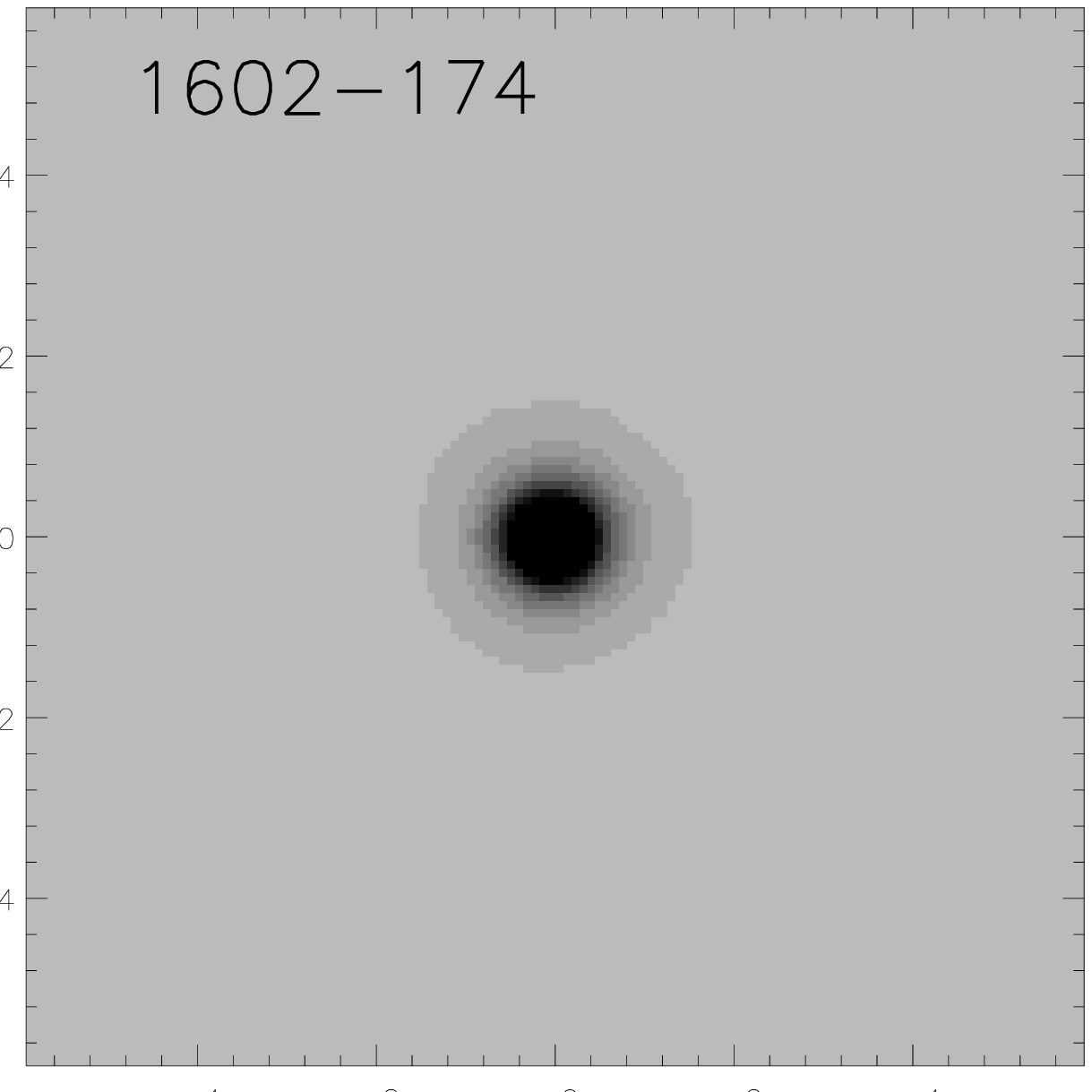,width=0.3\textwidth}&
\epsfig{file=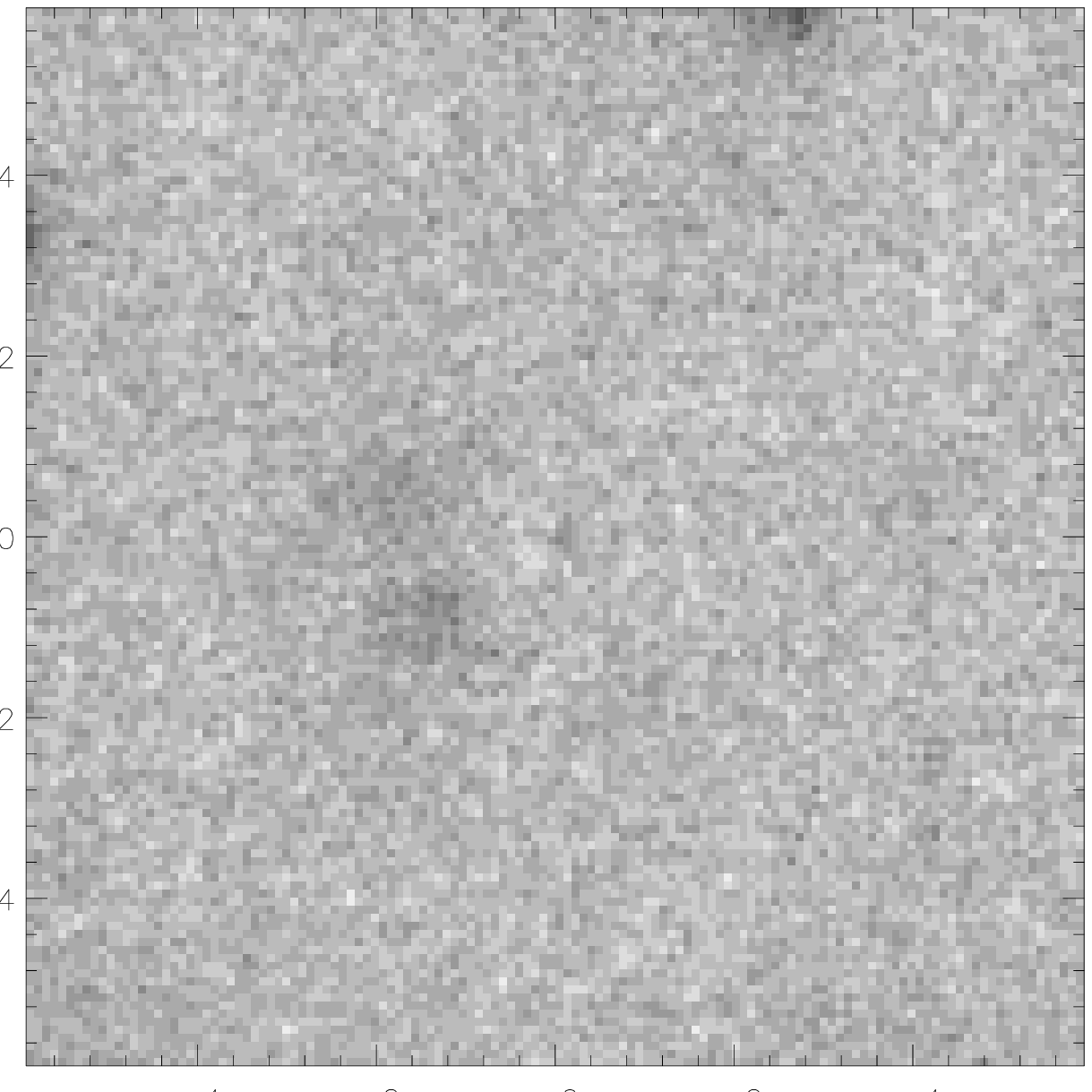,width=0.3\textwidth}\\
\end{tabular}
\addtocounter{figure}{-1}
\caption{- continued}
\label{ukirtmodel1}
\end{figure*}
\end{center}

\begin{center}
\begin{figure*}
\begin{tabular}{ccc}
\epsfig{file=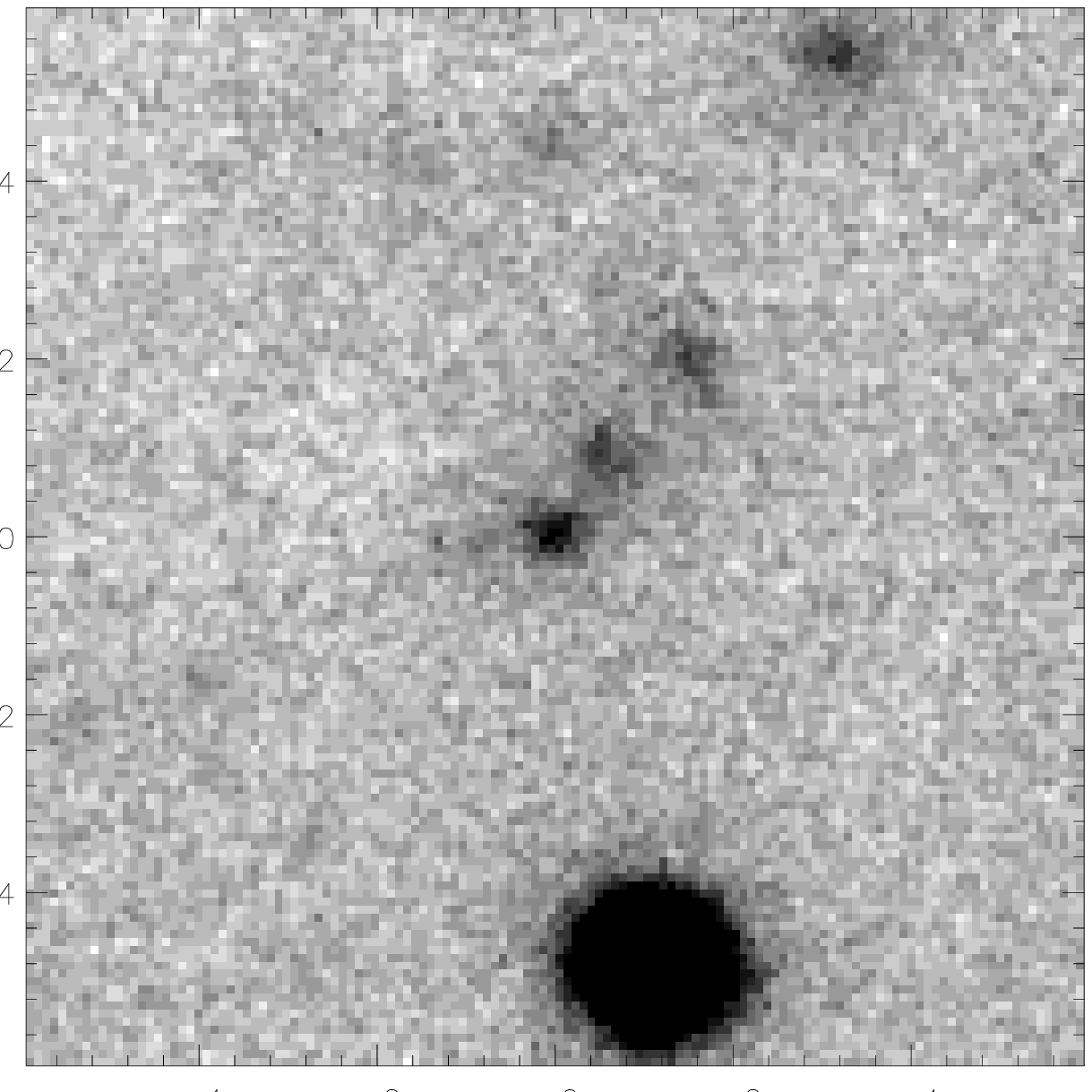,width=0.3\textwidth}&
\epsfig{file=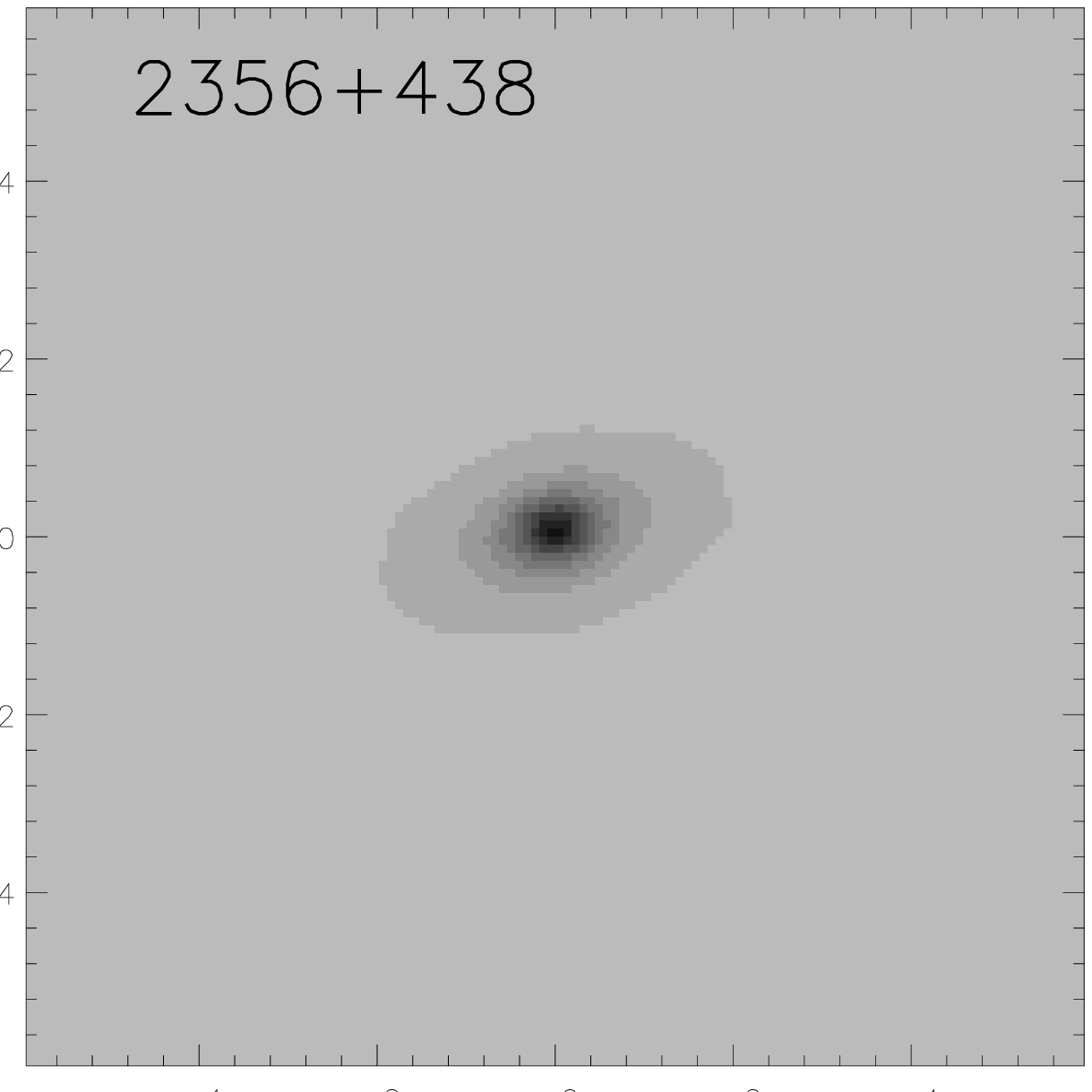,width=0.3\textwidth}&
\epsfig{file=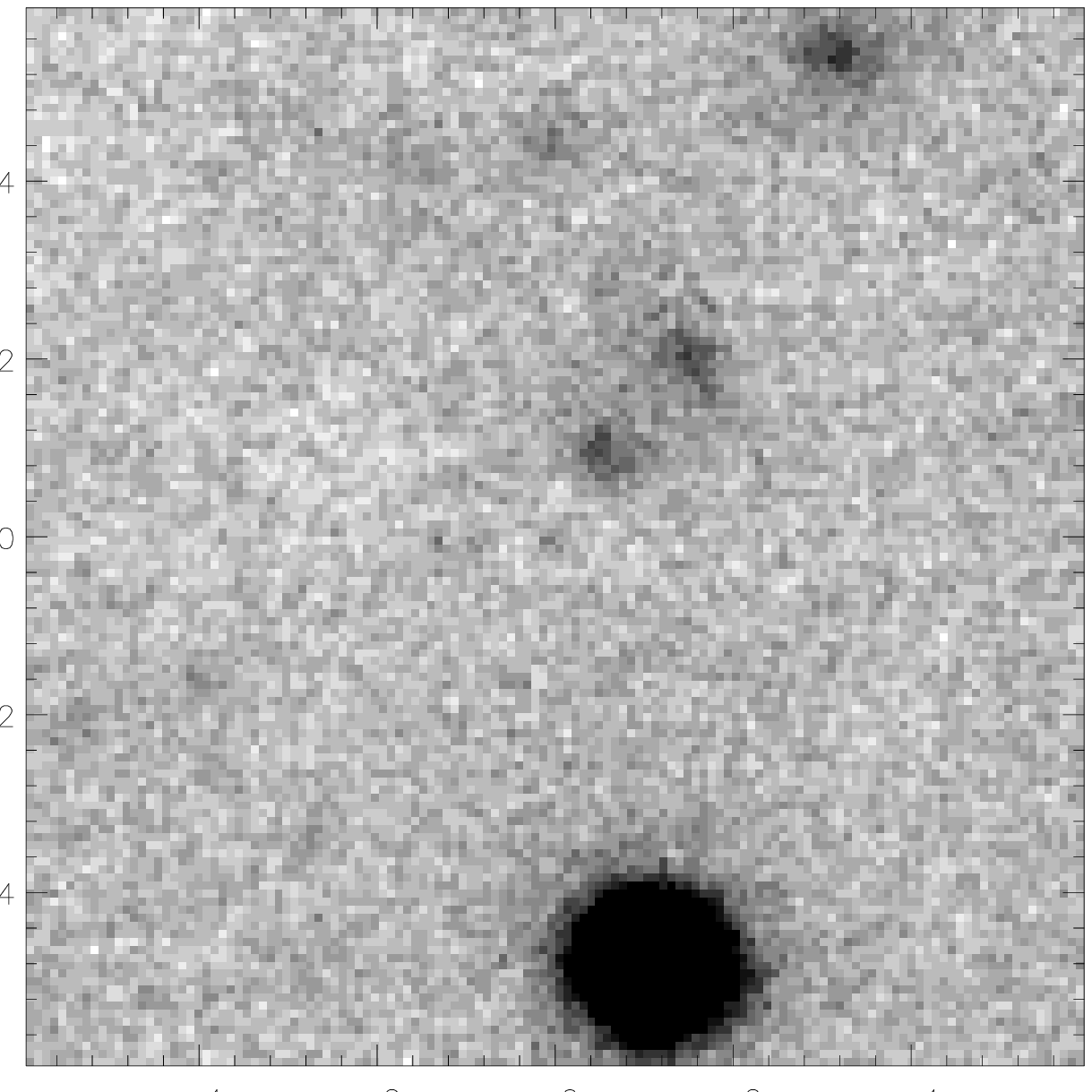,width=0.3\textwidth}\\
\end{tabular}
\addtocounter{figure}{-1}
\caption{- continued}
\label{ukirtmodel1}
\end{figure*}
\end{center}

\begin{center}
\begin{figure*}
\begin{tabular}{ccc}
\epsfig{file=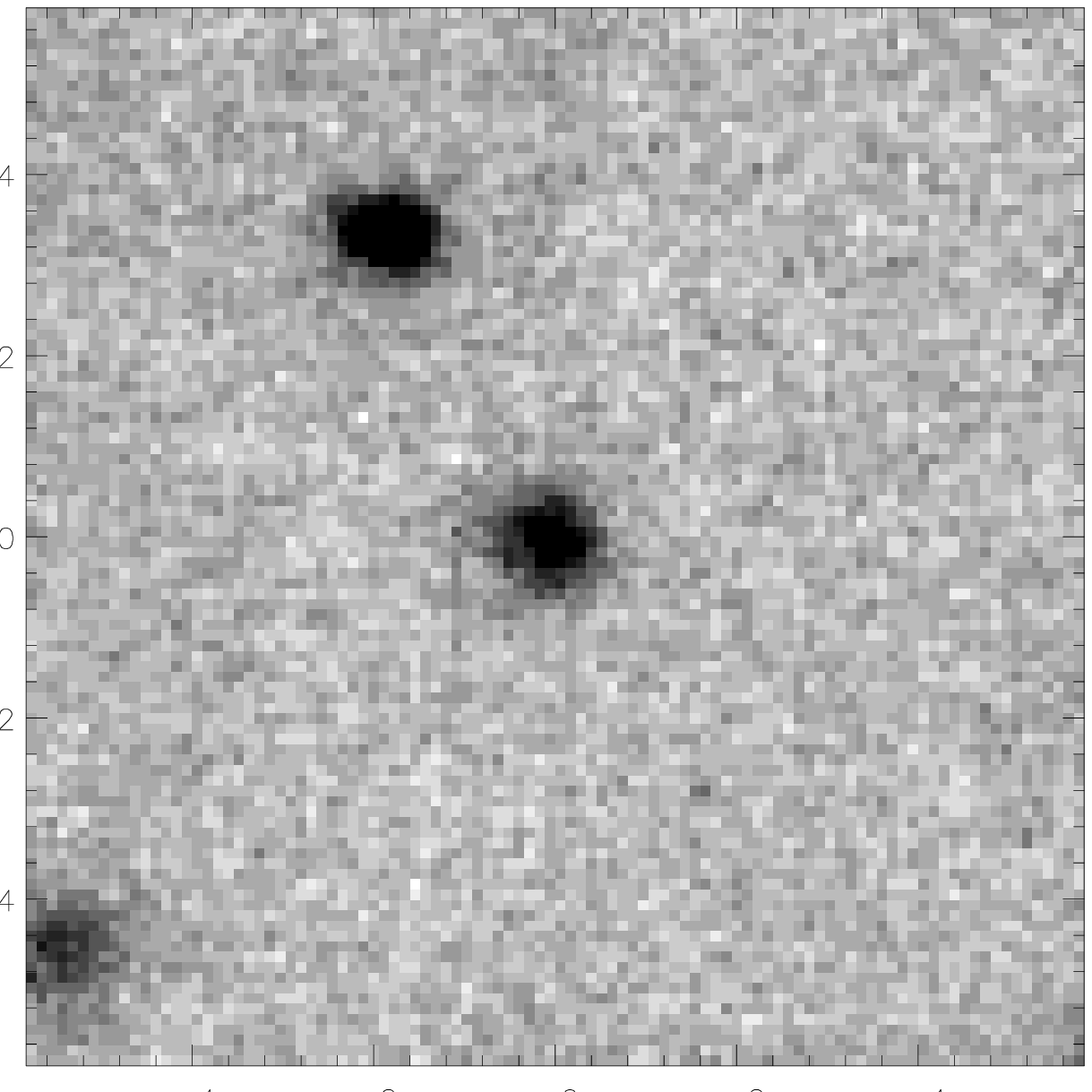,width=0.3\textwidth}&
\epsfig{file=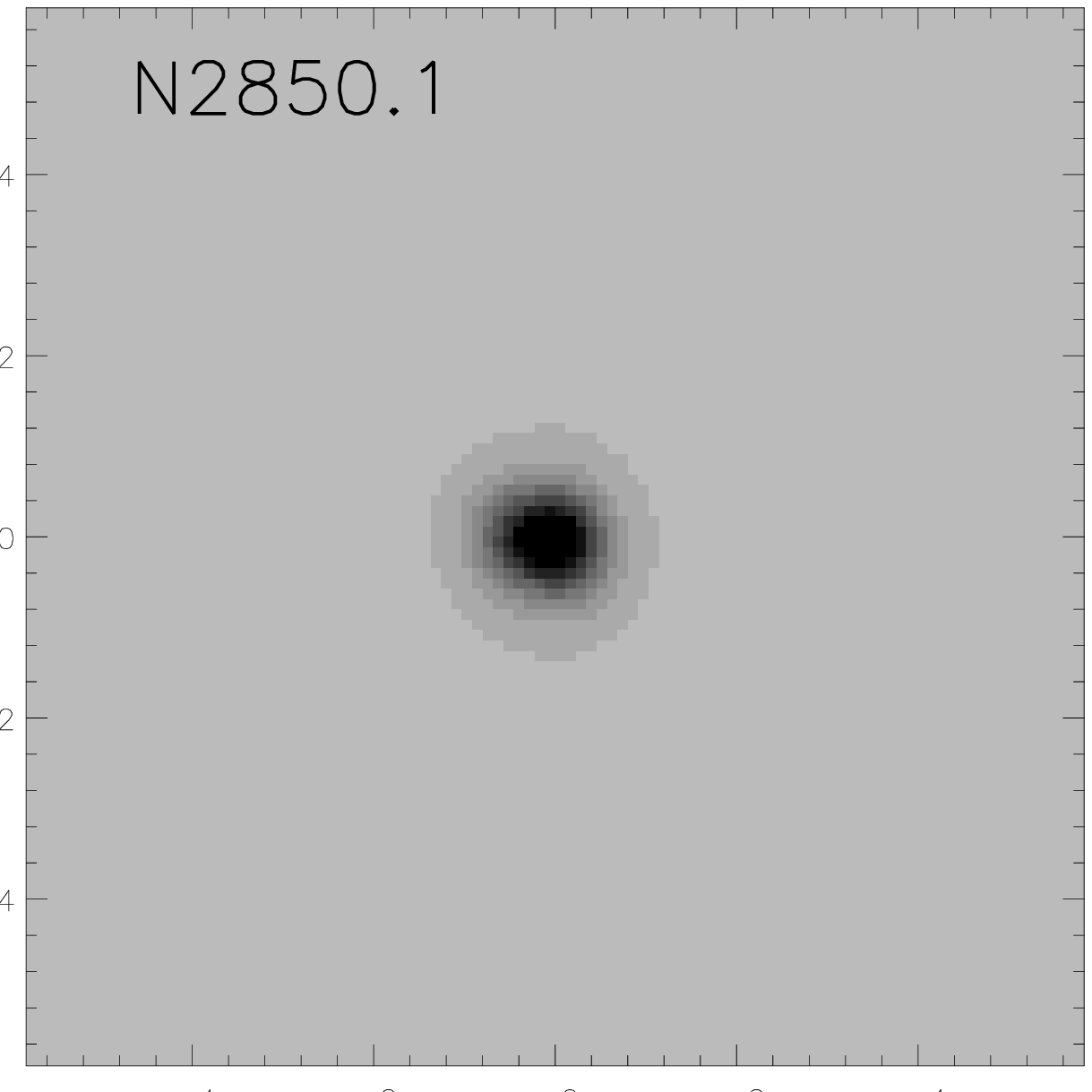,width=0.3\textwidth}&
\epsfig{file=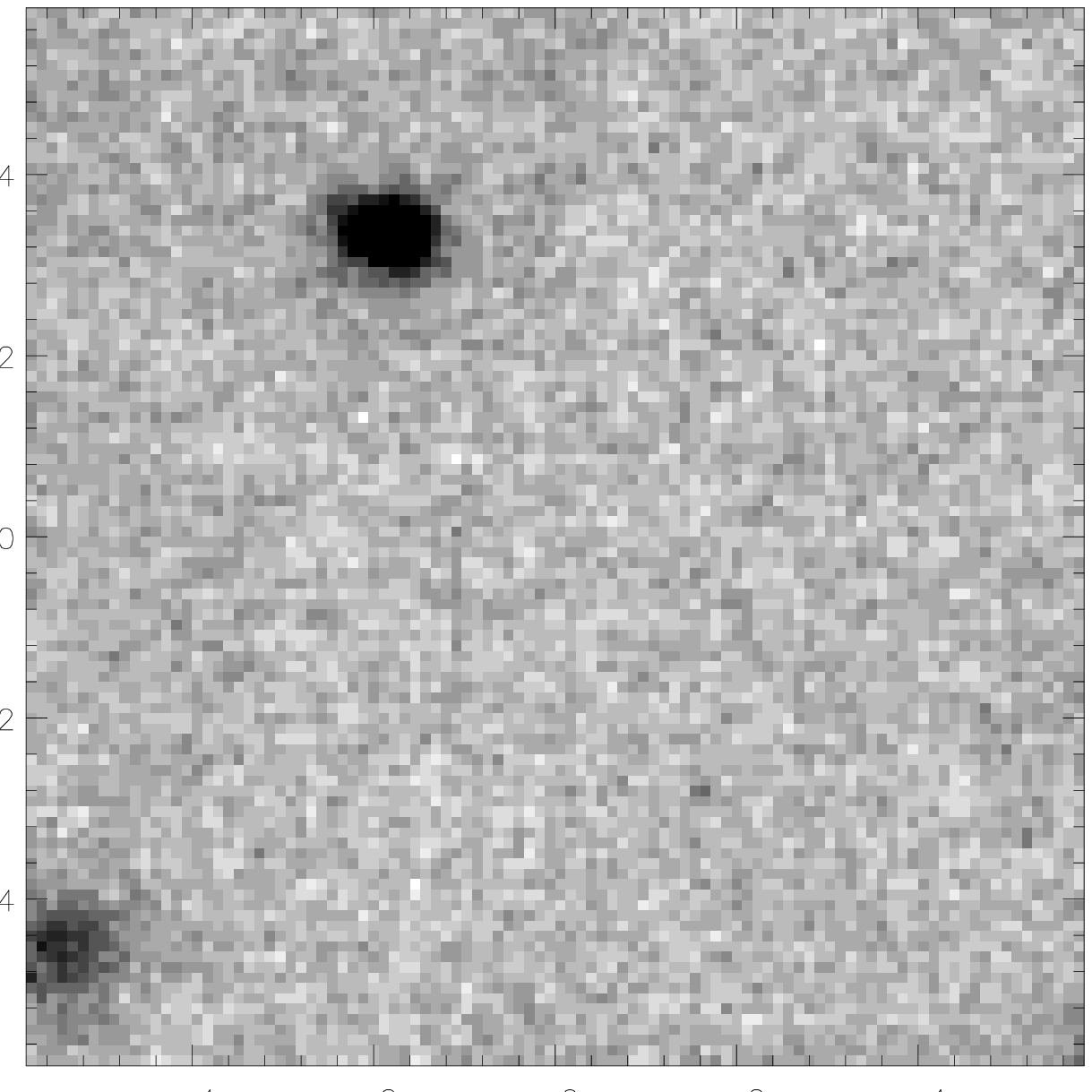,width=0.3\textwidth}\\
\\
\epsfig{file=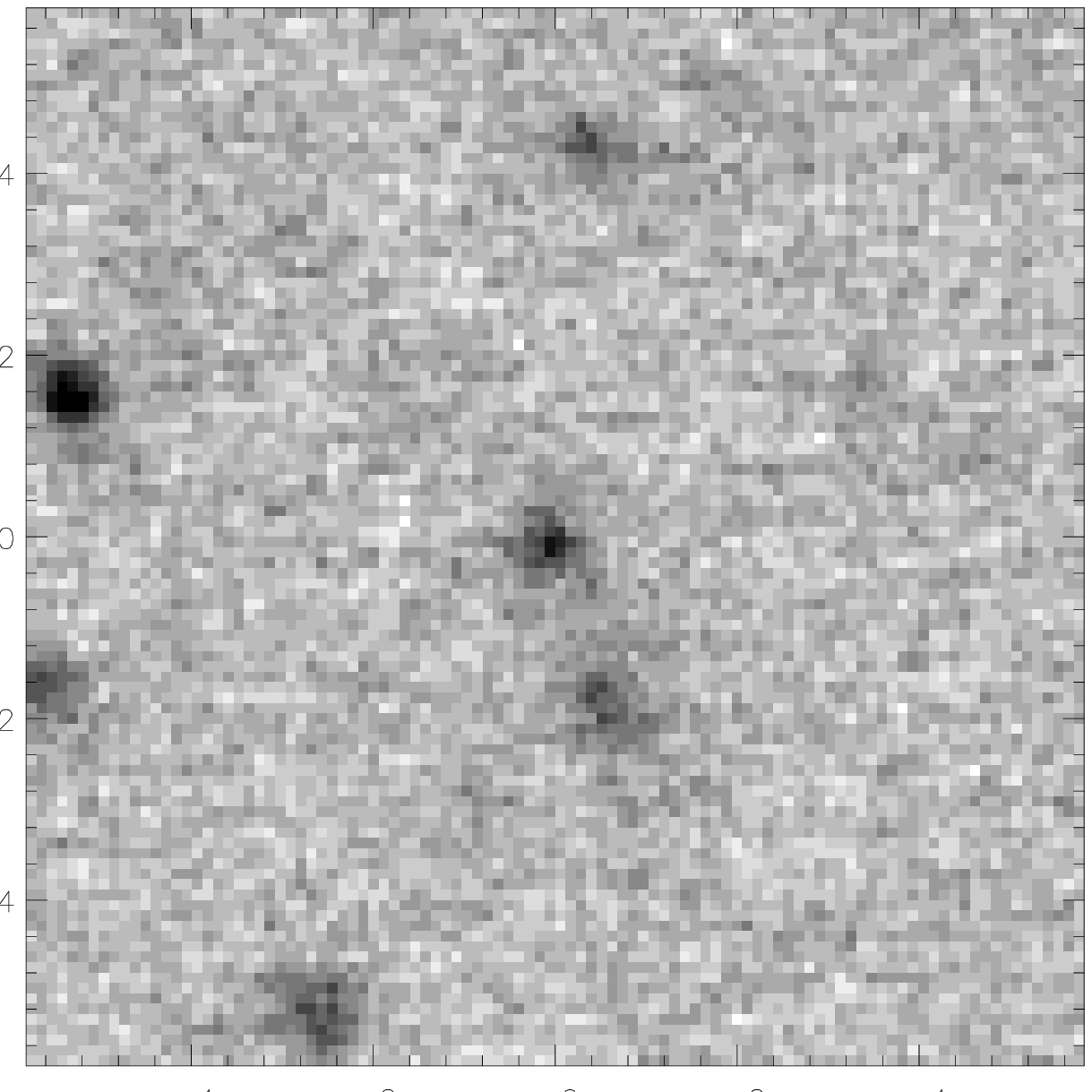,width=0.3\textwidth}&
\epsfig{file=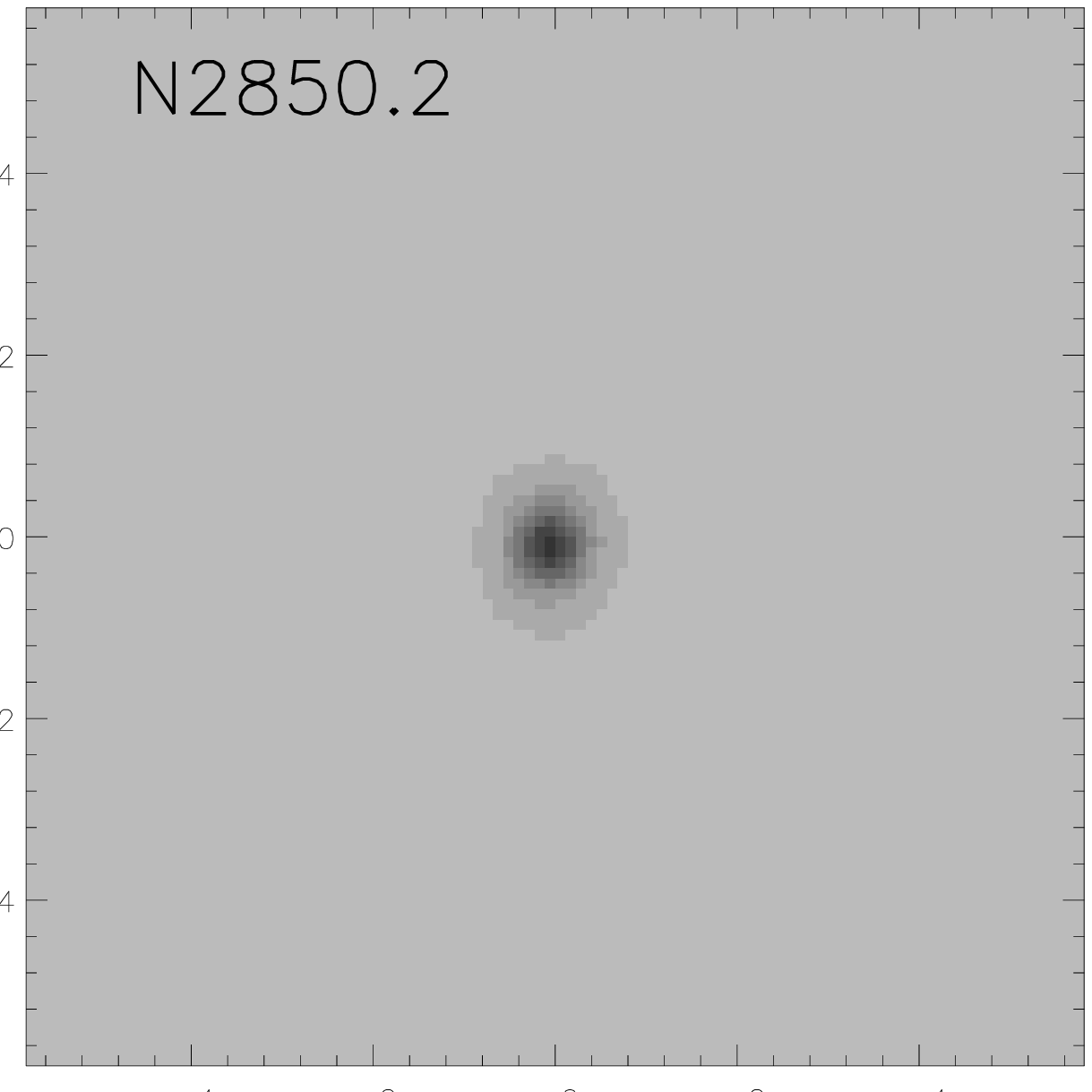,width=0.3\textwidth}&
\epsfig{file=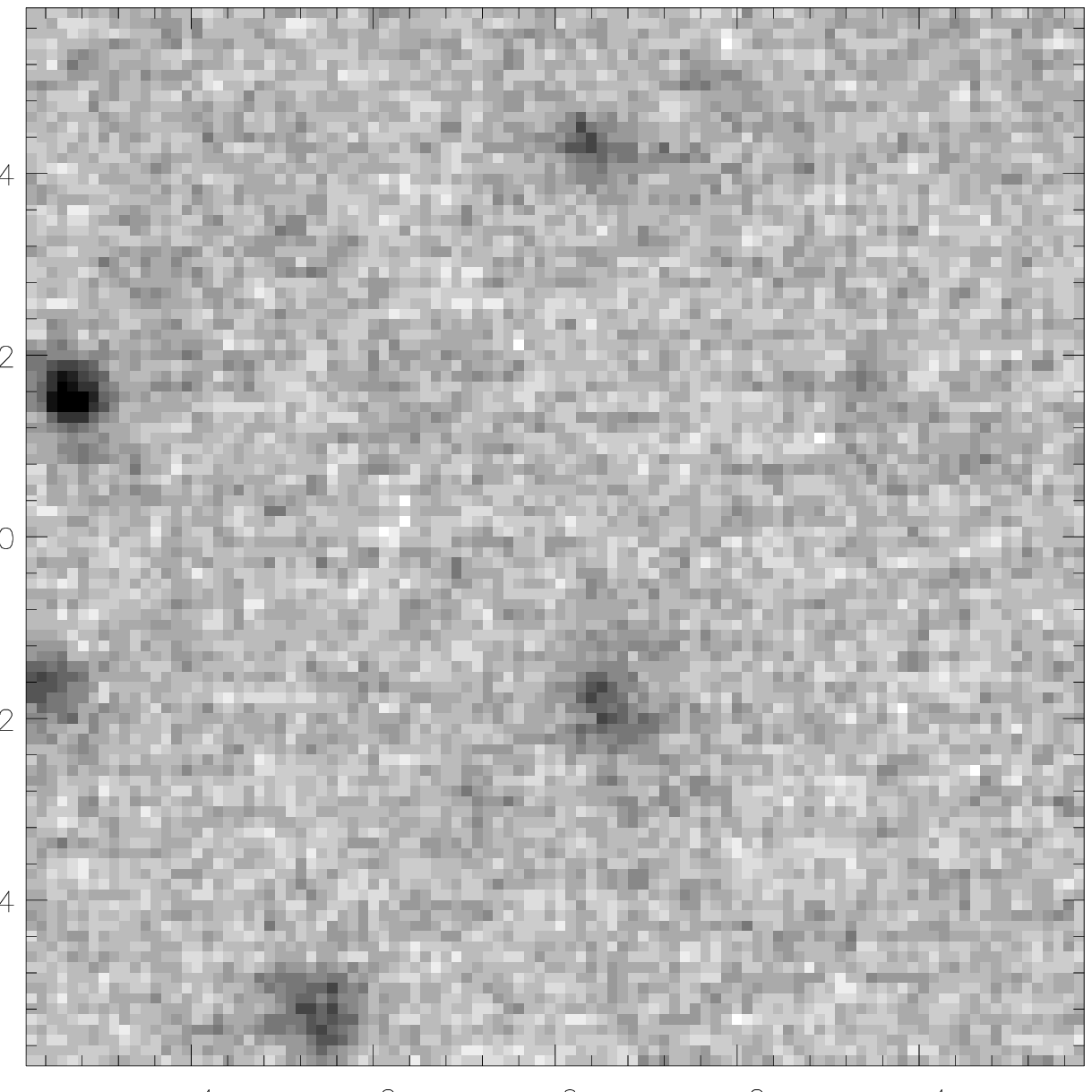,width=0.3\textwidth}\\
\\
\epsfig{file=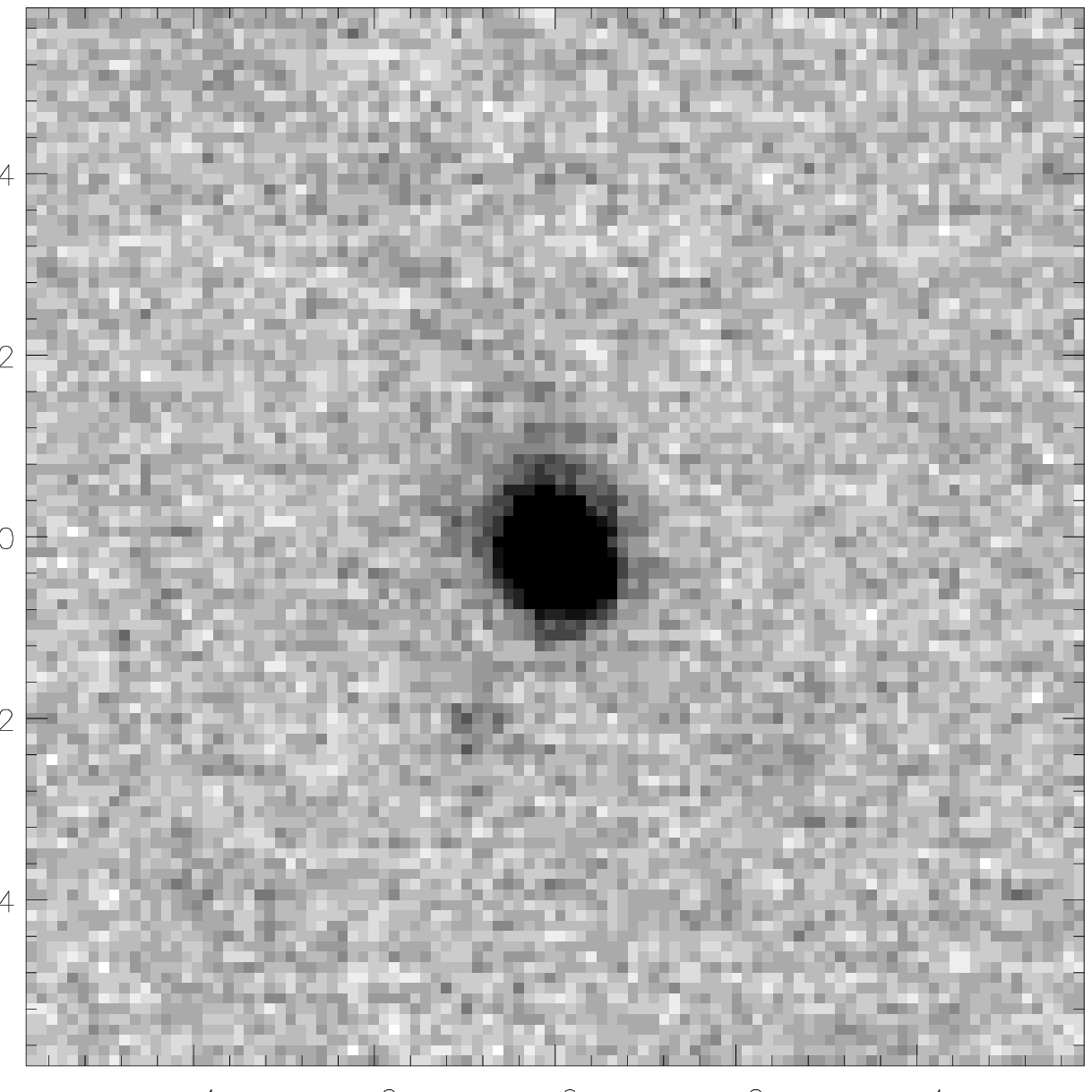,width=0.3\textwidth}&
\epsfig{file=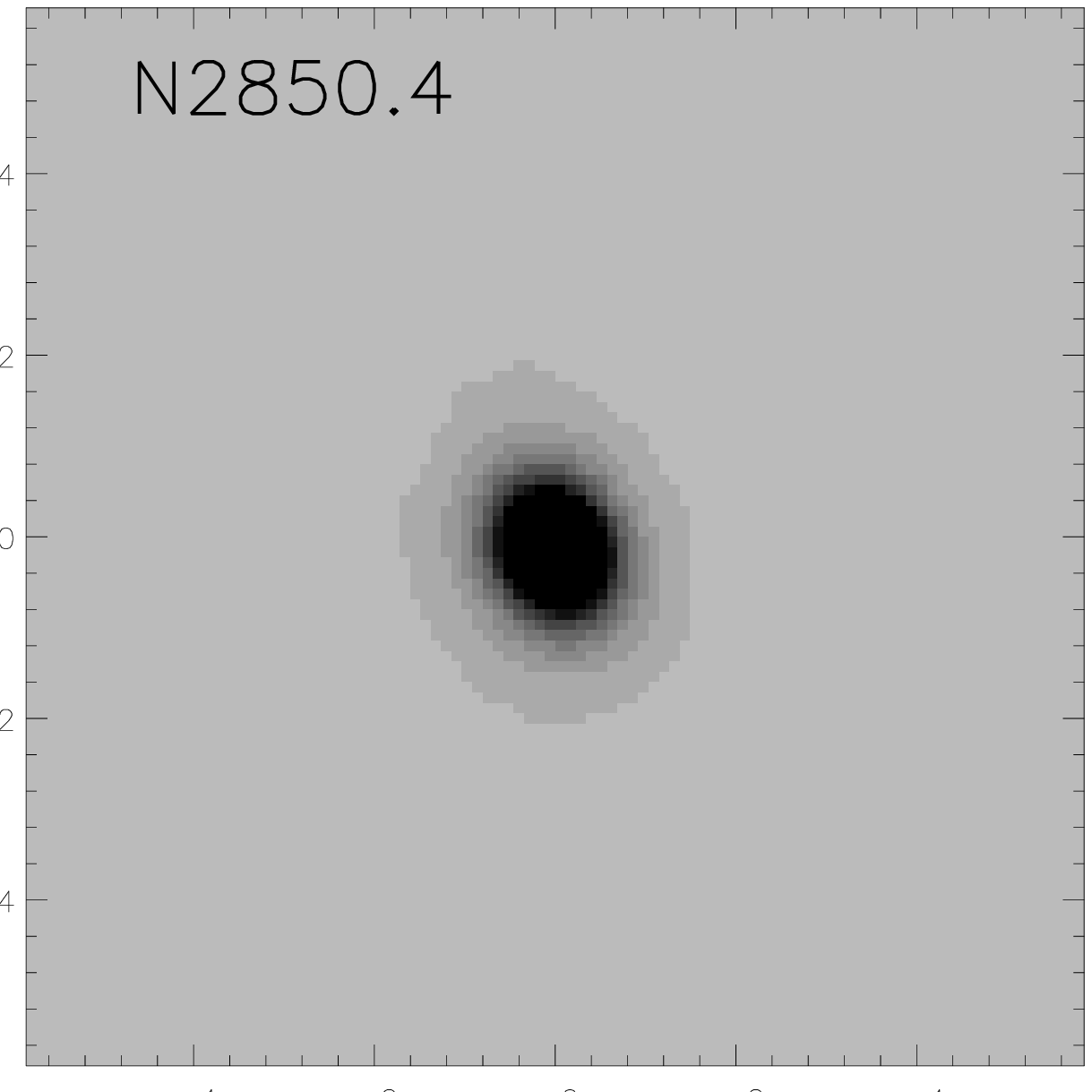,width=0.3\textwidth}&
\epsfig{file=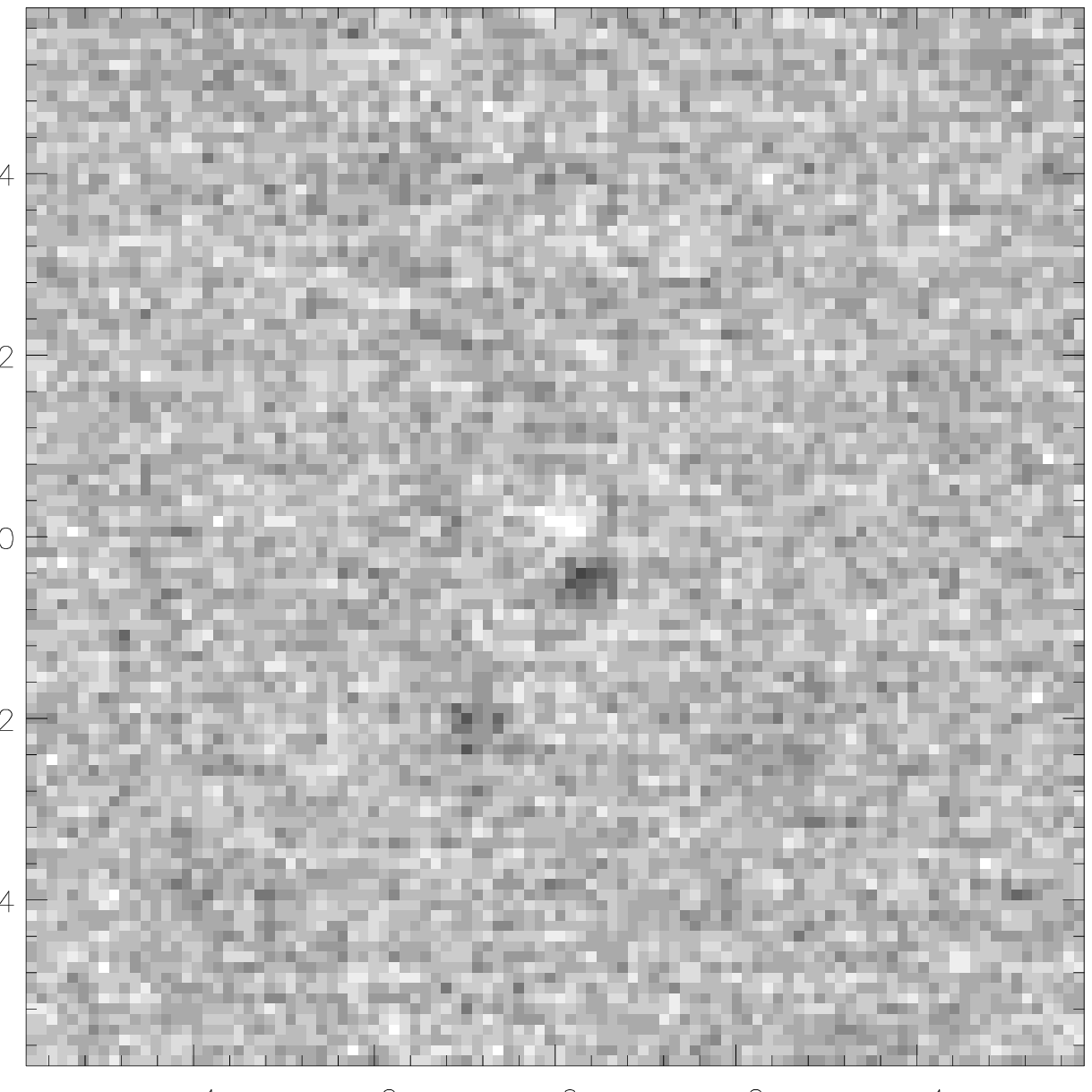,width=0.3\textwidth}\\
\\
\epsfig{file=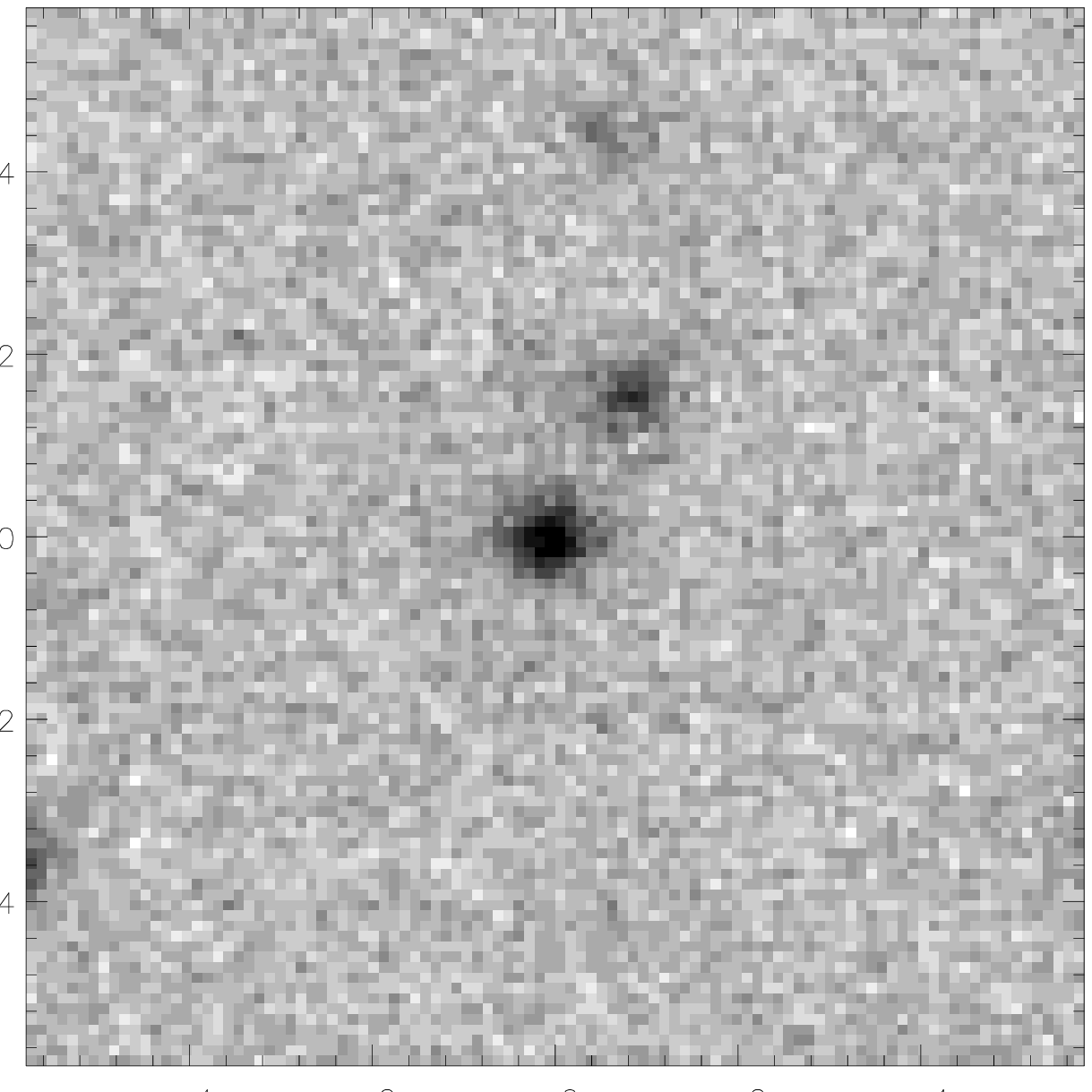,width=0.3\textwidth}&
\epsfig{file=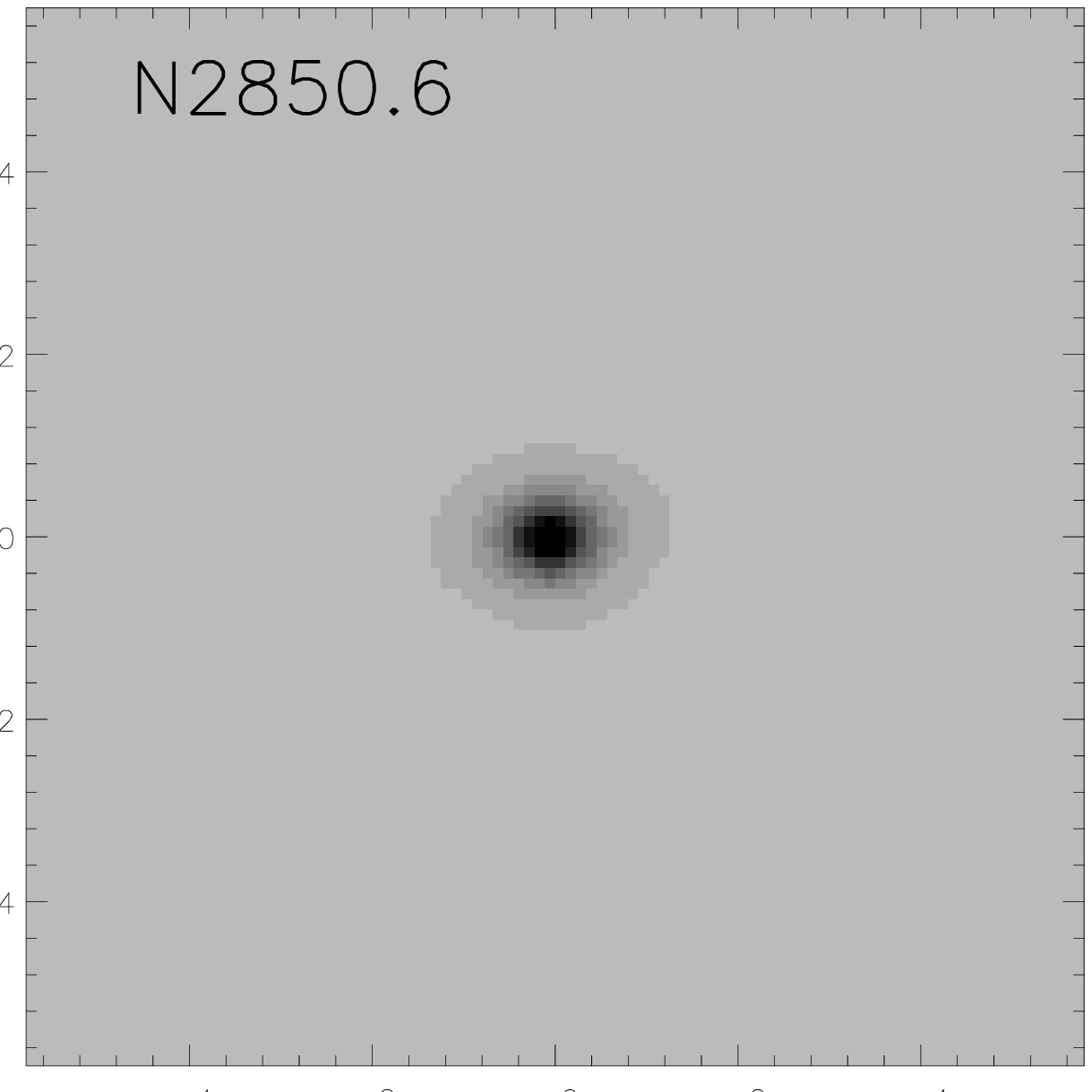,width=0.3\textwidth}&
\epsfig{file=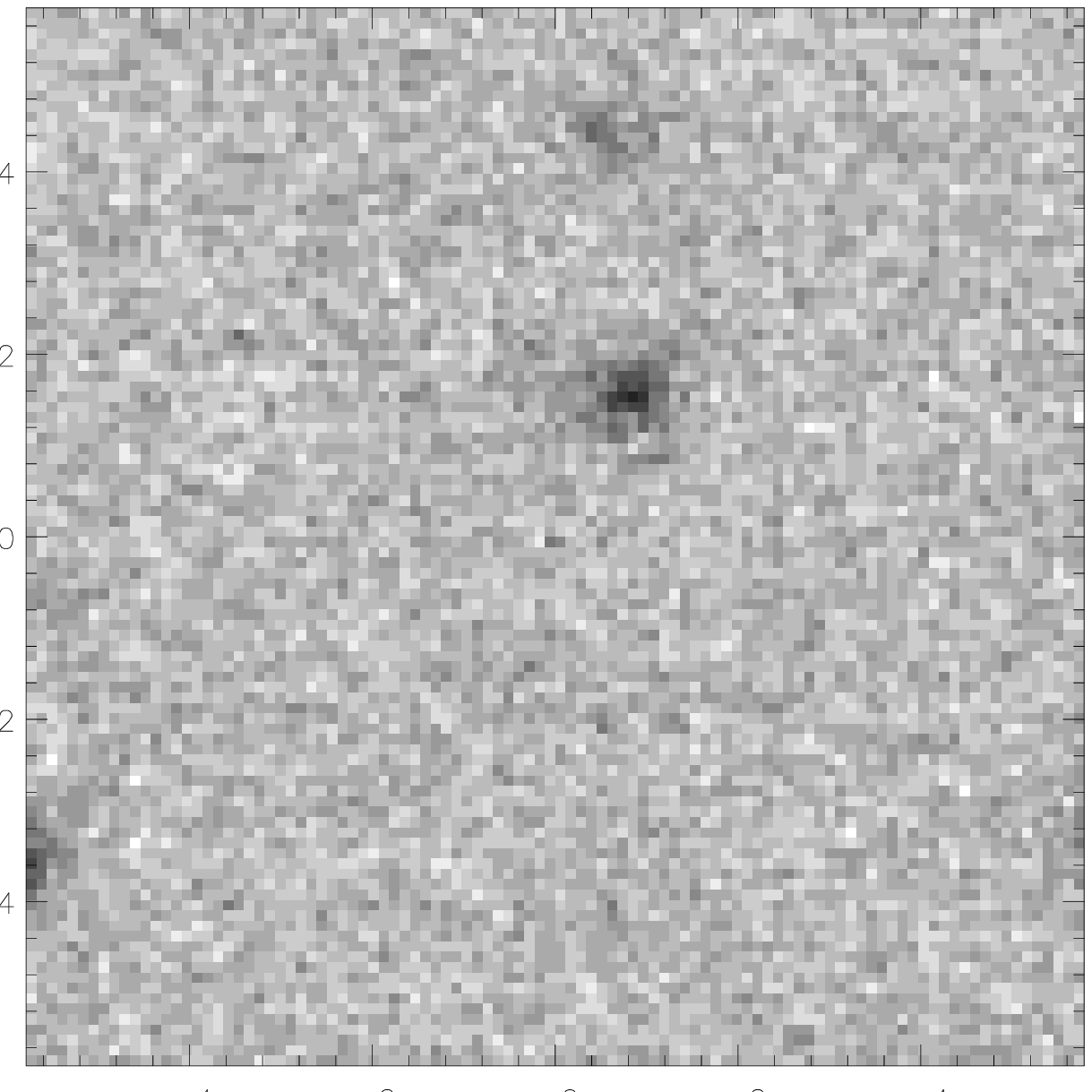,width=0.3\textwidth}\\
\end{tabular}
\caption[Two-dimensional modelling of the sub-millimetre galaxies]{Two-dimensional modelling of the sub-millimetre galaxies. The left-hand panel shows the $K$-band image centred on the sub-millimetre galaxy. The middle panel shows the best-fitting two-dimensional model. The right-hand panel shows the residual image after subtraction of the model from the data. All panels are 12.0$^{\prime \prime}$ $\times$12.0$^{\prime \prime}$, and the images are shown with a linear greyscale in which black corresponds to $2.5\sigma$ above, and white to $1\sigma$ below the median sky value.}
\label{geminimodel1}
\end{figure*}
\end{center}

\begin{center}
\begin{figure*}
\begin{tabular}{ccc}
\epsfig{file=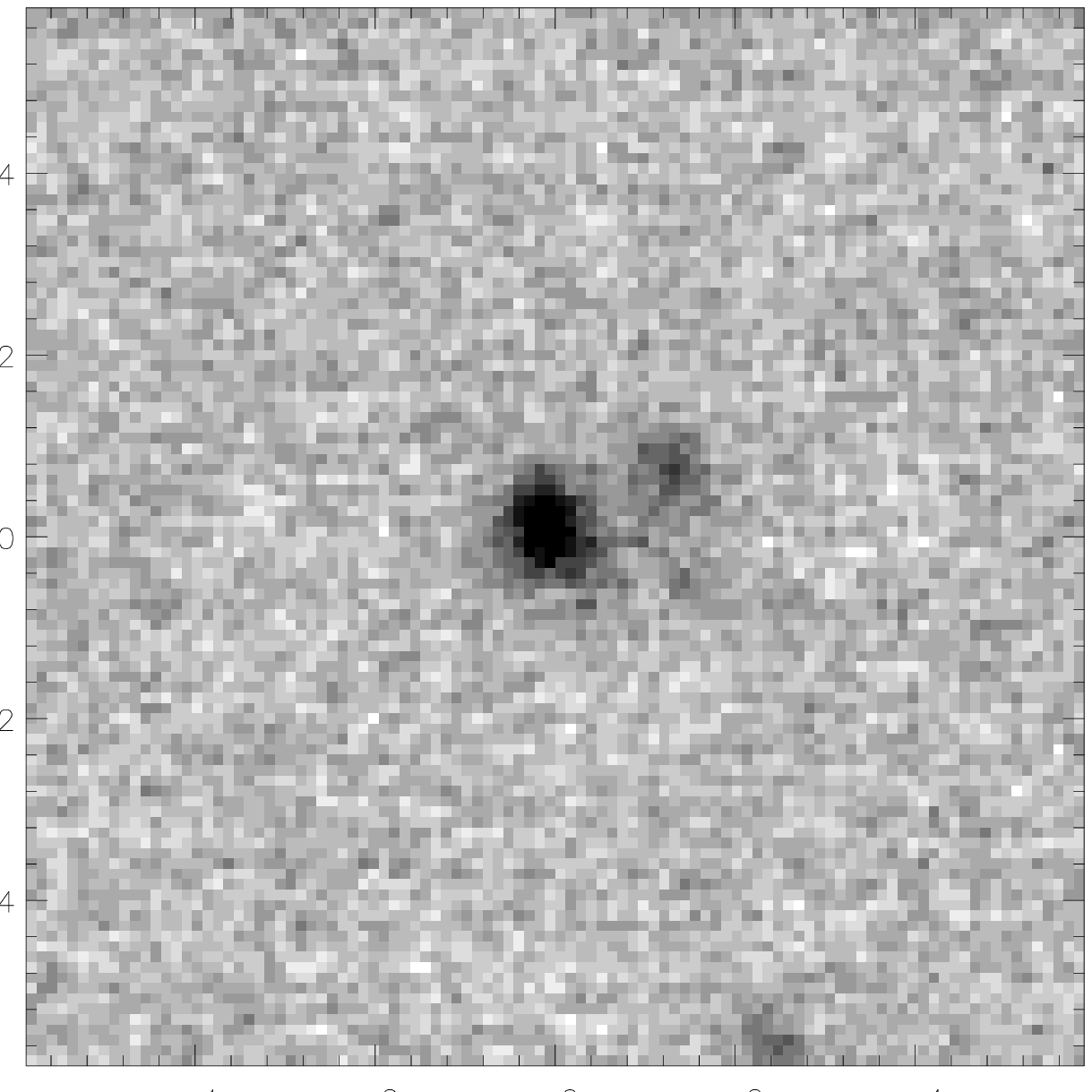,width=0.3\textwidth}&
\epsfig{file=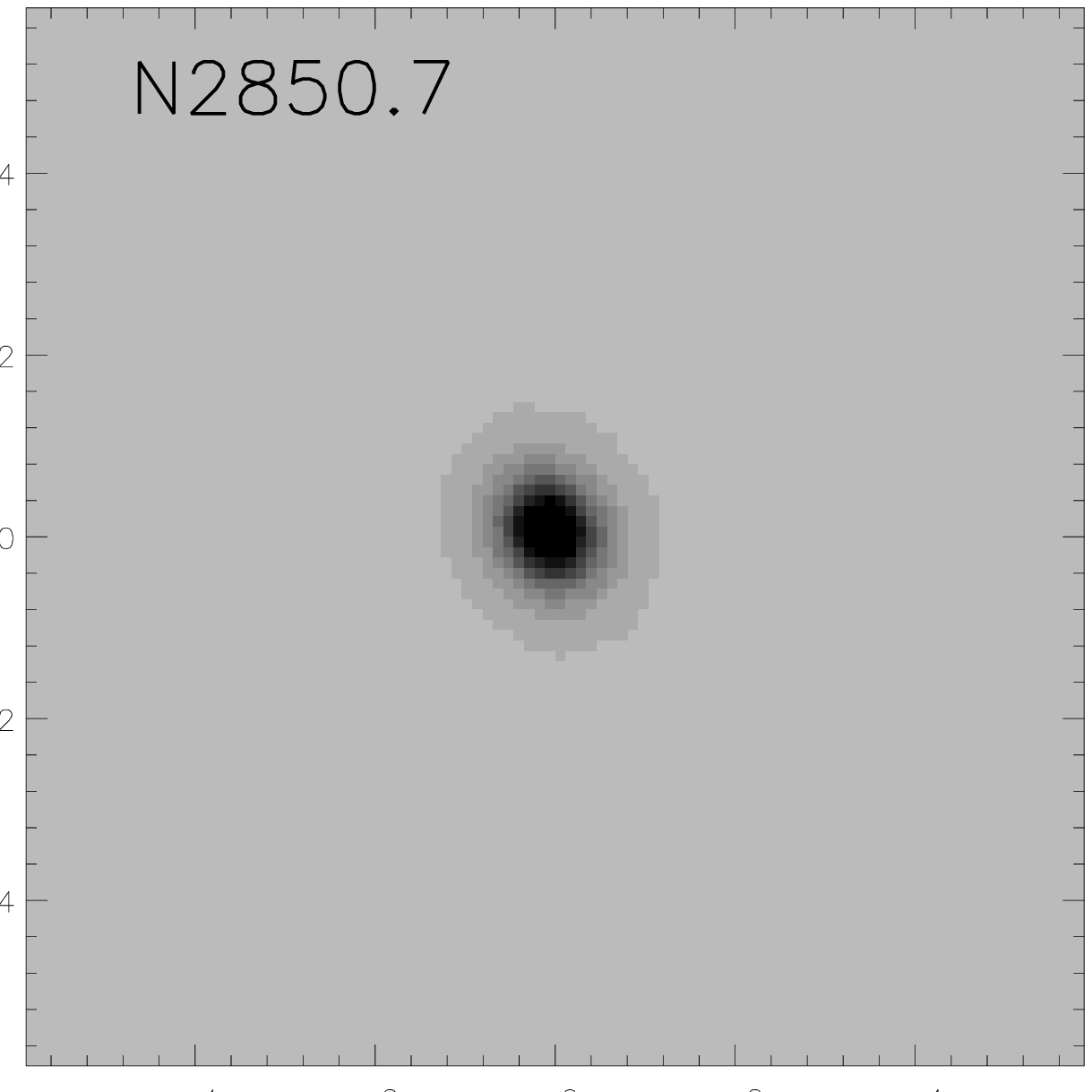,width=0.3\textwidth}&
\epsfig{file=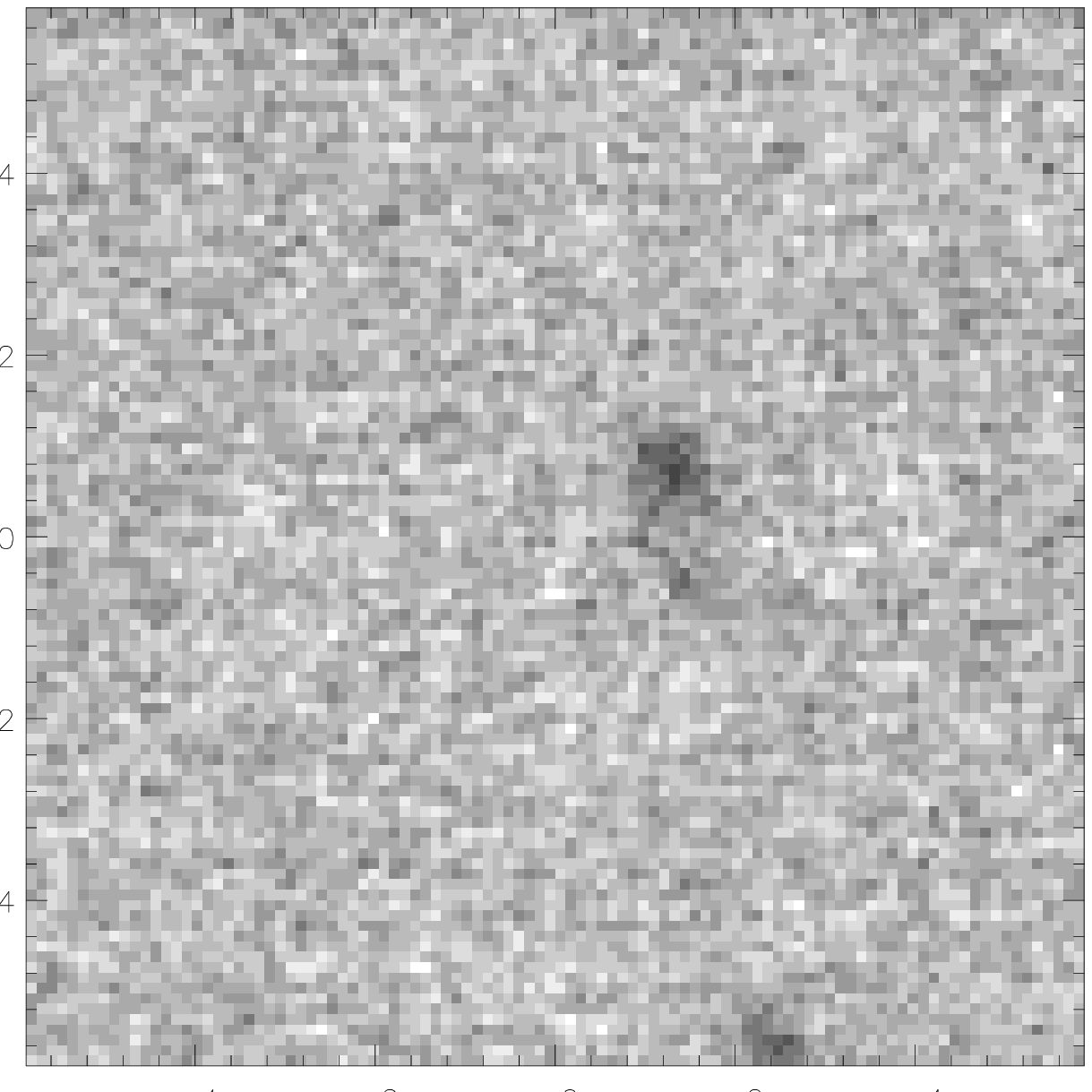,width=0.3\textwidth}\\
\\
\epsfig{file=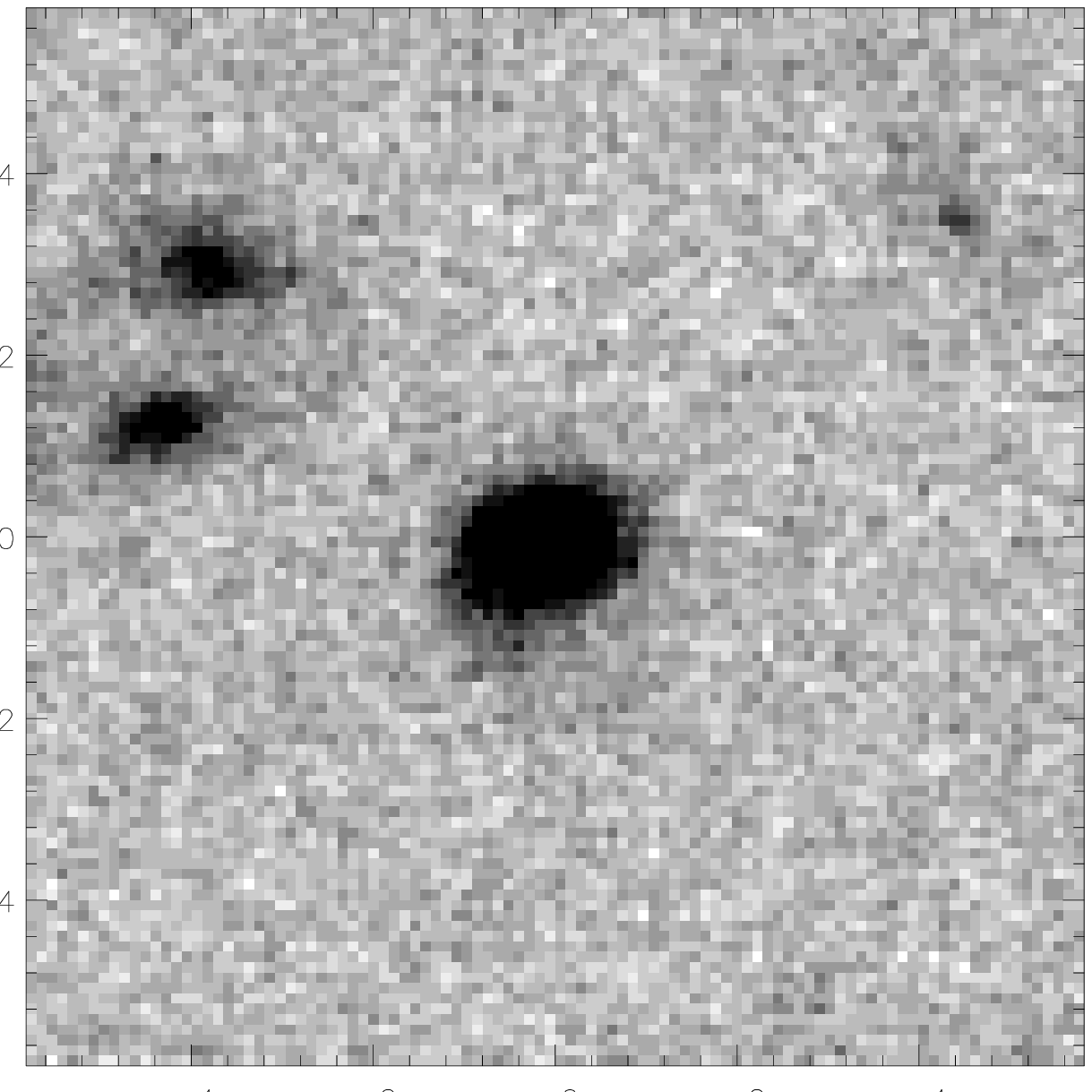,width=0.3\textwidth}&
\epsfig{file=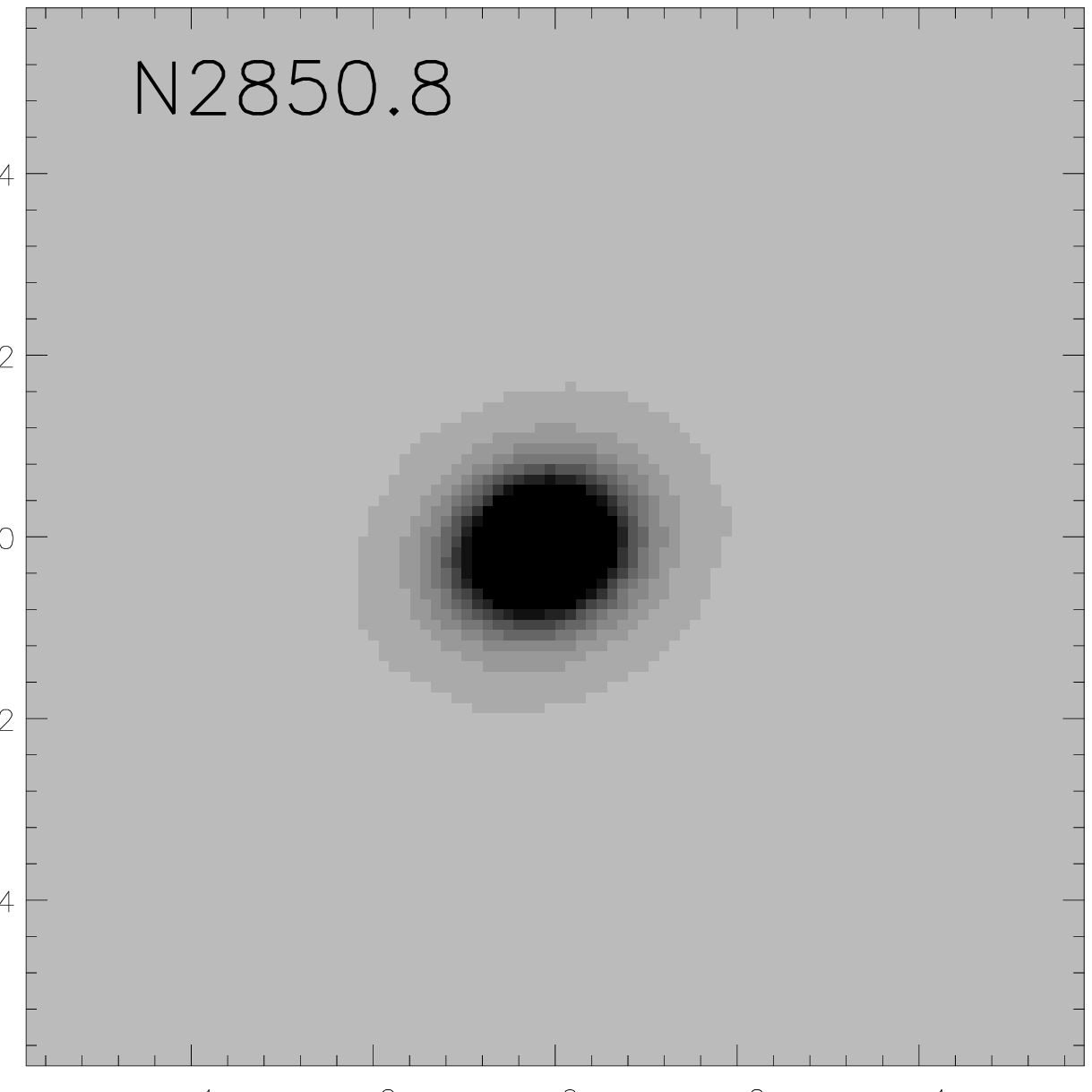,width=0.3\textwidth}&
\epsfig{file=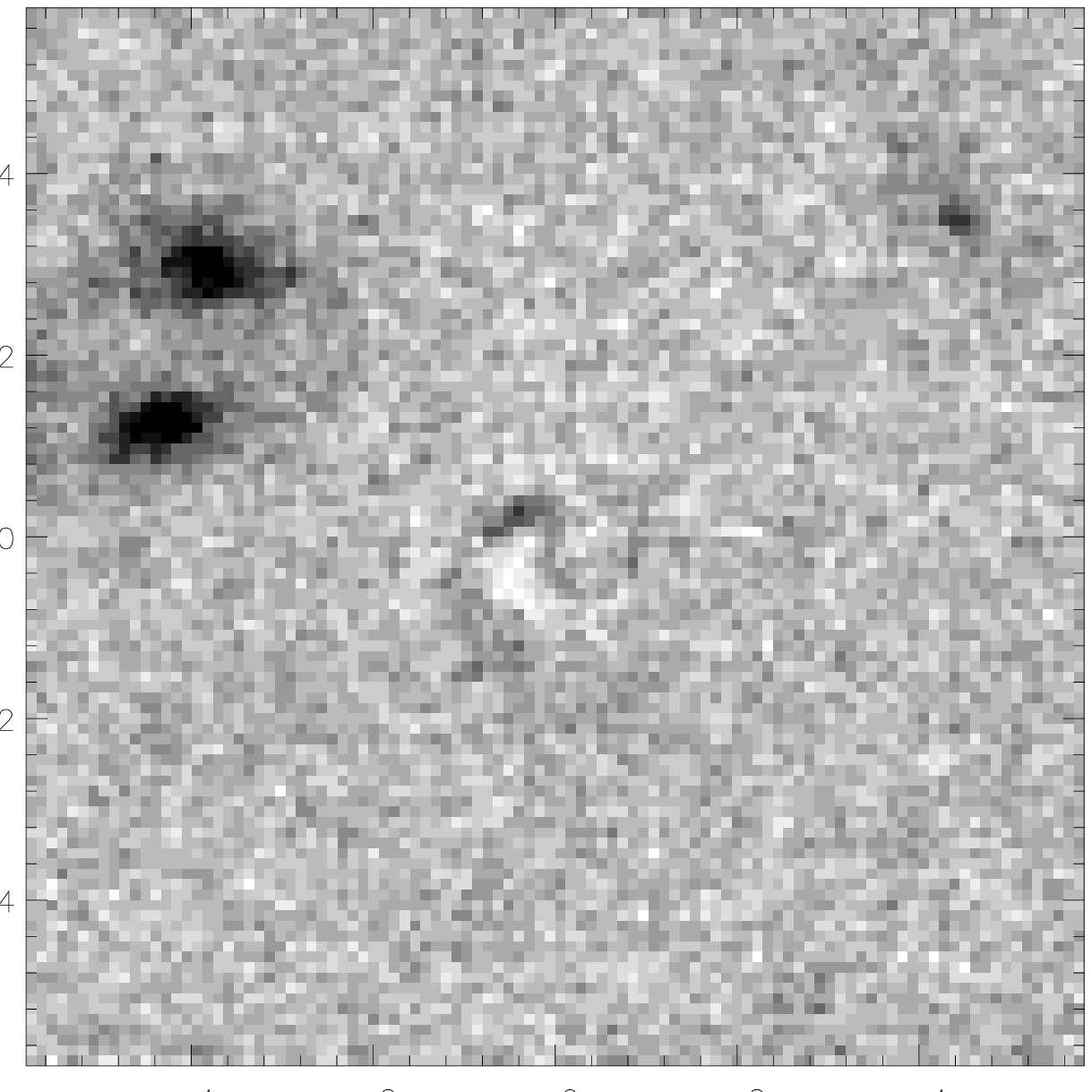,width=0.3\textwidth}\\
\\
\epsfig{file=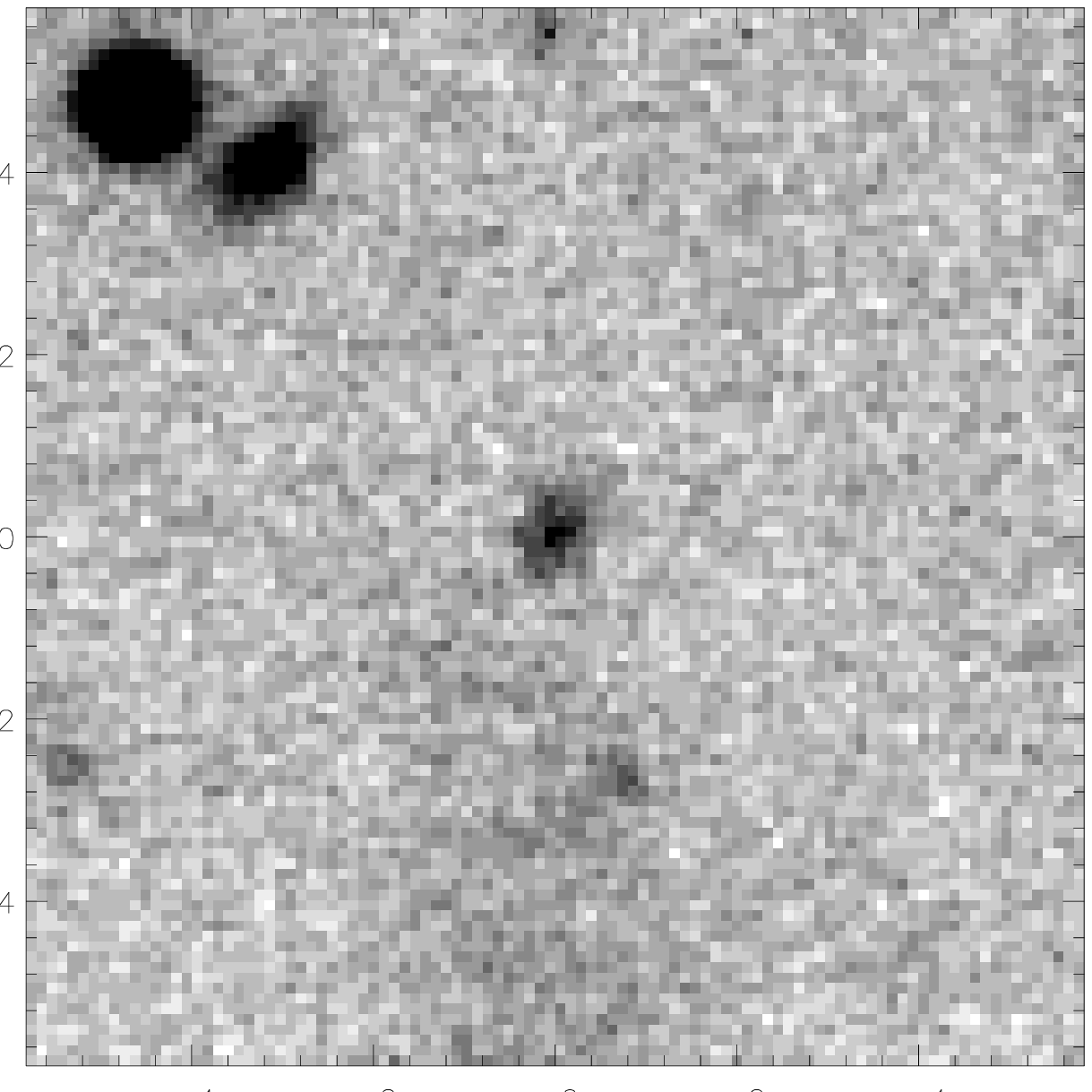,width=0.3\textwidth}&
\epsfig{file=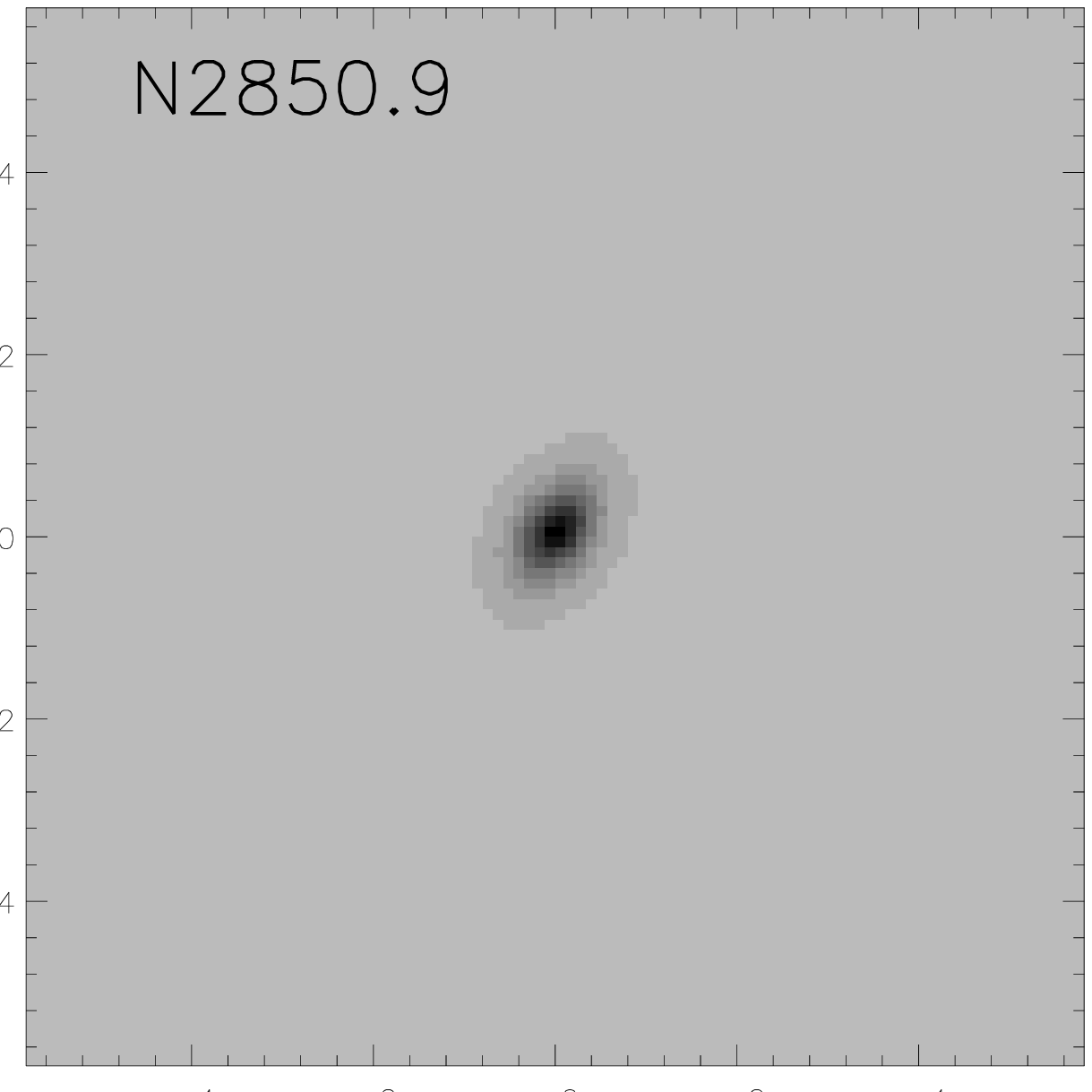,width=0.3\textwidth}&
\epsfig{file=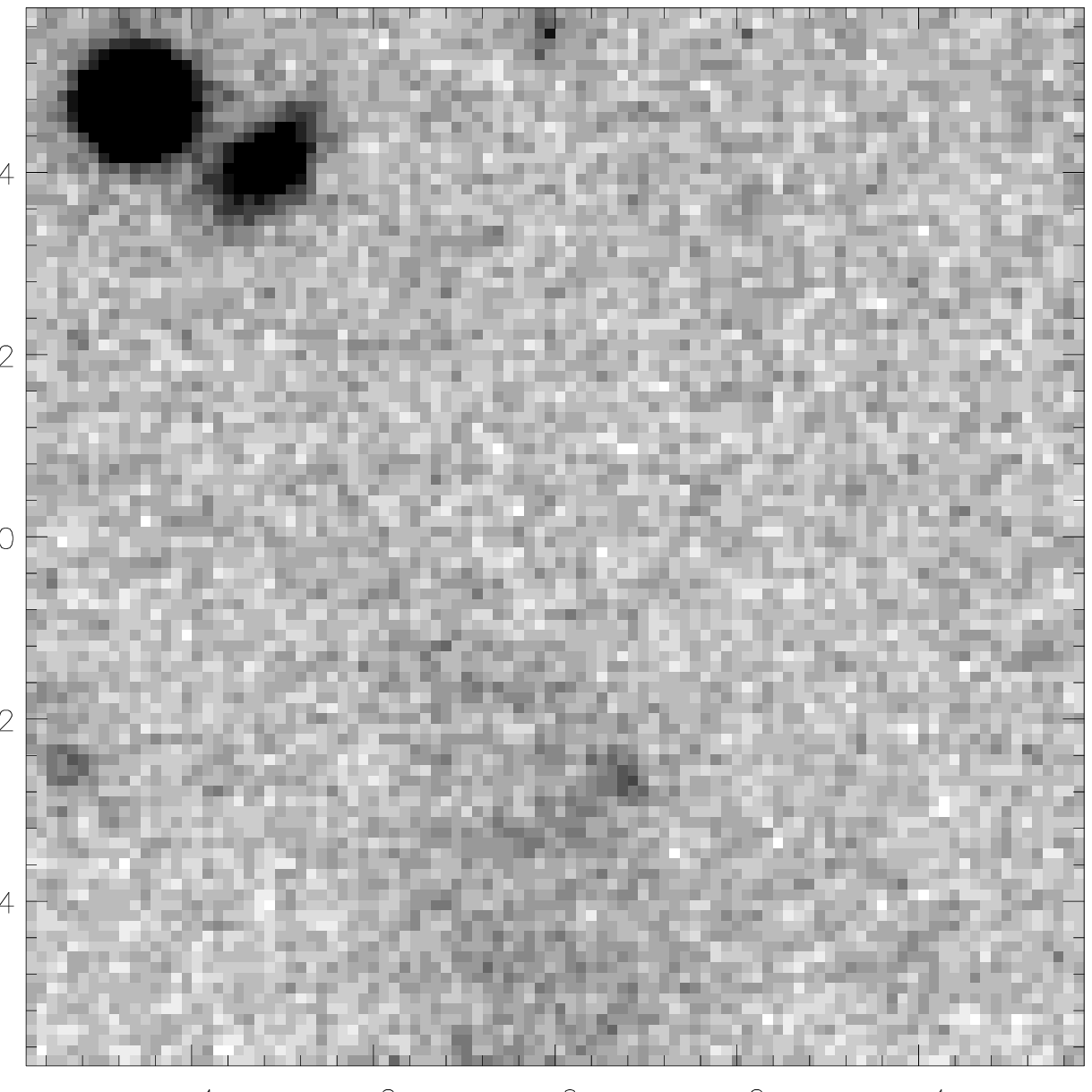,width=0.3\textwidth}\\
\\
\epsfig{file=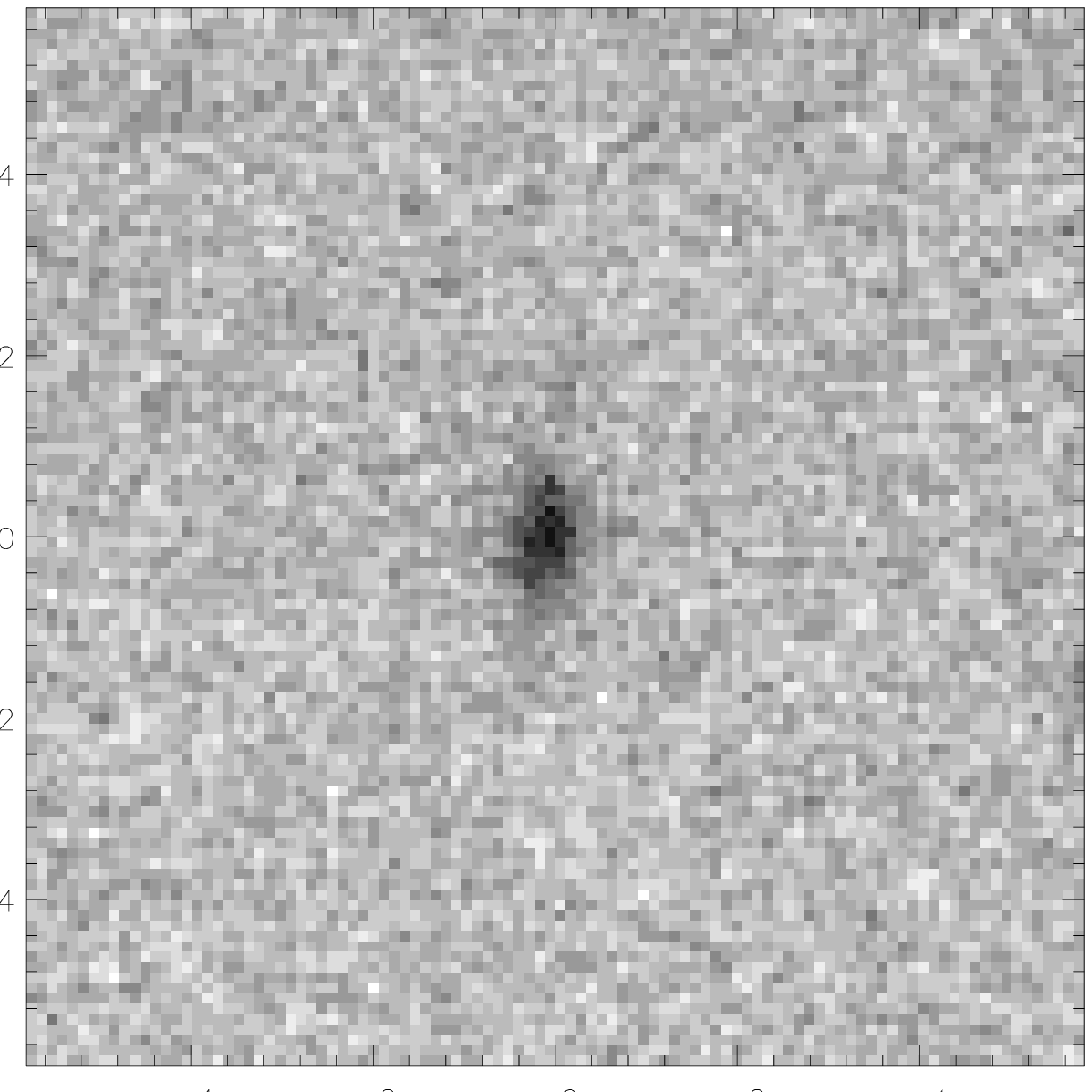,width=0.3\textwidth}&
\epsfig{file=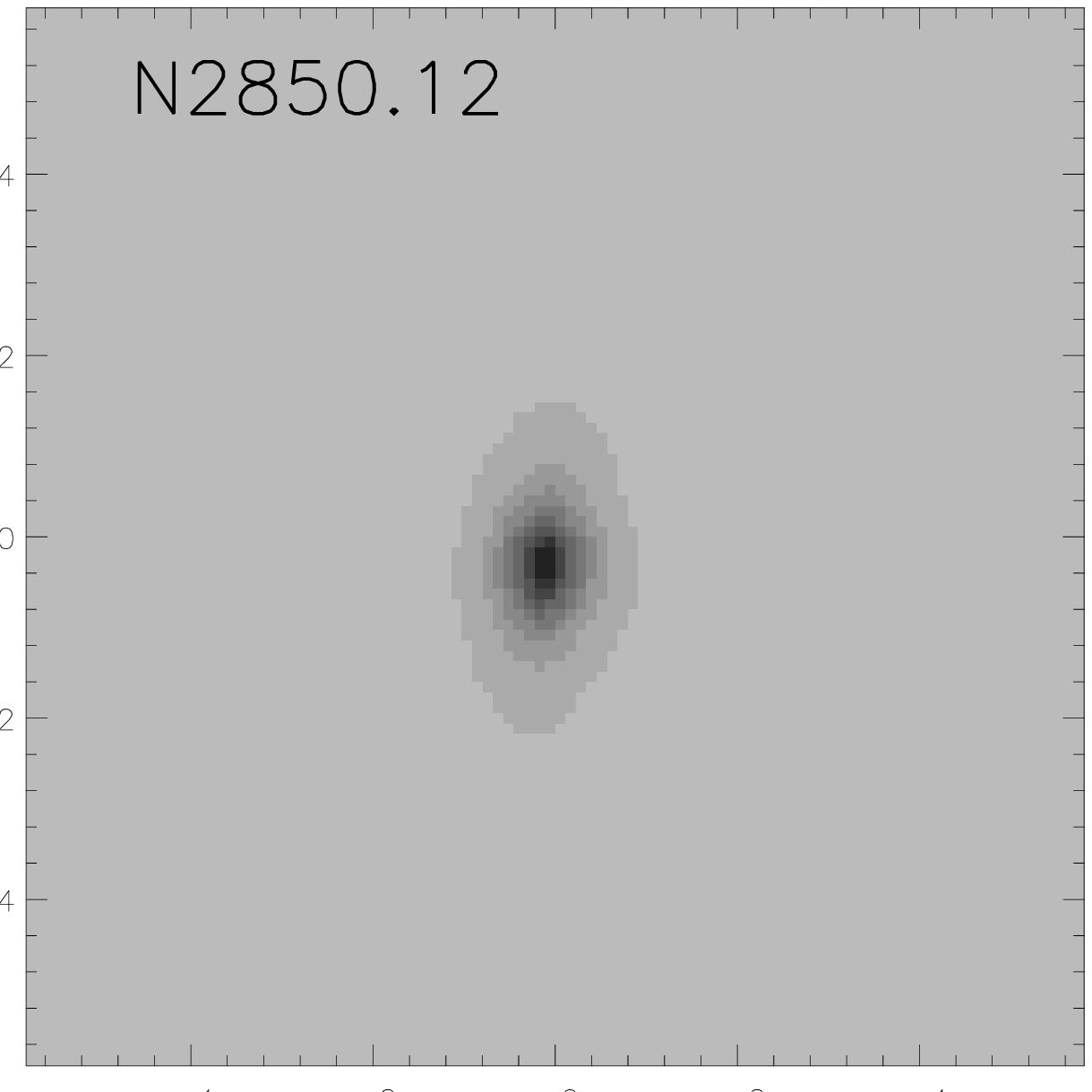,width=0.3\textwidth}&
\epsfig{file=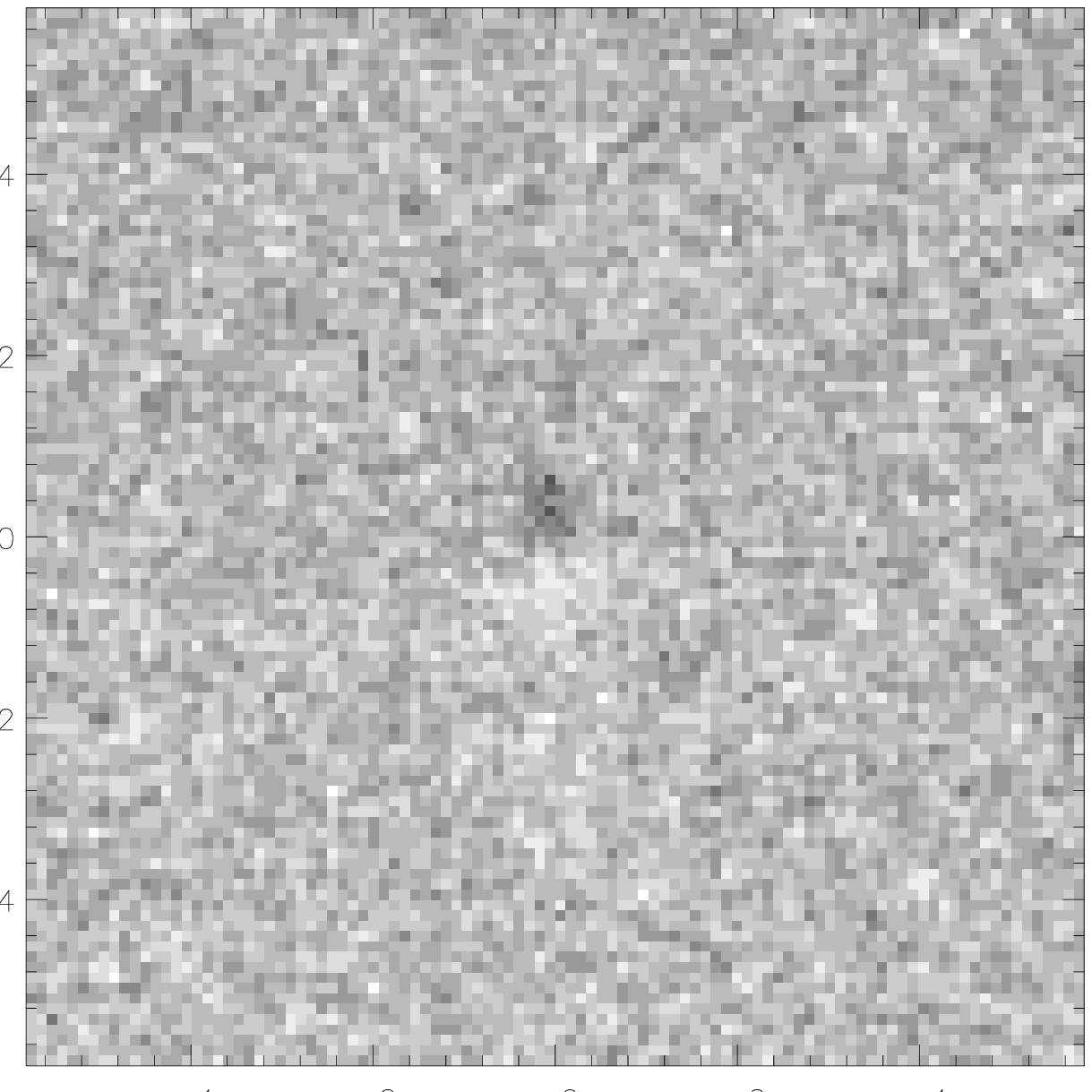,width=0.3\textwidth}\\
\end{tabular}
\addtocounter{figure}{-1}
\caption{- continued}
\label{geminimodel1}
\end{figure*}
\end{center}

\begin{center}
\begin{figure*}
\begin{tabular}{ccc}
\epsfig{file=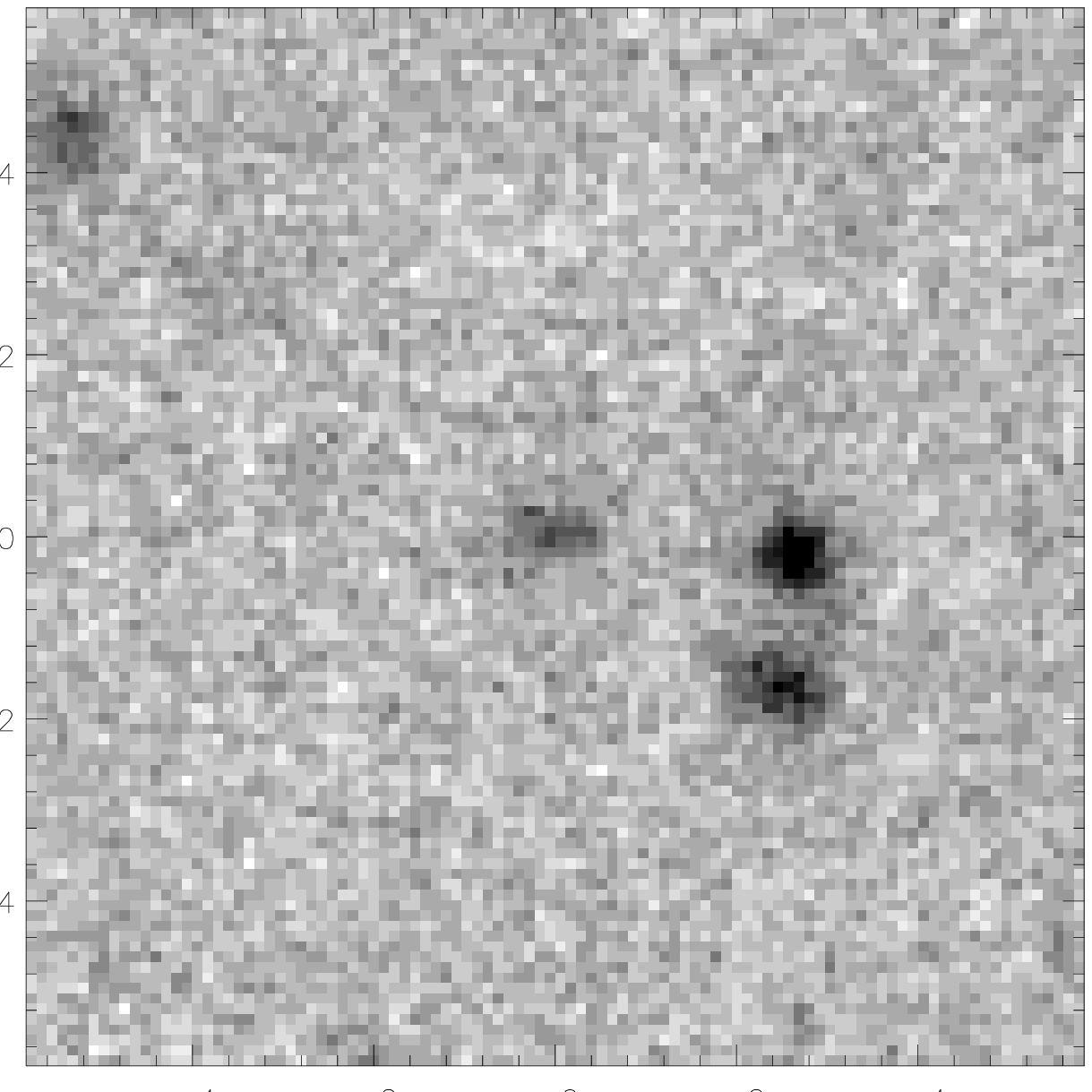,width=0.3\textwidth}&
\epsfig{file=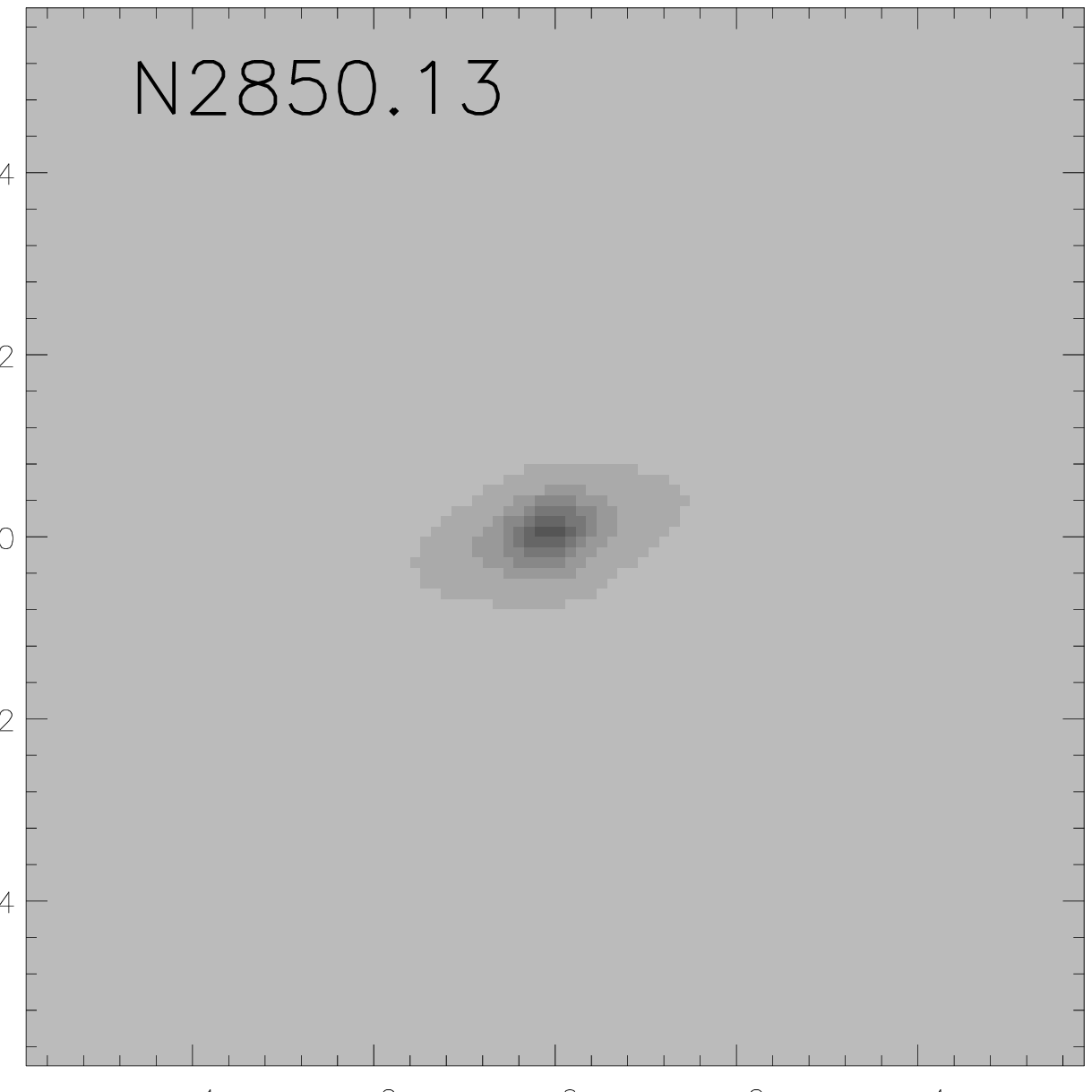,width=0.3\textwidth}&
\epsfig{file=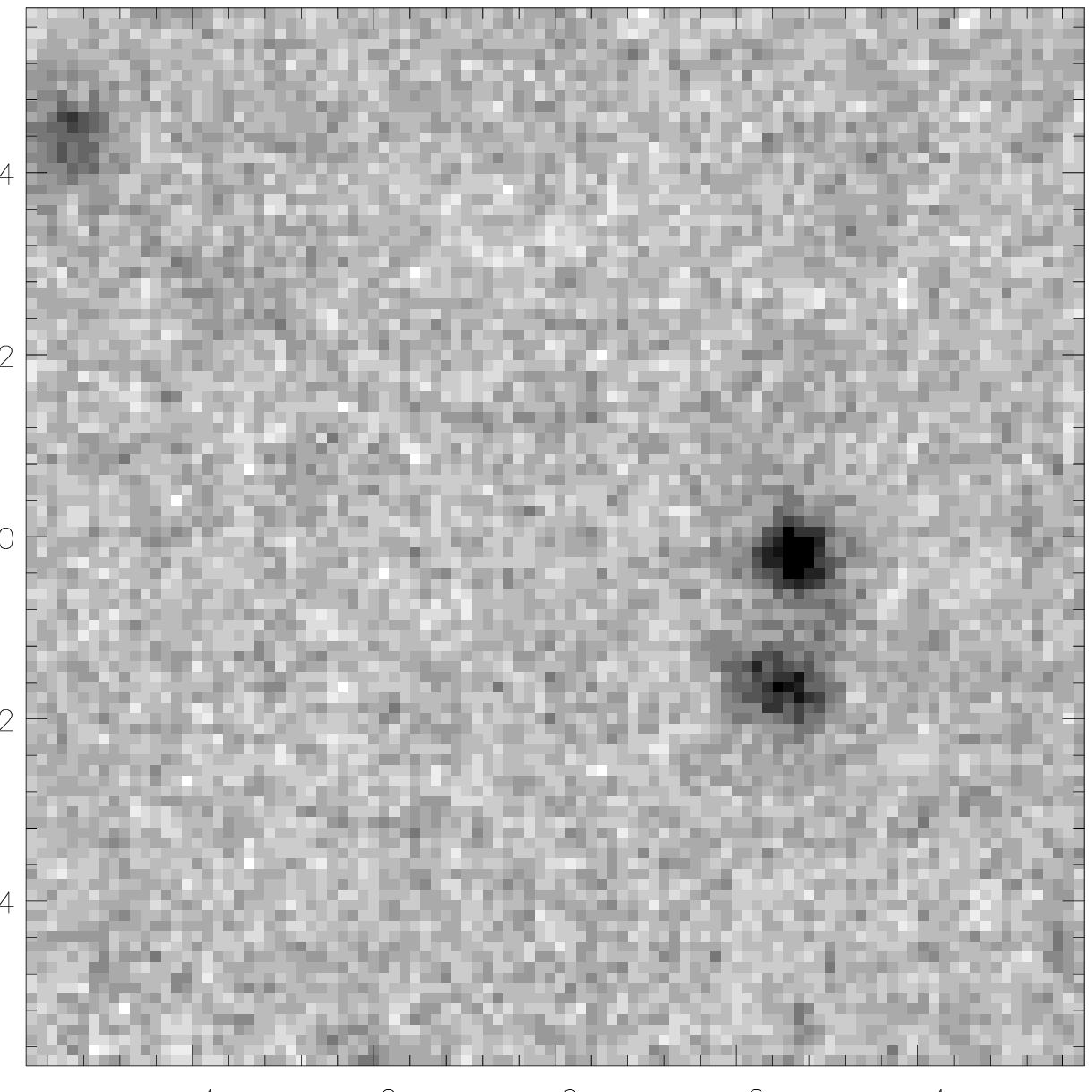,width=0.3\textwidth}\\
\\           
\epsfig{file=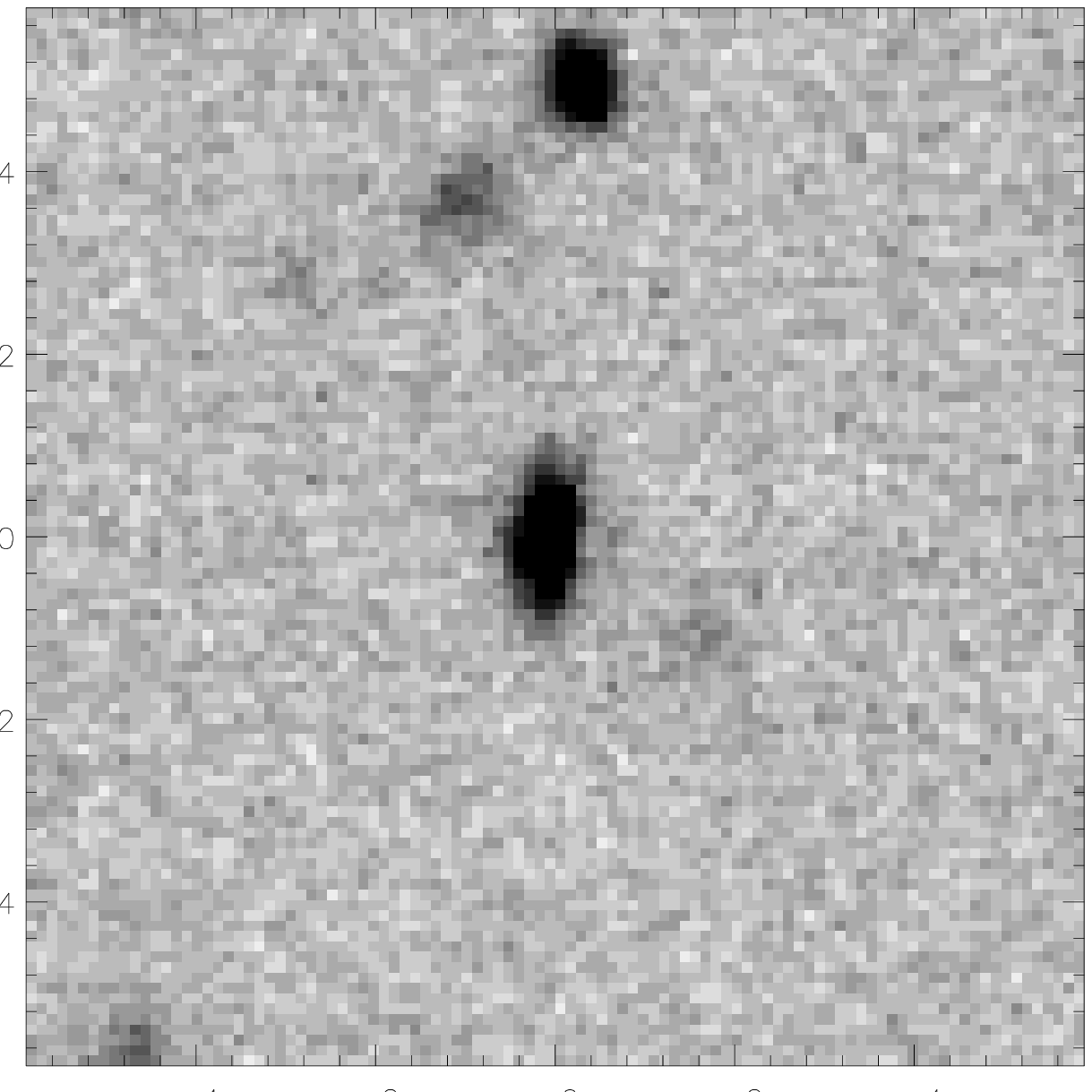,width=0.3\textwidth}&
\epsfig{file=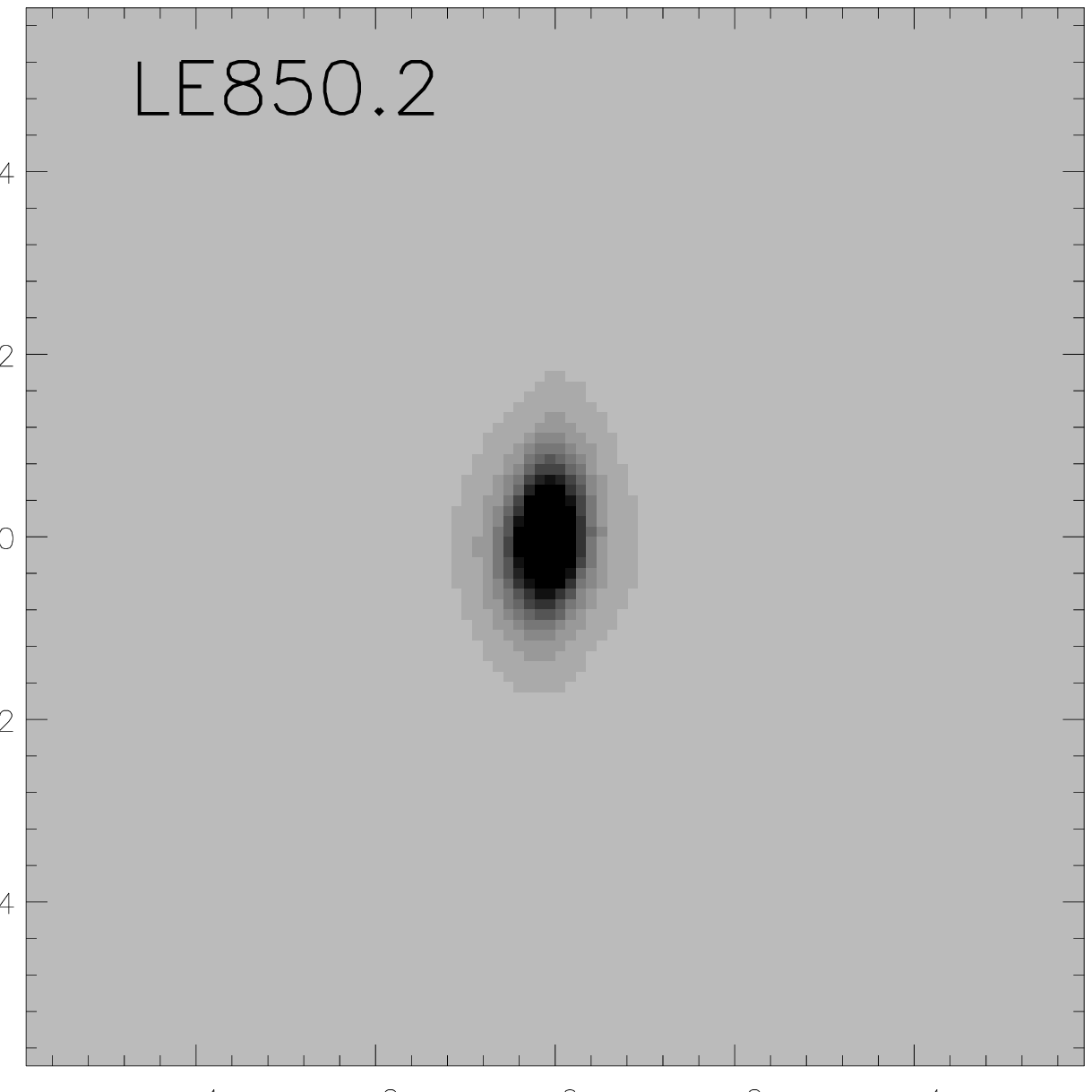,width=0.3\textwidth}&
\epsfig{file=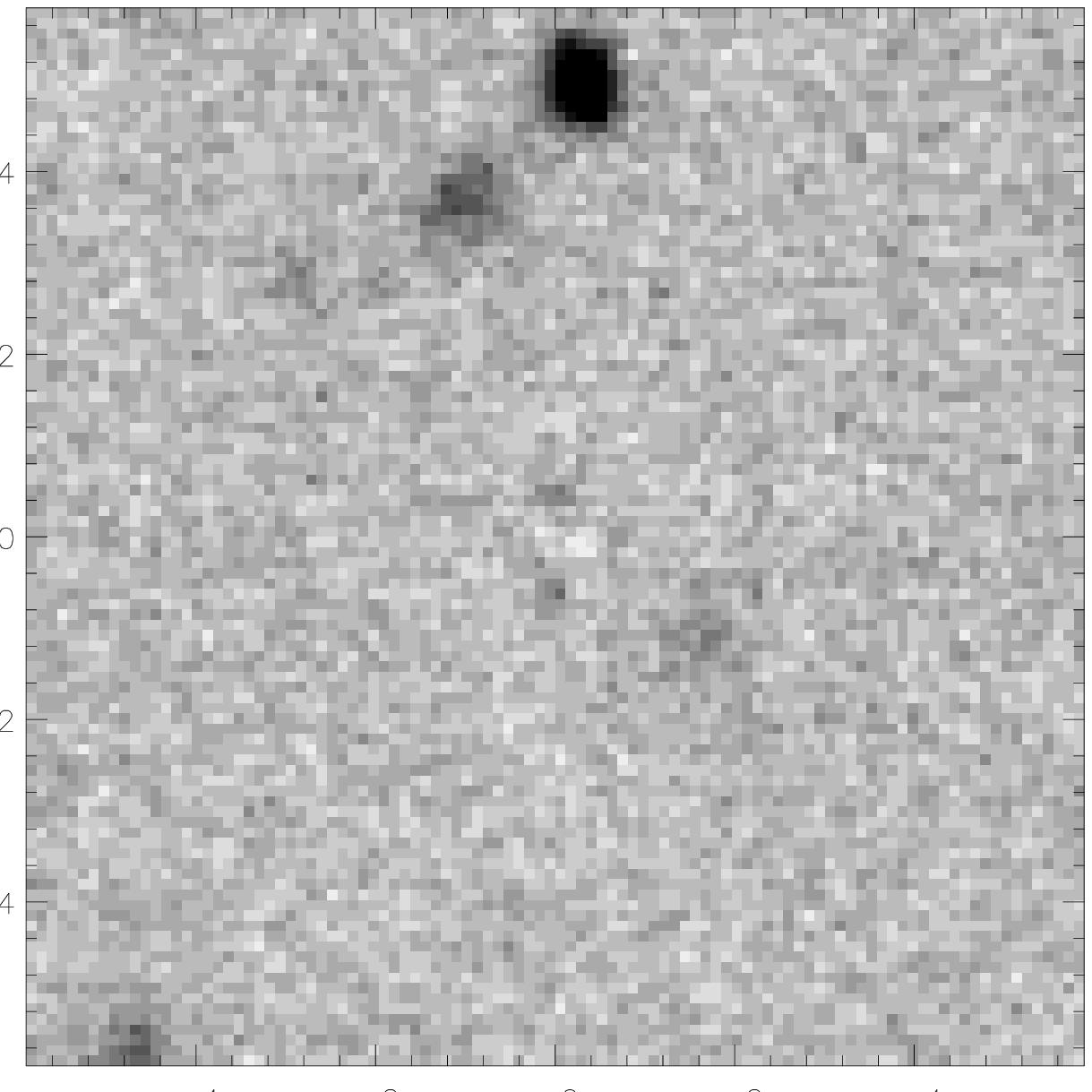,width=0.3\textwidth}\\
\\           
\epsfig{file=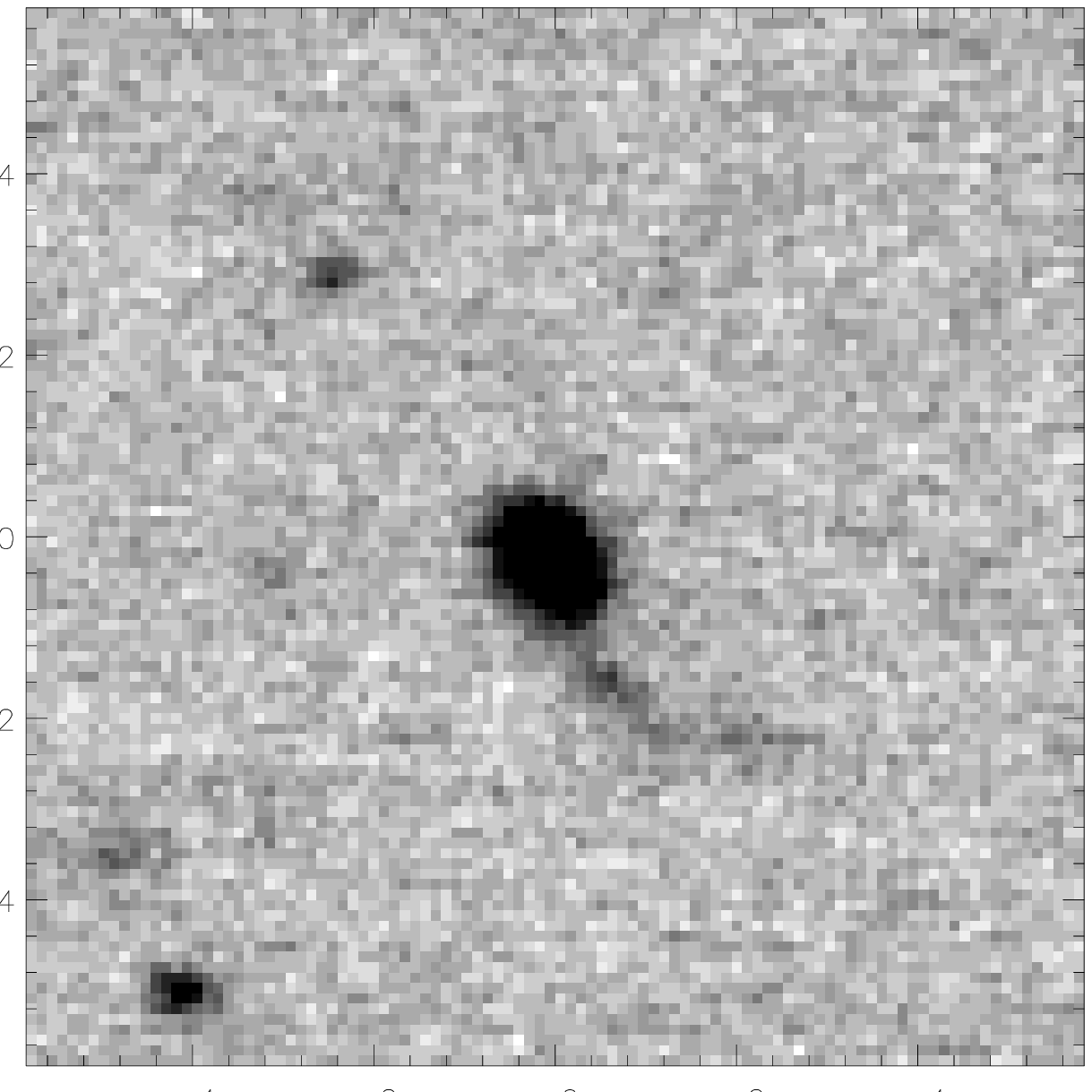,width=0.3\textwidth}&
\epsfig{file=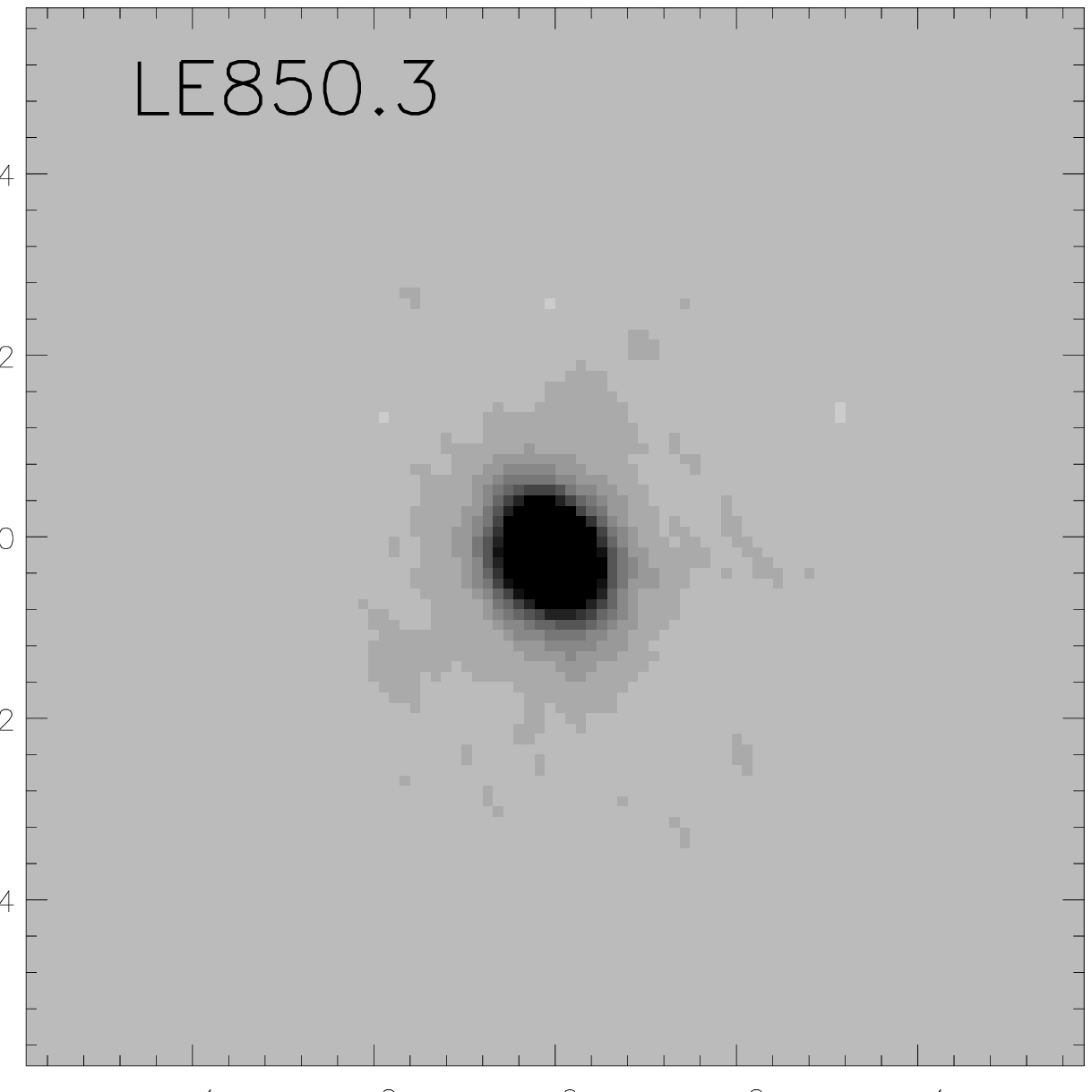,width=0.3\textwidth}&
\epsfig{file=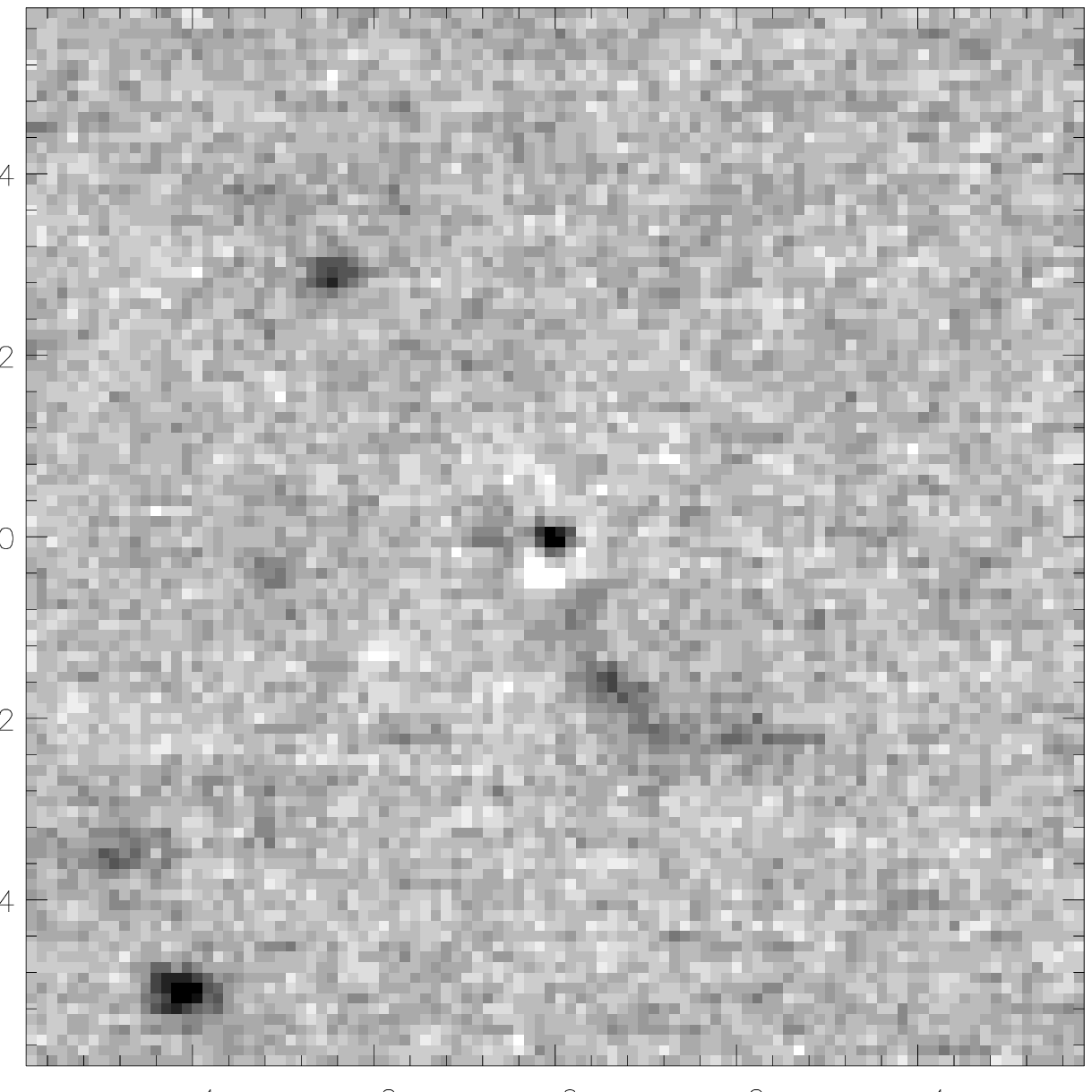,width=0.3\textwidth}\\
\\           
\epsfig{file=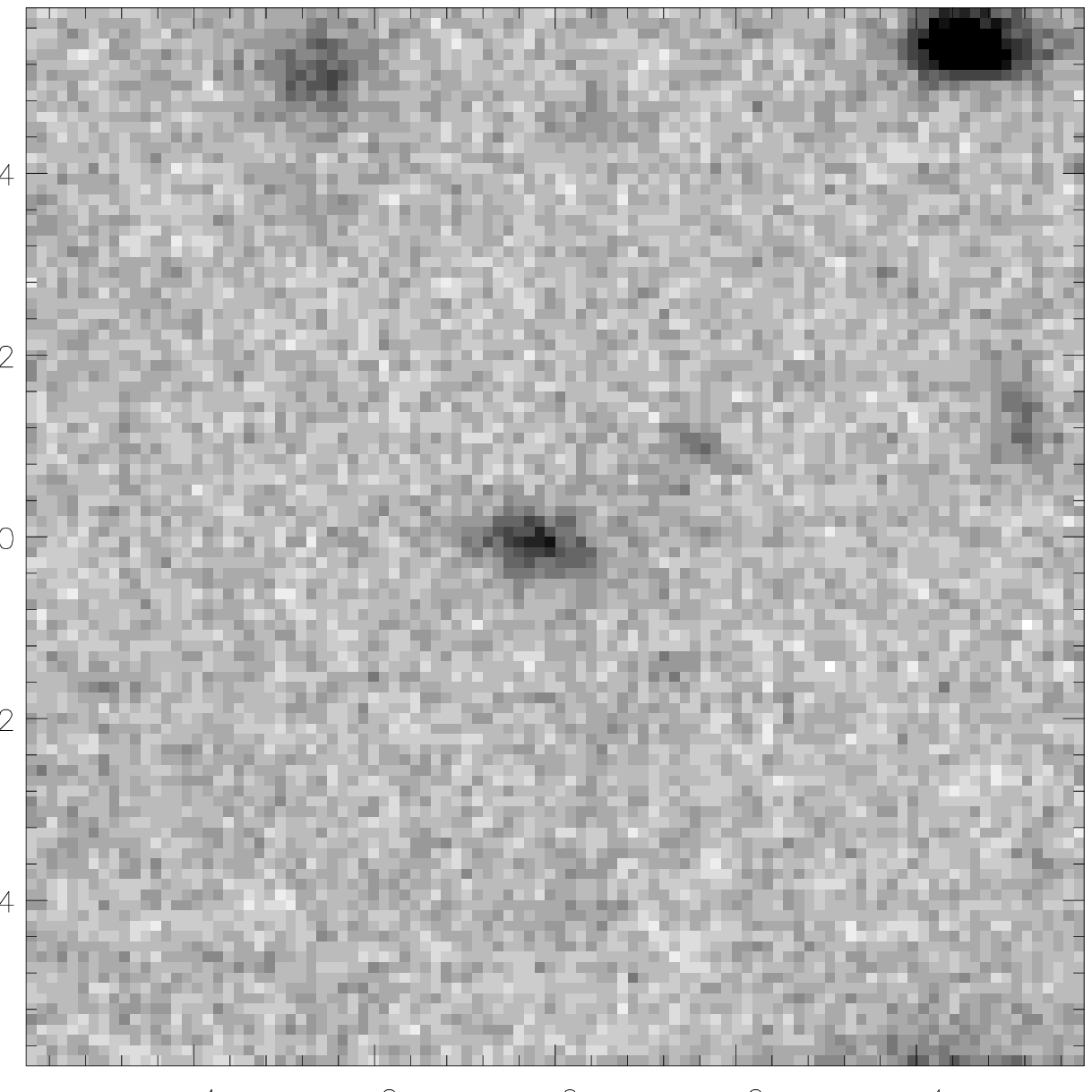,width=0.3\textwidth}&
\epsfig{file=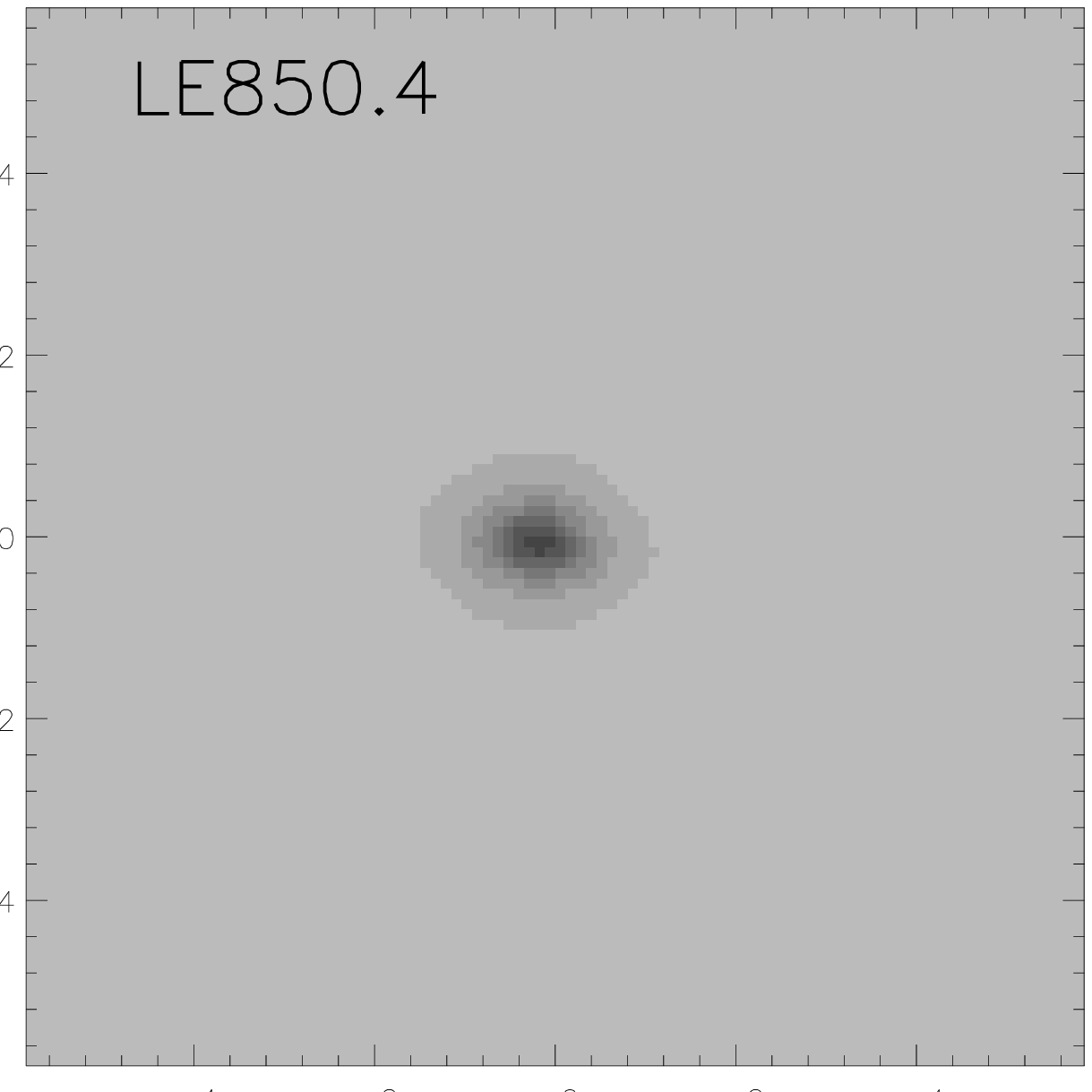,width=0.3\textwidth}&
\epsfig{file=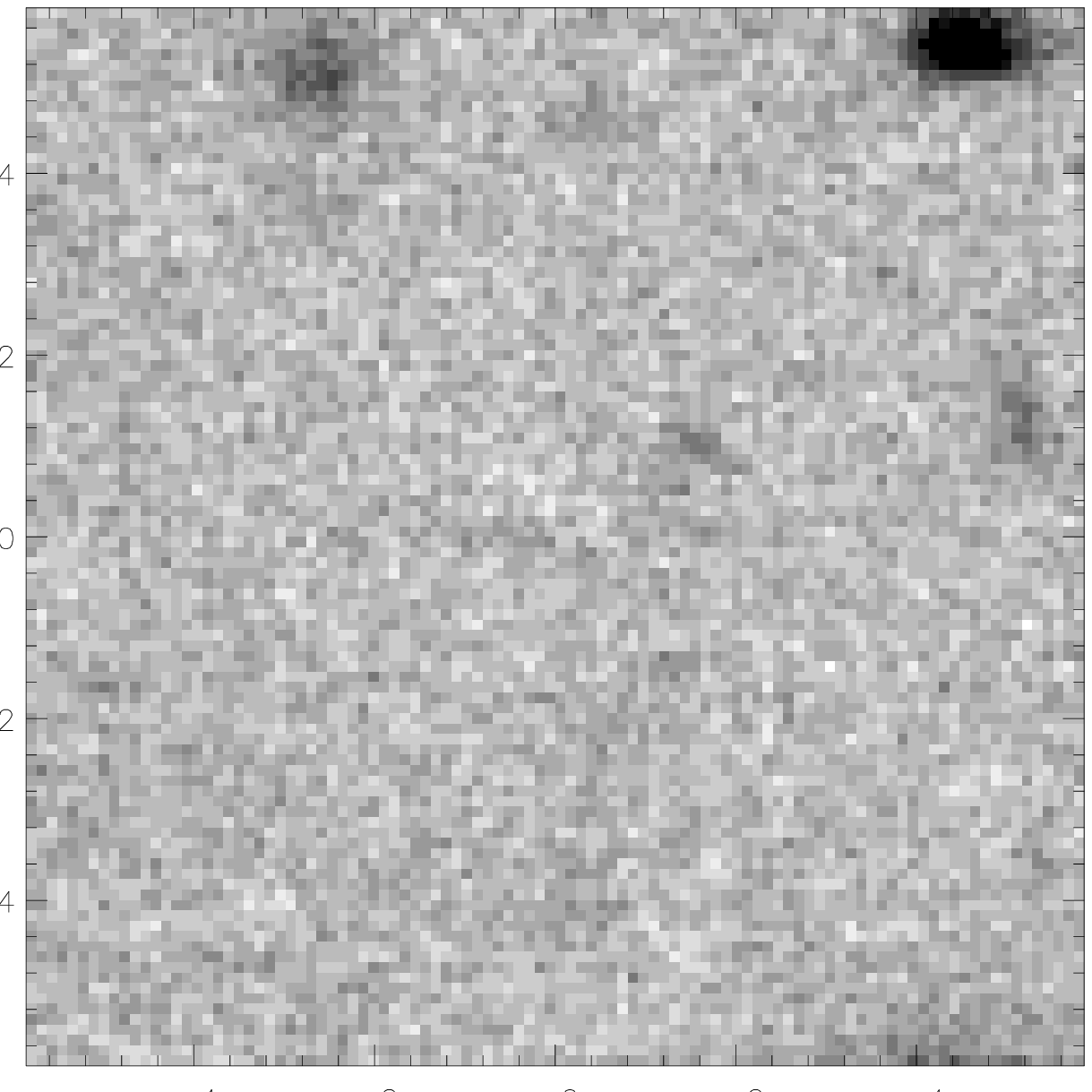,width=0.3\textwidth}\\
\end{tabular}
\addtocounter{figure}{-1}
\caption{- continued}
\label{geminimodel1}
\end{figure*}
\end{center}

\begin{center}
\begin{figure*}
\begin{tabular}{ccc}
\epsfig{file=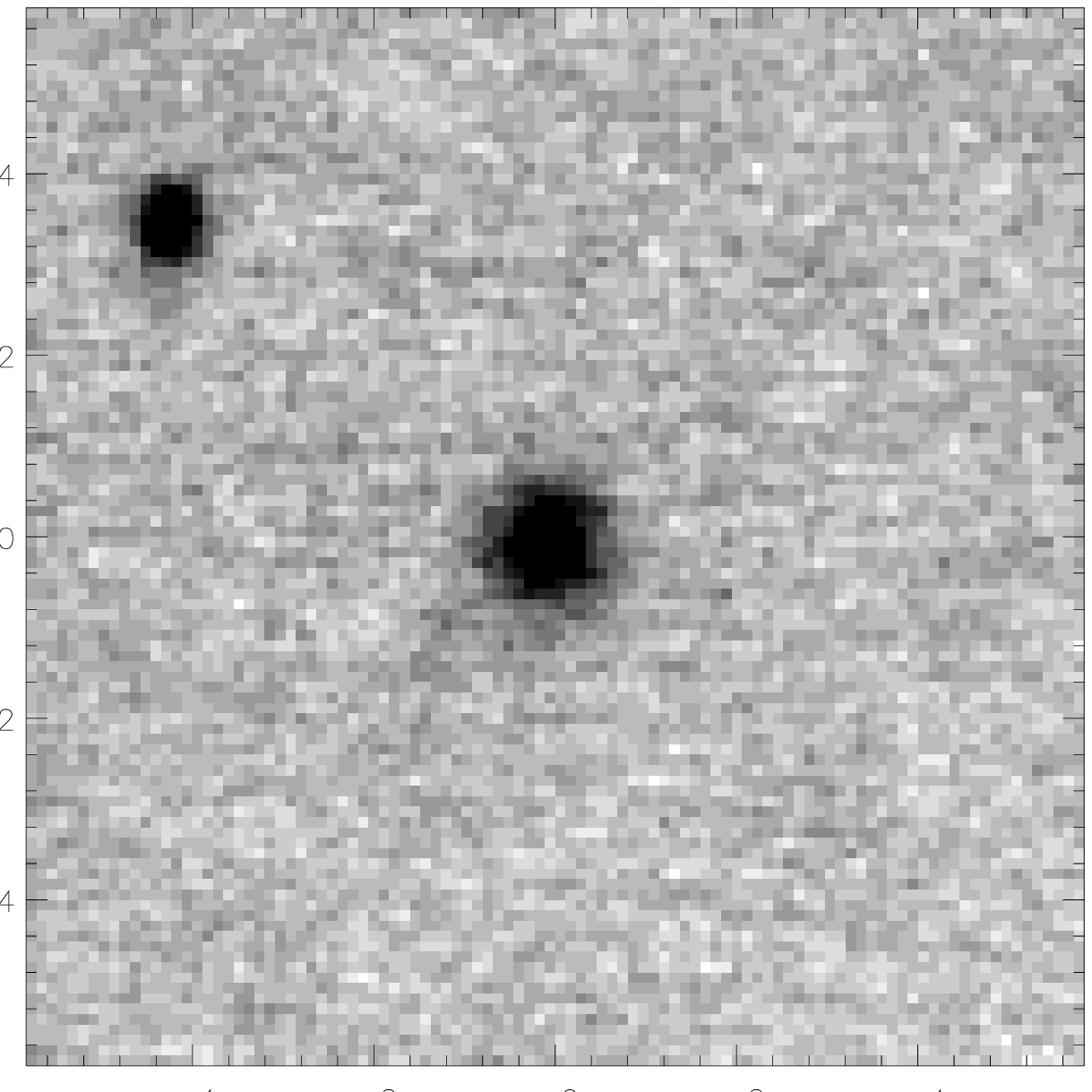,width=0.3\textwidth}&
\epsfig{file=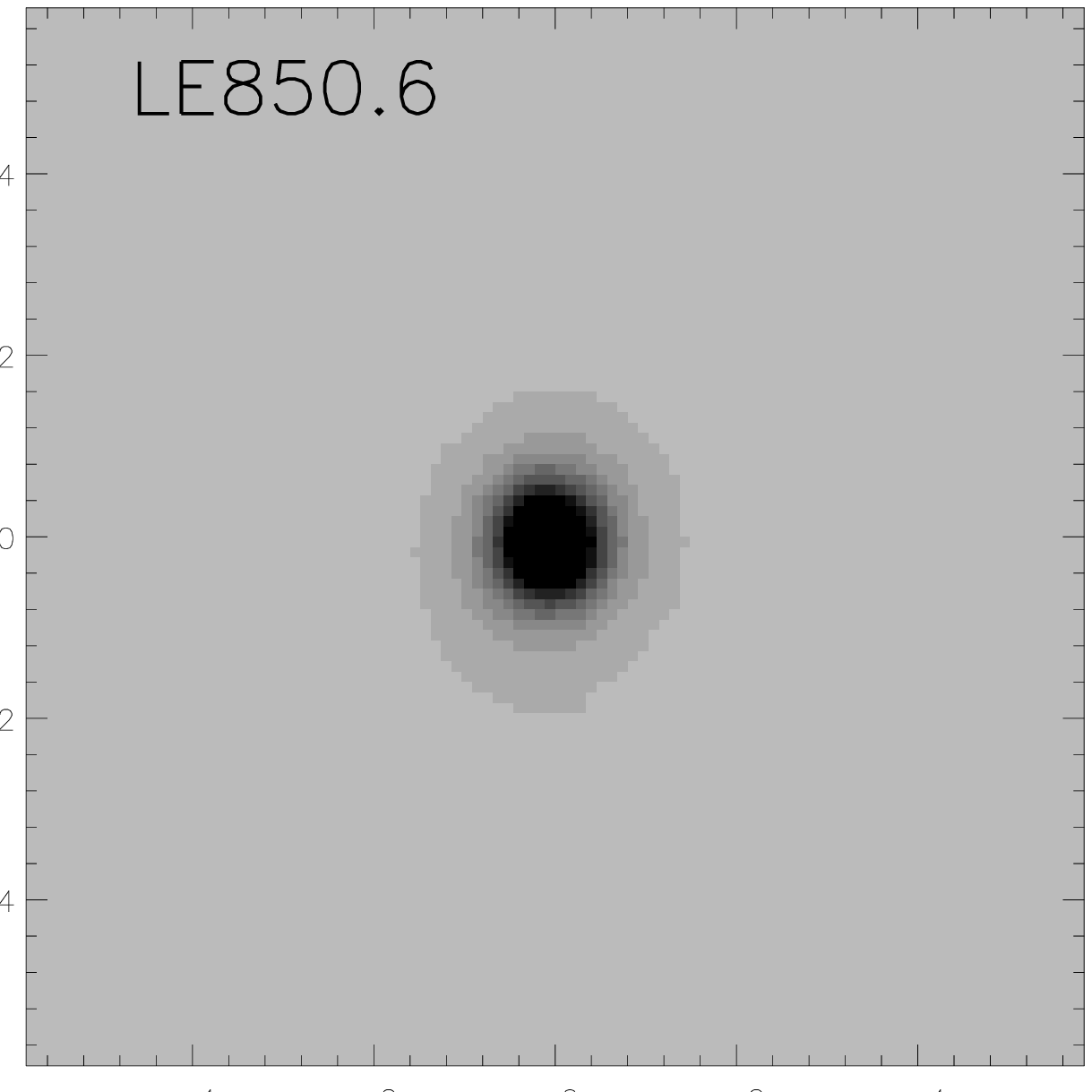,width=0.3\textwidth}&
\epsfig{file=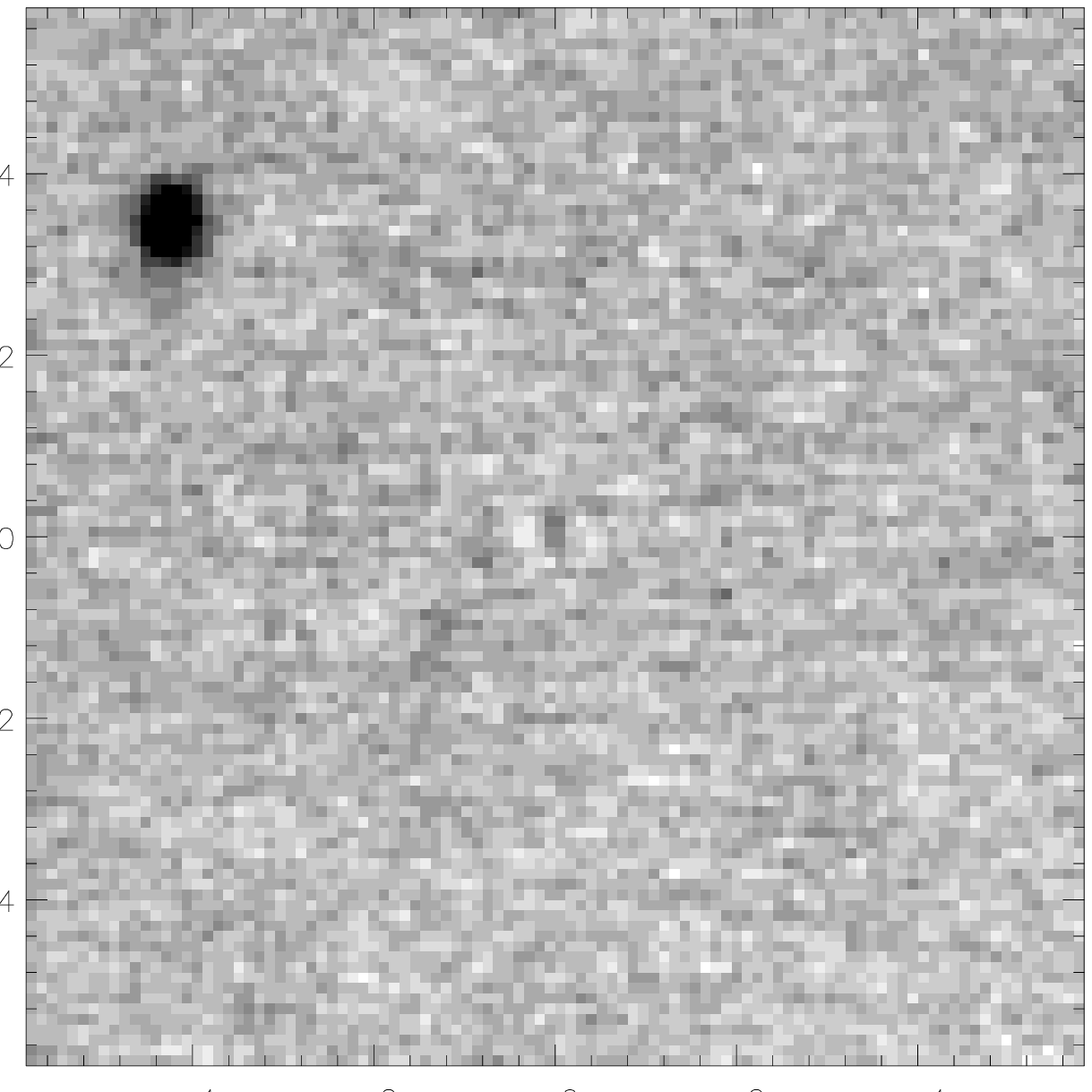,width=0.3\textwidth}\\
\\           
\epsfig{file=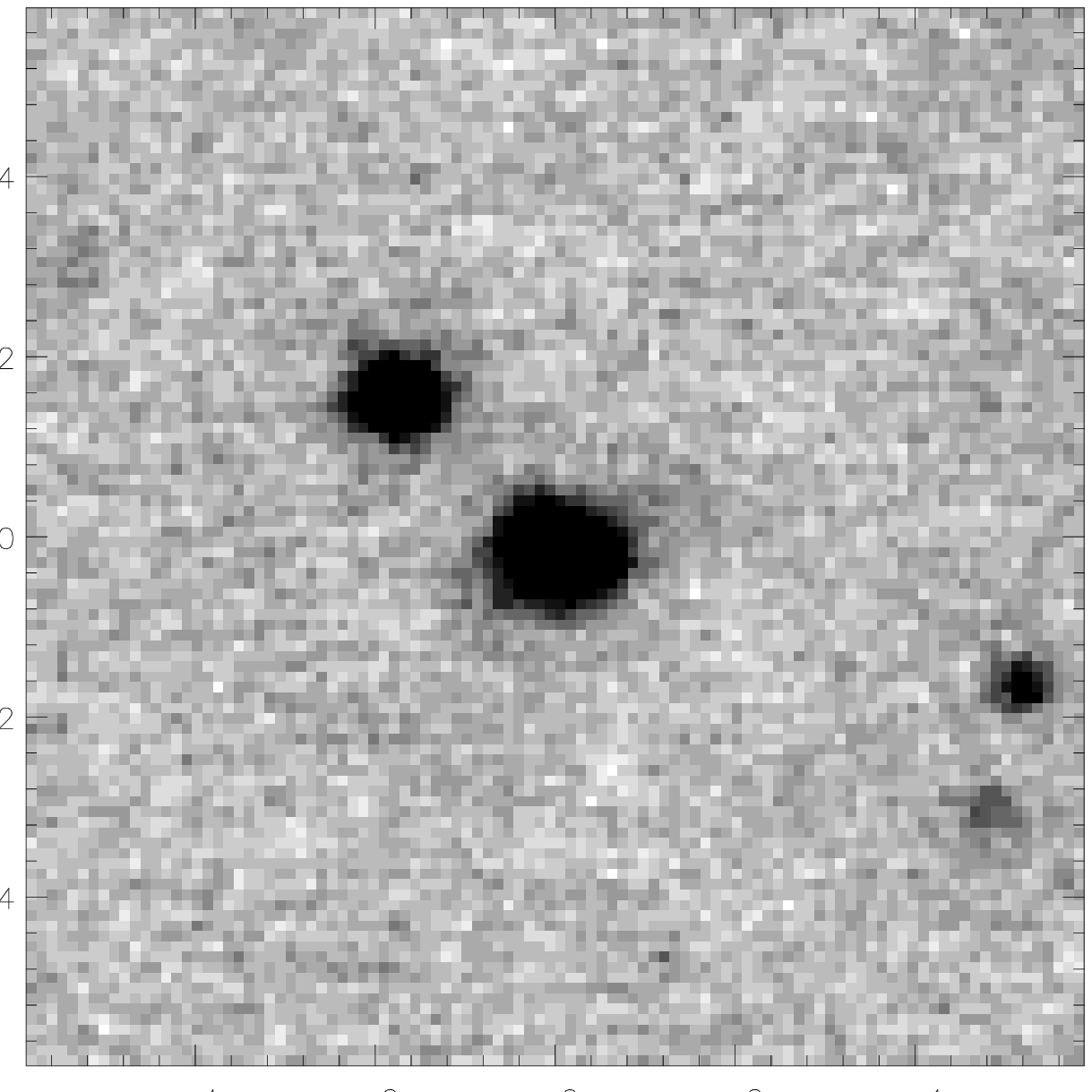,width=0.3\textwidth}&
\epsfig{file=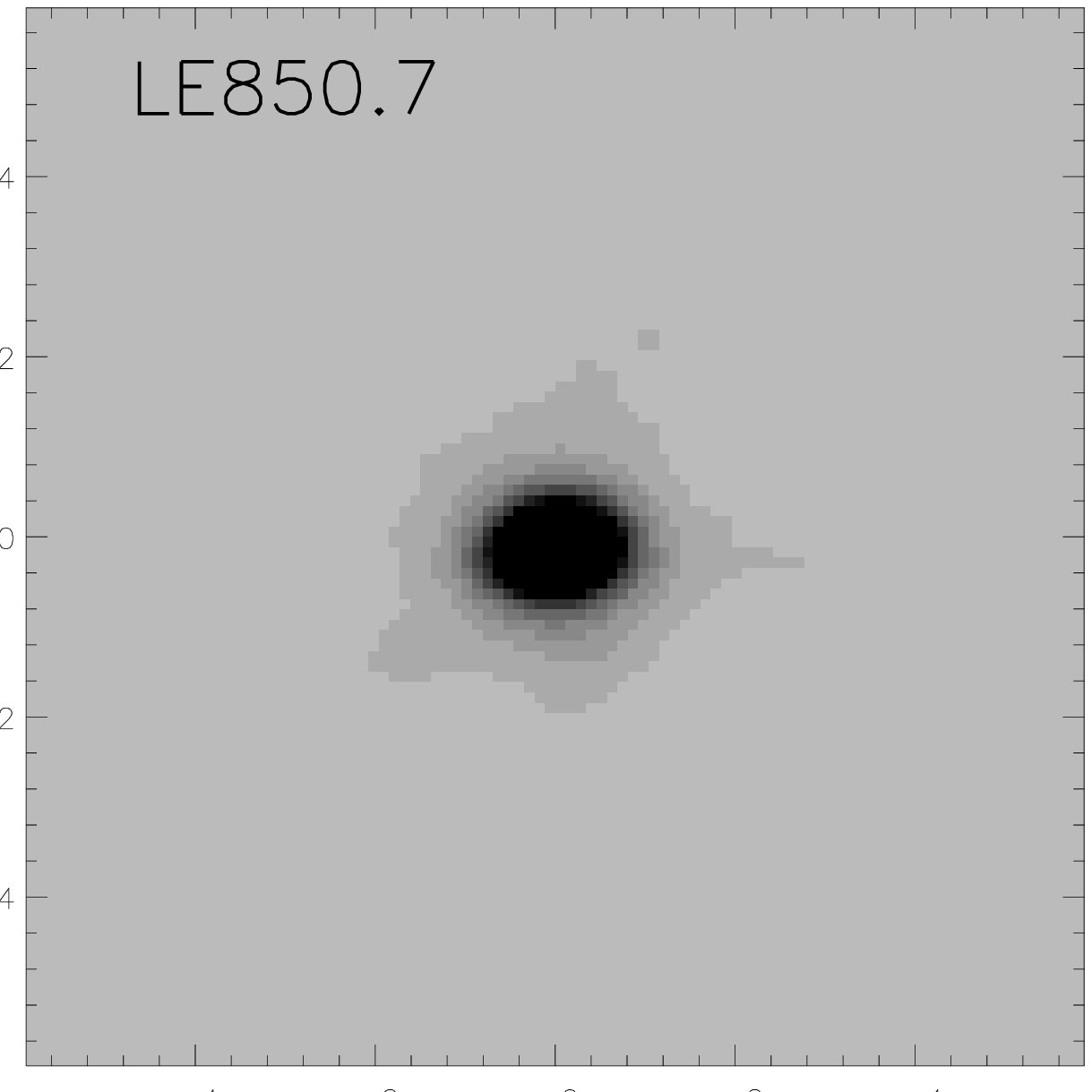,width=0.3\textwidth}&
\epsfig{file=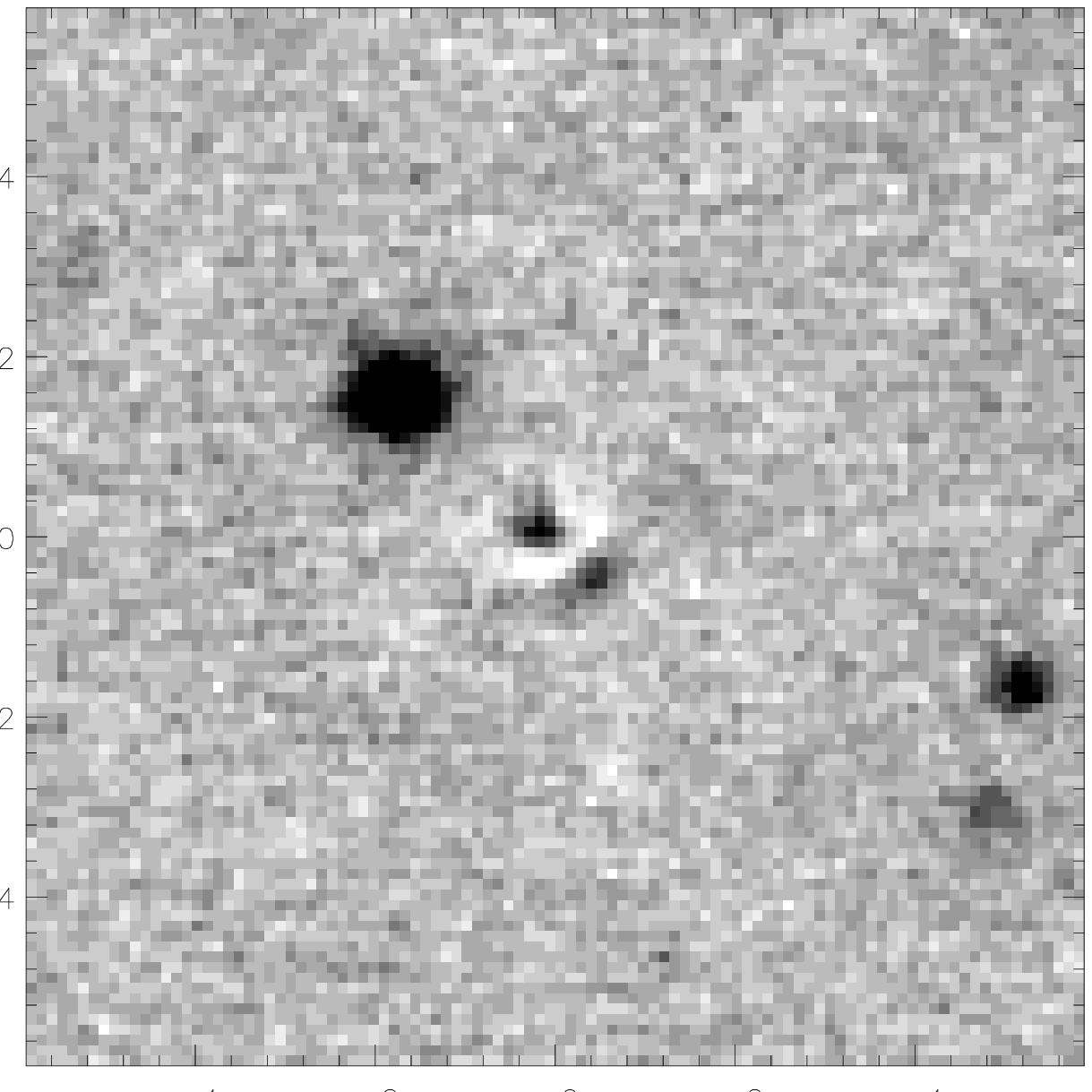,width=0.3\textwidth}\\
\\           
\epsfig{file=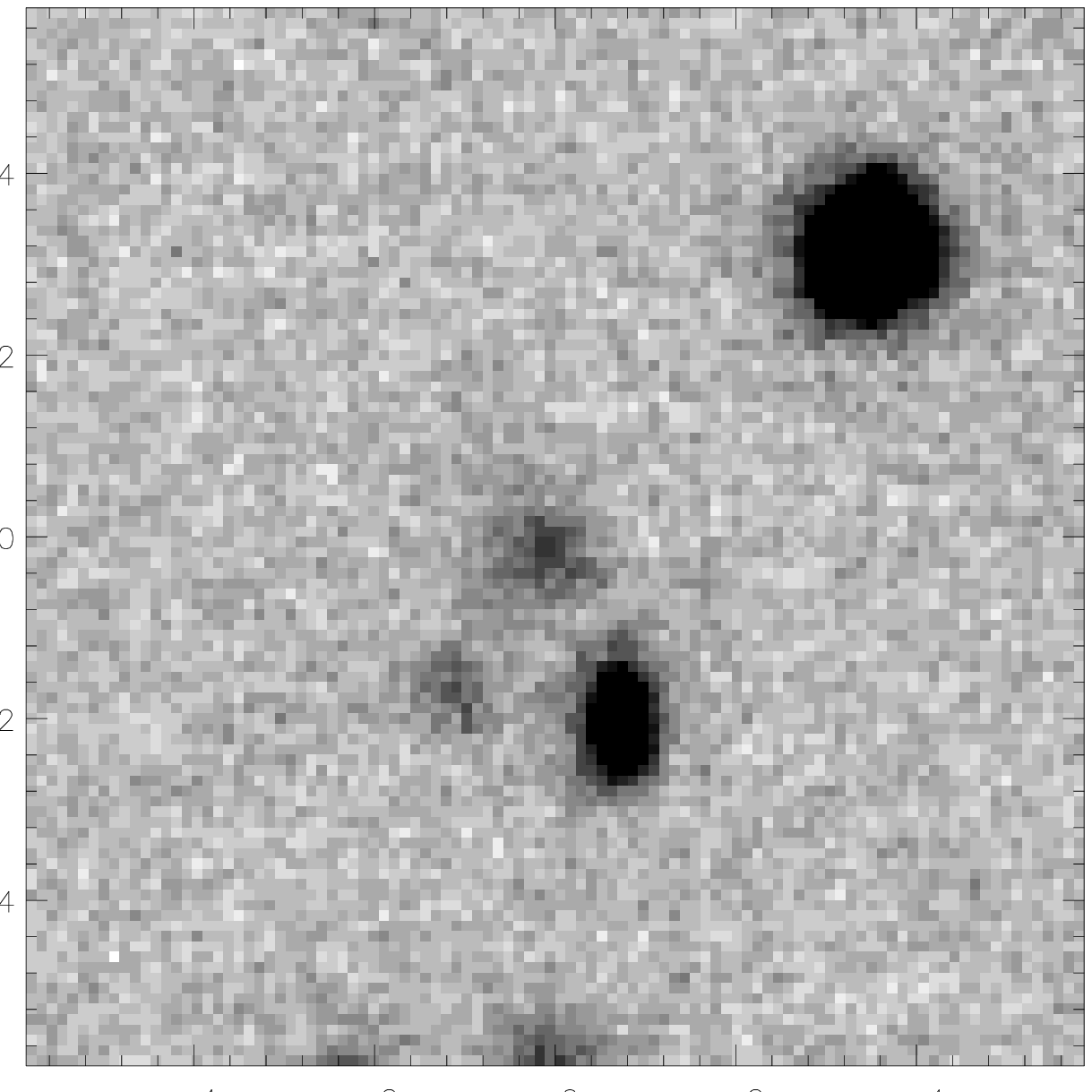,width=0.3\textwidth}&
\epsfig{file=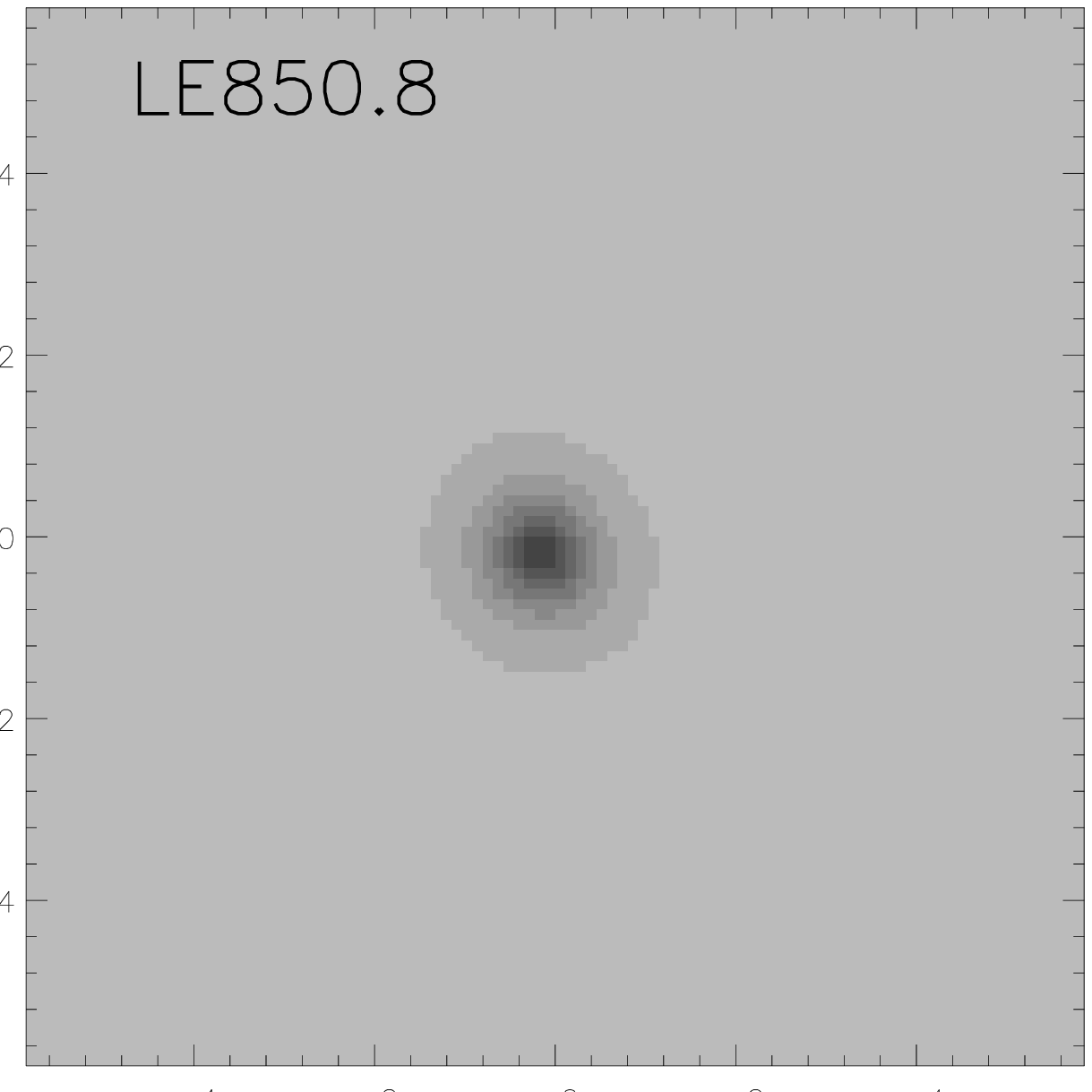,width=0.3\textwidth}&
\epsfig{file=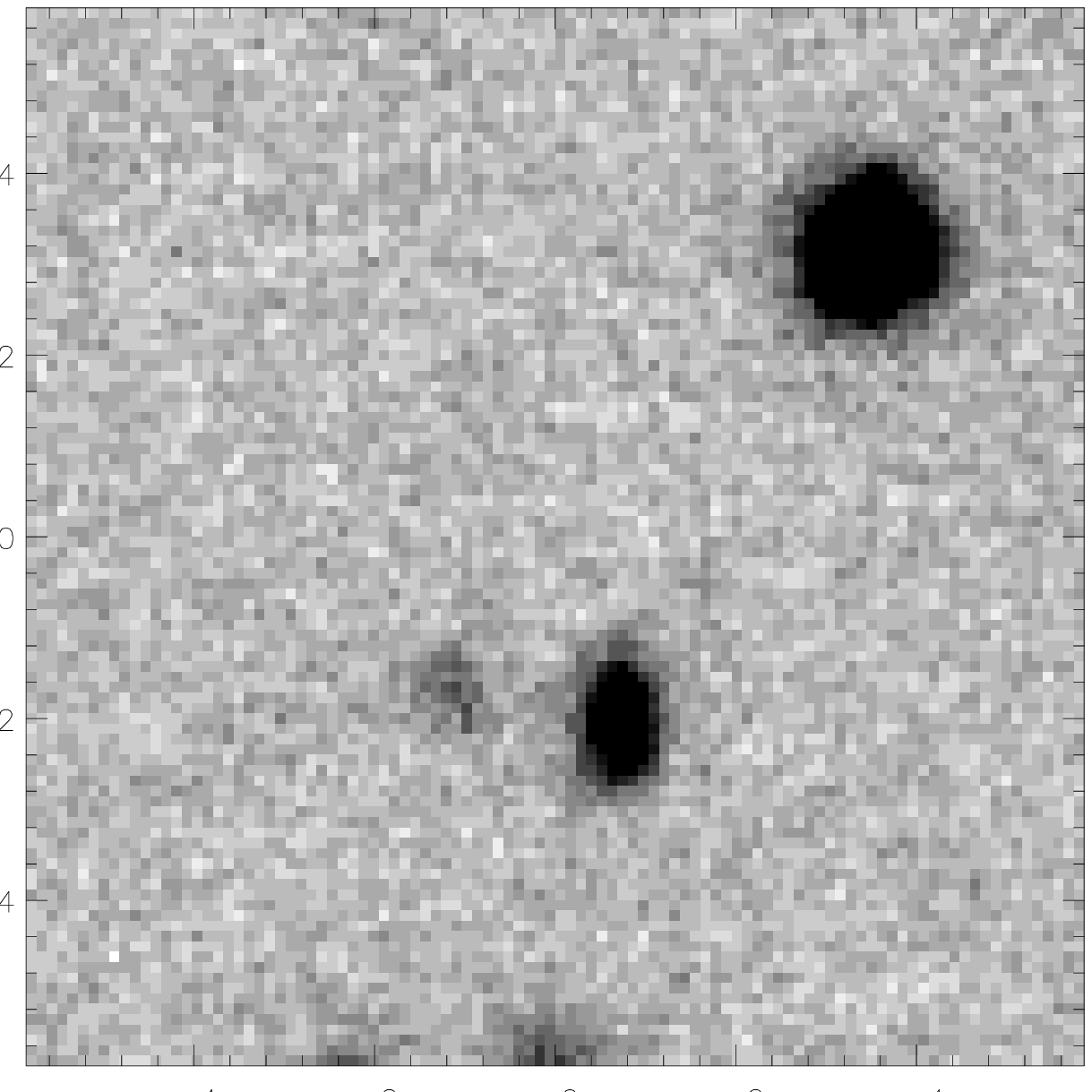,width=0.3\textwidth}\\
\end{tabular}
\addtocounter{figure}{-1}
\caption{- continued}
\label{geminimodel1}
\end{figure*}
\end{center}

\begin{figure*}
\begin{center}
\begin{tabular}{ccc}
\epsfig{file=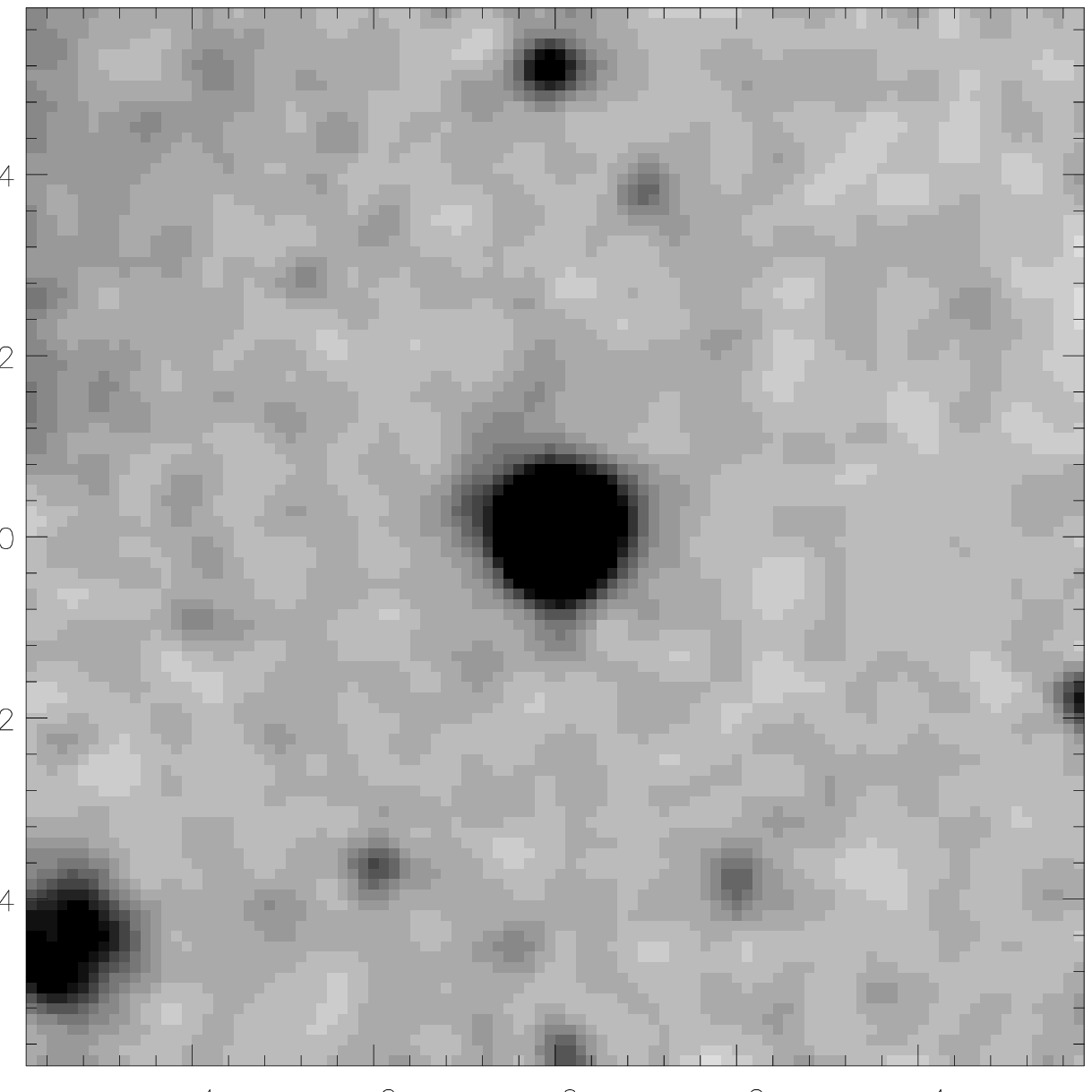,width=0.3\textwidth}&
\epsfig{file=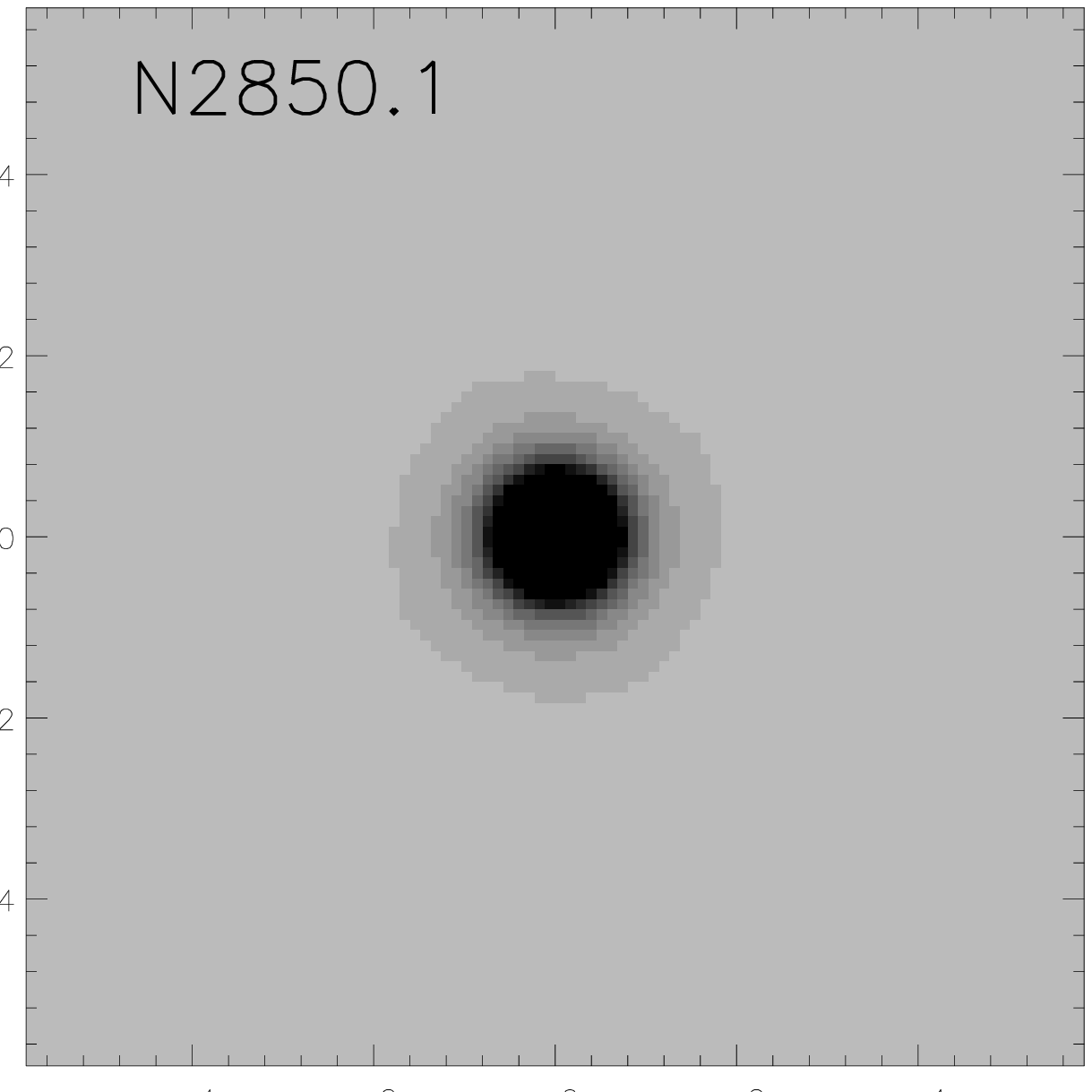,width=0.3\textwidth}&
\epsfig{file=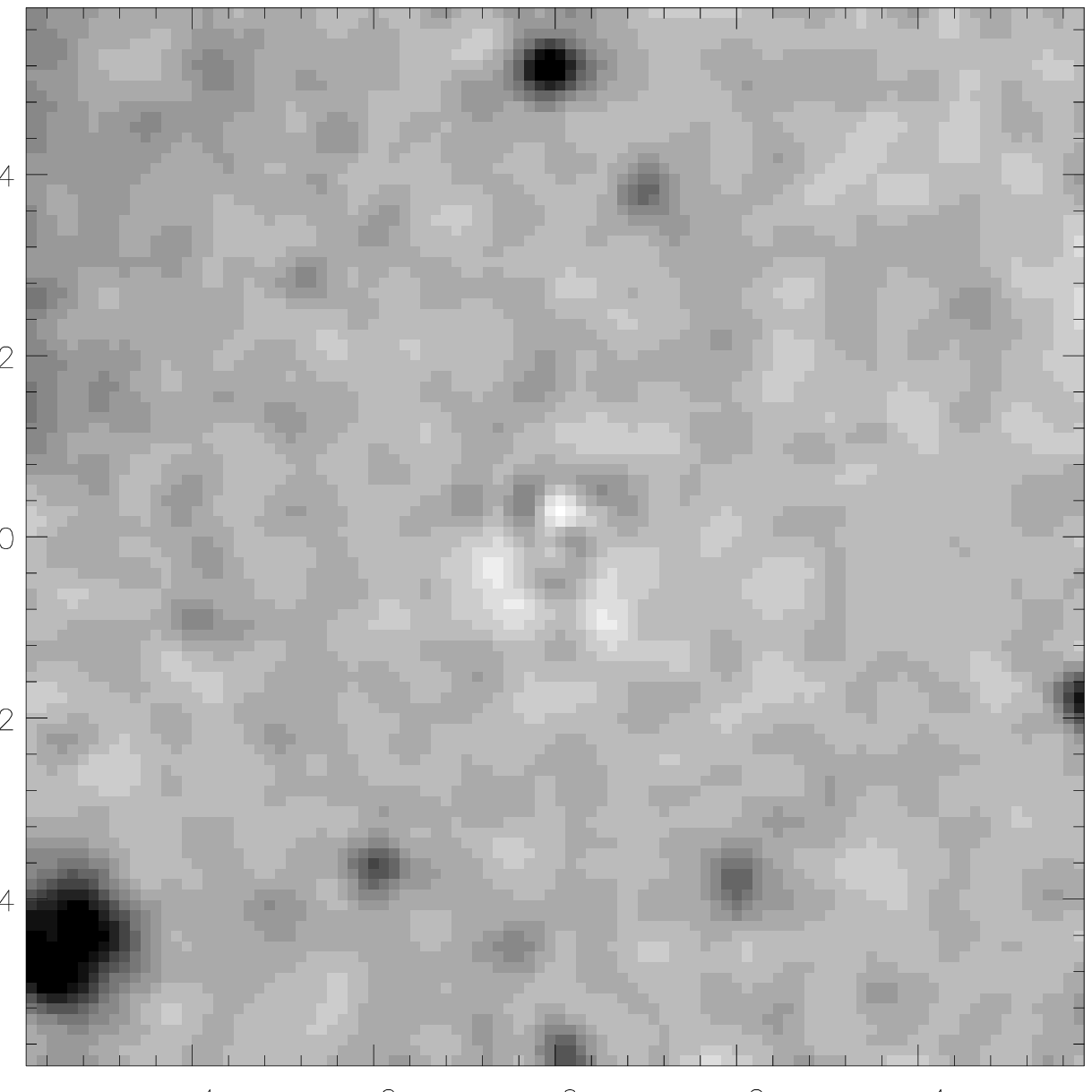,width=0.3\textwidth}\\
\\
\epsfig{file=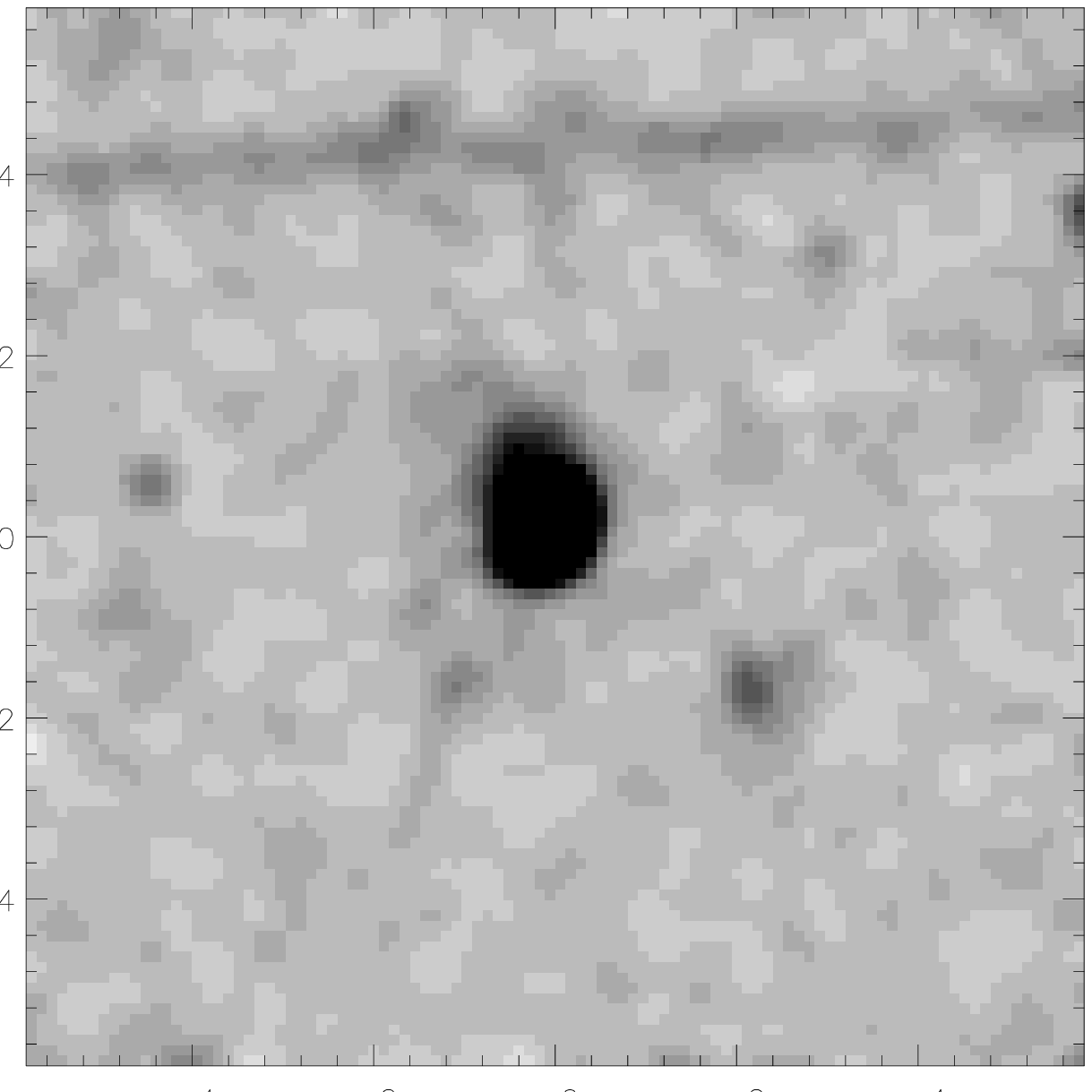,width=0.3\textwidth}&
\epsfig{file=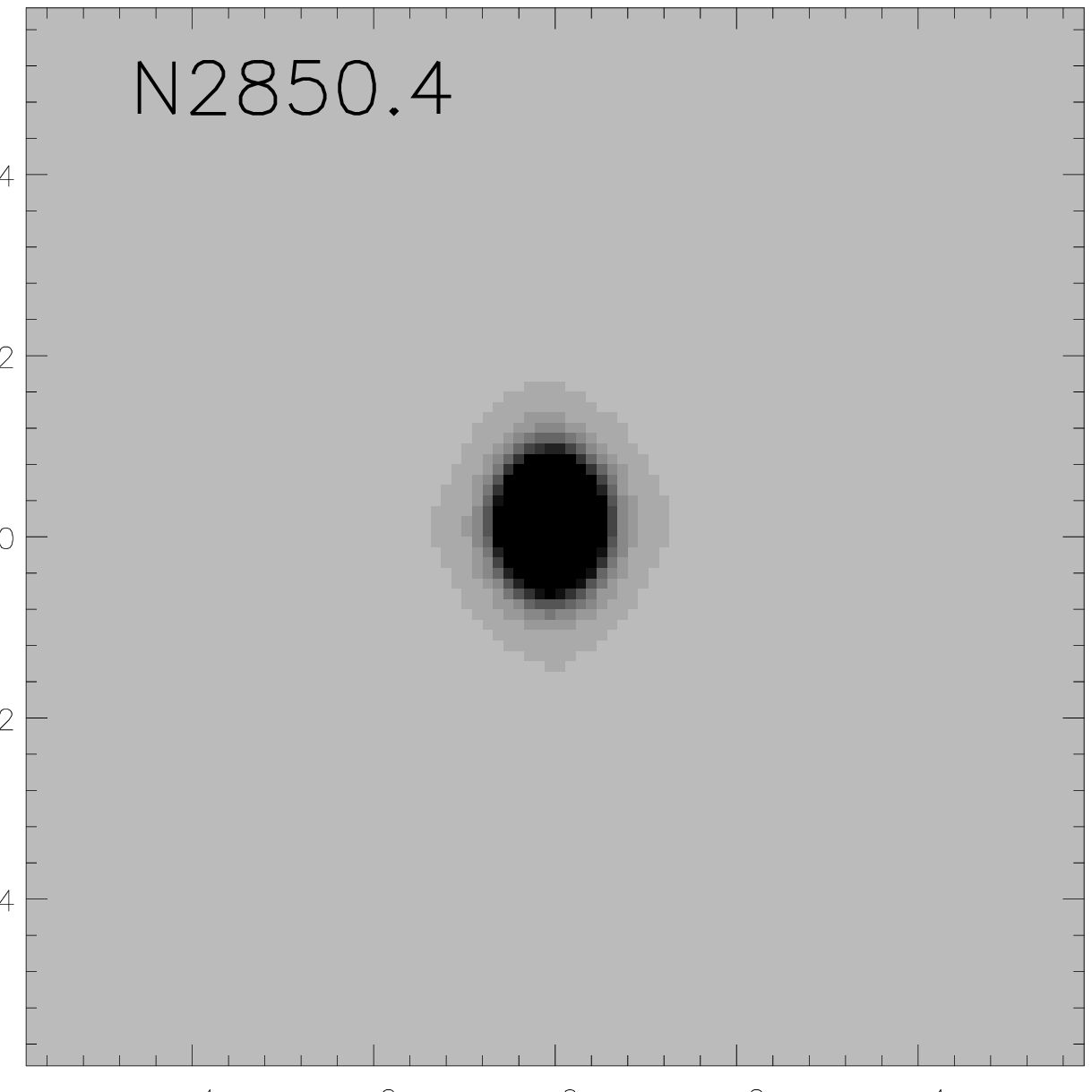,width=0.3\textwidth}&
\epsfig{file=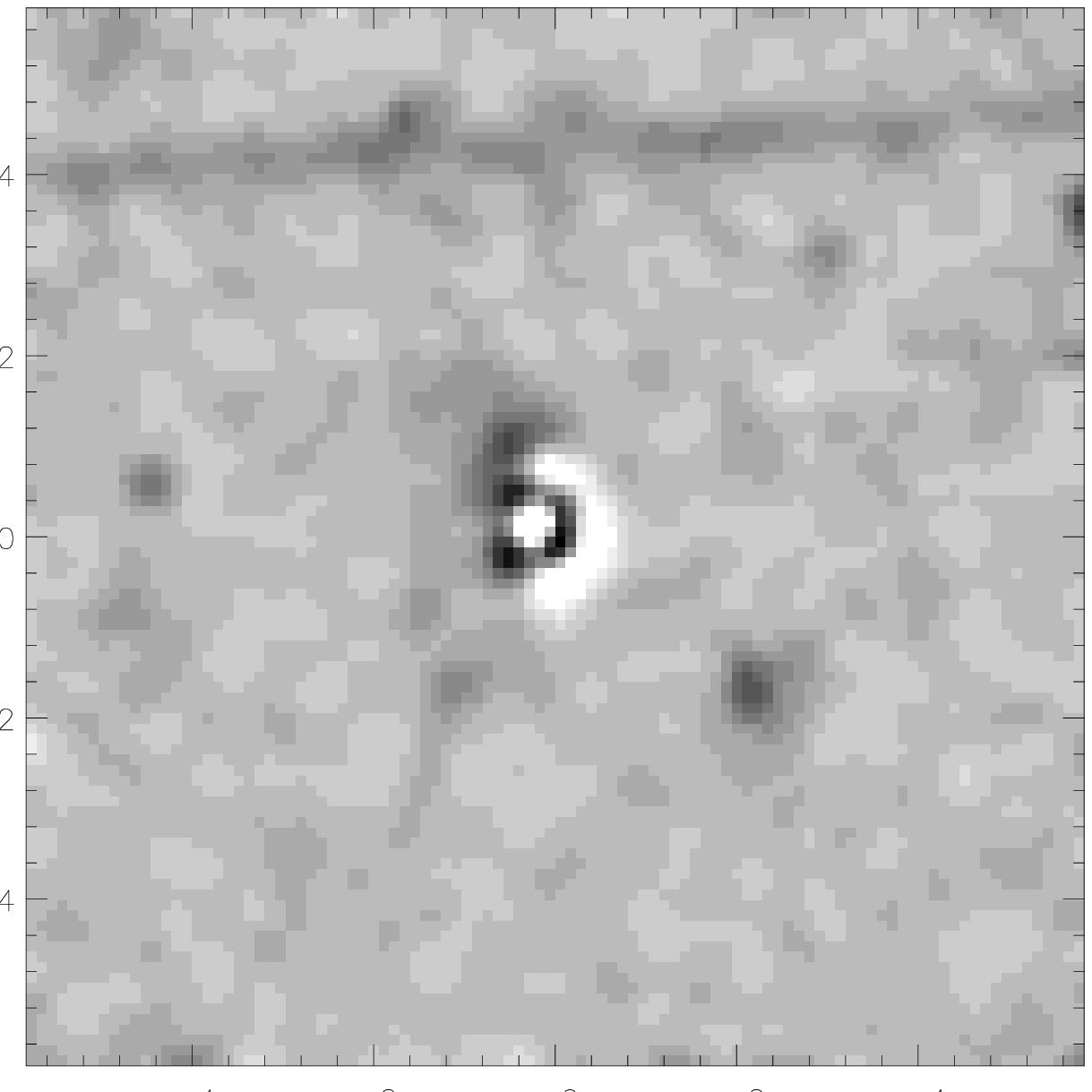,width=0.3\textwidth}\\
\\
\epsfig{file=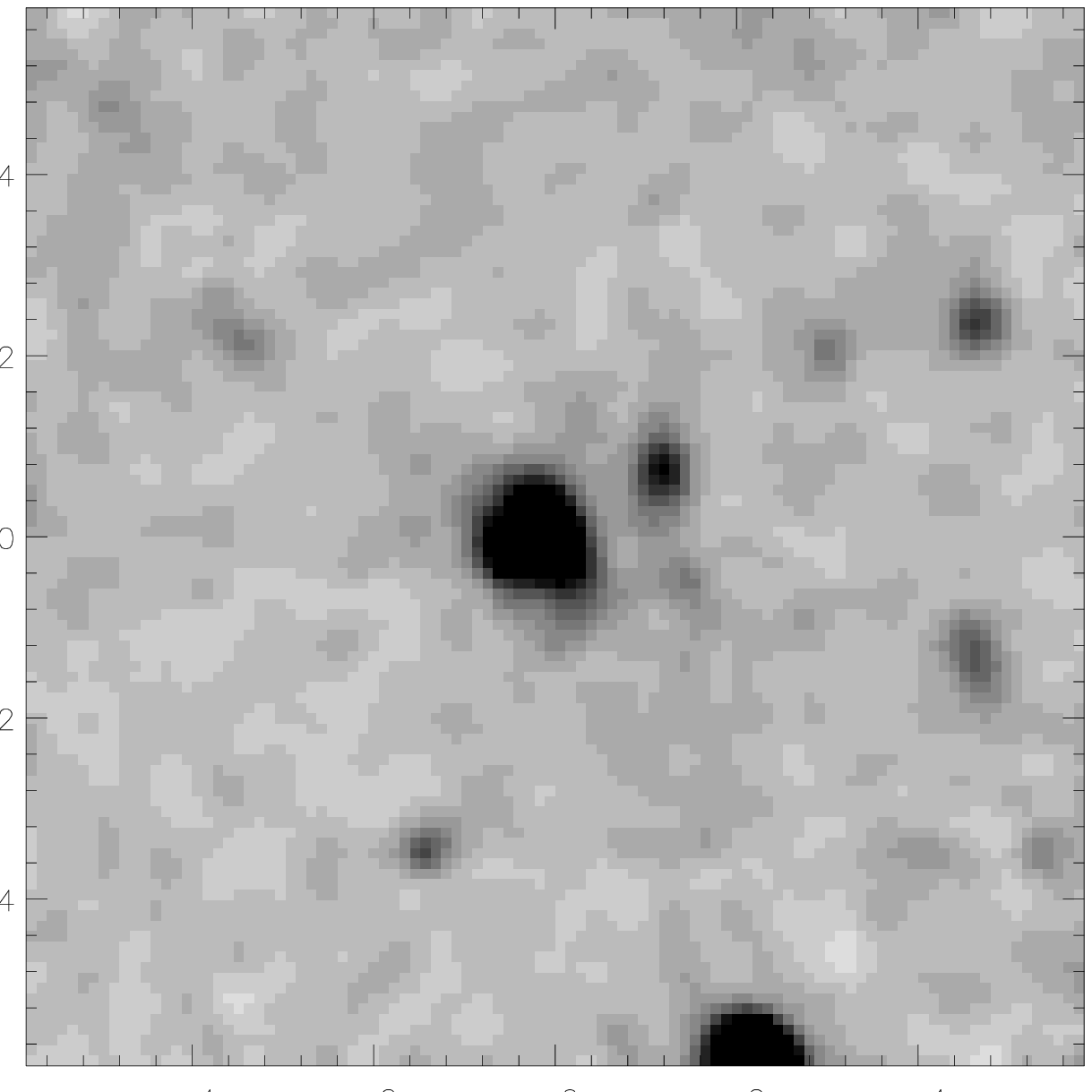,width=0.3\textwidth}&
\epsfig{file=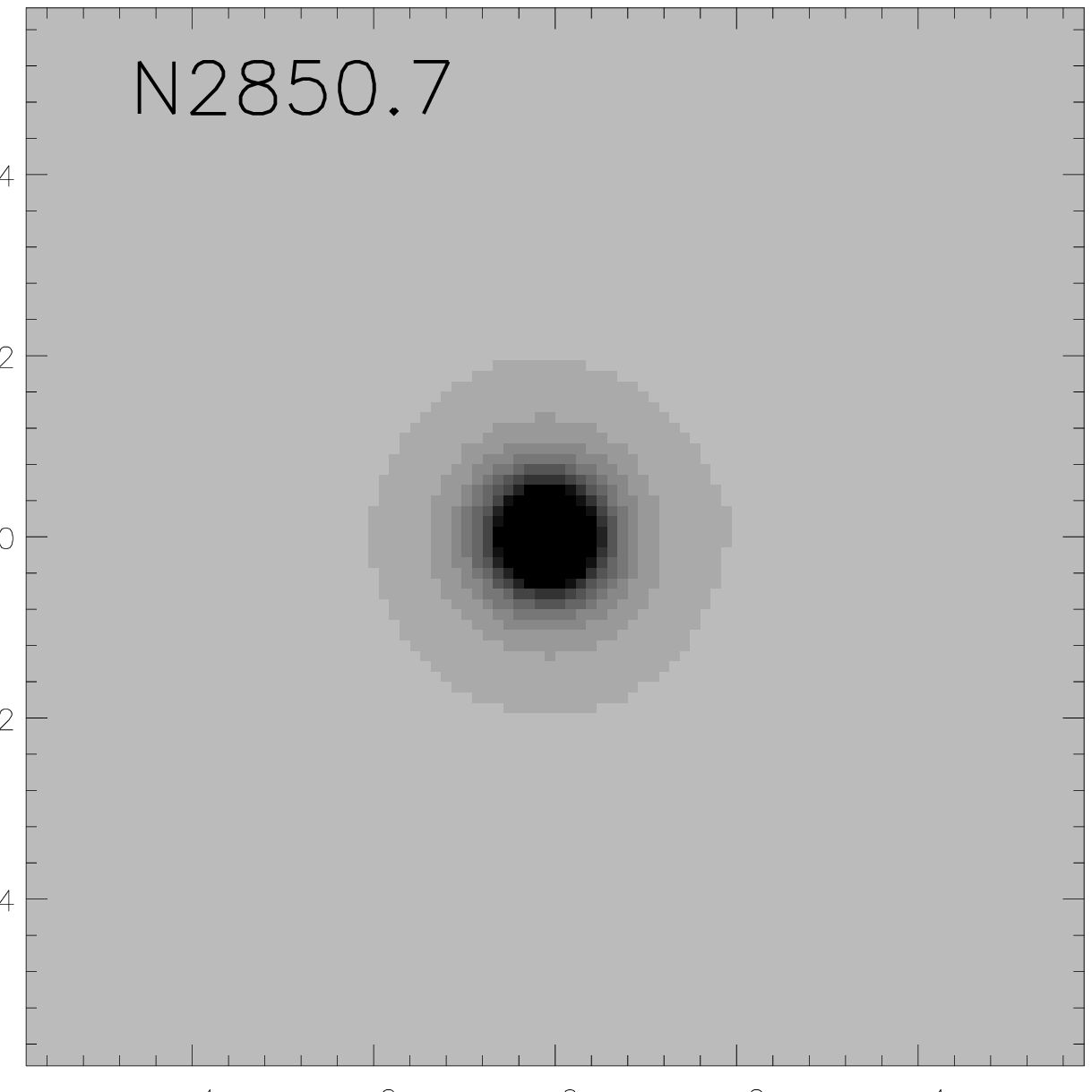,width=0.3\textwidth}&
\epsfig{file=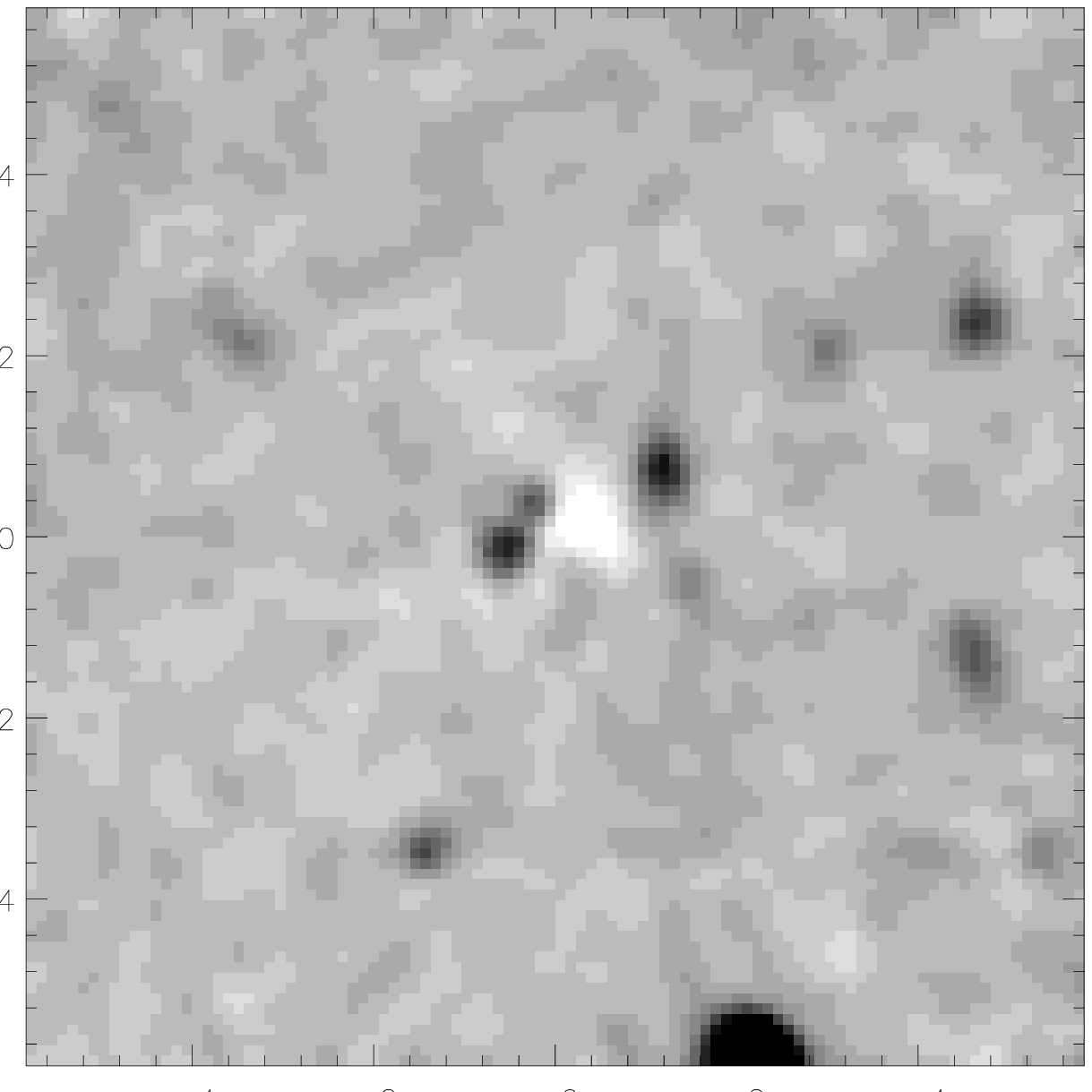,width=0.3\textwidth}\\
\\
\epsfig{file=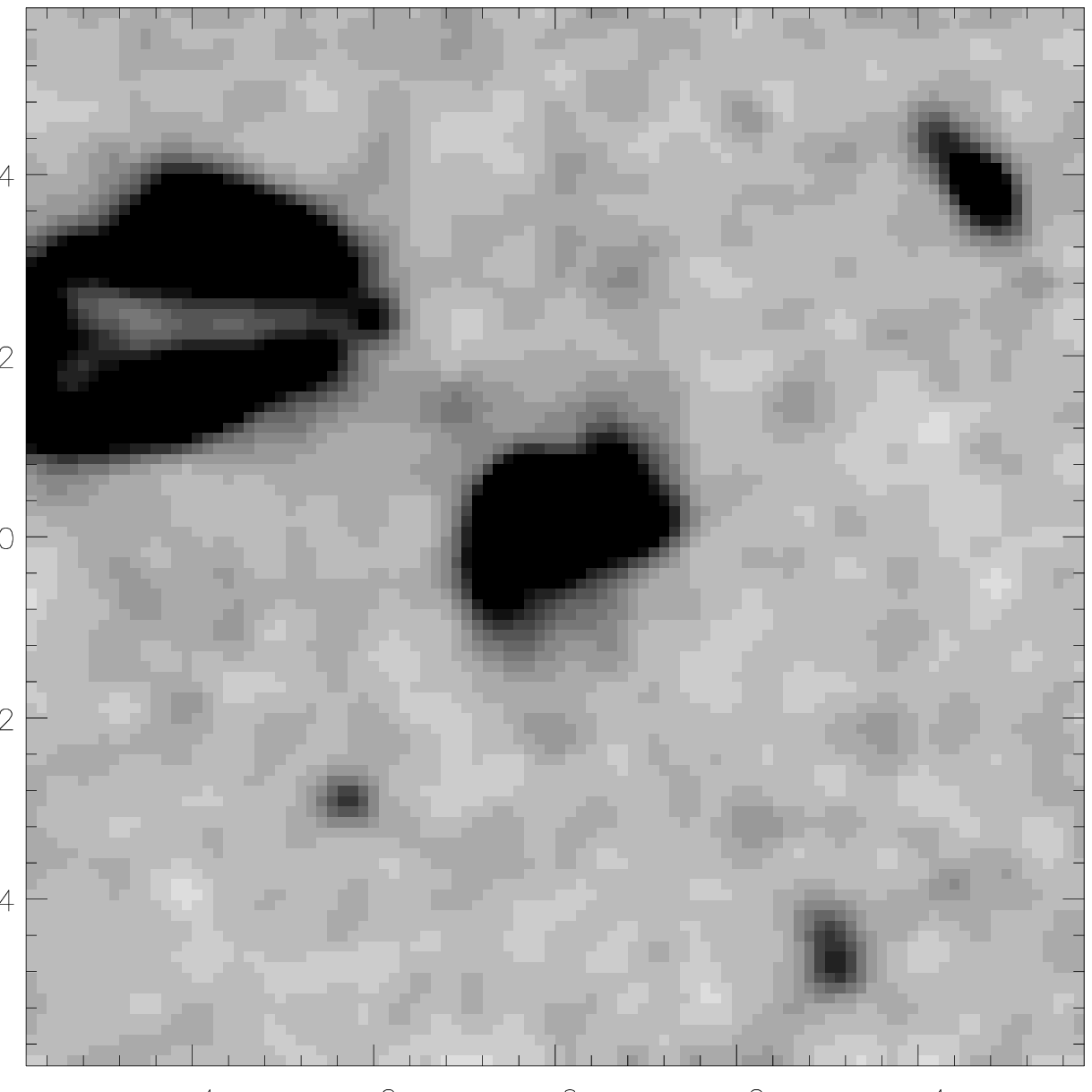,width=0.3\textwidth}&
\epsfig{file=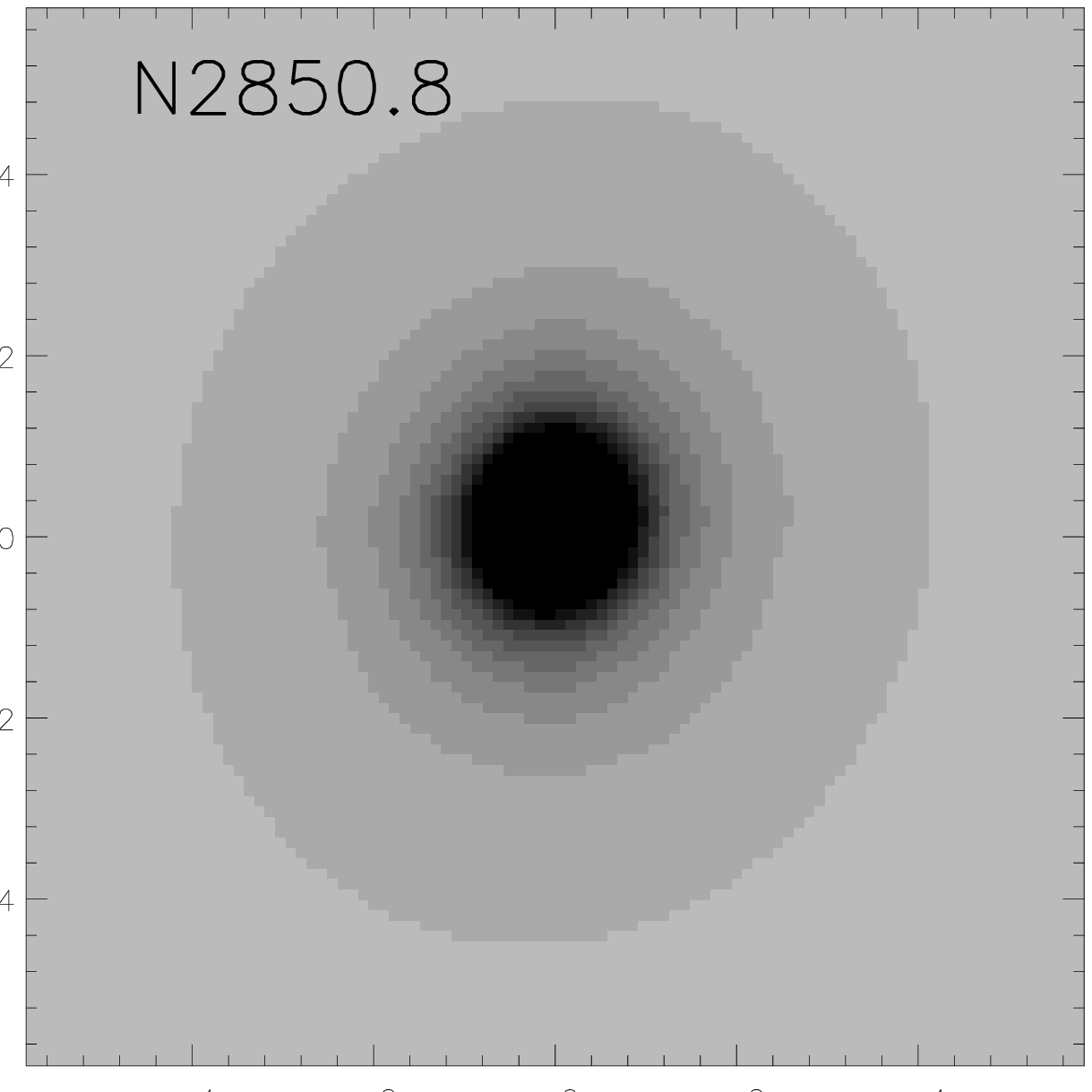,width=0.3\textwidth}&
\epsfig{file=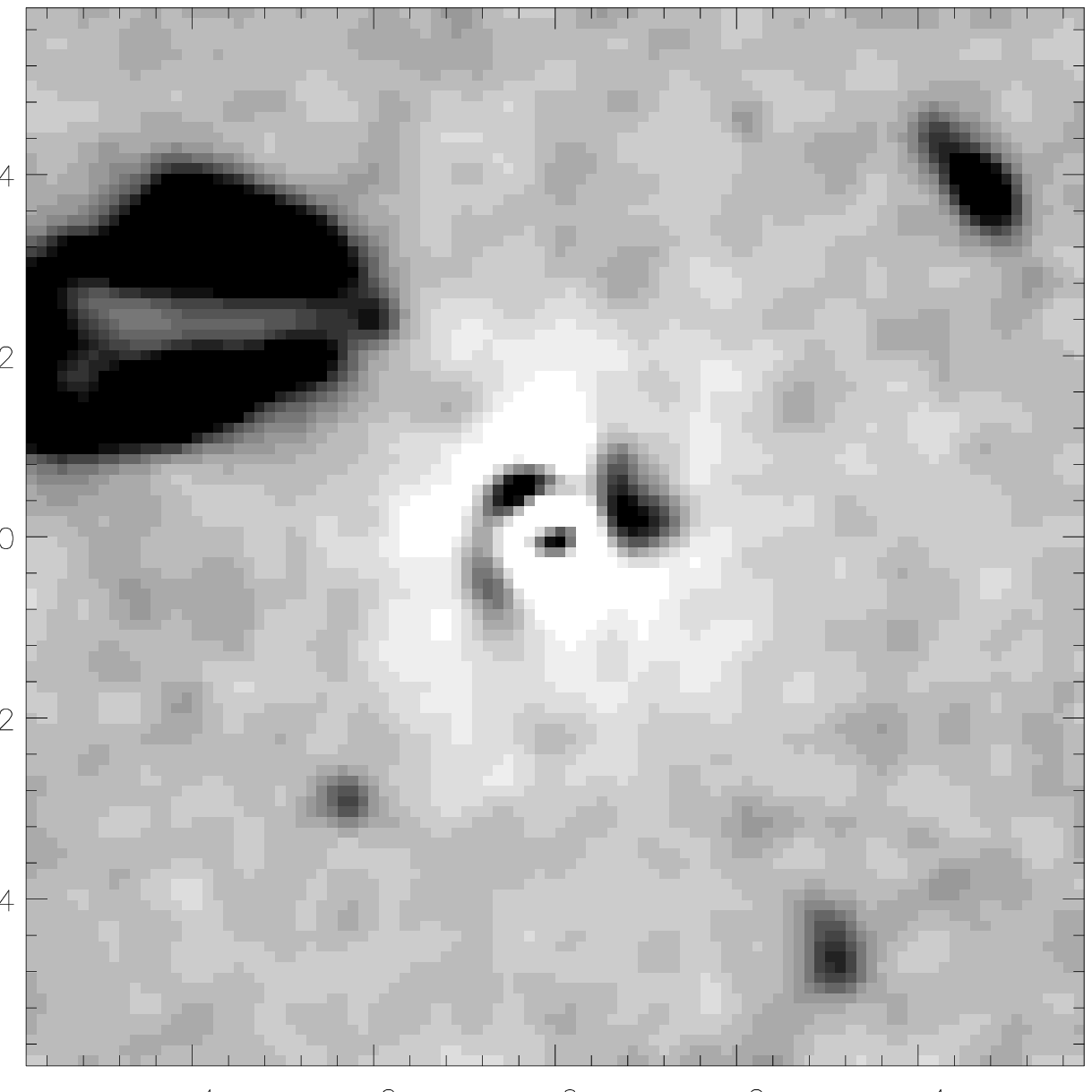,width=0.3\textwidth}\\

\end{tabular}
\caption[Two-dimensional modelling of the smmothed ACS data]{Two-dimensional modelling of the HST ACS $I$-band images of ELAIS N2\,850.1, 850.4, 850.7 and 850.8 (PI Almaini, PID: 9761), after degradation to ground-based seeing of $\simeq 0.5$ arcsec (FWHM). The left-hand panel shows the smoothed ACS image re-sampled to Gemini pixel-scale. The middle panel shows the best-fitting two-dimensional model. The right-hand panel shows the residual image after subtraction of the model from the data. All panels are 12.0$^{\prime \prime}$ $\times$12.0$^{\prime \prime}$, and the images are shown with a linear greyscale in which black corresponds to $2.5\sigma$ above, and white to $1\sigma$ below the median sky value.}
\label{smooth}
\end{center}
\end{figure*}

\end{document}